\documentstyle[12pt]{article}
\def\beq{\begin{eqnarray}}
\def\eeq{\end{eqnarray}}

\def\an{{({\rm an})}}
\begin{document}

\date{today}
\hskip 8cm {\small Review article}

\vskip 2cm

%\title{The Role of Density of States Fluctuations in the
% Normal State Properties of High $T_c$ Superconductors}
%\author{ A.A.Varlamov}
%\address{Forum: Institute of Solid State Theory of INFM, Pisa - 
%Florence, Dipartimento di Fisica, Universita di Firenze, L.go E. Fermi, 2 
%Firenze 50125, Italy\\
%and\\
%Department of Theoretical Physics,
%Moscow Institute for Steel and Alloys, 
%Leninski pr. 4, Moscow 117936, Russia}
%\author{ G.Balestrino, E.Milani}
%\address{Dipartimento di Scienze e Tecnologie Fisiche ed Energetiche, 
%II Universit\'a di Roma ``Tor Vergata", via Tor Vergata, 
%00133 Roma, Italy}
%\author{ D.V.Livanov}
%\address{Department of Theoretical Physics,
% Moscow Institute for Steel and Alloys, 
%Leninski pr. 4, Moscow 117936, Russia}
%\maketitle

\begin{center}
{\bf {\LARGE The Role of Density of States Fluctuations in the Normal State
Properties of High $T_c$ Superconductors}}
\end{center}

\vskip 1.5truecm

\begin{center}
{\bf A.A.Varlamov}
\end{center}

\vskip 0.5truecm

\begin{center}
{Forum: Institute of Solid State Theory of INFM, Pisa - Florence,
Dipartimento di Fisica, Universit\`{a} di Firenze, L.go E. Fermi, 2 Firenze
50125, Italy (e-mail: varlamov@fi.infn.it)\\and\\Department of Theoretical
Physics, Moscow Institute for Steel and Alloys, Leninski pr. 4, Moscow
117936, Russia}
\end{center}

\vskip 1.0truecm

\begin{center}
{\bf G.Balestrino, E.Milani}
\end{center}

\vskip 0.5truecm

\begin{center}
{Dipartimento di Scienze e Tecnologie Fisiche ed Energetiche,\\II
Universit\`a di Roma ``Tor Vergata", via Tor Vergata,\\00133 Roma, Italy
(e-mail: balestrino@uniroma2.it,  milani@uniroma2.it)}
\end{center}

\vskip 1.0truecm

\begin{center}
{\bf D.V.Livanov}
\end{center}

\vskip 0.5truecm

\begin{center}
{Department of Theoretical Physics, Moscow Institute for Steel and Alloys,
Leninski pr. 4, Moscow 117936, Russia (e-mail: livanov@trf.misa.ac.ru)}
\end{center}

\newpage

\begin{abstract}
During the last decade a lot of efforts have been undertaken to explain the
unusual normal state properties of high temperature superconductors (HTS) in
the framework of unconventional theories based on strongly interacting
electrons, pre-formed Cooper pairs, polaron mechanism of superconductivity
etc. A different approach to this problem would be to develop the
perturbation theory for interacting electrons in the normal phase of
strongly anisotropic superconductors without specifying the origin of this
interaction. The Cooper channel of interelectron interaction is equivalent
to the superconducting fluctuations which are unusually strong in HTS. We
show that the peculiarities of such systems not only lead to the increase of
the magnitude but are also frequently responsible for the change of the
hierarchy of different fluctuation effects and even of the sign of the total
corrections. As a result the fluctuation contributions can manifest
themselves in very unusual forms.

The first and well known result is that that now one has the ``pre-formed
Cooper pairs" automatically, from {\it ab initio} calculations: taking into
account thermal fluctuations leads to the appearance of a non-zero density
of fluctuating Cooper pairs (with finite lifetime) within layers without the
establishment of long range order in the system. The fluctuation Cooper pair
density decreases with temperature very slowly ($\sim \ln{\frac{T_c}{T-T_c}}$
in 2D case). The formation of these pairs of normal electrons leads to the
decrease of the density of one-electron states (DOS renormalization) at the
Fermi level, and this turns out to be the key effect in our discussion.

The DOS contribution to the most of characteristics is negligible for
traditional superconducting materials. However it becomes dominant when
highly anisotropic materials are discussed, and therefore is very important
in HTS, especially when transport along the c-axis is considered. We analyze
the role of the DOS fluctuations in the properties of HTS and show how,
taking into account this effect, many puzzling and long debated properties
of HTS materials (such as the steep increase of the electrical resistivity
along the c-axis just above $T_c$, the anomalous magnetoresistance, effects
of the magnetic field on the resistive transition along the c-axis, the
c-axis far infrared absorption spectrum, NMR characteristics around the
critical temperature etc.) can be understood leading to a simple, consistent
description in terms of the fluctuation theory.
\end{abstract}

\newpage

\tableofcontents

\newpage

\vskip 3.0truecm

\hskip 3.0truecm{\bf To the memory of our good friends and colleagues}

\hskip 3.0truecm{\bf Lev Aslamazov and Paolo Paroli}

\newpage

\section{Introduction}

There are no doubts that the puzzling anomalies of the normal state
properties of high temperature superconductors (HTS) are tightly connected
with the physical origin of superconductivity in these materials. Among them
are:

- a peak in the c-axis resistivity above $T_c$ followed by a decrease to
zero as temperature is decreased \cite{PHK88,MFF88};

- the giant growth of this peak in the presence of an external magnetic
field applied along the c-axis and its shift towards low temperatures \cite
{BCZ91};

- the giant magnetoresistance observed in a wide temperature range above the
transition \cite{YMHO95,HASH,AHE};

- the deviation from the Korringa law in the temperature dependence of the
NMR relaxation rate above $T_c$ \cite{PS90};

-the opening of a large pseudo-gap in the c-axis optical conductivity at
temperatures well above $T_c$ \cite{BTD94,BTD95} ;

- the anisotropic gap observed in the electron spectrum by angular resolved
photo-emission experiments \cite{PRB96}.

- the gap-like tunneling anomalies observed already above $T_c$ \cite
{tun,CNR,tao97,MSW97,SKN97,RRG97}.

- the anomalies in the thermoelectric power above $T_c$ \cite{H90,ZSY92}.

- the anomalies in the Hall effect above $T_c$ \cite{Ong92,Ch91,CWO91,S94}.

- the anomalies in the heat transport above $T_c$ \cite{Ong97,Ausloos96}.

\noindent These effects have been attributed by many authors to the opening
of a ``pseudo-gap''. Naturally this has led to numerous speculations about
the physical origin of such a gap.

During the last decade a lot of efforts have been undertaken to explain the
unusual normal state properties of HTS materials using unconventional
theories of superconductivity based on ideas of spin-charge separation,
pre-formed Cooper pairs, polaron mechanism of superconductivity, etc.(see
for instance \cite{PhysTod96}). They have been widely discussed and we will
not overview them here. In the case of HTS with a well developed Fermi
surface (i.e. in the optimally doped or overdoped part of the phase diagram) one
can approach this problem from another side, namely to develop the
perturbation theory for interacting electrons in the normal phase of a
strongly anisotropic superconductor. We will not specify the origin of this
interaction: for our purposes it is enough to assume that this interaction
is attractive and leads to the appearance of superconductivity with Cooper
pairs of charge $2e$ at temperatures below $T_c$. Of course, the smallness
of the effects magnitude (necessary for the applicability of the
perturbative approach) is a serious limitation of the proposed theory.
Nevertheless the current state of HTS investigations in some respects
reminds one of the situation which occurs in the study of metal-insulator
transitions in the 1970s. The weak localization theory did not describe
consistently the Anderson transition, but was successful in the explanation
of a set of anomalous properties of the disordered metal systems; besides
this it gave a hint to the development of the renormalization group approach
to the description of the metal-insulator transition.

The Cooper channel of interelectron interaction is equivalent to taking into
account superconducting fluctuations which are unusually strong in HTS \cite
{T94}. The reasons for this strength are the effective low dimensionally of
the electron spectrum, the low density of charge carriers and the high
values of critical temperature of HTS. We will show that these peculiarities
lead not only to the increase of the magnitude of the fluctuation effects,
but frequently change the hierarchy of the different fluctuation
contributions, leading to the appearance of competition among them and even
to the change of the habitual (in conventional superconductivity) sign of
the overall correction. As a result, the fluctuation effects can manifest
themselves in very unusual form, so that their origin cannot be identified
at first glance.

The first well-known result is that in the metallic phase one automatically
has the non-equilibrium analogue of pre-formed Cooper pairs above $T_c$.
Indeed, taking into account thermal fluctuations (or interelectron
interaction in the Cooper channel) leads to the appearance of some non-zero
density of fluctuation Cooper pairs (in contrast to pre-formed pairs with
finite lifetime) in the superconducting layers without the establishment of
the long range order in the system. It is important that in the 2D case,
typical for HTS materials, the density of Cooper pairs decreases with
temperature extremely slowly: $\sim \ln {\frac{T_c}{T-T_c}}$. One should
therefore not be surprised that precursor effects can often be detected in
the normal phase well above $T_c$ (especially in underdoped samples, see
section 5).

The formation of fluctuation Cooper pairs of normal electrons above $T_c$
has an important though usually ignored consequence: the decrease of the
density of one-electron states (DOS) at the Fermi level \cite{ARW70,CCRV90}.
This circumstance turns out to be crucial for the understanding of the
aforementioned effects and it will constitute the quintessence of this
review. In this way the following phenomena can be at least qualitatively (and in many cases
quantitatively as well) explained:

The behavior of the c-axis resistance \cite{BCZ91} which has been explained
in terms of the suppression of the one-electron DOS at the Fermi level and
the competition of this effect with the positive Aslamazov-Larkin (AL)
paraconductivity \cite{ILVY93,KG93,BMMVY93,BMV93,BMV96}.

The giant growth of the c-axis resistance peak in the presence of an
external magnetic field applied along the c-axis is explained using the same
approach \cite{BDKLV93} which was shown to fit well the experiments \cite
{BMV96,NBMLV96,BLM97}.

The anomalous negative magnetoresistance observed above $T_c$ in BSCCO
samples \cite{YMH95,HASH,L97} was again explained by the same DOS
fluctuation contribution \cite{BMV95}. Moreover, its competition with the
positive Aslamazov-Larkin magnetoresitance gave good grounds for the
prediction of a sign change in the magnetoresistance as temperature
decreases towards $T_c$ \cite{BMV95}. The latter effect was very recently
confirmed experimentally on YBCO samples \cite{AHE}.

The decrease of the thermoelectric power at the edge of transition turns out
to be the result of the DOS fluctuation contribution which dominates over
the AL term \cite{VLF97} previously assumed to play the crucial role \cite
{M74,VL90,RS94,MVF94}.

The temperature dependence of the NMR rate $\frac 1{T_1T}$ can be explained
as the result of the competition between the positive Maki-Thompson (MT)
correction to the Korringa law and the negative DOS contribution at the edge
of the transition \cite{RV94,CLRV96}.

The observed pseudo-gap-like structure in the far infra-red optical
conductivity along c-axis can also be attributed to the suppression of the
one-electron DOS at the Fermi level. This leads to the appearance of a
sizable negative contribution in optical conductivity, which shows up in a
wide range of frequencies (up to $\omega _{DOS}\sim \tau ^{-1}$), exceeding
the positive AL and MT contributions in magnitude and range of manifestation 
\cite{FV96}.

We believe that the fact that even the simple approach proposed here was
able to explain most of the anomalies of the normal state properties of HTS
mentioned above is not accidental. It shows the importance of the
interelectron interaction in the problem discussed and demonstrates that
even the way of ``up-grading'' the traditional BCS theory to include the HTS
peculiarities is creative in the explanation of the HTS properties. Further,
the approach considered provides clear results which can be compared with
those obtained from the alternative viewpoints and an attempt to match them
in the region of the intermediate strengths of interaction may be undertaken.

The review is organized in the following way. The sections 2 and 3 introduce
the reader to the short story and simplest notions of the fluctuation
theory. The existence of a finite non-equilibrium concentration of Cooper
pairs at temperatures above the critical one is shown and the effect of
fluctuations on the order parameter and critical temperature is discussed.
In the section 4, the renormalization of the one-electron density of states
is considered, and in section 5, its consequences on the tunneling
properties are discussed. This preliminary introduction prepares the reader
for the central section 6 where the different fluctuation effects are
discussed in their variety, first at a qualitative level and, then, within
the microscopic approach. The results obtained in section 6 are then applied
in section 7 to the analysis of the experimental data for the c-axis
electrical transport. Section 8 is devoted to the effect of an external
magnetic field on the c-axis transport. In section 9 we demonstrate that the
effect of DOS fluctuation renormalization causes the opening of a pseudo-gap
type structure in the c-axis optical conductivity. The importance of the DOS
fluctuation contribution for the thermoelectricity is shown in section 10.
The last two sections are devoted to the discussion of the effect of
fluctuations on the NMR characteristics at the edge of transition and the
possibility of application of the effect discussed as a tool to study the
order parameter symmetry of HTS.

\newpage

\section{ Excursus to superconducting fluctuation theory}

During the first half of the century, after the discovery of
superconductivity by Kammerlingh-Onnes, the problem of fluctuations smearing
the superconducting transition was not even considered. In bulk samples of
traditional superconductors the critical temperature $T_c$ sharply divides
the superconducting and the normal phases. It is worth mentioning that such
behavior of the physical characteristics of superconductors is in perfect
agreement both with the Ginzburg-Landau phenomenological theory (1950) and
the BCS microscopic theory of superconductivity (1957). However, at the
same time, it was well known that thermodynamic fluctuations can cause
strong smearing of other second -order phase transitions, such as the $%
\lambda $-point in liquid helium.

As already mentioned, the characteristics of high temperature and organic
superconductors, low dimensional and amorphous superconducting systems
studied today, differ strongly from those of the traditional superconductors
discussed in textbooks. The transitions turn out to be much more smeared
here. The appearance of thermodynamically nonequilibrium Cooper pairs
(superconducting fluctuations) above critical temperature leads to precursor
effects of the superconducting phase occurring while the system is still in
the normal phase, often far enough from $T_c$. The conductivity, the heat
capacity, the diamagnetic susceptibility, the sound attenuation etc. may
increase considerably in the vicinity of the transition temperature.

So, what is the principal difference between conventional and unconventional
superconductors with respect to fluctuation phenomena, and in general, what
determines the role and the strength of fluctuations in the vicinity of the
superconducting transition? How smeared out is the transition point in
existing superconducting devices, and how can one separate the fluctuation
contributions from the normal state ones? Which microscopic information can
be extracted from the analysis of the fluctuation corrections in different
physical characteristics of superconductors?

These questions, along with many others, find their answers in the theory of
fluctuation phenomena in superconductors. This chapter of the
superconductivity has been developed in the last 30 years.

The first numerical estimation of the fluctuation contribution to the heat
capacity of superconductors in the vicinity of $T_c$ was done by Ginzburg in
1960 \cite{G60}. In that paper he showed that superconducting fluctuations
increase the heat capacity even above $T_c$. In this way the fluctuations
smear the jump in the heat capacity which, in accordance with the
phenomenological Ginzburg-Landau theory of second order phase transitions
(see for instance \cite{Ab88}), takes place at the transition point itself.
The range of temperatures where the fluctuation correction to the heat
capacity of a bulk clean conventional superconductor is relevant was
estimated by Ginzburg as 
\begin{eqnarray}
\frac{\delta T}{T_c}\sim \left( \frac{T_c}{E_F}\right) ^4\sim \left( \frac
a\xi \right) ^4\sim 10^{-12}\div 10^{-14}  \label{Gi}
\end{eqnarray}
where $a$ is the interatomic distance, $E_F$ is the Fermi energy and $\xi $
is the superconductor coherence length at zero temperature \footnote{%
The same expression for the width of the strong fluctuations region was
obtained by Levanyuk in \cite{L59}. So in the modern theory of phase
transitions the relative width of fluctuation region is called the
Ginzburg-Levanyuk parameter $Gi_{(D)}$ and its value strongly depends on the
space dimensionality D and on the impurity concentration \cite{Ab88}.}. It
is easy to see that this is many orders of magnitude smaller than the
temperature range accessible in real experiments. This is why fluctuation
phenomena in superconductors were considered experimentally inaccessible for
a long time.

In the 1950s and 60s the formulation of the microscopic theory of
superconductivity, the theory of type-II superconductors and the search for
high-$T_c$ superconductivity attracted the attention of researchers to dirty
systems, and the properties of superconducting films and filaments began to
be studied. In 1968, in the well known paper of L. G. Aslamazov and A. I.
Larkin \cite{AL68}, the consistent microscopic theory of fluctuations in
the normal phase of a superconductor in the vicinity of the critical
temperature was formulated. This microscopic approach confirmed Ginzburg's
evaluation \cite{G60} for the width of the fluctuation region in a bulk
clean superconductor, but much more interesting results were found in \cite
{AL68} for low dimensional or dirty superconducting systems. The exponent $%
\nu $ of the ratio $(a/\xi _0)$, which enters in (\ref{Gi}), drastically
decreases as the effective dimensionality of the electron motion diminishes: 
$\nu =4$ for 3D, but $\nu =1$ for 2D electron spectrum (in the clean case)
which is the most appropriate for HTS materials.

Another source of the effective increase of the strength of fluctuation
effects is the decrease of the coherence length, which occurs in dirty
superconductors because of the diffusive character of the electronic motion.
This means that fluctuation phenomena are mainly observable in amorphous
materials with removed dimensionality, such as films and whiskers, where
both factors mentioned above come into play. HTS is of special interest in
this sense, because their electronic spectrum is extremely anisotropic and
their coherence length is very small. As a result the temperature range in
which the fluctuations are important in HTS may reach tens of degrees.

The manifestation of superconducting fluctuations above critical temperature
may be conveniently demonstrated considering the case of electrical
conductivity. In the first approximation there are three different effects.
The first one, a direct contribution, consists in the appearance of
nonequilibrium Cooper pairs with the characteristic lifetime $\tau _{GL}=\pi
\hbar /{8k_B(T-T_c)}$ in the vicinity of the transition. In spite of their
finite lifetime, a non-zero number of such pairs (which depends on the
proximity to $T_c$) is always present in the normal phase (below $T_c$ they
are in excess in comparison with the equilibrium value (see the next
section)). Their presence gives rise, for instance, to the appearance of the
precursor of the Meissner-Ochsenfeld anomalous diamagnetism in the normal
phase: that is \cite{TS75} the anomalous increase of the diamagnetic
susceptibility at the edge of the transition. As far as the conductivity is
concerned, one can say that above $T_c$, because of the presence of
nonequilibrium Cooper pairs, a new, non-dissipative, channel of charge
transfer has been opened. This direct fluctuation contribution to the
conductivity is called {\it paraconductivity} or the {\it Aslamazov-Larkin
(AL)} contribution \cite{AL68}.

Another consequence of the appearance of fluctuating Cooper pairs above $T_c$
is the decrease of the one-electron density of states at the Fermi level.
Indeed, if some electrons are involved in the pairing they can not
simultaneously participate in charge transfer and heat capacity as
one-particle excitations. Nevertheless, the total number of the electronic
states can not be changed by the Cooper interaction, and only a
redistribution of the levels along the energy axis is possible \cite
{ARW70,CCRV90} (see Fig. 2 in section 4). In this sense one can
speak about the opening of a fluctuation pseudo-gap at the Fermi level.

The decrease of the one-electron density of states at the Fermi level leads
to the reduction of the normal state conductivity. This, indirect,
fluctuation correction to the conductivity is called the {\it density of
states (DOS)} contribution and it appears side by side with the {\it %
paraconductivity}. It has an opposite (negative) sign and turns out to be
much less singular in $(T-T_c)^{-1}$ in comparison with the AL contribution,
so that in the vicinity of $T_c$ it was usually omitted. However, in
many cases \cite{ILVY93,BDKLV93,AL73,AV80,ARV83}, when for some special
reasons the main, most singular, corrections are suppressed, the DOS
correction becomes of major importance. Such a situation takes place in many
cases of actual interest (quasiparticle current in tunnel structures, c-axis
transport in strongly anisotropic high temperature superconductors, NMR
relaxation rate, thermoelectric power). In this context the study of the DOS
contribution will be our main goal in this review.

The third, purely quantum, fluctuation contribution is generated by the
coherent scattering of the electrons forming a Cooper pair on the same
elastic impurities. This is the so called {\it anomalous Maki-Thompson (MT)}
contribution \cite{M68,T70} which often turns out to be important in
conductivity and other transport phenomena at the edge of the transition.
Its temperature singularity near $T_c$ is similar to that of paraconductivity,
but this contribution turns out to be extremely sensitive to electron
phase-breaking processes. In HTS materials there are several sources of
strong pair-breaking above $T_c$ (such as localized magnetic moments,
thermal phonons etc.). So the MT contribution turns out to be depressed by
these phase-breaking processes and can usually be omitted in HTS fluctuation
analysis. Nevertheless in some special cases (like NMR relaxation rate) it
has to be taken into account even in its overdamped form.

Finally, for completeness, we have to mention the regular part of the
Maki-Thompson diagram which is much less singular and has an origin similar
to the DOS renormalization contribution.

We will see below that the strong anisotropy of the electron motion in HTS
makes the DOS contribution particularly important for c-axis transport and
related phenomena, changing significantly the hierarchy of the fluctuation
corrections in comparison with the conventional case.

\newpage

\section{The effect of fluctuations on the order parameter and critical
temperature}

\subsection{Introduction}

The order parameter temperature dependence and the critical temperature are
among the main characteristics of superconductors. The traditional BCS
approach gives simple expressions for both of them, but it turns
out to be valid for 3D systems only. The effect of fluctuations in the 2D
case becomes crucial \cite{H68} and leads to the break down of the
fundamental idea of BCS theory: the association of superconductivity with
long-range order in the system. The works of Berezinski \cite{B73}, Thouless
and Kosterlitz \cite{TK74} demonstrated that the requirement $\langle \Psi
(0)\Psi ({\vec{r}})\rangle _{|{\vec{r}}|\rightarrow \infty }=const$ for an
homogeneous 2D system is too rigid, actually the flow of supercurrent takes
place even if the system possesses some ``stiffness'' only. More precisely
the aforementioned correlator of the order parameter at different points
only has to decrease with distance $|{\bf r}|$ as some power law (in
contrast to its exponential decrease in the normal metal).

We will not discuss here the well known properties of the Berezinski-
Thouless-Kosterlitz state (see, for instance, the review \cite{BTKrev,BTKHTS}
) but will concentrate on the crucial, for our purposes, fact of the
nonequilibrium fluctuation Cooper pair formation above $T_c$ in
quasi-two-dimensional systems. As we will show below, in spite of the
exponential decrease of spatial superconducting correlations above $T_c$,
the density of nonequilibrium pairs in this case decreases only
logarithmically with temperature.

\subsection{Fluctuation Cooper pairs above $T_c$}

Let us start from the calculation of the density of fluctuation Cooper pairs
in the normal phase of a superconductor. We restrict ourselves to the region
of temperatures near the critical temperature, so we can operate in the
framework of the Landau theory of phase transitions \cite{Ab88}.

When we consider the system above the transition temperature, the order
parameter $\Psi (\vec{r})$ has a fluctuating origin (its mean value is equal
to zero) and it depends on the space variables even in the absence of
magnetic field. This is why we have to take into account the gradient term
in the Ginzburg-Landau functional for the fluctuation part of the
thermodynamical potential $\Omega _{(fl)}$: 
\begin{eqnarray}
\Omega _{(fl)}=\Omega _s-\Omega _n=\alpha \int dV\left\{ \varepsilon |\Psi (%
\vec{r})|^2+\frac b{2\alpha }|\Psi (\vec{r})|^4+\eta _D|\nabla \Psi (\vec{r}%
)|^2\right\}   \label{FP}
\end{eqnarray}
where $\varepsilon =\ln (T/T_c)\approx \frac{T-T_c}{T_c}\ll 1$ for the
temperature region discussed and $\alpha =\frac 1{4m\eta _D}$. The positive
constant $\eta _D$ of the phenomenological Ginzburg-Landau may be expressed
in terms of microscopic characteristics of the metal: 
\begin{eqnarray}
\eta _D &=&-\frac{v_F^2\tau ^2}D\left[ \psi \left( {\frac 12}+{\frac 1{{4\pi
\tau T}}}\right) -\psi \left( {\frac 12}\right) -{\frac 1{{4\pi \tau T}}}%
\psi ^{^{\prime }}\left( {\frac 12}\right) \right] \rightarrow   \label{d8}
\\
&&\frac{\pi v_F^2\tau }{8DT}\cases{1 &for $\tau T<<1$,\cr
7\zeta(3)/(2\pi^3\tau T) &for $\tau T>>1,$\cr}  \nonumber
\end{eqnarray}
where $v_F$ is the Fermi velocity $\tau $ is the quasiparticle scattering time,
D is the space dimensionality, $\psi (x)$ and $\psi ^{\prime }(x)$ are the
digamma function and its derivative respectively, and $\zeta (x)$ is the
Riemann zeta function \footnote{%
We will mostly use the system $\hbar = c = k_B =1$ everywhere, excluding the
situations where the direct comparison with experiments is necessary.}.
Dealing mostly with $2D$ case we will often use this definition omitting the
subscript ''2'': $\eta _2\equiv \eta .$

Dealing with the region of temperatures $\varepsilon >0$, in the first
approximation we can neglect the fourth order term in (\ref{FP}). Then,
carrying out the Fourier transformation of the order parameter 
\begin{eqnarray}
\Psi _{\vec{k}}={\frac 1{\sqrt{V}}}\int \Psi (\vec{r})\exp ^{-i\vec{k}\vec{r}%
}dV  \label{K}
\end{eqnarray}
one can easily write the fluctuation part of the thermodynamic potential as
a sum over Fourier components of the order parameter: 
\begin{eqnarray}
\Omega _{(fl)}=\alpha \sum_{\vec{k}}\left( \varepsilon +\eta _Dk^2\right)
\left| \Psi _{\vec{k}}\right| ^2.  \label{FPK}
\end{eqnarray}
Here 
\[
\vec{k}={\frac{{2\pi }}{{L_x}}}n_x\vec{i}+{\frac{{2\pi }}{{L_y}}}n_y\vec{j}+{%
\frac{{2\pi }}{{L_z}}}n_z\vec{l}, 
\]
where $L_{x,y,z}$ are the sample dimensions in appropriate directions; $\vec{%
i},\vec{j},\vec{l}$ are unit vectors along the axes; $n_{x,y,z}$ are integer
numbers; $V$ is the volume of the sample.

In the vicinity of the transition the order parameter $\Psi $ undergoes
equilibrium fluctuations. The probability of the fluctuation realization of
a given configuration $\Psi (\vec{r})$ is proportional to \cite{Ab88}: 
\begin{eqnarray}
{\cal P}\propto \exp \left[ -\frac \alpha T\sum_{\vec{k}}\left( \varepsilon
+\eta _Dk^2\right) \left| \Psi _{\vec{k}}\right| ^2\right] ,  \label{prob}
\end{eqnarray}
Hence the average equilibrium fluctuation of the square of the order
parameter Fourier component $\left| \Psi _{\vec{k}}^{(fl)}\right| ^2$ may be
calculated as 
\begin{eqnarray}
\langle \left| \Psi _{\vec{k}}^{(fl)}\right| ^2\rangle  &=&\displaystyle{\ 
\frac{\int \left| \Psi _{\vec{k}}\right| ^2\exp \left[ -\frac \alpha T\left(
\varepsilon +\eta _D{k^2}\right) \left| \Psi _{\vec{k}}\right| ^2\right]
d\left| \Psi _{\vec{k}}\right| ^2}{\int \exp \left[ -\frac \alpha T\left(
\varepsilon +\eta _D{k^2}\right) \left| \Psi _{\vec{k}}\right| ^2\right]
d\left| \Psi _{\vec{k}}\right| ^2}}  \nonumber  \label{aef} \\
&& \\
&=&\displaystyle{\ \frac T{\alpha \left( \varepsilon +\eta _Dk^2\right) }}, 
\nonumber
\end{eqnarray}

The concentration of Cooper pairs ${\cal N}_{c.p.}$ is determined by the
average value of the square of the order parameter modulus \cite{Ab88}. For
the two-dimensional case, which is of most interest to us, one finds:

\begin{eqnarray}
{\cal N}_{c.p.}^{(2)}=\langle |\Psi _{(fl)}|^2\rangle &=&\int \frac{d^2{\ \ 
\vec{k}}}{(2\pi )^2}|\Psi _{\vec{k}}^{(fl)}|^2\exp {i({\vec{k}}\cdot {\ \vec{%
r}})}|_{{\ \vec{r}}\rightarrow 0}=  \nonumber  \label{5} \\
&=&\frac T\alpha \int \frac 1{{\ \varepsilon }+\eta _2{\vec{k}}^2}\frac{d^2{%
\ \vec{k}}}{(2\pi )^2}=2{\cal N}_e^{(2)}\frac{T_c}{E_F}\ln {\ \frac
1\varepsilon }  \label{del}
\end{eqnarray}
where ${\cal N}_e^{(2)}=\frac m{2\pi }E_F$ is the one-electron concentration
in 2D case, and $\eta _2$ is defined by the expression (\ref{d8}) \footnote{%
One can notice that the number of Cooper pairs in (\ref{del}) surprisingly
does not depend on the concentration of electrons even when {\it ad absurdum}
the number of electrons tends to zero. In this relation it is worth to
mention two circumstances. The first one consists in the fact that we used
the degenerate Fermi gas model from the very beginning, so the electron
density cannot tend to zero. The second comment concerns the very special
role of the dimensionality "2" for electron systems ( disordered, or
superconducting). The well known universality of the paraconductivity
expression in this case is directly related with the discussed property of ( 
\ref{del}) (see section 6).}.

We see that in the 2D case the density of fluctuation Cooper pairs decreases
very slowly as the temperature increases : logarithmically only. Of course these
are nonequilibrium pairs, their lifetime being determined by the
Ginzburg-Landau time $\tau _{GL}=\frac \pi {8(T-T_c)}$ and there is no long
range order in the system. Nevertheless, one can see that even in the normal
phase of a superconductor at each moment there is a non-zero density of such
pairs which may participate in charge transfer, anomalous diamagnetism, heat
capacity increase near transition. In this sense we can speak about the
existence of the average modulus of the order parameter (which is defined as
the square root of the average square of modulus (\ref{del})).

The participation of normal electrons in nonequilibrium Cooper pairing above 
$T_c$ is an inelastic process leading to some decay of the phase coherence
between initial and final quasiparticle states. This means that fluctuations
themselves act as a source of some phase-breaking time $\tau _\phi
(\varepsilon )$ side by side with paramagnetic impurities and thermal
phonons. The consequence of this fact is the shift of the transition
temperature toward lower temperatures with respect to its mean field value $%
T_{c0}$.

This shift is easy to estimate by taking into account the next order
correction ($\sim |\Psi |^4$) in the Ginzburg-Landau functional. We make the
Hartree approximation by replacing the $|\Psi (\vec{r})|^4$ term in (\ref{FP}%
) by $\langle |\Psi _{(fl)}|^2\rangle |\Psi (\vec{r})|^2$. This leads to the
renormalization of the reduced critical temperature value 
\begin{eqnarray}
\varepsilon ^{*}=\varepsilon +\frac b{2\alpha }\langle |\Psi
_{(fl)}|^2\rangle   \label{THF}
\end{eqnarray}
and to appropriate reduction of the critical temperature $T_c^{*}$ with
respect to its BCS value $T_{c0}$. Using the microscopic values of $\alpha $
and $b$ and also the results (\ref{aef})-(\ref{5}) for $\langle |\Psi
_{(fl)}|^2\rangle $, one can easily find within logarithmic accuracy: 
\begin{eqnarray}
\frac{\delta T_c}{T_c}=\frac{T_c^{*}-T_{c0}}{T_{c0}}\sim -Gi_{(2)}\ln {\
\frac 1{Gi_{(2)}}}  \label{dr}
\end{eqnarray}
where $Gi_{(2)}^{(d)}\sim \frac 1{p_F^2ld}$ for dirty, $Gi_{(2)}^{(cl)}\sim
\frac 1{p_Fd}\frac{T_c}{E_F}$ for clean film of thickness $d$ ($l$ is the
electron mean free path). In the case of HTS single crystal $%
Gi_{(2)}^{(cl)}\sim \frac{T_c}{E_F}$.

The consistent description of fluctuations in superconductors above $T_c$ is
possible only in the framework of the microscopic approach based on BCS
theory and it will be presented later (section 6).

\subsection{Fluctuations below $T_c$}

The description of fluctuations is much more complicated at temperatures
below $T_c$. In contrast to the simple fluctuation picture of the normal
phase, where nonequilibrium pairs appear and decay, below $T_c$ the
consequent construction of the fluctuation theory requires one to go beyond
the simple BCS picture of superconductivity and to take into account
correctly the different channels of interelectron interaction.

The electron-electron interaction in a normal metal may be considered as the
sum of the dynamically screened Coulomb interaction and virtual phonon
exchange. Generally speaking both of them involve momentum transfers in the
range of $|\Delta \vec{p}|\leq 2p_F$. However, in normal phase, it is
usually possible to consider the interaction as taking place in just two
channels: low-momentum and high-momentum transfer.

The first one is appropriate for the momentum transfers $\Delta p\rightarrow
0$ and is called the diffusion channel or the dynamically screened Coulomb
interaction. Virtual phonon exchange is neglected in it. The role of this
interaction becomes really pronounced in amorphous systems where the delay
of the screening, due to the diffusive character of the electron motion, is
considerable. It leads to the renormalization of the density of states and
other thermodynamical and transport properties of a metal (see \cite{AA85}
and section 5).

The second, so-called Cooper, channel is appropriate for the values of the
momentum transfers $\Delta p\approx 2p_F$, while the total momentum of two
particles $\vec{p}_1+\vec{p}_2\rightarrow 0$. Here both Coulomb and virtual
phonon exchange interactions are important and it is this part of
interaction which leads to superconductivity. In the BCS model it is
described by the effective constant of the electron-electron interaction
only.

Above $T_c$ these two channels of interaction do not mix, so the interaction
corrections from both channels can be considered separately \cite{AA85}.
Accounting for interaction in the diffusion channel leads to corrections
important in the description of the disordered systems properties. An
effective attraction in the Cooper channel, on the other hand, leads to the
reconstruction of the electronic ground state below $T_c$ to form the
superconducting state. Taking into account the Cooper channel above $T_c$ is
equivalent to treating superconducting fluctuations.

Below $T_c$ the interelectron interaction, in the framework of the BCS
theory, shows up itself via $T_c$ only. Fluctuations are not taken into
account here and the equilibrium coexistence of Cooper pairs condensate with
one-particle excitations is supposed. Fluctuations may be taken into
account by developing the perturbation series in the interaction beyond the
BCS picture. It is important to stress that the interaction in a
superconductor cannot be reduced to separate diffusion (screened Coulomb
interaction) and Cooper channels only, as was done above $T_c$: the presence
of the condensate mixes up the channels leading to a variety of collective
processes. These include: the appearance and decay of non-equilibrium Cooper
pairs, different types of quasiparticle scattering processes involving the
condensate, and scattering of Cooper pairs between themselves \cite{D78,VD86}%
.

Another way of considering fluctuation phenomena in the superconducting
phase is to speak in terms of fluctuations of the modulus and phase of the
order parameter, to calculate their correlators in different points of the
superconductor and to study the physically measurable values \cite
{VD86,O73,VDS88}. In this scheme, the inseparability of scalar potential
(dynamically screened Coulomb interaction) and order parameter phase
fluctuations below $T_c$ becomes evident. Indeed, above $T_c$ the dynamical
screening of the charge fluctuation originates by the space and time
redistribution of electrons only. Below the critical temperature such
quasiparticle currents cause charge redistribution by means of supercurrent
flows too, i.e. the appearance of gradients of the phase of order parameter.
One can see that this means the linking of scalar potential and phase
fluctuations, i.e. the appearance of off-diagonal elements in the matrix
correlation function.

We can discuss the effect of fluctuations on $T_c$ and the modulus of the
order parameter $\Psi (T)$ starting from the superconducting phase (in
contrast to the previous section). Evidently fluctuations suppress both of
these with respect to their BCS values and, as we will see below, this
effect is very similar to the effect of paramagnetic impurities.
However, while in the case of paramagnetic impurities, it is possible,
at least in principle, to clean up the sample and to determine the value of $%
T_{c0}$, it is impossible to do this with fluctuations. They reduce the
values of the critical temperature and the order parameter modulus with
respect to their BCS values, but there is no way to ``switch them off'' at
finite temperatures and especially near the critical temperature \footnote{%
The effect of fluctuations in principle can be suppressed by external
magnetic field but this will change $T_c$ with respect to the BCS value
itself.}.

We confine ourselves here to the discussion of the most interesting case of
2D fluctuations (films with thickness $d\ll \xi (T)$ or strongly anisotropic
layered superconductors). The renormalization of the order parameter modulus
may be calculated directly from the order parameter self-consistent equation
by including the fluctuation corrections to Gorkov F-function. In the
vicinity of $T_c$ one finds for a thin film \cite{VD86}: 
\begin{eqnarray}
\langle |\Psi _{(fl)}|^2(T,T_{c0})\rangle &=&\Psi _0^2(T,T_{c0})-  \nonumber
\label{delta2} \\
&& \\
&-&\frac{9\pi }{4p_Fd}\left( \frac{T_c}{E_F}\right) {\cal N}_e\left\{ \ln {\ 
\frac{T_{c0}}{T_{c0}-T}}+2\ln {\frac L{L_T}}+\frac{8\pi ^2}{63\zeta (3)}\ln
^3{\frac{L_T}d}\right\}  \nonumber
\end{eqnarray}
where $L_T=\sqrt{\frac{{\bf D}}T}$ is the diffusion length and ${\bf D}$ is
diffusion coefficient. The inclusion of $T_{c0}$ amongst the arguments of $%
\Psi _{(fl)}$ and $\Psi _0$ is done on purpose to underline that this
parameter can vary due to the effect of fluctuations too. For a layered
superconductor the analogous expression can be found by replacing the film
thickness $d$ by the interlayer spacing $s$ in the coefficient, and replacing $%
d\rightarrow \max \{s,\xi _{\perp }\sim \frac{Js}T\}$ in the argument of the
last logarithm (here $J$ is a hopping integral describing the Josephson
interaction between layers, see section 6).

The first correction to $\langle \Psi _{(fl)}^2\rangle $ is due to
fluctuations of the modulus of the order parameter and is primarily
responsible for the temperature dependence. One can see that this term is
analogous to $\langle \Psi _{(fl)}^2(T)\rangle $ from above the transition.

The second correction comes from the fluctuations of the phase of order
parameter and depends on some longitudinal cut-off parameter $L$ which has
to be chosen in accordance with the problem considered. Its origin is
connected with the destruction of long range order by phase fluctuations in
low dimensional systems \cite{H68} \footnote{%
Let us mention that whilst above $T_c$ both degrees of freedom (modulus and
phase of the order parameter) fluctuate in the same way, below the critical
temperature their behavior is quite different. Namely, as one can see from
the result (\ref{delta2}), the modulus fluctuates qualitatively in the same
way as above $T_c$, while the phase fluctuations lead to the destruction of
the long range order. If the electron Fermi liquid were uncharged, one could
associate phase fluctuations with the appearance of the Goldstone mode. In
superconductors, due to the electro-neutrality condition, the massless
Goldstone boson cannot propagate, but some traces of this phenomenon can be
observed at finite frequencies \cite{ArVol79,Kul80}.}. Thus in the framework
of such an approach the longitudinal dimension of the specimen $L$ has to be
taken as a cut-off parameter in (\ref{delta2}) and the divergence which
appears there indicates the absence of long-range order for infinite 2D
superconductors. However, as mentioned above, a supercurrent can flow in the
system even when the average value of the order parameter is not
well-defined. The only requirement is that the correlation function of the
order parameter behaves as a power of $r$ in the long-range limit $%
r\rightarrow \infty $. In the calculation of thermodynamical functions in
such a state, the presence of vortices in 2D superconductors gives rise to
another cut-off parameter for the contribution of phase fluctuations: the
characteristic distance between vortices.

The last term on the right-hand side of (\ref{delta2}) represents the
contribution both of the scalar potential fluctuations and off-diagonal
phase -- scalar potential interference terms \cite{VD86,O73}. One can see
that it becomes important for very thin films with $d\ll L_T$, and for
strongly anisotropic layered superconductors.

Such considerations lead to the renormalization of the critical temperature
which can be obtained from (\ref{delta2}) \cite{VD86,O73,F87}. For not too
thin (but with $d\ll \xi (T)$) superconducting films: 
\begin{eqnarray}
\frac{\delta T_c}{T_{c0}}\sim -Gi_{(2)}^{}\ln {\frac 1{Gi_{(2)}^{}}}
\label{T_c}
\end{eqnarray}
in accordance with (\ref{dr}).

In the important for HTS case of a clean strongly anisotropic layered
superconductor the consideration of the first two terms in (\ref{delta2}) is
analogous but the last term $\ln ^3{\frac{\xi_{ab}}{\max\{s,\xi_c\}}} \sim
\ln^3{\frac{E_F}{\max\{J,T_c\}}}$, generally speaking, cannot be omitted. As
a result the shift of the critical temperature due to the interelectron
interaction (which includes fluctuations of the modulus and phase of the
order parameter side by side with fluctuations of the scalar potential) in
this case is determined by the formulae

\begin{eqnarray}
\frac{\delta T_c}{T_{c0}}\sim -\frac{T_c}{E_F}\left[ \ln {\frac{E_F}{T_c}}+%
\frac{4\pi ^2}{63\zeta (3)}\ln ^3{\frac{E_F}{\max \{J,T_c\}}}\right]
\label{Tln3}
\end{eqnarray}
and the last term can even dominate in the case of extreme anisotropy.

\subsection{Discussion}

We discuss here the full picture of the temperature dependence of the order
parameter modulus $\sqrt{\langle |\Psi _{(fl)}|^2(T)\rangle }$ renormalized
by fluctuations for quasi-two-dimensional systems (see Fig. 1):

We start from the unperturbed BCS curve $\Psi _0(T,T_{c0})$ (dashed-dot line
at the Fig. 1). The effect of fluctuations is described by
equation (\ref{delta2}) and results in the deviation of $\sqrt{\langle |\Psi
_{(fl)}|^2(T)\rangle }$ below $\Psi _0(T,T_{c0})$ (solid line) due to the
growth of order parameter modulus fluctuations with the increase of
temperature.

The second effect of fluctuations is the decrease of the critical
temperature with respect to $T_{c0}$ (see (\ref{T_c})), so we have to
terminate our consideration based on (\ref{delta2}) at the renormalized
transition temperature $T_c^{*}$. One can see that the value of $\langle
|\Psi _{(fl)}|^2(T_c^{*},T_{c0})\rangle $ is of the order of $N_eGi_{(2)}\ln
\frac 1{Gi_{(2)}}$ and it matches perfectly with the logarithmic tail
calculated for the temperatures above $T_c$ in section 3.1. The full curve $%
\sqrt{\langle |\Psi _{(fl)}|^2(T)\rangle }$ is presented in Fig. 1
by the solid line.

One further comment should be made at this point. In practice the
temperature $T_{c0}$ is a formal value only, $T_c^{*}$ is measured by
experiments. So instead of $\Psi _0(T,T_{c0})$ the curve $\Psi _0(T,T_c^{*})$
more naturally has to be plotted (dashed line at Fig. 1). It
starts at $T_c^{*}$ and, in accordance with \cite{SRW95}, finishes at zero
temperature a little bit below the BCS value $\Psi _0(0,T_{c0})$. The shift $%
\delta \Psi _0(0)=\Psi _0(0,T_{c0})-\Psi _0(0,T_c^{*})$ due to quantum
fluctuations turns out to be have the same relative magnitude as $\delta T_c$%
: 
\begin{eqnarray}
\frac{\delta \Psi (0)}{\Psi (0)}\sim \frac{\delta T_c}{T_{c0}}\sim Gi\ln {Gi}
\label{dd}
\end{eqnarray}
One can see that the renormalized by fluctuations curve $\sqrt{\langle |\Psi
_{(fl)}|^2(T)\rangle }$ (\ref{delta2}) passes above to this, more natural
parametrization of the BCS $\Psi _0(T,T_c^{*})$ temperature dependence.

The renormalization of the critical temperature by fluctuations (\ref{T_c})
was first obtained by Yu.N. Ovchinnikov in 1973 \cite{O73} by means of the
expansion of the Eilenberger equations in the fluctuations of the order
parameter and scalar potential degrees of freedom. Then it was reproduced in
different approaches in following studies \cite{VD86,F87}. Nevertheless, the
studies of the effect of fluctuations on the superconducting transition
properties are still in progress. For instance, the authors of \cite{CUK94}
have revised the BCS theory including Goldstone bosons fluctuations. They
have stated that this procedure led them to find the {\it increase} of the
critical temperature due to fluctuations. Other authors \cite{VD86,F87,KT94}
(including the authors of this review), are much less optimistic: it seems
today evident that fluctuations can only reduce the critical temperature,
and even suppress the zero-temperature value of the order parameter \cite
{SRW95}. The growth of the critical temperature discovered in \cite{CUK94}
is very likely related to a lack of the correct account of the ''phase --
scalar potential'' correlations which leads to the exact cancelation of the
logarithmic contribution from boson-like Goldstone phase fluctuations (see 
\cite{O73,VD86}).

The decrease of the critical temperature is specially pronounced in the case
of bad metals such as synthetic metals and optimally doped or slightly underdoped
HTS compounds \cite{EK95,D96}. We will see in the following that as oxygen
concentration decreases the $Gi$ number grows and the role of fluctuation
effects (Cooper channel interaction) becomes more and more important, making
inapplicable the perturbation theory methods far enough from the optimal
oxygen concentration. The first example of this process is the strong
suppression of $T_c$ seen from the formula (\ref{T_c}).

\newpage

\section{The effect of superconducting fluctuations on the one-electron
density of states.}

As was already mentioned, the appearance of non-equilibrium Cooper pairing
above $T_c$ leads to the redistribution of the one-electron states around
the Fermi level. A semi-phenomenological study of the fluctuation effects on
the density of states of a dirty superconducting material was first carried
out while analyzing the tunneling experiments of granular $Al$ in the
fluctuation regime just above $T_c$ \cite{CA68}. The second metallic
electrode was in the superconducting regime and its well developed gap gave
a bias voltage around which a structure, associated with the superconducting
fluctuations of $Al$, appeared. The measured density of states has a dip at
the Fermi level \footnote{%
Here we refer the energy $E$ to the Fermi level, where we assume $E=0$.},
reaches its normal value at some energy $E_0(T),$ show a maximum at an
energy value equal to several times $E_0$, finally decreasing towards its
normal value at higher energies (Fig. 2). The characteristic
energy $E_0$ was found to be of the order of the inverse of the
Ginzburg-Landau relaxation time $\tau _{GL}\sim \frac
1{T-T_c}=T_c\varepsilon ^{-1}$ introduced above.

The presence of the depression at $E=0$ and of the peak at $E\sim (1/{\ \tau
_{GL}})$ in the density of states above $T_c$ are the precursor effects of
the appearance of the superconducting gap in the quasiparticle spectrum at
temperatures below $T_c$. The microscopic calculation of the fluctuation
contribution to the one-electron density of states is a nontrivial problem
and can not be carried out in the framework of the phenomenological
Ginzburg-Landau theory. It can be solved within the diagrammatic technique
by calculating the fluctuation correction to the one-electron temperature
Green function with its subsequent analytical continuation to the real
energies \cite{ARW70,CCRV90}. We omit here the details of the cumbersome
calculations and present only the results obtained from the first order
perturbation theory for fluctuations. They are valid near the transition
temperature, in the so-called Ginzburg-Landau region, where the deviations
from the classical behavior are small. The theoretical results reproduce the
main features of the experimental behavior cited above. The strength of the
depression at the Fermi level is proportional to different powers of $\tau
_{GL}$, depending on the effective dimensionality of the electronic spectrum
and the character of the electron motion (diffusive or ballistic). In a
dirty superconductor for the most important cases of dimensions D=3,2 one
can find the following values of the relative corrections to the density of
states at the Fermi level \cite{ARW70}:

\begin{equation}
\delta N_{fl}^{(d)}(0)\sim -\left\{ 
\begin{array}{ll}
\displaystyle{\frac{T_c^{1/2}}{{\bf D}^{3/2}}(T_c\tau _{GL})^{3/2}}, & D=3
\\ 
\displaystyle{\frac 1{{\bf D}}(T_c\tau _{GL})^2}, & D=2
\end{array}
\right.  \label{dirtyds}
\end{equation}
where ${\bf D}=\frac{v_Fl}D$ is the diffusion coefficient. At large energies 
$E\gg \tau _{GL}^{-1}$ the density of states recovers its normal value,
according to the same laws (\ref{dirtyds}) but with the substitution $\tau
_{GL}\rightarrow {E}^{-1}$.

It is interesting that in the case of the density of states fluctuation
correction the critical exponents change when moving from a dirty to a clean
superconductor \cite{CCRV90}:

\begin{equation}
\delta N^{(cl)}_{fl}(0) \sim - \left\{ 
\begin{array}{ll}
\displaystyle{\frac{1}{T_c \xi_0^3}(T_c \tau_{GL})^{1/2}}, & D=3 \\ 
\displaystyle{\frac{1}{T_c \xi_0^2}(T_c \tau_{GL})}, & D=2
\end{array}
\right.  \label{cleands}
\end{equation}
(the subscripts {\em (cl)} and {\em (d)} stand here for clean and for dirty
cases respectively). Nevertheless, as it will be seen below, due to some
specific properties of the corrections obtained, this difference between
clean and dirty systems does not manifest itself in the physically
observable quantities (tunneling current, NMR relaxation rate etc.) which
are associated with the density of states by means of some convolutions.

Another important respect in which the character of the density of states
renormalization in the clean and dirty cases differs strongly is the energy
scale at which this renormalization occurs. In the dirty case ($\xi _0\gg l$%
) this energy turns out to be \cite{ARW70}: 
\begin{equation}
E_0^{(d)}\sim T-T_c\sim \tau _{GL}^{-1},  \label{direnerg}
\end{equation}
while in the clean one ($\xi _0\ll l$) \cite{CCRV90}: 
\begin{equation}
E_0^{(cl)}\sim \sqrt{T_c(T-T_c)},  \label{clenerg}
\end{equation}
To understand this important difference one has to study the character of
the electron motion in both cases discussed \cite{CCRV90}. We recall
that the size of the fluctuating Cooper pair is determined by the coherence
length 
\begin{equation}
\xi (T)=\xi _0\left( \frac{T_c}{T-T_c}\right) ^{1/2}  \label{cohGL}
\end{equation}
of the Ginzburg-Landau theory. The zero-temperature coherence length $\xi _0$
differs considerably for the clean and dirty cases: 
\begin{equation}
\xi _{0,cl}^2=\frac{7\zeta (3)}{12\pi ^2T_c^2}\frac{E_F}{2m}  \label{cohcl}
\end{equation}
\begin{equation}
\xi _{0,d}^2=\frac{\pi {\bf D}}{8T_c}  \label{cohd}
\end{equation}
To pass from the dirty to the clean case one has to make the substitution 
\begin{equation}
{\bf D}\sim \frac{p_Fl}m\sim \frac{E_F\tau }m\rightarrow \frac{E_F}{mT_c}.
\label{subst}
\end{equation}
The relevant energy scale in the dirty case is the inverse of the time
necessary for the electron to diffuse over a distance equal to the coherence
length $\xi (T)$. This energy scale coincides with the inverse relaxation
time $\tau _{GL}$: 
\begin{equation}
t_\xi ^{-1}={\bf D}\xi ^{-2}\sim \tau _{GL}^{-1}\sim T-T_c.  \label{td}
\end{equation}
In the clean case, the ballistic motion of the electrons gives rise to a
different characteristic energy scale 
\begin{equation}
t_\xi ^{-1}\sim v_F\xi ^{-1}\sim (T_c\tau _{GL}^{-1})^{1/2}\sim \sqrt{%
T_c(T-T_c)}.  \label{tcl}
\end{equation}

The fluctuation corrections to the density of states may be presented as a
function of the energy and the temperature in a general form, for any
dimensionality of the isotropic electron spectrum and any impurity
concentration \cite{CCRV90}, but the relevant expressions are very
cumbersome and we restrict ourselves to report the 2D clean case only

\begin{eqnarray}
\delta N_{fl(2)}^{cl}(E) &=&-N_{(2)}\frac{8aT_c}{\pi E_F}\frac{T_c^2}{%
(4E^2+aT_c\tau _{GL}^{-1})}\times  \nonumber  \label{ds2Dc} \\
&& \\
&&\times \left\{ 1-\frac{2E}{(4E^2+aT_c\tau _{GL}^{-1})^{1/2}}\ln \left[ 
\frac{2E+(4E^2+aT_c\tau _{GL}^{-1})^{1/2}}{(aT_c\tau _{GL}^{-1})^{1/2}}%
\right] \right\} .  \nonumber
\end{eqnarray}
where $N_{(2)}=\frac m{2\pi }$ is the 2D density of electron states in the
normal metal and $a$ is some number of the order of unity. One can check
that the integration of this expression over all positive energies gives
zero. This is a consequence of conservation of the number of particles: the
number of quasiparticles is determined by the number of cells in the crystal
and cannot be changed by the interaction. So the only effect which can be
produced by the interelectron interaction is the redistribution of energy
levels near the Fermi energy. This statement can be written as the ``sum
rule'' for the fluctuation correction to the density of states: 
\begin{equation}
\int_0^\infty \delta N_{fl}(E)dE=0  \label{N}
\end{equation}

This sum rule plays an important role in the understanding of the
manifestation of the fluctuation density of states renormalization in the
observable phenomena. As we will see in the next section the singularity in
tunneling current (at zero voltage), due to the density of states
renormalization, turns out to be much weaker than that in the density of
states itself ($\ln {\varepsilon }$ instead of $\varepsilon ^{-1}$ or $%
\varepsilon ^{-2}$, see (\ref{dirtyds})-(\ref{cleands})). The same features
occur in the opening of the pseudo-gap in the c-axis optical conductivity,
in the NMR relaxation rate etc. These features are due to the fact that we
must always form the convolution of the density of states with some slowly
varying function: for example, a difference of Fermi functions in the case
of the tunnel current. The sum rule then leads to an almost perfect
cancellation of the main singularity at low energies. The main non-zero
contribution then comes from the high energy region where the DOS correction
has its `tail'.

Another important consequence of the conservation law (\ref{N}) is the
considerable increase of the characteristic energy scale of the fluctuation
pseudo-gap opening with respect to $E_0$: this is $eV_0=\pi T$ for tunneling
and $\omega \sim \tau ^{-1}$ for c-axis optical conductivity.

\newpage

\section{ The effect of fluctuations on the tunnel current}

\subsection{Introduction}

One of the most currently discussed problems of the physics of high
temperature superconductivity is the observation of pseudo-gap type
phenomena in the normal state of these materials \footnote{%
It is worth mentioning that some confusion with the concept of
''pseudo-gap'' takes place in literature. It is used often as synonym for
the spin gap (even being far from the anti-ferromagnetic phase), the same
definition is used in the description of a variety of HTS normal state
anomalies observed by means of transport and photo-emission measurements,
NMR relaxation, optical conductivity and tunneling at temperatures above $%
T_c $ in all range of oxygen concentrations without any proof that it has
the same origin in the different experiments.}. In this Chapter we deal with
the ''pseudo-gap'' type behaviour of the tunneling characteristics of HTS
materials in the metallic phase (slightly under-, optimally or over-doped
compounds, where the Fermi surface is supposed to be well developed). By
this we mean the observation at temperatures above $T_c$ critical one of
non-linear $I-V$ characteristics usual for the superconducting phase where a
real gap in the quasiparticle spectrum occurs.

First of all we would like to attract the readers attention to a possible,
relatively old, scenario of a ''pseudo-gap'' opening in the tunneling
resistance \cite{VD83,CCRV90}. This is caused by the fluctuation
renormalization of the one-electron density of states in a very narrow
interval of energies near the Fermi surface as discussed in the previous
section. It turns out that, due to the conservation of particle number, this
sharp renormalization manifests itself in the tunneling conductance by means
of the appearance of an unexpectedly wide ''pseudo-gap'' type structure with
a central minimum and two lateral maxima at a characteristic bias $eV=\pm
\pi T$ and a strong temperature dependence of its magnitude in the vicinity
of the transition. The appearance of a characteristic ''kink'' (of the same
origin as the fluctuation growth of the c-axis resistance \cite
{ILVY93,BMV93,BMV96}) in the temperature dependence of the zero-bias
conductance $G(V=0,T)$ is also predicted at temperatures close to $T_c$.
This theory was checked and confirmed long ago on the conventional
superconducting junction $Al-I-Sn$ \cite{Khachat} and we now apply it to
analysis of the very recent tunneling experiments on HTS materials.

We start with a discussion of the effect of the inter-electron interaction
on the tunneling properties of superconductors. In the following subsection
these results are then used to explain the anomaly in the zero-bias
tunneling conductance $G(V=0,T)$ of a $YBaCuO/Pb$ junction above the $YBaCuO 
$ transition temperature \cite{tun,AMC91,AMC94,AMC96}. Finally, we discuss
the results of traditional electron tunneling spectroscopy \cite{tao97}, STM
measurements \cite{MSW97} and interlayer tunneling spectroscopy \cite{SKN97}%
, where pseudo-gap structures in the tunnel conductance of $BiSrCaCuO$ were
found in a wide range of temperatures above $T_c$.

\subsection{Preliminaries}

It is quite evident that the renormalization of the density of states near
the Fermi level, even of only one of the electrodes, will lead to the
appearance of anomalies in the voltage-current characteristics of a tunnel
junction. So-called zero-bias anomalies, which are the increase of the
differential resistance of a junction with amorphous electrodes at zero
voltage and low temperatures, have been observed for a long time. They have
been explained in terms of a density of states depression in an energy range
of the order of ${E}_{am}\sim \tau ^{-1}$ around the Fermi level due to the
electron-electron interaction in the diffusion channel.

The quasiparticle current flowing through a tunnel junction may be presented
as a convolution of the densities of states with the difference of the
electron Fermi distributions in each electrode (L and R):

\begin{eqnarray}
I_{qp} &=&\frac 1{eR_nN_L(0)N_R(0)}\times  \label{ambar} \\
&&\int_{-\infty }^\infty \left( \tanh {\frac{E+eV}{2T}}-\tanh {\frac E{2T}}%
\right) N_L(E)N_R(E+eV)dE,  \nonumber
\end{eqnarray}
where $R_n$ is the Ohmic resistance per unit area and $N_L(0)$, $N_R(0)$ are
the densities of states at the Fermi levels in each of electrodes in the
absence of interaction. One can see that for low temperatures and voltages
the expression in parenthesis is a sharp function of energy near the Fermi
level. The characteristic width of it is $E_{\ker }\sim \max {\{T,V\}}\ll
E_F $. Nevertheless, depending on the properties of densities of states
functions, the convolution (\ref{ambar}) may exhibit different properties.

If the energy scale of the density of states correction is much larger than $%
E_{\ker }$ , the expression in parenthesis in (\ref{ambar}) acts as a
delta-function and the zero-bias anomaly in the tunnel conductivity strictly
reproduces the anomaly of the density of states around the Fermi level:

\begin{equation}
\frac{\delta G(V)}{G_n(0)}=\frac{\delta N(eV)}{N(0)},  \label{cond}
\end{equation}
where $G(V)$ is the differential tunnel conductance and $G_n(0)$ is the
background value of the Ohmic conductance supposed to be bias independent, $%
\delta G(V)=G(V)-G_n(0)$.

This situation occurs in a junction with amorphous electrodes \cite{AA79a}.
In the amorphous metal, the electron-electron interaction with small
momentum transfer (diffusion channel) is retarded and this fact leads to a
considerable suppression of the density of states in the vicinity of the
Fermi level, within an energy range ${E}_{am}{\sim }$ $\tau ^{-1}\gg T\sim $ 
$E_{\ker }.$ At zero temperature for the 2D case one has:

\begin{eqnarray}
\delta N_2(E)=\frac \lambda {4\pi ^2{\bf D}}\ln ({E\tau ),}  \label{2DAA}
\end{eqnarray}
where the constant $\lambda $ is related to the Fourier transform of the
interaction potential. In the 3D case the correction to the density of
states turns out to be proportional to $|E|^{1/2}$.

In the framework of this approach Altshuler and Aronov \cite{AA79a,AA79c}
analyzed the experimental data obtained in studies of the tunneling
resistance of $Al-I(O_2)-Au$ junctions and showed it to be proportional to $%
|\ V|^{1/2}$ at $eV\ll T$. The identification of the ``wings'' in the $I-V$
characteristics of such junctions with (\ref{2DAA}) was a key success of the
theory of the electron-electron interaction in disordered metals \cite{AA85}.

It is worth stressing that the proportionality between the tunnel current
and the electron DOS of the electrodes is widely accepted as an axiom, but
generally speaking this is not always so. As one can see from the previous
section, the opposite situation occurs in the case of the DOS
renormalization due to the electron-electron interaction in the Cooper
channel: in this case the DOS correction varies strongly already in the
scale of $E_0\ll T$ $\sim E_{\ker }$ and the convolution in (\ref{ambar})
with the density of states (\ref{ds2Dc}) has to be carried out without the
simplifying approximations assumed to obtain (\ref{cond}). This statement
represents the central point of this section. We will see in the following
that this fluctuation induced pseudo-gap like structure in the tunnel
conductance differs drastically from the anomaly of the density of states ( 
\ref{ds2Dc}), both in its temperature singularity near $T_c$ and in the
energy range of manifestation.

\subsection{The effect of fluctuations on the tunnel current}

Let us first discuss the effect of the fluctuation suppression of the
density of states on the properties of a tunnel junction between a normal
metal and a superconductor above $T_c$. The effect under discussion turns
out to be most pronounced in the case of thin superconducting films ($d\ll
\xi (T)$) and layered superconductors like HTS cuprates.

We now derive the explicit expression for the fluctuation contribution to
the differential conductance of a tunnel junction with one thin film
electrode close to its $T_c$. To do this we differentiate (\ref{ambar}) with
respect to voltage, and insert the density of states correction given in (%
\ref{ds2Dc}). This gives (see \cite{VD83}): 
\begin{eqnarray}
\frac{\delta G_{fl}(V,\varepsilon )}{G_n(0)} &=&\frac 1{2T}\int_{-\infty
}^\infty \frac{dE}{\cosh ^2\displaystyle{\frac{E+eV}{2T}}}\delta
N_{fl}^{(2)}(E,\varepsilon )=  \nonumber \\
&&  \label{flcon1} \\
&=&Gi_{(2)}(4\pi T\tau )\ln \left( \frac 1\varepsilon \right) {\rm Re}\psi
^{\prime \prime }\left( \frac 12-\frac{ieV}{2\pi T}\right) ,  \nonumber
\end{eqnarray}
where $\psi (x)$ is the digamma function, and $\tau $ is the electron's
elastic scattering time. We have introduced the Ginzburg-Levanyuk parameter $%
Gi_{(2)}(4\pi T\tau )$, which characterizes the strength of fluctuations in
the 2D case with arbitrary impurity concentration, and is given by :

\begin{eqnarray}
Gi_{(2)}(x)|_{x=4\pi T\tau } &=&  \label{GiLe} \\
\frac 2{x^2[\psi (\frac 12)-\psi (\frac 12+x)+\frac 1x\psi ^{\prime }(\frac
12)]}_{x=4\pi T\tau }\left( \frac{T_c}{E_F}\right) &=&\cases{ \frac1{\pi^3
E_F \tau} &for $4 \pi T \tau \ll 1$ \cr \frac{T_c}{14 \zeta(3)E_F} & for $4
\pi T \tau \gg 1$}.  \nonumber
\end{eqnarray}

It is important to emphasize several nontrivial features of the result
obtained. First, the sharp decrease ($\varepsilon ^{-2(1)}$) of the density
of the electron states generated by the inter-electron interaction in the
immediate vicinity of the Fermi level surprisingly results in a much more
moderate growth of the tunnel resistance at zero voltage ($\ln {\frac
1\varepsilon }$). Second, in spite of the manifestation of the density of
states renormalization at the characteristic scales $E_0^{(d)}\sim T-T_c$ or 
$E_0^{(cl)}\sim \sqrt{T_c(T-T_c)}$, the energy scale of the anomaly
development in the $I-V$ characteristic is much larger: $eV=\pi T\gg $ $E_0$
(see Fig. 3).

This departure from the habitual idea of the proportionality between the
tunnel conductance and the so-called tunneling density of states (\ref{cond}%
) is a straightforward result of the convolution calculated in ( \ref{ambar}%
) with the difference of Fermi-functions as a kernel. As already explained
in the previous section the physical reason is that the presence of
inter-electron interaction cannot create new electron states: it can only
redistribute the existing states.

In the inset of Fig. 3 the measurements of the differential
resistance of the tunnel junction $Al-I-Sn$ at temperatures slightly above
the critical temperature of $Sn$ electrode are presented. This experiment
was accomplished by Belogolovski, Khachaturov and Chernyak in 1986 \cite
{Khachat} with the purpose of checking the theory proposed by Varlamov and
Dorin \cite{VD83} which led to the result (\ref{flcon1}). The non-linear
differential resistance was precisely measured at low voltages which
permitted the observation of the fine structure of the zero-bias anomaly.
The reader can compare the shape of the measured fluctuation part of the
differential resistance (the inset in Fig. 3 with the
theoretical prediction. It is worth mentioning that the experimentally
measured positions of the minima are $eV\approx \pm 3T_c$, while the
theoretical prediction following from (\ref{flcon1}) is $eV=\pm \pi T_c$.
Recently similar results on an aluminium film with two regions of different
superconducting transition temperatures were reported \cite{PIP95}.

We will now consider the case of a symmetric junction between two
superconducting electrodes at temperatures above $T_c$. In this case,
evidently, the correction (\ref{flcon1}) has to be multiplied by a factor
''two'' because of the possibility of fluctuation pairing in both
electrodes. Furthermore, in view of the extraordinarily weak ($\sim \ln {({%
\frac 1\varepsilon )}}$) temperature dependence of the first order
correction, different types of high order corrections may manifest
themselves on the energy scale $eV\sim T-T_c$ or $\sqrt{T_c(T-T_c)}$.

The first type of higher order correction appears in the first order of
barrier transparency but in the second of fluctuation strength ($\sim Gi^2$) 
\cite{VD83}. Such corrections are generated by the interaction of
fluctuations through the barrier and they can be evaluated directly from ( 
\ref{ambar}) applied to a symmetric junction \cite{VD83}. T he second order
correction in $Gi$ can be written as:

\begin{equation}
\delta G_{fl}^{(2)}\sim \int_{-\infty }^\infty \frac{dE}{\cosh ^2{\ %
\displaystyle{\frac{E+eV}{2T}}}}\left[ \delta N_{fl}^{(2)}(E)\right] ^2.
\label{ds2ord}
\end{equation}
Because of the evident positive sign of the integrand, the condition (\ref{N}%
) does not lead to the cancellation observed in the first order correction $%
\delta G_{fl}^{(1)}$. As a result this correction turns out to be small ($%
\sim Gi^2$), but it is strongly singular in temperature and has the opposite
sign to $\delta G_{fl}^{(1)}$. Apparently it leads to the appearance of a
sharp maximum at zero voltage in $G(V)$ with a characteristic width $eV\sim
T-T_c$ in the immediate vicinity of $T_c$ (one can call this peak the
hyperfine structure). To our knowledge such corrections were never observed
in tunneling experiments.

The second type of correction is related to the coherent tunneling of the
Cooper pairs through barrier. These appear in the second order in
transparency but in the first order of fluctuation strength $Gi,$ so they
can only be expected to be observed in transparent enough junctions. These
corrections are similar to the {\em paraconductivity} and {\em anomalous
Maki-Thompson} contributions to electric conductivity. Recently, some of
them have been found to play an important role in the case of transverse
resistance of strongly anisotropic layered superconductors, where the
conducting layers are connected in the Josephson way \cite
{ILVY93,BDKLV93,K94}. The following sections will be devoted to the detailed
discussion of the interplay of all these contributions.

\subsection{Zero-bias anomaly studies in HTS junctions}

The tunneling study of HTS materials is a difficult task for many reasons.
For instance, the extremely short coherence length requires mono-layer-level
perfection at surfaces which are subject to long oxygen anneals; in general
they do not satisfy this stringent requirement. Another problem is the
intrinsic, (unrelated to superconductivity), bias dependence of the normal
tunnel conductance $G_n(V)$ in voltage ranges as wide as $\ \sim 100\ mV,$
in which it is necessary to scan for the study of $I-V$ characteristics of
HTS junctions. Nevertheless in recent years high quality junctions were
realized and $I-V$ characteristics can be measured today through a wide
temperature and voltage range by means of different techniques \cite
{tao97,MSW97,SKN97}).

Much attention in the scientific literature is devoted to the temperature
behaviour of the zero-bias conductance. The appearance of a kink around $T_c$
has been often observed \cite{AMC91}) and in this section we demonstrate
that quantitative fitting of $G(V=0,\varepsilon )$ vs temperature can be
obtained if the fluctuation contribution to the DOS in (\ref{ambar}) is
taken into account.

For the junction with one thin film (2D) superconducting electrode in the
vicinity of $T_c$ we have obtained the result (\ref{flcon1}); in the case of
HTS tunnel junction it has to be modified to take into account the
quasi-two-dimensional spectrum. In section 6, where the microscopic theory
of fluctuation conductivity will be presented, it is demonstrated that the
generalization of the logarithmic dependence on reduced temperature $%
\varepsilon $ (see (\ref{flcon1})) from 2D to the quasi-2D spectrum of the
corrugated cylinder type (see (\ref{d1})) is trivial: it is enough to
replace 
\begin{eqnarray}
\ln \left( \frac 1\varepsilon \right) \rightarrow 2\ln \left( {{\frac 2{{\ 
\sqrt{\varepsilon }+\sqrt{\varepsilon +r}}}}}\right) .  \label{q2D}
\end{eqnarray}
Here $r$ is the anisotropy parameter (Lawrence-Doniach crossover
temperature) which is defined precisely in the following section in terms of
the microscopic parameters of the HTS material. Physically this
dimensionless parameter determines the reduced temperature at which the
c-axis Ginzburg-Landau coherence length reaches the interlayer distance: $%
\xi _c(\varepsilon =r)\simeq $ $s$ .

Experiments have been performed on $YBaCuO/Pb$ planar junctions obtained by
chemically etching the degraded surfaces of $YBaCuO$ single crystals by a $%
1\%Br$ solution in methanol. Careful quality controls \cite{AMC94} have been
carried out on these junctions in order to assure that a pure tunneling
process without any interaction in the barrier takes place. 

In Fig. 4 we show the theoretical fitting (full line) for the
normalized fluctuation part of the tunneling conductance at zero bias, $%
\delta G_{fl}(0,\varepsilon )/G_n(0,T=140K)$ to formula (\ref{flcon1}) and (%
\ref{q2D}) with $V=0$. In the inset the complete set of experimental data
between 30 K and 180 K is shown, as reported in \cite{AMC96}. The junction's
critical temperature and the magnitude of the Ginzburg-Levanyuk parameter,
have been taken as fitting parameters \cite{CCV97}.

It is worth noticing that the tunneling spectroscopy probes regions of the
superconducting electrodes to a depth of $(2\div 3)\xi $ in contrast with
resistive measurements which sense the bulk percolation length. The
transition temperatures determined by the two types of measurement can be
quite different In our case the resistive critical temperature measured on
the $YBaCuO$ single crystal was $T_c=91~K$, while a ''junction's $T_c"=89K$
was obtained from the fitting procedure. This indicates that a slightly
oxygen deficient $YBaCuO$ layer in the junction area is probed by the
tunneling measurements. However the sample turns out to be still in the
proper (metallic) region of the phase diagram.

The value of Lawrence-Doniach crossover reduced temperature $r=0.07$ was
taken from the analysis of the crossover between 2D and 3D regimes in the
in-plane conductivity measurements \cite{FPFV97}. Another independent measurement 
of $r$ from the analysis of the non-linear
fluctuation magnetization \cite{Buz96,BBH95} leads to the very similar value 
$r=0.057$.

The magnitude of the fluctuation correction $|\psi ^{\prime \prime }(\frac
12)|Gi_{(2)}=0.029\simeq T_c/E_F$ for the sample under discussion leads to
the value of $E_F\simeq 0.3~eV$ which is in the lower range of the existing
estimates ($0.2\div 1.0~eV$).

The temperature range in which (\ref{flcon1}) satisfactorily reproduces the
behaviour of the zero-bias conductance, extends from the "junction $T_c$" up
to 110 K.

\subsection{Analysis of the pseudo-gap observation experiments}

As we have mentioned above the fluctuation renormalization of the DOS at the
Fermi level leads to the appearance of a pseudo-gap type structure in the
tunnel conductance at temperatures above $T_c$ which is very similar to the
one in superconducting phase, with the maximum position being determined by
the temperature $T$ instead of the superconducting gap value: $eV_m=\pi T$.
For the HTS samples this means a scale of $20-40~meV$, considerably larger
than in the case of conventional superconductors.

The observations of the pseudo-gap type anomalies of $G(V,T)$ at
temperatures above $T_c$ obtained by a variety of experimental techniques on 
$BiSrCaCuO-2212$ samples were reported very recently \cite{tao97,MSW97,SKN97}%
. Let us discuss their results based on the theoretical discussion presented
above.

1. The tunneling study of $BiSrCaCuO-2212/Pb$ junctions (as grown single
crystal samples) where the pseudo-gap type conductance nonlinearities were
observed in the temperature range from $T_c=87~K-89~K$ up to $110K$ (see
Fig. 5 \cite{tao97}. The maximum position moves to higher
voltages with the growth of temperature and at $T=100~K$ it turns out to be
of the order of $30-35~mV$ (depending on the sample). Both the magnitude of
the effect and the measured maximum position are in qualitative agreement
with the theoretical predictions ($eV_m(100~K)=\pm \pi T=27~mV$). 

2. The STM study of $Bi_{2.1}Sr_{1.9}CaCu_2O_{8+\delta }$ vacuum tunneling
spectra \cite{MSW97}, demonstrates the appearance of pseudo-gap type maxima
in $G(V,T)$ starting from temperatures far above $T_c$ ($T^{*}=260~K$ for
optimum and $T^{*}=180~K$ for overdoped samples). The energy scale
characterizing the pseudo-gap dip was estimated by the authors to be of the
order of $100~mV$ which again is consistent with the predicted maximum
position $eV_m(100~K)=\pm \pi T=\pm 27mV$.

3. Impressive results on tunneling pseudo-gap observations are presented in 
\cite{SKN97}. They were carried out by interlayer tunneling spectroscopy
using very thin stacks of intrinsic Josephson junctions fabricated on the
surface of $Bi_2SrCaCu_2O_8$ single crystal. The opening of the pseudo-gap
was found in $dI/dV$ characteristics at temperatures below $180K$. The data
presented can be fitted (Fig. 6) by means of formula (\ref{flcon1}) with $%
Gi=0.008$ and tunneling critical temperature T$_c=87~K$ (in view of the
strong anisotropy of the BSCCO spectrum $r=0$ may be assumed)\cite{CCV97}. 

One can see that this fit reproduces not only the magnitude of the effect
and the maxima positions, but also the shape of the experimental curves
(especially at low voltages). At higher voltages the theoretical curves,
being background voltage independent (Ohm law), overestimate the
experimental results.

\subsection{Discussion}

We have demonstrated that the idea of relating the pseudo-gap type phenomena
observed in tunneling studies with the fluctuation renormalization of
one-electron DOS in the immediate vicinity of the Fermi surface permits us
to fit the experimental data available with values of microscopic parameters
($E_F,r$) consistent with those obtained from independent measurements.

Two important comments are necessary. Both of them concern the limits of
applicability of the approach proposed.

The first concerns the magnitude of the effect. It is clear that (\ref{flcon1}%
), being a perturbative result, has to be small, so the parameter $Gi\ln {%
1/\varepsilon }$ has to be restricted. The temperature range is evidently
restricted by $\varepsilon \geq Gi$, so the criterion for applicability of
the theory is $Gi\ln {1/Gi}\ll 1$. The analysis of experimental data
obtained on samples with different oxygen concentration gives us grounds to
believe that $Gi$ increases as the oxygen concentration is decreased from
optimal doping. It follows that the importance of the electron-electron
interaction increases in the underdoped region of the phase diagram. The
qualitative extension of our perturbative results to this poor metal -
insulator region permits us to attribute the huge growth of the anomalies
observed to the effect of the strong e-e interaction.

Secondly it is very important to discuss the temperature range in which
pseudo-gap phenomena are expected to be observed following the approach
proposed. The result (\ref{flcon1}) is obtained in the mean field region $%
\varepsilon \ll 1$, neglecting the contribution of the short wave-length
fluctuations. Nevertheless, characteristic for 2D very slow (logarithmic)
dependence of the fluctuation correction on $~\varepsilon =\ln {\ \frac
T{T_c}}$ permits us to believe that the result (\ref{flcon1}) can be
qualitatively extended on wider temperature range up to several $T_c$. The
study of high temperature asymptotics ($\ln {\frac T{T_c}}\gg 1$) for the
e-e interaction in the Cooper channel \cite{ARV83} demonstrates the
appearance of an extremely slow $\ln {\ln {\frac T{T_c}}}$ dependence which
fits $\ln {1/\varepsilon }$ in the intermediate region and shows the
importance of the interaction effects up to high temperatures. In such a way
one can understand the reported gap-opening temperature $T^{*}\sim 200-300~K$
in the underdoped part of the phase diagram as the temperature where a
noticeable concentration of short-lived ($\tau \sim \frac \hbar {k_BT}$)
fluctuation Cooper pairs first manifests itself.

\newpage

\section{ Theory of fluctuation conductivity in a layered superconductor}

\subsection{Introduction}

Among all unconventional properties of the high temperature superconductors,
the transport properties are the most puzzling. The transition in the
``in-plane'' resistivity is usually smeared for tens of Kelvins and then a
mysterious linear increase with temperature takes place. If the smearing of
the transition finds a satisfactory explanation in the framework of the
fluctuation theory, the linear increase of the resistance is still the
subject of much speculation and discussion.

The temperature dependence of the transverse resistivity turns out to be
even more complicated. For not too underdoped samples at high temperatures
(considerably higher than $T_c$) $R_c(T)$ diminishes linearly with the
temperature in an analogous way to the in-plane resistivity. Nevertheless as
temperature decreases, for many samples this moderate decrease is followed
by a precipitous growth, the resistance passing through a maximum and then
abruptly decreasing to zero as the sample passes into the superconducting
phase. Such behavior in some degree has been observed in all high $T_c$
compounds \cite{BCZ91,BMV93,V93} and even in conventional layered
superconductors \cite{K94} .

As noted by Anderson \cite{AZ88}, the difference in temperature dependence
between the transverse and in-plane resistivities is extremely difficult to
explain in the framework of conventional Fermi liquid theory. Following the
general ideology of this review we show that such an explanation is possible
by taking into account the electron-electron interaction in the Cooper
channel. Discussing in detail the peculiarities of the fluctuation
conductivity tensor of a layered superconductor we will identify at least
one source of the difference between its transverse and longitudinal
components: the interplay of the suppression of positive paraconductivity
along the c-direction by the square of the interlayer transparency with the
growth of the normal resistance due to the fluctuation depression of the
density of states at the Fermi level.

In this section after a short review of the theory of electrical
conductivity in a layered metal (section 6.2) the main fluctuation
contributions to conductivity (AL, MT, DOS) will be examined, first
qualitatively (section 6.3) and then quantitatively within the full
microscopic theory (section 6.4). Section 6.5 is devoted to the analysis of
the role of the MT contribution as the precursor of the Josephson effect.
The results obtained are summarized in section 6.6.

\subsection{ Normal conductivity tensor of a layered metal. Electron
interlayer hopping time}

The goal of this subsection is to discuss the charge transport mechanisms in
a normal layered metal with an arbitrary impurity concentration and more
specifically the $c$-axis transport. Interest in this problem of classical
theory of metals was recently revived and the problem was discussed in
various models \cite{kosh,KLS90,rain,levin,V94} in connection with the study
of the normal properties of high temperature layered superconductors. Below
we present the traditional results of anisotropic diffusion theory which
will be necessary for the following analysis. In addition we will
demonstrate, on the basis of simple ideas of band motion, how the crossover
between the normal Drude regime and the hopping regime can appear \cite{V94}.

We assume that the electronic spectrum of a layered metal has the form: 
\begin{eqnarray}
\xi (p)=E({\bf p})+J\cos (p_zs)-E_F,  \label{d1}
\end{eqnarray}
where $E({\bf p})={\bf p}^2/(2m)$, $p\equiv ({\bf p},p_z)$, ${\bf p}\equiv
(p_x,p_y)$ is a two-dimensional intralayer wave-vector, and $J$ is the
effective nearest-neighbor interlayer hopping energy for quasiparticles. We
note that $J$ characterizes the width of the band in the $c$-axis direction
taken in the strong-coupling approximation. The Fermi surface defined by $%
\xi (p_F)=0$ is a corrugated cylinder (see Fig. 7), and $E_F$ is
the Fermi energy.

We assume that electrons are scattered by an elastic random potential
of arbitrary origin, and that the scatterers are located in the layers
and the scattering amplitude $U$ is isotropic.

We start from the Einstein relation for conductivity expressed through the
diffusion constant ${\bf D}_{\alpha \beta }$%
\begin{eqnarray}
\sigma _{\alpha \beta }=N(0)e^2{\bf D}_{\alpha \beta }.  \label{Drud}
\end{eqnarray}
The dirty case ($\tau \ll J^{-1}$) is trivial, since the usual anisotropic diffusion 
takes place. The
conductivity tensor is determined by the formula (\ref{Drud}) with a
diffusion coefficient 
\begin{eqnarray}
{\bf D}_{\alpha \beta }^{(d)}=\langle v_{\alpha \beta }^2(p_{\perp })\rangle
\tau =\tau \left( 
\begin{array}{cc}
\frac 12v_F^2 & 0 \\ 
0 & \frac{J^2s^2}2
\end{array}
\right)  \label{Dd}
\end{eqnarray}
where $<...>$denotes angular averaging over the Fermi surface. This leads to
a Drude conductivity with the substitution of $v_F$ by its c-axis analogue
for the transverse component: 
\begin{eqnarray}
\left( 
\begin{array}{l}
\sigma _{\parallel }^{(d)} \\ 
\sigma _{\perp }^{(d)}
\end{array}
\right) =\frac 12N(0)e^2\tau \left( 
\begin{array}{c}
v_F^2 \\ 
J^2s^2
\end{array}
\right)  \label{Dru}
\end{eqnarray}

Before starting the discussion of the clean case we recall the old story
of Zener (or Bloch) oscillations \cite{AKL}. In the case of a normal
isotropic metal it seems that there are no restrictions on the applicability
of the Drude formula for large $\tau $, and at first glance one might say
that the conductivity of a perfect crystal in the absence of any type of
scattering processes is infinite. However, more careful consideration shows
that when the work produced by the electric field on the acceleration of the
quasiparticle is accelerated by the electric field to the edge of the
Brillouin zone, it will be Bragg reflected. Consequently the electrons
oscillate with an amplitude which is determined by the inverse value of the
electric field and is huge on a microscopic scale (of the order of
centimeters for a field $\sim 1\frac V{cm}$). So no charge transfer takes
place in this ideal case and the conductivity is zero. Naturally any
scattering event destroys such a localized state and leads to a finite
conductivity.

The same situation must occur in an ideal clean layered superconductor. The
important point here is that the band along the $c$-axis direction is
extremely narrow (for HTCS it is estimated that $J\sim 10-200K$) so the
conditions on the mean free path are much less rigid than for the normal
Zener oscillations. So one can believe that for an ideal layered crystal the 
$c$-axis conductivity has to be zero. Now we consider what happens if a
small number of impurities are introduced (but $\tau \gg {\cal T}{\sim J^{-1}%
}$, where ${\cal T}$ is the period of oscillations and $\tau $ is the
elastic scattering time introduced above). In this case Zener oscillations
of electrons along $c$-axis will be destroyed from time to time by the
scattering events. In order to estimate the transverse resistance in this
clean case one can use formula ( \ref{Drud}) with the diffusion coefficient
describing the process of charge transfer discussed above. The hopping of an
electron from one plane to the next is possible only as the result of
scattering on an impurity which dephases the Zener oscillations. So the
diffusion coefficient in this case may be estimated as 
\begin{eqnarray}
{\bf D}_{\perp }^{(cl)}=\frac{\overline{{\bf r}_{\perp }^2(t)}}\tau \sim 
\frac{s^2}\tau  \label{strange}
\end{eqnarray}
(the time average here is taken over an interval much larger then $\tau $).
The transverse conductivity in this case is 
\begin{eqnarray}
\sigma _{\perp }^{(cl)}\sim N(0)e^2\frac{s^2}\tau
\end{eqnarray}
and one can see that it really vanishes in the clean limit in accordance
with the Zener statement. This value evidently matches with transverse
component of (\ref{Dru}) in the crossover region $J\tau \sim 1$.

Let us try to understand the physical meaning of the result obtained. The
diffusion in the case of a strongly anisotropic superconductor should not be
thought of as the continuous traveling of an electron in the insulating
space between metal layers. Even in the dirty case, the c-axis motion has
the character of wavepacket propagation, but the impurity scattering
destroying this state occur too often: $\tau \ll J^{-1}$(we remind that
the $J$ determines the period of oscillations). In a weak enough electric
field (in practice the only reachable in real experiment) the displacement
of the wavepacket during the time $\tau $ is of the order of many interlayer
distances and all above consideration concerning hopping to a neighbor plane
remains the same as in the case of full oscillation.

We can estimate the characteristic time of anisotropic diffusion from one
layer to a neighboring one ($\tau _{hop}$) on the basis of the Einstein
relation. The average displacement $s$ in the direction of the field
(c-axis) takes place in the time ($\tau _{hop}$), so 
\begin{eqnarray}
s=\sqrt{{\bf D}_{\perp }\tau _{hop}^{(d)}}  \label{A}
\end{eqnarray}
and hence 
\begin{eqnarray}
\tau _{hop}^{(d)}\sim \frac 1{J^2\tau },  \label{P}
\end{eqnarray}
Similarly we find 
\begin{eqnarray}
\tau _{hop}^{(cl)}\sim \tau .
\end{eqnarray}

The last result seems much more natural that the previous one. In the clean
case each scattering process leads to a hop along the c-axis, while in the
dirty case the hopping time turns out to be proportional to $\tau ^{-1}$.
Nevertheless replacing $\tau $ in (\ref{strange}) by (\ref{P}) reproduces
the Drude formula.

In conclusion, the results of this subsection can be summarized as follows:

1. In the dirty limit ($J\tau \ll 1$) both components of the layered metal
conductivity have the Drude form: 
\begin{eqnarray}
\sigma_{\perp} \propto \sigma_{\parallel} \propto \tau.
\end{eqnarray}

2. In the clean case ($J\tau \gg 1$) along the $c$-axis direction Zener
oscillations with infrequent dephasing take place, the conductivity $\sigma
_{\perp }\propto \tau ^{-1}$ has a hopping character, and the components of
the conductivity tensor satisfy the relation \cite{kosh,KLS90,rain,levin} 
\begin{eqnarray}
\sigma _{\perp }\sigma _{\parallel }=const.
\end{eqnarray}

It is worth mentioning, that formally in the extremely clean case Zener
oscillations can take place also in the $ab$-planes, but this possibility
may be excluded from consideration for any real experimental situation.

\subsection{ Qualitative consideration of different fluctuation contributions
}

Let us start with the qualitative discussion of the various fluctuation
contributions to conductivity \cite{ILVY94}.

The first effect of the appearance of fluctuation Cooper pairs above
transition is the opening of a new channel for charge transfer. Cooper pairs
can be treated as carriers with charge $2e$ and lifetime $\tau _{GL}=\frac
\pi {8(T-T_c)}$. This lifetime has to play the role of scattering time in
the Drude formula because, as is evident from its derivation, carrier
scattering or annihilation are essentially equivalent.

Finally we should replace the electron concentration ${\cal N}_e$by the
Cooper pair concentration ${\cal N}_{c.p.}$ in the Drude formula. This gives
the paraconductivity

\begin{eqnarray}
\delta \sigma _{AL}^{(2)}\sim \frac{{\cal N}_{c.p.}(2e)^2\tau _{GL}}{2m}=%
\frac{\pi e^2}{4m}\frac 1{(T-T_c)}{\cal N}_{c.p.}  \label{ALq}
\end{eqnarray}
In the 2D case ${\cal N}_{c.p.}^{(2)}=\frac{p_F^2}{2\pi d}\frac{T_c}{E_F}\ln 
{\ \frac 1\varepsilon }$. Substituting this result in (\ref{ALq}) one
reproduces with logarithmic accuracy the well known result of microscopic
calculations \cite{AL68}: 
\begin{eqnarray}
\delta \sigma _{AL}^{(2)}=\frac{e^2}{16d}~\frac{T_c}{T-T_c}  \label{AL2}
\end{eqnarray}

The 2D result (\ref{AL2}) is rather surprisingly identical for clean and
dirty cases. For electronic spectra of other dimensions this universality is
lost, and the paraconductivity becomes dependent on the electronic mean free
path via the $\eta _D$ parameter given in (\ref{d8}): 
\begin{eqnarray}
\delta \sigma _{AL}^{(D)}=\left\{ 
\begin{array}{ll}
\displaystyle{\frac 1{8\pi }\left( \frac e\hbar \right) ^2\left( \frac
1{4\eta _3\epsilon }\right) ^{1/2}} & {\rm three-dimensional\,case}, \\ 
\displaystyle{\frac 1{16}\frac{e^2}{\hbar d\epsilon }} & {\rm film,thickness:%
}d\ll \xi , \\ 
\displaystyle{\frac{\pi \eta _1^1/2}{16}\frac{e^2}{\epsilon ^{3/2}S}} & {\rm %
wire,cross\, section:}S\ll \xi ^2.
\end{array}
\right.  \label{parac}
\end{eqnarray}
Nevertheless one can see that the paraconductivity critical exponent $\nu
=(D-4)/2$ depends only on the effective dimensionality of the electronic
spectrum, and not on the nature of the scattering process.

This universality of the critical exponent obtained in the framework of the
simple mean field theory of fluctuations turns out to be very robust. The
recent revisions of the theory of paraconductivity for the models of
superconductivity with Eliashberg's strong coupling \cite{N94},
self-consistent treatment of the coexistence of superconductivity and
localization \cite{BSV86,S97} and a derivation of TDGL theory using
stochastic differential equations \cite{DM97} resulted in the same critical
exponents for $\delta \sigma _{AL}.$ Only the overall coefficient depends on
the characteristics of the model.

The correction to the normal state conductivity above the transition
temperature related with the one-electron density of states renormalization
can be reproduced in analogous way. The fact that some electrons participate
in fluctuation Cooper pairing means that the effective number of carriers
taking part in one-electron charge transfer diminishes leading to a decrease
of conductivity: 
\begin{eqnarray}
\delta \sigma _{DOS}=-\frac{\Delta {\cal N}_ee^2\tau _{imp}}m=-\frac{2{\cal N%
}_{c.p.}e^2\tau _{imp}}m  \label{DOSq}
\end{eqnarray}
which in the 2D case yields 
\begin{eqnarray}
\delta \sigma _{DOS}^{(2)}=-\frac{e^2}{2\pi d}(T_c\tau )\ln {\ \frac
1\varepsilon }  \label{DOS2}
\end{eqnarray}
The exact diagrammatic consideration of the DOS fluctuation effect on
conductivity agrees with this estimate obtained in its sign and temperature
dependence. The impurity scattering time dependence of (\ref{DOS2}) is
correct in the dirty case ($\xi \ll l$) but in the clean case the exact
calculations show a stronger dependence on $\tau $ ($\tau ^2$instead of $%
\tau $, see the next subsection).

Finally, we discuss the Maki-Thompson contribution. This anomalous term
has the same singularity in $\varepsilon $ as the AL one (within logarithmic
accuracy), but has a purely quantum nature and does not appear in the usual
TDGL approach at all \footnote{%
Recently it was reported \cite{G95} that the account of the interference
between superfluid and normal motions of charge carriers in TDGL scheme
permits to derive the MT contribution}. Its physical nature has remained
mysterious since 1968, when it was calculated in the diagrammatic approach
by Maki. This contribution is related to coherent electron scattering,
manifest itself only in transport properties, and is strongly phase
sensitive. These facts suggest that the MT contribution should be treated in
the same way as Altshuler and Khmelnitskii ( \cite{Ab88}) have treated weak
localization and interaction corrections to conductivity.

Let us consider possible types of Cooper pairing above $T_c$ in real space.
The simplest one is the appearance of Cooper correlation between two
electrons with momenta $\vec{p}$ and $-\vec{p}$ moving along straight lines
in opposite directions. (see Fig. 8(a)). Such pairing does not
have the characteristics mentioned above and has to be attributed to the AL
process.

Nevertheless, another, much more sophisticated pairing process can occur:
one electron with spin up and momentum $\vec{p}$ can move along some
self-intersecting trajectory; simultaneously another electron with spin down
and the opposite momentum $-\vec{p}$ can move in the opposite direction
along the same trajectory (see Fig. 8(b)). The interaction of such
a pair of electrons during their motion along the trajectory leads to the
appearance of some special contribution similar to the localization and
Coulomb interaction corrections to conductivity \cite{AA85}, but evidently
singular in the vicinity of $T_c$. One can easily see that such pairing is
possible only in the case of diffusive motion (necessary for the realization
of a self-intersecting trajectory). Finally any phase-breaking mechanism
leads to the loss of coherence and destruction of the Cooper correlation. So
all properties of the Maki-Thompson contribution coincide with the
properties of the process proposed. The contribution to the conductivity of
such process must be proportional to the ratio of the number of interfering
Cooper pairs $\delta {\cal N}_{s.i.}$ to the full concentration of
fluctuation Cooper pairs. In the 2D case 
\begin{eqnarray}
\frac{\delta {\cal N}_{s.i.}}{{\cal N}_{c.p.}}\sim \int_{\tau _{GL}}^{\tau
_\phi }\frac{lv_Fdt}{({\bf D}t)}=\ln {\frac \varepsilon {\gamma _\varphi }}
\label{MT2}
\end{eqnarray}
where $\tau _\phi $ is the one-electron phase-breaking time and $\gamma
_\varphi =\frac \pi {8T_c\tau _\phi }$ is the appropriate phase-breaking
rate.

The denominator of this integral, as in \cite{Ab88}, describes the volume
available for the diffusive electronic motion with the coefficient {\bf D}
during the time $t$, $({\bf D}t)^{d/2}$. The numerator describes the
volume element of the tube in which superconducting correlation of two
electron states of opposite momenta, can occur. The width of the tube is
determined by $l$ (mean free path) while the element of arc length is $v_Fdt$
(see Fig. 8). The lower limit of the integral is chosen so that at
least one Cooper pair occurs along the trajectory. The upper limit reflects
the fact that for times $t>\tau _\phi $ an electron loses its phase and
coherent Cooper pairing of two electron states above $T_c $ is impossible.

The contribution of Cooper pairs generated by coherent electrons moving
along self-intersecting trajectories is therefore 
\begin{eqnarray}
\Delta \sigma _{s.i.}^{(2)}={\cal N}_{c.p.}^{(2)}\ln {(\frac \varepsilon
{\gamma _\varphi })}\frac{(2e)^2\tau _{GL}}{2m}\sim \sigma _{AL}^{(2)}\ln {%
(\frac \varepsilon {\gamma _\varphi }).}  \label{MTq2}
\end{eqnarray}
One finds that this result coincides with the result of microscopic
calculations of the anomalous Maki-Thompson contribution \cite{T70}: 
\begin{eqnarray}
\sigma _{MT}^{(2)}=\frac{e^2}{8d}\frac 1{\varepsilon -\gamma _\varphi }\ln {%
\frac \varepsilon {\gamma _\varphi }}
\end{eqnarray}

In a similar way the correct temperature dependence of all contributions in
all dimensions can be obtained.

We will now try to understand the effect of fluctuations on the transverse
resistance of a layered superconductor in the same qualitative manner. As we
have demonstrated above, the in-plane component of paraconductivity is
determined by the Aslamazov-Larkin formula (\ref{AL2}). To modify this
result for c-axis paraconductivity one has to take into account the hopping
character of the electronic motion in this direction. One can easily see
that, if the probability of one-electron interlayer hopping is ${\cal P}_1$,
then the probability of coherent hopping for two electrons during the
virtual Cooper pair lifetime $\tau _{GL}$ is the conditional probability of
these two events: 
\begin{eqnarray}
{\cal P}_2={\cal P}_1\cdot ({\cal P}_1\cdot \tau _{GL}).  \label{3}
\end{eqnarray}
The transverse paraconductivity may thus be estimated as 
\begin{eqnarray}
\sigma _{\perp }^{AL}\sim {\cal P}_2\cdot \sigma _{\parallel }^{AL}\sim 
{\cal P}_1^2\frac 1{\varepsilon ^2}.  \label{4}
\end{eqnarray}
We see that the temperature singularity of $\sigma _{\perp }^{AL}$ turns out
to be stronger than that in $\sigma _{\parallel }^{AL}$ because of the
hopping character of the electronic motion (the critical exponent ''2'' in
the conductivity is characteristic of zero-dimensional band motion), However
for a strongly anisotropic layered superconductor $\sigma _{\perp }^{AL}$,
is considerably suppressed by the square of the small probability of
inter-plane electron hopping which enters in the prefactor.

It is this suppression which leads to the necessity of taking into account
the DOS contribution to the transverse conductivity. The latter is less
singular in temperature but, in contrast to paraconductivity, manifests
itself at the first, not the second, order in the interlayer transparency.
One can estimate it in the same way as above by multiplying the in plane
result (\ref{DOS2}) by the one-electron hopping probability: 
\begin{eqnarray}
\Delta \sigma _{\perp }^{DOS}\sim -{\cal P}_1\ln {\ \frac 1\varepsilon }.
\label{6}
\end{eqnarray}

It is important that, in contrast to the paraconductivity, the DOS
fluctuation correction to the one-electron transverse conductivity is
obviously negative and, being proportional to the first order of ${\cal P}_1$%
, can completely change the traditional picture of fluctuations just
rounding the resistivity temperature dependence around transition. Excluding
temporarily from consideration the anomalous Maki-Thompson contribution
(which is strongly suppressed in HTS by strong pair-breaking effects \cite
{HVS94}), one can say that the shape of the temperature dependence of the
transverse resistance is determined by the competition of two contributions
of the opposite sign: the paraconductivity, which is strongly temperature
dependent but is suppressed by the square of the barrier transparency ($\sim
J^4$) and the DOS contribution which has a weaker temperature dependence but
depends only linearly on the barrier transparency ($\sim J^2$), 
\begin{eqnarray}
\sigma _{fl}^{\perp }\sim k_1{\cal P}_1^2\frac 1{\varepsilon ^2}-k_2{\cal P}%
_1\ln {\ \frac 1\varepsilon .}  \label{41}
\end{eqnarray}
Here $k_1$ and $k_2$ are coefficients which will be calculated in the
framework of the exact microscopic theory presented in the next subsection.
It is this competition which leads to the formation of a maximum in the
c-axis resistivity.

\subsection{Microscopic theory of fluctuation conductivity in layered
superconductor}

\subsubsection{The model}

In a layered superconductor the zero-field resistivity is a diagonal tensor (%
$\rho _{xx}\equiv \rho _a$, $\rho _{yy}\equiv \rho _b$, $\rho _{zz}\equiv
\rho _c$), so the calculation of its various components is required.
Neglecting weak in-plane anisotropy, we further assume that the most
appropriate model for high-temperature superconductors has isotropic
in-plane electronic motion ($\rho _{xx}\equiv \rho _{yy}$). Therefore, we
will evaluate the fluctuation corrections to the two remaining independent
components of the resistivity tensor, $\rho _{xx}$ and $\rho _{zz}$.

We begin by discussing the quasiparticle normal state energy spectrum. While
models with several conducting layers per unit cell and with either
intralayer or interlayer pairing have been considered \cite{KL91}, it has
recently been shown \cite{LK93} that all of these models give rise to a
Josephson pair potential that is periodic in $k_z$, the wave-vector
component parallel to the $c$-axis, with period $s$, the $c$-axis repeat
distance. While such models differ in their superconducting densities of
states, they all give rise to qualitatively similar fluctuation propagators,
which differ only in the precise definitions of the parameters and in the
precise form of the Josephson coupling potential \cite{LK93}. Ignoring the
rather unimportant differences between such models in the Gaussian
fluctuation regime above $T_c(H)$, we therefore consider the simplest model
of a layered superconductor, in which there is one layer per unit cell, with
intralayer singlet $s$-wave pairing \cite{K74,KBL73}. These assumptions lead
to the simple spectrum (\ref{d1}) and hence to a Fermi surface having the
form of a corrugated cylinder.\ Experiments \cite{HER95} confirm the
assumption about the rough isotropy of the Fermi surface in the ab-plane for
metallic part of the HTS phase diagram.

Some remarks regarding the normal-state quasiparticle momentum relaxation
time are necessary. In the ''old'' layered superconductors such as $TaS_2$%
(pyridine)$_{1/2}$, the materials were generally assumed to be in the dirty
limit \cite{KBL73}. In the high-$T_c$ cuprates, however, both single
crystals and epitaxial thin films are nominally in the ''clean'' limit, with 
$l/\xi _{ab}$ values generally exceeding unity, where $l$ and $\xi _{ab}$
are the intralayer mean-free path and BCS coherence length, respectively.
However, as $l/\xi _{ab}\approx 2-5$ for most of the cuprates, these
materials are not extremely clean. In addition, the situation in the
cuprates is complicated by the presence of phonons for $T\simeq T_c\simeq
100K$, the nearly localized magnetic moments on the Cu$^{2+}$ sites, and by
other unspecified inelastic processes. In the following, we assume simple
elastic intralayer scattering \cite{KBL73}, keeping the impurity
concentration $n_i$ and the resulting mean-free path arbitrary with respect
to $\xi _{ab}$ \footnote{%
As it will be clear from the following assumptions for the impurity vertex
calculations the necessity to deal with the anisotropic spectrum (\ref{d1})
restricts us in this section by the requirement $l < \xi_{ab}(T) = \frac{
\xi_{ab}}{\sqrt{\varepsilon}}$}. The phase-breaking time $\ \tau _\phi $ is
taken to be much larger than $\tau $.

We now consider the various diagrams for the electromagnetic response
operator $Q_{\alpha \beta }(\omega _\nu )$, ( $\omega _\nu =(2\nu +1)\pi T$%
are the Matsubara frequencies) which contribute to the fluctuation
conductivity of layered superconductors.{\ }The diagrams corresponding to
the first order of perturbation theory in the fluctuation amplitude are
shown in Fig. 9. In this notation, the subscripts $\alpha ,\beta $
refer to polarization directions and thus to the conductivity tensor
elements according to 
\begin{eqnarray}
\sigma _{\alpha \beta }=-\lim_{\omega \rightarrow 0}{\frac 1{{i\omega }}}%
[Q_{\alpha \beta }]^R(\omega )\ \ \ ,
\end{eqnarray}

Intralayer quasiparticle scattering is included in the Born approximation,
giving rise to a scattering lifetime $\tau $ and resulting in a
renormalization of the single quasiparticle normal state Green's function to 
\begin{eqnarray}
G(p,\omega _n)={\frac 1{{i\tilde{\omega}_n-\xi (p)}}},  \label{d2}
\end{eqnarray}
where $\tilde{\omega}_n=\omega _n[1+1/(2|\omega _n|\tau )]$. Such
renormalizations are indicated in the vertices of Fig. 9 by
shadowing. The resulting expression for the triangle vertex, valid for
impurity concentration $n_i$ with the restrictions mentioned above, was
calculated in \cite{AV80} (see Appendix A): 
\begin{eqnarray}
\lambda (q,\omega _n,\omega _{n^{\prime }})=\biggl[1-{\frac{{\Theta (-\omega
_n\omega _{n^{\prime }})}}{{\tau (\tilde{\omega}_n-\tilde{\omega}_{n^{\prime
}})}}}\Bigl(1-{\frac{{\langle [\xi (p)-\xi (q-p)]^2\rangle }}{{(\tilde{\omega%
}_n-\tilde{\omega}_{n^{\prime }})^2}}}\Bigr)\biggr]^{-1},  \label{d3}
\end{eqnarray}
where $\Theta (x)$ is the Heaviside step function, and $\langle \cdots
\rangle $ denotes an average over the Fermi surface. Performing the Fermi
surface average, we find, 
\begin{eqnarray}
\langle [\xi (p)-\xi (q-p)]^2\rangle ={\frac 12}\Bigl(v_F^2{\bf q}%
^2+4J^2\sin ^2(q_zs/2)\Bigr)\equiv \tau ^{-1}\hat{{\bf D}}q^2,  \label{d4}
\end{eqnarray}
where $v_F=|{\bf p}_F|/m$ is the magnitude of the Fermi velocity parallel to
the layers.

For (\ref{d3}) to be valid we have to make the assumption $\tau \hat{{\bf D}}%
q^2\ll 1$. This will prevent us from discussing non-local corrections in
this section. We will explain what this means at the end of this subsection.

With each electromagnetic field component $eA_\alpha $ we associate the
external vertex $ev_\alpha (p)$ where $e$ is the quasiparticle electronic
charge, and $v_\alpha (p)={\frac{{\partial \xi (p)}}{{\partial p_\alpha }}}$%
. For longitudinal conductivity tensor elements (parallel to the layers, for
which $\alpha =x,y$), the resulting vertex is simply $ep_\alpha /m$. For the 
$c$-axis conductivity, the vertex $ev_z(p)$ is given by \cite{K74} 
\begin{eqnarray}
v_z(p)={\frac{{\partial \xi (p)}}{{\partial p_z}}}=-Js\sin (p_zs).
\label{d5}
\end{eqnarray}

Each wavy line in the diagrams represents a {\it fluctuation propagator} $%
L(q,\omega _\mu )$, which is a chain of superconducting bubble diagrams.
This object, which is the two-particle Green's function of the
fluctuation Cooper pair, was introduced in \cite{AL68,M68} and is calculated
from the Dyson equation in the ladder approximation (see Appendix B): 
\begin{eqnarray}
L^{-1}(q,\omega _\mu )=g^{-1}-\Pi (q,\omega _\mu ).  \label{Dyson}
\end{eqnarray}
Here $g$ is the effective constant of the electron-electron interaction in
the Cooper channel and the {\it polarization operator} $\Pi (q,\omega _\mu )$
consists of the correlator of two one-electron impurity Green functions: 
\begin{eqnarray}
\Pi (q,\omega _\mu )==T\sum_{\omega _n}\int {\frac{{d^3p}}{{(2\pi )^3}}}%
\lambda (q,\omega _{n+\mu },\omega _{-n})G(p+q,\omega _{n+\mu })G(-p,\omega
_{-n}),  \label{polar}
\end{eqnarray}

In the absence of a magnetic field and in the vicinity of $T_c$, the inverse
of $L(q,\omega _\mu )$ has the form 
\begin{eqnarray}
L^{-1}(q,\omega _\mu )=-N(0)\left[ \varepsilon +\psi \left( {\frac 12}+{\ 
\frac{{\omega _\mu }}{{4\pi T}}}+\alpha _q\right) -\psi \left( {\frac 12}%
\right) \right] ,  \label{d6}
\end{eqnarray}
where $\varepsilon =\ln (T/T_c)$ and 
\begin{eqnarray}
\alpha _q={\frac{{4\eta \hat{{\bf D}}q^2}}{{\pi ^2v_F^2\tau }}},  \label{d7}
\end{eqnarray}
(we remind, that for simplicity we omit the subscript ''2'' in $\eta
_2:\eta _2\equiv \eta $).

We integrate over the internal momenta $q$ and sum over the internal
Matsubara frequencies $\omega _\mu $, with momentum and energy conservation
at each internal vertex (fluctuation propagator endpoint) in the analytical
expressions for the diagrams presented at Fig. 9.

It is worthwhile making some comment there As it
was already mentioned above, the assumption $\tau \hat{{\bf D}}q^2<<1$ is
necessary in order to derive an explicit expression for $\lambda (q,\omega
_n,\omega _{n^{\prime }})$ valid for an arbitrary spectrum. This expression
will be used for the further treatment of fluctuation conductivity of
layered superconductor. It is therefore important to understand, which are
the effective momenta involved in the following integrations. It turns out
that in the vicinity of $T_c$ the convergence of integrals is determined by
the fluctuation propagator $L(q,0)$, which means 
\begin{eqnarray}
\alpha _q={\frac{{4\eta \hat{{\bf D}}q^2}}{{\pi ^2v_F^2\tau }}}\sim
\varepsilon \ll 1,
\end{eqnarray}

This does not imply any restrictions on $q_z$ but ${\bf q}_{eff}\sim \frac{%
\sqrt{\varepsilon }}{\xi _{ab}}$. It follows that the vertex part (\ref{d3})
does not permit us to treat the non-local limit $l\gg \frac{\xi _{ab}}{\sqrt{%
\varepsilon }}=\xi _{ab}(\varepsilon )$ appropriate to extremely clean
systems or in the immediate vicinity of $T_c$. Since we want to apply our
result to the analysis of real HTS compounds with $l\sim 2-5\xi _{ab}$ we
sacrifice the non-local limit here for the possibility of treating the
anisotropic spectrum (\ref{d1}). We stress again that the most important
arbitrary (in sense of the relation between $l$ and $\xi _{ab}$) case is
accessible to our consideration. The treatment of the non-local limit will
be presented in section 11, where its consequences on the MT contribution
will be discussed in detail for the example of NMR relaxation rate.

After these necessary introductory remarks and definitions we pass to the
microscopic calculation of the different fluctuation contributions
represented by the diagrams of Fig. 9.

\subsubsection{Aslamazov-Larkin contribution}

We first examine the AL paraconductivity (diagram 1 of Fig. 9).
The proper contribution to the electromagnetic response tensor has the form: 
\begin{eqnarray}
Q_{\alpha \beta }^{AL}(\omega _\nu )=2e^2T\sum_{\omega _\mu }\int {\frac{{%
d^3q}}{{(2\pi )^3}}}B_\alpha (q,\omega _\mu ,\omega _\nu )L(q,\omega _\mu
)L(q,\omega _\mu +\omega _\nu )B_\beta (q,\omega _\mu ,\omega _\nu ),
\label{d9}
\end{eqnarray}
where $\int d^3q\equiv \int d^2{\bf q}\int_{-\pi /s}^{\pi /s}dq_z$ is the
appropriate momentum space integral for a layered superconductor \cite
{K74,ILVY93,BPBHA92}, and the Green functions block is given by 
\begin{eqnarray}
B_\alpha (q,\omega _\mu ,\omega _\nu ) &=&T\sum_{\omega _n}\int {\frac{{d^3p}%
}{{\ (2\pi )^3}}}v_\alpha (p)\lambda (q,\omega _{n+\nu },\omega _{\mu
-n})\lambda (q,\omega _n,\omega _{\mu -n})\times  \nonumber  \label{d10} \\
&&\times G(p,\omega _{n+\nu })G(p,\omega _n)G(q-p,\omega _{\mu -n}),
\end{eqnarray}
($\lambda (q,\omega _n,\omega _{n^{\prime }})$ is defined by (\ref{d3}), and 
$\omega _{n\pm \nu }=\omega _n\pm \omega _\nu $, etc).

In the vicinity of $T_c$, the leading contribution to the response $%
Q_{\alpha \beta }^{AL}$ arises from the fluctuation propagators in (\ref{d9}%
) rather than from the frequency dependences of the vertices $B_\alpha $, so
we can to neglect the $\omega _\mu $- and $\omega _\nu $-dependences of $%
B_\alpha $. This approximation leads to \cite{AV80,BDKLV93} 
\begin{eqnarray}
B_\alpha ({\bf q},0,0) &=&-2\rho \frac \eta {v_F^2}\frac \partial {\partial
q_\alpha }\langle [\xi ({\bf p})-\xi ({\bf q}-{\bf p})]^2\rangle =  \nonumber
\label{d11} \\
&& \\
&=&-2\rho \frac \eta {v_F^2}\cases{ sJ^2sin(q_zs) &for $\alpha=z$ \cr v_{
F}^2q_{\alpha} & for $\alpha=x,y $}  \nonumber
\end{eqnarray}
Notice, that since $v_z(p)$ is odd in $p_z$ from (\ref{d5}), $B_z(q,\omega
_\mu ,\omega _\nu )$ is proportional to $J^2$ to leading order in $J$.

Using the expression in (\ref{d9}) followed by analytic continuation of the
external Matsubara frequencies to the imaginary axis (to obtain the
appropriate retarded response $Q^R(\omega )$) and integration over $q_z$,
the zero-frequency AL contribution to the in-plane fluctuation conductivity
response was found in the static limit to be 
\begin{eqnarray}
\sigma _{xx}^{AL} &=&={\frac{{\pi ^2e^2\eta ^2}}{{s}}}\int {\frac{{d^2{\bf q}%
}}{{(2\pi )^2}}}{\frac{{\bf q}^2}{{\bigl[(\eta {\bf q}^2+\varepsilon )(\eta 
{\bf q}^2+\varepsilon +r)\bigr]^{3/2}}}}  \nonumber  \label{d12} \\
&=&\frac{e^2}{16s}\frac 1{[\varepsilon (\varepsilon +r)]^{1/2}}\rightarrow {%
\ \frac{{e^2}}{{16s}}}\cases{(1/(\varepsilon r)^{1/2},&for
$\varepsilon<<r$,\cr 1/\varepsilon, &for $\varepsilon>>r$,\cr}.
\end{eqnarray}
Here 
\begin{eqnarray}
r=4\eta _2J^2/v_F^2\rightarrow \cases{{{\pi J^2\tau}\over{4T}} &for $\tau
T<<1$,\cr {{7\zeta(3)J^2}\over{8\pi^2T^2}} &for $\tau T>>1$},  \label{d13}
\end{eqnarray}
where $r(T_c)=4\xi _{\perp }^2(0)/s^2$ is the usual Lawrence-Doniach
anisotropy parameter \cite{LD70} characterizing the dimensional crossover
from the 2D to the 3D regimes in the thermodynamic fluctuation behavior at $%
T_{LD}$, and $\xi _{\perp }(0)$ is the zero-temperature Ginzburg-Landau
coherence length in the $c$-axis direction.

In the same way one can evaluate the AL contribution to the $c$-axis
fluctuation conductivity \cite{K74,ILVY93,BPBHA92} 
\begin{eqnarray}
\sigma _{zz}^{AL} &=&{\frac{{\pi e^2sr^2}}{{32}}}\int {\frac{{d^2{\bf q}}}{{%
\ (2\pi )^2}}}{\frac 1{{\bigl[(\eta {\bf q}^2+\varepsilon )(\eta {\bf q}%
^2+\varepsilon +r)\bigr]^{3/2}}}}  \nonumber  \label{d12_2} \\
&=&{\frac{{e^2s}}{{32\eta }}}\biggl({\frac{{\varepsilon +r/2}}{{\
[\varepsilon (\varepsilon +r)]^{1/2}}}}-1\biggr) \rightarrow {\frac{{e^2s}}{{%
\ 64\eta }}}\cases{(r/\varepsilon)^{1/2},&for $\varepsilon<<r$,\cr
[r/(2\varepsilon)]^2, &for $\varepsilon>>r$,\cr}
\end{eqnarray}
Note, that contrary to the case of in -plane conductivity, for $\sigma _{zz}$
the crossover occurs from 0D to 3D at $T_{LD}$.

In the region $\xi _{\perp }(T)<<s/2$ of two-dimensional fluctuation
behavior, $\sigma _{zz}^{AL}$ is smaller than the static in-plane
fluctuation conductivity $\sigma _{xx}^{AL}$ by the factor $[2\xi _{\perp
}(T)/s]^2\allowbreak (\sigma _{zz}^N/\sigma _{xx}^N)$, so that the other
contributions to the transverse fluctuation conductivity need to be
considered as well. The normal state conductivity tensor components in this
model are $\sigma _{xx}^N=N(0)e^2v_F^2\tau /2=E_F\tau e^2/(2\pi s)$, and $%
\sigma _{zz}^N/\sigma _{xx}^N=J^2s^2/v_F^2$ is the square of the ratio of
effective Fermi velocities in the parallel and perpendicular directions,
respectively.

\subsubsection{Contributions from fluctuations of the density of states}

The specific forms of the AL and, as shown below, MT contributions to the
fluctuation conductivity, which are suppressed for small interlayer
transparency, suggest that one should compare these terms with those arising
from other, less divergent, diagrams which are of lower order in the
transmittance \cite{ILVY93}. Such diagrams are pictured in diagrams 5-10 of
Fig. 9. These (DOS) diagrams arise from corrections to the
normal quasiparticle density of states due to fluctuations of the normal
quasiparticles into the superconducting state. In the dirty limit, the
calculation of the contributions to the longitudinal fluctuation
conductivity $\sigma _{xx}$ from such diagrams was discussed previously \cite
{ARV83}. Diagrams 9 and 10 arise from averaging diagrams 5 and 6 over
impurity positions. It was shown \cite{ARV83} that, for $\sigma _{xx},$
diagrams 9 and 10 are less temperature dependent than diagrams 5 and 6, and
can therefore be neglected. In the dirty limit, diagrams 7 and 8 were shown 
\cite{ARV83} to be equal to $-{\frac 13}$ times diagrams 5 and 6, which are
evidently equal to each other. In the clean limit, diagrams 7 and 8 can be
neglected relative to diagrams 5 and 6. For general impurity scattering, the
ratio of these diagrams depends upon $\tau $. As we are interested in the
results for arbitrary impurity concentration, we shall evaluate all these
diagrams separately. Calculations show that contrary to the case of the AL
contribution, the in-plane and out-of-plane components of DOS contribution
differ only in the square of the ratio of effective Fermi velocities in the
parallel and perpendicular directions. This allows us to calculate both
components simultaneously. The contribution to the fluctuation conductivity
due to diagram 5 is 
\begin{eqnarray}
Q_{\alpha \beta }^5(\omega _\nu ) &=&2e^2T\sum_{\omega _\mu }\int {\frac{{%
d^3q}}{{(2\pi )^3}}}L(q,\omega _\mu )T\sum_{\omega _n}\int {\frac{{d^3p}}{{%
(2\pi )^3}}}v_\alpha (p)v_\beta (p)\lambda ^2(q,\omega _n,\omega _{\mu
-n})\times  \nonumber  \label{d14} \\
&&\times G^2(p,\omega _n)G(q-p,\omega _{\mu -n})G(p,\omega _{n+\nu
}),~~~~~~~~~~~~
\end{eqnarray}
and diagram 6 gives an identical contribution. Evaluation of the
integrations over the internal momenta ${\bf p}$ and the summation over the
internal frequencies $\omega _n$ are straightforward. Treatment of the other
internal frequencies $\omega _\mu $ is less obvious, but in order to obtain
the leading singular behavior in $\varepsilon <<1$ of $Q^5$, it suffices to
set $\omega _\mu =0$. After integration over $q_z$, we have 
\begin{eqnarray}  \label{d15}
\sigma^{5+6}_{\left[xx \atop zz \right]} &=&-{\frac{{e^2s\pi r_1}}{{4}}}
\left[(v_F/sJ)^2 \atop 1 \right] \int_{|{\bf q}|\le q_{{\rm max}}}{%
\frac{{d^2{\bf q}}}{{(2\pi)^2}}}{\frac{1}{{\bigl[(\varepsilon+\eta{\bf q}
^2)(\varepsilon+r+\eta{\bf q}^2)\bigr]^{1/2}}}}  \nonumber \\
&=&-{\frac{{e^2sr_1}}{{8\eta}}} \left[(v_F/sJ)^2 \atop 1 \right]
\ln \biggl({\frac{{(\varepsilon+\eta q_{{\rm max}}^2)^{1/2}+(\varepsilon+r+%
\eta q^2_{{\rm max}})^{1/2}}}{{\varepsilon^{1/2}+(\varepsilon+r)^{1/2}}}}%
\biggr) \\
&\approx&-{\frac{{e^2sr_1}}{{8\eta}}} \left[(v_F/sJ)^2 \atop 1 \right]
\ln \Bigl({\frac{2}{{\varepsilon^{1/2}+(\varepsilon+r)^{1/2}}}}%
\Bigr),  \nonumber
\end{eqnarray}
where 
\begin{eqnarray}  \label{d16}
r_1={\frac{{2(J\tau)^2}}{{\pi^2}}}\biggl[\psi^{^{\prime}}\Bigl({\frac{1}{2}}%
+ {\frac{1}{{4\pi T\tau}}}\Bigr)-{\frac{3}{{4\pi T\tau}}}\psi^{^{\prime
\prime}}\Bigl({\frac{1}{2}}\Bigr)\biggr].
\end{eqnarray}
In the clean limit, $\sigma_{zz}$ in (\ref{d15}) reduces to that obtained in 
\cite{ILVY93}. In (\ref{d15}), we have introduced a cutoff in the integral
at $|{\bf q}|=q_{{\rm max}}$, where $\eta q_{{\rm max}}^2\approx1$, as in 
\cite{ILVY93,BMV93,BMMVY93}. This cutoff arises from the $q$-dependence of
the vertices and of the Green's functions, which had been neglected in
comparison with the contribution from the propagator, and is appropriate for
both the clean and dirty limits.

In a similar manner, the equal contributions from diagrams 7 and 8 sum to 
\begin{eqnarray}  \label{d17}
\sigma^{7+8}_{\left[xx \atop zz \right]} &=&-{\frac{{e^2s\pi r_2}}{{4}%
}} \left[(v_F/sJ)^2  \atop 1 \right] \int_{|{\bf q}|\le q_{{\rm max}%
}}{\frac{{d^2{\bf q}}}{{(2\pi)^2}}}{\frac{1}{{\bigl[(\varepsilon+\eta{\bf q}
^2)(\varepsilon+r+\eta{\bf q}^2)\bigr]^{1/2}}}}  \nonumber \\
&\approx&-{\frac{{e^2sr_2}}{{8\eta}}} \left[(v_F/sJ)^2 \atop 1 \right]
\ln\Bigl({\frac{2}{{\varepsilon^{1/2} +(\varepsilon+r)^{1/2}}}}%
\Bigr),
\end{eqnarray}
where 
\begin{eqnarray}  \label{d18}
r_2={\frac{{J^2\tau}}{{2\pi^3 T}}}\psi^{^{\prime\prime}}\Bigl({\frac{1}{2}}
\Bigr).
\end{eqnarray}
Comparing (\ref{d15}) and (\ref{d17}), we see that in the clean limit, the
main contributions from the DOS fluctuations arise from diagrams 5 and 6. In
the dirty limit, diagrams 7 and 8 are also important, having -1/3 the value
of diagrams 5 and 6, for both $\sigma_{xx}$ and $\sigma_{zz}$. Diagrams 9
and 10 are not singular in $\varepsilon<<1$ in the 2D regime, and can be
neglected. The total DOS contributions to the in-plane and $c$-axis
conductivity are therefore 
\begin{eqnarray}  \label{d19}
\sigma^{DOS}_{\left[xx \atop zz \right]} =-\frac{e^2}{2s} \kappa
\left[1 \atop (sJ/v_F)^2 \right] \ln\Bigl({\frac{2}{{%
\varepsilon^{1/2}+( \varepsilon+r)^{1/2}}}}\Bigr),
\end{eqnarray}
where 
\begin{eqnarray}  \label{d20}
\kappa={\frac{{r_1+r_2}}{{r}}}={\frac{{-\psi^{^{\prime}}\Bigl({\frac{1}{2}}+{%
\ \frac{1}{{4\pi\tau T}}}\Bigr)+{\frac{1}{{2\pi\tau T}}}\psi^{^{\prime%
\prime}} \Bigl({\frac{1}{2}}\Bigr)}}{{\pi^2\Bigl[\psi \Bigl({\frac{1}{2}}+{%
\frac{1}{{\ 4\pi\tau T}}}\Bigr)-\psi\Bigl({\frac{1}{2}}\Bigr)-{\frac{1}{{%
4\pi\tau T}}} \psi^{^{\prime}}\Bigl({\frac{1}{2}}\Bigr)\Bigr]}}}  \nonumber
\\
\rightarrow \cases{56\zeta(3)/\pi^4\approx0.691,&for $T\tau<<1$,\cr
8\pi^2(\tau T)^2/[7\zeta(3)]\approx9.384(\tau T)^2, &for $T\tau>>1$\cr}
\end{eqnarray}
is a function of $\tau T$ only.

\subsubsection{Maki-Thompson contribution}

We now consider the Maki-Thompson (MT) contribution (diagram 2 of Fig. 9) to fluctuation conductivity. The contributions from the two other
diagrams of the MT type (diagrams 3 and 4 of Fig. 9) are
negligible, because they are less singular in $\varepsilon $. Although the
MT contribution to in-plane conductivity is expected to be important in the
case of low pair-breaking, experiments on high-temperature superconductors
have shown that excess in-plane conductivity can usually be explained in
terms of the fluctuation paraconductivity alone. Two possible explanations
can be found for this fact. The first one is that pair-breaking in these
materials is not weak. The second is connected with the possibility of $d-$%
wave pairing which does not permit the anomalous Maki-Thompson process at
all (see section 12 for details).

Concerning the out-of-plane MT contribution, as was stated in \cite{ILVY93},
even when the pair lifetime $\tau _\phi $ is short, this contribution is
proportional to $J^4$ for small $J$ above $T_{LD}$, but is less singular
above $T_{LD}$ than the AL diagram, and was therefore excluded from that
treatment. As we shall show in the following, neglecting the MT diagram is
usually justified for the out-of-plane component of fluctuation conductivity
in layered materials. Nevertheless, there are a variety of situations where
the MT diagram may be important, depending upon the material parameters.
Because of its dependence on $\tau _\phi $, the MT contribution to the
transverse conductivity can have different temperature dependencies, and its
order in the interlayer transmittance can vary. For completeness, we
consider the scattering lifetime $\tau $ and the pair-breaking lifetime $%
\tau _\phi $ to be arbitrary, but satisfying $\tau _\phi >\tau $. The MT
contribution to the electromagnetic response tensor is then 
\begin{eqnarray}
Q_{\alpha \beta }^{MT}(\omega _\nu )=2e^2T\sum_{\omega _\mu }\int {\frac{{%
d^3q}}{{(2\pi )^3}}}L(q,\omega _\mu )I_{\alpha \beta }(q,\omega _\mu ,\omega
_\nu ),  \label{d21}
\end{eqnarray}
where 
\begin{eqnarray}
I_{\alpha \beta }(q,\omega _\mu ,\omega _\nu )
&=&~~~~~~~~~~~~~~~~~~~~~~~~~~~~~~~  \nonumber \\
&=&T\sum_{\omega _n}\int {\frac{{d^3p}}{{(2\pi )^3}}}v_\alpha (p)v_\beta
(q-p)\lambda (q,\omega _{n+\nu },\omega _{\mu -n-\nu })\lambda (q,\omega
_n,\omega _{\mu -n})\times \nonumber
\\
&&\times G(p,\omega _{n+\nu })G(p,\omega _n)G(q-p,\omega _{\mu -n-\nu
})G(q-p,\omega _{\mu -n})~.
\label{d22}
\end{eqnarray}
In the vicinity of $T_c$, it is possible to take the static limit of the MT
diagram simply by setting $\omega _\mu =0$ in (\ref{d21}). Although dynamic
effects can be important for the longitudinal fluctuation conductivity well
above $T_{LD}$, the static limit is correct very close to $T_c$, as shown in 
\cite{K90,RVV91}.

In evaluating the sums over the Matsubara frequencies $\omega_n$ in (\ref
{d22}), it is useful to break up the sum into two parts. In the first part, $%
\omega_n$ is in the domains $]-\infty, -\omega_{\nu}[$ and $[0,\infty[$.
This gives rise to the {\it regular} part of the MT diagram. The second ( 
{\it anomalous}) part of the MT diagram arises from the summation over $%
\omega_n$in the domain $]-\omega_{\nu},0[$. In this domain, analytic
continuation leads to an additional diffusive pole in the integration over $%
q $, with a characteristic pair-breaking lifetime $\tau_{\phi}$.

We start with the MT contribution to in-plane conductivity. Since in this
case the regular part of the MT diagram is completely similar to the DOS
contribution \cite{BDKLV93}, we list the result only: 
\begin{eqnarray}
\sigma _{xx}^{MT(reg)}=-\frac{e^2}{2s}\tilde{\kappa}\ln \Bigl({\frac 2{{\
\varepsilon ^{1/2}+(\varepsilon +r)^{1/2}}}}\Bigr)
\end{eqnarray}
where 
\begin{eqnarray}
\tilde{\kappa} &=&{\frac{{-\psi ^{^{\prime }}\Bigl({\frac 12}+{\frac 1{{4\pi
T\tau }}}\Bigr)+\psi ^{^{\prime }}\Bigl({\frac 12}\Bigr)+{\frac 1{{4\pi
T\tau }}}\psi ^{^{\prime \prime }}\Bigl({\frac 12}\Bigr)}}{{\pi ^2\Bigl[\psi
\Bigl({\frac 12}+{\frac 1{{4\pi \tau T}}}\Bigr)-\psi \Bigl({\frac 12}\Bigr)-{%
\frac 1{{4\pi \tau T}}}\psi ^{^{\prime }}\Bigl({\frac 12}\Bigr)\Bigr]}}} 
\nonumber  \label{d25} \\
&\rightarrow &\cases{28\zeta(3)/\pi^4\approx0.3455 &for $T\tau<<1$,\cr
\pi^2/[14\zeta(3)]\approx0.5865 &for $T\tau>>1$\cr}
\end{eqnarray}
is another function only of $\tau T$. We note that this regular MT term is
negative, as is the overall DOS contribution.

For the anomalous part of in-plane MT contribution we have: 
\begin{eqnarray}
\sigma _{xx}^{MT(an)} &=&8e^2\eta T_c\int {\frac{{d^3q}}{{(2\pi )^3}}}\frac
1{[1/\tau _\phi +\hat{{\bf D}}q^2][\varepsilon +\eta {\bf q}^2+{\frac r2}%
(1-\cos q_zs)]}  \nonumber \\
&=&\frac{e^2}{4s(\varepsilon -\gamma _\varphi )}\ln \left( \frac{\varepsilon
^{1/2}+(\varepsilon +r)^{1/2}}{\gamma _\varphi ^{1/2}+(\gamma _\varphi
+r)^{1/2}}\right)  \label{d25_1}
\end{eqnarray}
where 
\begin{eqnarray}
\gamma _\varphi ={\frac{{2\eta }}{{v_F^2\tau \tau _\phi }}}\rightarrow {%
\frac{{\pi }}{{\ 8T\tau _\phi }}}\cases{1 &for $T\tau<<1$,\cr
7\zeta(3)/(2\pi^3 T\tau) &for $T\tau>>1$.\cr}  \label{d27}
\end{eqnarray}
Equation (\ref{d25_1}) indicates that in the weak pair-breaking limit, the
MT diagram makes an important contribution to the longitudinal fluctuation
conductivity: it is of the same order as the AL contribution in the 3D
regime, but is larger than the AL contribution in the 2D regime above $%
T_{LD} $. For finite pair-breaking, however, the MT contribution is greatly
reduced in magnitude.

We now consider the calculation of the MT contribution to the transverse
conductivity. From the forms of $v_z(p)$ and $v_z(q-p)$in (\ref{d22})
obtained from (\ref{d5}), the bare electromagnetic vertices are proportional
to $\sin (p_zs)\sin (q_z-p_z)s$. After integration over the momentum $p=(%
{\bf p},p_z)$, the non-vanishing contribution is proportional to $\cos q_zs$%
. We take the limit $J\tau <<1$ in evaluating the remaining integrals, which
may then be performed exactly. The normal part of the MT contribution to
transverse conductivity is 
\begin{eqnarray}
\sigma _{zz}^{MT({\rm reg})} &=&-{\frac{{e^2s^2\pi r\tilde{\kappa}}}{{4}}}%
\int {\ \frac{{d^3q}}{{(2\pi )^3}}}{\frac{{\cos q_zs}}{{\varepsilon +\eta 
{\bf q}^2+{\ \frac r2}(1-\cos q_zs)}}}  \nonumber  \label{d24} \\
&=&-{\frac{{e^2s\pi \tilde{\kappa}}}{{2}}}\int {\frac{{d^2{\bf q}}}{{(2\pi
)^2}}}\biggl( {\frac{{\varepsilon +\eta {\bf q}^2+r/2}}{{[(\varepsilon +\eta 
{\bf q}^2)(\varepsilon +\eta {\bf q}^2+r)]^{1/2}}}}-1\biggr) \\
&=&-{\frac{{e^2sr\tilde{\kappa}}}{{16\eta }}}\biggl({\frac{{\ (\varepsilon
+r)^{1/2}-\varepsilon ^{1/2}}}{{r^{1/2}}}}\biggr)^2  \nonumber \\
&\rightarrow &-{\frac{{e^2sr\tilde{\kappa}}}{{16\eta }}}\cases{1 &for
$\varepsilon<<r$,\cr r/(4\varepsilon) &for $\varepsilon>>r$,\cr}  \nonumber
\end{eqnarray}
This term is smaller in magnitude than is the DOS one, and therefore makes a
relatively small contribution to the overall fluctuation conductivity and to
the temperature at which the $c$-axis resistivity is a maximum. In the 3D
regime below $T_{LD}$, it is proportional to $J^2$, and in the 2D regime
above $T_{LD}$, it is proportional to $J^4$.

The anomalous part of the MT diagram gives rise to a contribution to the
transverse conductivity of the form 
\begin{eqnarray}
\sigma _{zz}^{MT({\rm an)}} &=&{\frac{{\pi e^2J^2s^2\tau }}{{4}}}\int {\frac{%
{d^3q}}{{(2\pi )^3}}}{\frac{{\cos q_zs}}{{[1/\tau _\phi +\hat{{\bf D}}%
q^2][\varepsilon +\eta {\bf q}^2+{\frac r2}(1-\cos q_zs)]}}}  \nonumber
\label{d26} \\
&=&{\frac{{\pi e^2s}}{{4(\varepsilon -\gamma _\varphi )}}}\int {\frac{{d^2%
{\bf q}}}{{\ (2\pi )^2}}}\biggl[{\frac{{\gamma _\varphi +\eta {\bf q}^2+r/2}%
}{{\bigl[(\gamma _\varphi +\eta {\bf q}^2)(\gamma _\varphi +\eta {\bf q}^2+r)%
\bigr]^{1/2}}}}-{\frac{{\varepsilon +\eta {\bf q}^2+r/2}}{{\bigl[%
(\varepsilon +\eta {\bf q}^2)(\varepsilon +\eta {\bf q}^2+r)\bigr]^{1/2}}}}%
\biggr] \nonumber
\\
&=&{\frac{{e^2s}}{{16\eta }}}\biggl({\frac{{\gamma _\varphi +r+\varepsilon }%
}{{\ [\varepsilon (\varepsilon +r)]^{1/2}+[\gamma _\varphi (\gamma _\varphi
+r)]^{1/2}}}}-1\biggr),\ \ \ \ \ \ \ \ \ \ \ \ \ \ \ \ \ \   
\end{eqnarray}

In examining the limiting cases of (\ref{d26}), it is useful to consider the
cases of weak ($\gamma _\varphi <<r$, $\equiv $ $J^2\tau \tau _\phi >>1/2$)
and strong ($\gamma _\varphi >>r$, $\equiv $ $J^2\tau \tau _\phi <<1/2$)
pair-breaking separately. For weak pair-breaking, we have 
\begin{eqnarray}
\sigma _{zz}^{MT({\rm an})}\rightarrow {\frac{{e^2s}}{{16\eta }}}\cases{
(r/\gamma_\varphi)^{1/2} &for $\varepsilon<<\gamma_\varphi<<r$,\cr (r/\varepsilon)^{1/2} &
for $\gamma_\varphi<<\varepsilon<<r$,\cr r/(2\varepsilon) &if
$\gamma_\varphi<<r<<\varepsilon$.\cr}  \label{d28}
\end{eqnarray}
In this case, there is the usual 3D to 2D dimensional crossover in the
anomalous MT contribution at $T_{LD}$, for which $\varepsilon (T_{LD})=r$.
There is an additional crossover at $T_1$(where $T_c<T_1<T_{LD}$),
characterized by $\varepsilon (T_1)=\gamma _\varphi $, below which the
anomalous MT term saturates. Below $T_{LD}$, the MT contribution is
proportional to $J$, but in the 2D regime above $T_{LD}$, it is proportional
to $J^2$.

For strong pair-breaking, 
\begin{eqnarray}
\sigma _{zz}^{MT({\rm an})}\rightarrow {\frac{{e^2s}}{{32\eta }}}%
\cases{r/\gamma_\varphi &for $\varepsilon<<r<<\gamma_\varphi$,\cr
r^2/(4\gamma_\varphi \varepsilon)
&for $r<<{\rm min}(\varepsilon,\gamma_\varphi)$.\cr}  \label{d29}
\end{eqnarray}
In this case, the 3D regime (below $T_{LD}$) is not singular, and the
anomalous MT contribution is proportional to $J^2$, rather than $J$ for weak
pair-breaking. In the 2D regime, it is proportional to $J^4$ for strong
pair-breaking, as opposed to $J^2$ for weak pair-breaking. In addition, the
overall magnitude of the anomalous MT contribution with strong pair-breaking
is greatly reduced from that for weak pair-breaking.

Let us now compare the regular and anomalous MT contributions. Since these
contributions are opposite in sign, it is important to determine which will
dominate. For the in-plane resistivity, the situation is straightforward:
the anomalous part always dominates over the regular and the latter can be
neglected. The case of $c-$axis resistivity requires more discussion. Since
we expect $\tau _\phi \geq \tau $, strong pair-breaking is likely in the
dirty limit. When the pair-breaking is weak, the anomalous term is always of
lower order in $J$ than the regular term, so the regular term can be
neglected. This is true for both the clean and dirty limits. The most
important regime for the regular MT term is the dirty limit with strong
pair-breaking. In this case, when $\tau _\phi T\sim 1$, the regular and
anomalous terms are comparable in magnitude. In short, it is usually a good
approximation to neglect the regular term, except in the dirty limit with
relatively strong pair-breaking and only for out-of-plane conductivity.
However, we include it for generality.

As discussed in greater detail in the previous section, when $J\tau <<1$,
the effective interlayer tunneling rate is of the order of $J^2\tau $. When $%
1/\tau _\phi <<J^2\tau <<1/\tau $, the quasiparticles scatter many times
before tunneling to the neighboring layers \cite{KBL73}, and the pairs live
long enough for them to tunnel coherently. When $J^2\tau <<1/\tau _\phi $,
the pairs decay before both paired quasiparticles tunnel.

It is interesting to compare the anomalous MT contribution with the DOS and
AL contributions to transverse fluctuation conductivity. This comparison is
best made in the 2D regime above $T_{LD}$. For weak pair-breaking, the
anomalous MT and DOS terms are proportional to $J^2$, but opposite in sign,
and the former has a stronger temperature dependence than the latter. With
strong pair-breaking, the anomalous MT and AL contributions are proportional
to $J^4$, but the former is less singular in $\varepsilon <<1$ than is the
latter. Hence, the transverse MT contribution is in some sense {\it %
intermediate} between the transverse DOS and AL contributions. Nevertheless,
as we will show in the following, the MT contribution can be important in
the overall temperature dependence of the transverse resistivity,
eliminating the peak for weak pair-breaking.

\subsection{MT anomalous contribution as the precursor phenomenon of
Josephson effect}

Now, after the cumbersome calculations, let us speculate about the
nontrivial results obtained for the MT anomalous contribution to the c-axis
current.

Comparing the results of section 5.2 (\cite{VD83,CCRV90}) and the section
6.4 one can notice the richness of the approach based on spectrum (\ref{d1})
with that based on tunneling Hamiltonian without the momentum conservation 
\cite{VD83}. The latter does not take into account such delicate effects as
the fluctuation pairing through the barrier, coherent hopping etc., while
the former provides these possibilities. One can ask: does a precursor
phenomenon of the Josephson effect exist and, if so, with fluctuation
process is it?

I.O.Kulik replied positively on this question \cite{Kul69,Kul70}. He
demonstrated that although the average Josephson current above $T_c$ in the
absence of the an applied voltage is zero, precursor radiation of the
Josephson junction can be expected above $T_c$ at the frequency $\Omega \sim
T-T_c$. He associated this radiation with the AL paraconductivity.

The analysis of the expression for $\sigma _{zz}^{AL}$ contradicts to this
association. In spite of the fact that the related current is due to the
motion of Cooper pairs, it is evidently proportional to the square of
transparency of the junction ($\sim J^4$) and can in no way be matched with
the Josephson component of current flowing through the junction at
temperatures below $T_c$. The DOS contribution occurs in the first order of
the barrier transparency , but is evidently due to the quasiparticle branch
of the tunnel current below $T_c$(\cite{ILVY93,KG93}). So the last candidate
is the MT contribution and it is easy to see that under specified conditions
it really represents the precursor phenomenon of the Josephson effect.

One can see that in the weak pair-breaking limit ($\tau _\phi \gg t_{hop}$)
the $\sigma _{zz}^{MT}$ is proportional to the first order of the barrier
transparency and so the hypothetical current associated with the MT process
can be matched in this parameter with the Josephson current below the
transition. In the opposite case ($\tau _\phi \ll t_{hop}$) there are no
traces of the Josephson effect above $T_c$.

These results can easily be understood from the picture of self-intersecting
trajectories proposed in section 6.3. The hopping character of the electron
motion along the c-axis leads to the minimal self-intersecting trajectory in
this case consisting of intralayer diffusion, followed by scattering to a
neighboring layer, diffusion in this layer, and hopping back to the same
point in the original layer (see Fig. 10). The Josephson current
below $T_c$ is related to the phase coherence of the Cooper pairs condensate
of both electrodes. Since the MT contribution appears as the pairing of two
electrons moving along the trajectory described one can see that the
condition ($\tau _\phi \gg t_{hop}$) is evidently necessary to permit the
required phase coherence along the minimal self-intersecting trajectory.

And there is a final argument, formal but convincing. The Josephson current
is represented diagrammatically as the correlator of two Gorkov's
F-functions (\cite{BP82,VD86}). Approaching $T_c$ from below the F-functions
can be expended in powers of $\Psi $ (see Fig. 11). Above $T_c$ $%
\langle \Psi \rangle =0$ but the correlator $\langle \Psi \Psi ^{*}\rangle $
is nothing else as the fluctuation propagator and we pass to the MT diagram.

The message of these speculations is the following: the layered
superconductor with low phase-breaking could, in principle, irradiate or
react to external irradiation at a frequency $\sim T-T_c$\cite{Kul69} in the
vicinity of $T_c$. This radiation can be related both to order parameter and
pancake vortex fluctuations (at $T<T_c$). Nevertheless, the estimate for HTS
compounds are very pessimistic: the necessary condition $\tau _\phi \gg
\frac \tau {(J\tau )^2},J\tau \ll 1$ is unrealistic.

\subsection{The total fluctuation conductivity}

From the previous considerations, the total zero-field in-plane and
out-of-plane fluctuation conductivities are found to be 
\begin{eqnarray}
\sigma _{xx}^{fl} &=&\sigma _{xx}^{AL}+\sigma _{xx}^{DOS}+\sigma _{xx}^{MT(%
{\rm reg})}+\sigma _{xx}^{MT({\rm an})}  \nonumber  \label{d30_1} \\
&=&{\frac{{e^2}}{{16s}}}\left[ \frac 1{[\varepsilon (\varepsilon
+r)]^{1/2}}-8(\kappa +\tilde{\kappa})\ln \left( \frac 2{\varepsilon
^{1/2}+(\varepsilon +r)^{1/2}}\right) \right.  \nonumber \\
&&\left. +\frac 4{\varepsilon -\gamma _\varphi }\ln \left( \frac{\varepsilon
^{1/2}+(\varepsilon +r)^{1/2}}{\gamma _\varphi ^{1/2}+(\gamma _\varphi
+r)^{1/2}}\right) \right]
\end{eqnarray}
and 
\begin{eqnarray}
\sigma _{zz}^{fl} &=&\sigma _{zz}^{AL}+\sigma _{zz}^{DOS}+\sigma _{zz}^{MT(%
{\rm reg})}+\sigma _{zz}^{MT({\rm an})}  \nonumber  \label{d30} \\
&=&{\frac{{e^2s}}{{16\eta }}}\left[ -r\kappa \ln \left( {\frac 2{{\
\varepsilon ^{1/2}+(\varepsilon +r)^{1/2}}}}\right) ^2+\left[ (\varepsilon
+r)^{1/2}-\varepsilon ^{1/2}\right] ^2\left( {\frac 1{{\ 4[\varepsilon
(\varepsilon +r)]^{1/2}}}}-\tilde{\kappa}\right) \right.  \nonumber \\
&&\left. +\left( {\frac{{\varepsilon +\gamma _\varphi +r}}{{[\varepsilon
(\varepsilon +r)]^{1/2}+[\gamma _\varphi (\gamma _\varphi +r)]^{1/2}}}}%
-1\right) \right] .
\end{eqnarray}
We note that the second term in (\ref{d30}) contains both the AL and the
regular MT contributions. Writing the AL term in this fashion, it is easy to
see that the AL term is generally larger in magnitude than the regular MT
term, except when $r<<\varepsilon \approx 1$.

Although equations (\ref{d30_1}) and (\ref{d30}) contain both positive (AL
and anomalous MT) and negative (DOS and regular MT) contributions, the
behavior of the overall in-plane and out-of-plane fluctuation conductivities
is qualitatively different.

In fact, for $\sigma _{xx}^{fl}$, the negative contributions are less than
the positive ones in the entire temperature range above the transition,
leading to a total correction which is always positive. On the other hand,
in the case of $\sigma _{zz}^{fl}$, both positive terms (AL and anomalous
MT) are suppressed by interlayer transparency, leading to a competition
between positive and negative terms. This can lead to a maximum in the
c-axis fluctuation resistivity (minimum in the c-axis fluctuation
conductivity), whilst the in-plane resistivity is expected to be monotonous.
Since the temperature dependencies of $r$, $\kappa $, and $\tilde{\kappa}$
are weak compared with that present in $\varepsilon $, to obtain the
position of the $\rho _c(T)$ maximum, it suffices to extremise (\ref{d30})
with respect to $\varepsilon $. Using the restriction $J\tau <<1$ for the
validity of our theory, it is sufficient to consider the cases in which the
resistive maximum occurs in the 2D regime. Setting $\varepsilon =\varepsilon
_m$ at $T=T_m$, we have 
\begin{eqnarray}
\varepsilon _m/r\approx {\frac 1{{(8r\kappa )^{1/2}}}}-{\frac 1{{8\kappa }}}%
\Bigl[\tilde{\kappa}-{\frac 1{{2\gamma _\varphi }}}\Bigr],  \label{d31}
\end{eqnarray}
which is valid for $r\kappa <<1$and $\gamma _\varphi \kappa >1$.

We see that the regular MT term decreases the position of the maximum
somewhat, but the anomalous term increases it somewhat. It is then
qualitatively correct to neglect the MT terms altogether (as confirmed e.g.
by \cite{SM97}), but quantitatively, they can change the overall shape of
the fluctuation resistivity. We remark that there may be cases in which $T_m$
could occur in the 3D regime, but such cases cannot be addressed by our
theory, as they would require a proper treatment of non-local effects, as
well as the removal of the restriction $J\tau <<1$.

In Figs. 12 and 13, we have plotted $\rho _{xx}/\rho
_{xx}^N$ and $\rho _{zz}/\rho _{zz}^N$ versus $T/T_{c0}$ for various values
of the scattering lifetime $\tau T_{c0}$ and the pair-breaking lifetime $%
\tau _\phi T_{c0}$. We have taken $\sigma _{xz}^N={\ \frac 12}%
N(0)e^2v_F^2\tau $ and $\sigma _{zz}^N={\frac 12}N(0)J^2e^2s^2\tau $ here.

In these figures, the solid curves are plots of $\rho _{zz}/\rho _{zz}^N$.
The dashed curves are plots of $\rho _{xx}/\rho _{xx}^N$, for the same sets
of parameters. In Fig. 12, we have chosen $\tau T_{c0}=1$, which
is relevant for the high-$T_c$cuprates. In each of these figures, curves for 
$\tau _\phi T_{c0}=1,10,100$ are shown. We have $r(T_{c0})=0.1,0.01,$
and 0.001 in Figs.12(a),(b),(c), respectively. These values
correspond roughly to those expected for YBCO, BSCCO, and Tl$_2$Ba$_2$CaCu$%
_2 $O$_{8+\delta }$ (TBCCO), respectively. In order that the overall
fluctuation conductivity not give a large correction to the normal state
conductivity at temperatures well above the transition (i. e., for $%
T>1.03T_{c0}$), we have chosen $E_F/T_{c0}=300$ for $r(T_{c0})=0.1$and 0.001
[Figs. 12(a),(b)], but $E_F/T_{c0}=500$for $r(T_{c0})=0.001$%
(Fig. 12(c)). As can be seen from each of these figures, for a
fixed amount of pair-breaking and intralayer scattering, strong
pair-breaking (e. g., $\tau _\phi T_{c0}=1$) gives rise to a peak, or
maximum in $\rho _{zz}/\rho _{zz}^N $. A small, broad peak in $\rho
_{xx}/\rho _{xx}^N$ can also occur, but only for weak anisotropy ($%
r(T_{c0})=0.1$) and for such strong pair-breaking ($\tau _\phi T_{c0}=1$).
Increasing the anisotropy (or decreasing $r(T_{c0})$) greatly enhances the
magnitude of the peak in $\rho _{zz}/\rho _{zz}^N$. Decreasing the amount of
pair-breaking decreases the amplitude of the peak, as seen in each figure.

In Fig. 13, plots with the same parameters as in Fig. 
12(b) are shown, except that the intralayer scattering lifetime has
been decreased to $\tau T_{c0}=0.1$, which is in the dirty limit. It can be
seen that the magnitude of the peak in $\rho _{zz}/\rho _{zz}^N$ is reduced
by interlayer hopping, by interlayer scattering, and by pairbreaking. For
highly anisotropic materials, no peak in $\rho _{xx}/\rho _{xx}^N$ is
expected for any amount of pair-breaking shown in these figures.

Another important issue which should be discussed is the role of the DOS
term in the interpretation of in-plane resistivity data within fluctuation
theory. As we have mentioned above, the DOS term in $\sigma _{xx}^{fl}$
cannot result in a change of sign, but is able to change the magnitude of
the total correction which is important for the quantitative comparison with
experimental data. The DOS correction gives rise to the term proportional to 
$\kappa $ in (\ref{d30}). In the dirty limit, since $\tilde{\kappa}%
\rightarrow 0.5\kappa $, the DOS contribution is not much larger than the
(relatively small) regular MT term, and was therefore neglected by all
previous workers. However, when $\tau T_{c0}=1$, $\kappa (T_{c0})=14.3123$,
which is much larger than $\tilde{\kappa}(T_{c0})=0.5578$, and the DOS
contribution {\it cannot} be neglected relative to the other terms. Hence,
fits to data in which the DOS term has been neglected can only be trusted
for systems which are in the dirty limit. Since the cuprates are thought to
have $\tau T_{c0}\approx 1$, it is necessary to include the DOS contribution
in the fits. This term dramatically alters the shape of the overall parallel
resistivity, even for zero magnetic field. This change in $\rho _{xx}$ due
to the inclusion of the DOS contribution is pictured in Fig. 14.
In this figure, we have plotted $\rho _{xx}/\rho _{xx}^N$ for $\tau T_{c0}=1$%
, $\tau _\phi T_{c0}=10$, $E_F/T_{c0}=300$, for both $r(T_{c0})=0.1$ (dashed
curves), and $r(T_{c0})=0.01$ (solid curves), both with and without the DOS
contribution. As is easily seen from Fig. 14, the DOS
contribution greatly alters the overall resistivity, with the main aspect of
the alteration being an overall increase in the resistivity.

However, as we will discuss in section 8.3.1, no experiment revealed up to
now a significant contribuion of the DOS fluctuations. Indeed, this
small contribution can be masked by a change in the parameters of fits
which do not include the DOS term.

\newpage

\section{Experimental observations of fluctuation conductivity in HTS}

\subsection{Introduction}

We have already remarked that the normal state electrical transport
properties of HTS are very peculiar in many aspects. Particularly
interesting are the differences between the in-plane and c-axis
resistivities. While the quantitative differences among them (up to a factor 
$10^4$) are obviously due to the layered nature of these compounds, the
explanation of their qualitative behavior is related to the quasi-2D
character of HTS in a much less straightforward way. The apparently opposite
behaviors of $\rho _{ab}(T)$ and $\rho _c(T)$ close to $T_c$ in HTS
(decrease of resistivity for the former, increase for the latter as
temperature is decreased) could not be explained for a long time. Although
the peak in $\rho _c(T)$ observed in all high $T_c$ compounds \cite
{PHK88,MFF88,BCZ91,BMV93,M89,M90} appeared to be strictly connected with the
anisotropy of the sample, being very pronounced in highly anisotropic
samples (BSCCO, YBCO annealed in reducing atmosphere) but almost absent in
samples with low anisotropy (fully oxygenated BSCCO or YBCO), every attempt
to explain it through normal state conductivity models in highly anisotropic
systems failed to give satisfactory results \cite
{KLS90,AND,CLZ89,CZ91,KJ92,LV92,KLEMM}. This failure extends to recent
models \cite{ABR,YMH95} which, as we will see, do not satisfactorily match
the experimental behavior in the vicinity of $T_c$.

However, it has been shown in section 6 how the different behaviors of $\rho
_{ab}(T)$ and $\rho _c(T)$ close to $T_c$ can be explained by a single
physical mechanism (namely fluctuations), provided that all fluctuation
contributions are taken into account, the hierarchy of the various
fluctuation contributions being different for in-plane and c-axis
conductivity. As we have seen, the suppression of the positive fluctuation
paraconductivity along the c-direction by the square of the interlayer
transparency together with the decrease of the normal state conductivity due
to the fluctuation decrease of the density of states at the Fermi level
leads to an increase of resistivity in c-axis measurements in samples having
a sufficiently high anisotropy.

In the following sections we will analyze experimental data in the framework
of the theory presented in section 6 and show how the theory can
quantitatively describe the experimental data. Section 7.1 will be devoted
to a brief review of the role of fluctuations in the in-plane conductivity,
where the AL paraconductivity dominates, section 7.2 will deal with the
origin of the resistivity peak in transverse measurements, where competition
between the positive AL paraconductivity and the negative DOS contribution
to conductivity takes place. A comparison between the two methods employed
to explain the transverse resistivity peak (normal state conductivity or
fluctuation conductivity) concludes this section.

\subsection{In-plane resistance: Crossover phenomena observations}

Soon after the discovery of HTS superconductivity in the YBCO compound, the
observation of a large in-plane excess conductivity above the
superconducting transition in measurements was reported by Freitas et al. 
\cite{Freitas87} and Dubson et al. \cite{Dubson87}. This excess conductivity
was attributed to thermodynamic fluctuations, which in HTS are expected to
be much larger than in conventional superconductors because of their short
coherence length and high transition temperature. An early review of the
fluctuation effects on the electrical transport properties was given in \cite
{APC92}. As pointed out in \cite{VAM93,MPD94} earlier measurements,
especially when carried out on bulk single crystals, were sometimes affected
by $T_c$ (oxygen-content) inhomogeneities. Such difficulties were surmounted
mostly by using monocrystalline films with well defined geometries. A
typical $\rho _{ab}(T)$ curve in HTS is shown in Fig. 15.

These measurements generated great interest, especially because the nature
of thermodynamic fluctuations is related to the important topic of the
dimensionality of superconductivity in HTS \cite
{Aus88,Ong88,Oh88,Vidala88,Vidalb88,Veira88,HR95,BD94}. In fact, since the
cuprate superconductors have layered structure, both 2D and 3D behavior can
be observed, depending on the relative values of the temperature-dependent
coherence length perpendicular to the layers $\xi _{{\rm c}}(T)$ and of the
spacing $s$ between superconducting layers. When $\xi _{{\rm c}}(T)\gg $ $s$
(i.e. close to $T_{{\rm c}}$) the behavior is 3D, while as the temperature
is increased and $\xi _{{\rm c}}(T)$ decreases, a Lawrence-Doniach crossover
between the 3D and 2D regimes occurs at the temperature $T_{{\rm LD}}$
defined by the condition $\xi _{{\rm c}}(T_{{\rm LD}})\approx s$.

For the YBCO compound \cite{Freitas87} the excess conductivity has been
described by the 3D AL fluctuation contribution ($\Delta \sigma \sim
\varepsilon ^{-1/2}$) in a wide temperature range $-4\leq \ln \varepsilon
\leq -2$, while at higher temperatures ($\ln \varepsilon >-2$) a breakdown
of this simple theory was observed as shown in Fig. 16. These
conclusions were confirmed by later experiments \cite{Aus88, Ong88, Oh88,
Vidala88, Vidalb88, Veira88}

On the other hand, the much more anisotropic BSCCO\ compound showed a 3D
fluctuation behavior only in the close proximity of $T_{{\rm c}}$, while at
higher temperatures a clear 2D behavior ($\Delta \sigma \sim \varepsilon
^{-1}$) was observed \cite{BNVM89}.

At sufficiently high temperatures the experimental behavior of fluctuation
conductivity deviates from the simple AL theory, which is indeed valid only
for $T-T_c\ll T_c$. Reggiani et al. \cite{RVV91} generalized the 2D AL
theory for the high temperature region by taking into account the short
wavelength fluctuations and obtained the following universal formula for
paraconductivity 
\begin{eqnarray}
\sigma _{{\rm fl}}^{{\rm 2D}}=\frac{e^2}{16\hbar s}f(\varepsilon ).
\label{al}
\end{eqnarray}
Here $f(\varepsilon )$ is a function calculated in \cite{RVV91} which, for
clean 2D superconductors, tends to $f(\varepsilon )=1/\varepsilon $ in the
GL region of temperature ($\varepsilon \ll 1$), so that the result coincides
with the well known AL one, while in the opposite case ($\varepsilon \gg 1$
), tends to the asymptotic $f(\varepsilon )\sim 1/\varepsilon ^3=1/\ln
^3(T/T_{{\rm c}})$.

In \cite{BMMRVV92} the validity of this formula was carefully verified.
Clearly, when dealing with the small fluctuation effects measured far from $%
T_{{\rm c}}$ extreme care must be taken in the method of measurement. Small
sample-dependent deviations of the $R(T)$ curves from the optimal,
''intrinsic'' behavior could severely affect the fluctuation effects deduced
from the $R(T)$ measurements. Therefore, from a batch of 20 epitaxial $%
Bi_2Sr_2CaCu_2O_{8+x}$ films, only three were selected after checking their
compositional homogeneity, structural quality and electrical transport
properties (low extrapolated resistivity at 0 $K$, narrow transition). The $%
R(T)$ curves of these three films were directly compared by plotting their
resistance normalized to a reference value (namely $R(1.33T_c)$) vs. the
normalized critical temperature $T/T_c$ as shown in Fig. 15. The
curves for all three films superimpose very well, so that they can be
assumed to represent the ''intrinsic'' resistive transition for the 2212
BSCCO\ compound in spite of the small spread of their critical temperature
and resistivity values. For these films the excess conductivity was
analyzed. A very good fitting with the formula (\ref{al}) was found in the
temperature region $0.02\le \varepsilon \le 0.14$ (i.e. $1.5$ $K<T-T_c<11$ $%
K $), while the original AL theory fits the data only in a much narrower
temperature range (Fig. 17). For temperatures below $T_c+1.5$ $K$
Eq. \ref{al} fails because the sample is no longer in the 2D region, while
for $T>T_c+11$ $K$ the usual choice of the normal state resistance as the
linear extrapolation from the high-temperature behavior artificially forces
the extracted fluctuation conductivity to go to zero as the temperature
approaches the range used for the linear extrapolation.

As already mentioned, formula (\ref{al}) predicts the asymptotic behavior $%
f(\varepsilon )\sim 1/\varepsilon ^3=1/\ln ^3(T/T_{{\rm c}})$ at high enough
temperatures. In a recent paper \cite{FPFV97} a careful analysis of the
higher temperature region (above the edge of the region investigated in \cite
{BMMRVV92}) permitted observation of this asymptotic regime, although at a
surprisingly low reduced temperature ($\varepsilon ^{*}\sim \ln (T^{*}/T_{%
{\rm c}})\sim 0.23$) in YBCO, 2212 BSCCO and 2223 BSCCO\ samples. The
background of the normal state conductivity $\sigma _{{\rm N}}=1/\rho _{{\rm %
N}}$ was evaluated with particular accuracy by starting the linear
extrapolation at temperatures higher than about $150$ $K$ and checking that
in the range from $150$ $K$ to $330$ $K$ $\rho _{{\rm N}}$ did not change by
shifting the interpolation temperature region. Therefore the upper limit of $%
\varepsilon $ at which the excess conductivity could be analyzed was $%
\varepsilon _{{\rm up}}\approx \ln (160/92)=0.55$ for YBCO, $\varepsilon _{%
{\rm up}}\approx 0.46$ for 2212 BSCCO and $\varepsilon _{{\rm up}}\approx
0.51$ for 2223 BSCCO.

In Fig. 18 $(16\hbar s/e^2)\ \sigma _{{\rm fl}}$ is plotted
for the three samples as a function of $\varepsilon $; the solid line
represents $1/\varepsilon $, the dashed line $1/\varepsilon ^3$ and the
dotted line $3.2/\sqrt{\varepsilon }$. The value of the interlayer distance $%
s$ is adjusted so that the experimental data follow the $1/\varepsilon $
behavior in the temperature region where the AL behavior is expected.
Obviously, the extension of the region where the 2D AL behavior ($%
1/\varepsilon $) is followed depends on the sample anisotropy. The less
anisotropic YBCO compound asymptotically tends to the 3D behavior ($%
1/\varepsilon ^{1/2}$) for $\varepsilon <0.1$, showing the LD crossover at $%
\varepsilon \approx 0.07$; the 2223 BSCCO curve starts to bend for $%
\varepsilon <0.03$ while the most anisotropic 2212 BSCCO shows a 2D behavior
in the whole temperature range investigated. All three compounds show a
universal high temperature behavior of in-plane conductivity in the 2D
regime, above the LD crossover. At $\varepsilon \approx 0.24$ all the curves
bend down and follow the same asymptotic $1/\varepsilon ^3$ behavior.
Finally at the value $\varepsilon \approx 0.45$, close to the values of $%
\varepsilon _{{\rm up}}$ reported above, all the curves fall down indicating
the end of the observable fluctuation regime.

A further refinement of the theory of the in-plane fluctuation conductivity
was carried out considering two different interlayer distances and different
strengths of the tunneling coupling between adjacent layers for YBCO \cite
{PDR93,RPV96} and BSCCO 2212 \cite{PRM96}. This approach led to the same
qualitative results as those obtained by the conventional single layer
approach, but to a better quantitative agreement between the theory and
experiments.

In conclusion, the analysis of in-plane fluctuation conductivity has shown
that in the whole temperature range from $T_c$ to temperatures high enough
that the fluctuation contribution becomes lower than the experimental
resolution, the AL theory, corrected to take into account short wavelength
fluctuations, is able to explain all the experimentally observed features in
HTS, correctly describing the crossovers between different regimes as the
temperature is increased. Therefore in most cases there is no need to
introduce other contributions to fluctuation conductivity (DOS, regular and
anomalous MT), since they are small with respect to the AL one as predicted
by (\ref{d30_1}). However very accurate measurements have shown that for
YBCO, a logarithmic contribution to $\sigma _{ab}^{fl}$ exists at
temperatures as high as $180~K$ which could originate from in-plane DOS
fluctuations \cite{AGL91}.

\subsection{Out-of-plane resistance: fluctuation origin of the c-axis peak}

The quantitative agreement of the transverse fluctuation theory outlined in
section 6.3 with the experimental data was proved shortly after its proposal 
\cite{BMMVY93,BMV93, V93} by fitting the resistivity peaks of BSCCO and YBCO
samples. This shows good metallic behavior far from the transition, and thus
had relatively small resistivity peak (Fig. 19). For strongly oxygen
deficient samples, the increase in the c-axis resistivity begins so far from 
$T_c$ and the peak has such a large magnitude \cite{V93} that it cannot be
due to fluctuation effects only: in this case the effect is probably due to
some metal-insulator transition.

In \cite{BMMVY93} the theory was fit to data from 2212 BSCCO\ films grown on
misaligned substrates (to allow measurement of c-axis resistivity on
epitaxial films) in the temperature region $93-110$ $K$. The fit used only
the DOS contribution since the temperatures are far enough from $T_c$ for
the AL contribution to be negligible. In this way, using a single fitting
parameter representing the amplitude of the DOS fluctuation correction to
conductivity, the agreement of the logarithmic increase of resistivity with
experimental data was proven. This analysis was later completed \cite{BMV93}
by enlarging the fitting region to include the peak and considering the AL
contribution besides the DOS one. Their different temperature dependencies
allow them to be separated and therefore to extract the values of the
physical parameters involved. Two fitting parameters were necessary in this
case, the Fermi velocity $v_F=1.4\cdot 10^7$ $cm/s$ and the electronic
elastic scattering time $\tau \simeq 5\cdot 10^{-14}$ $s$. In the same paper
the carrier concentration and the anisotropy of a BSCCO film were changed by
means of reducing and oxidizing annealing treatments. As the AL contribution
is heavily dependent on the interlayer coupling than the DOS one, a more
pronounced peak is expected for materials with higher anisotropy. The
carrier density also affects the magnitude of the peak, since a higher
carrier concentration means a lower fluctuation contribution as compared to
the normal-state conductivity. These facts strongly reduce the relative
change in conductivity for samples having high oxygen content. The evolution
of the resistivity peak under redox treatments confirmed these predictions.

We have just seen that the main features of the resistivity peak in HTS can
be explained by the fluctuation theory including the DOS contribution.
However, this is not the only possible approach. Indeed, both before and
after this theory was proposed and experimentally checked, many attempts
were carried out to describe such a peak by means of non metallic normal
state conduction mechanisms. Although early attempts in this direction
failed to give a satisfactory description of the steep increase of
resistivity just above $T_c$ \cite{AND, CLZ89, KLS90, CZ91, KJ92, LV92},
recent approaches involving an ''activated'' behavior of the normal state
resistivity seem more promising. It is therefore important to discuss these
mechanisms in order to see whether the less traditional description in terms
of fluctuations is really necessary. We will here analyze the resonant
tunneling model proposed by Abrikosov \cite{ABR} and a phenomenological
model proposed by Yan et al. \cite{YMH95}.

According to the model of Abrikosov the ratio between c-axis ad ab-plane
resistivity is 
\begin{equation}
\frac{\rho _c(T)}{\rho _{ab}(T)}=A\frac{\cosh \left( T_0/T\right) +\cosh
(T_1/T)}{\sinh \left( T_1/T\right) }  \label{AB1}
\end{equation}
where $T_0$ and $T_1$ are respectively the mean energy and the half width of
the energy spread of the resonant defects relative to Fermi level. The
agreement of this formula with experimental data for both YBCO \cite{ABR}
and BSCCO \cite{BMV96} is very good in the whole temperature range between
the temperature $T_m$ at which the ratio of the resistivities shows a
maximum, and room temperature. Even when the oxygen concentration in the
samples is varied by annealing over a fairly wide range the quality of the
fit remains good. However this large-scale agreement of the theory with
experiment is not conclusive: really, the important temperature region for
the comparison of normal-state and fluctuation theories of the resistivity
peak is very close to $T_c$. It is here, indeed, that the two approaches are
fundamentally different, since the fluctuation theory predicts a weak
(logarithmic) divergence of $\rho _c(T)$ at $T_c$, while {\it any }normal
state theory can at most provide a divergence at $T=0$ $K$. If the $\rho
_c(T)$ peak is fluctuation induced, therefore, a normal state theory should
not be able to reproduce the divergence close to $T_c$, whilst we know that
our fluctuation theory can. The comparison of the two approaches must
therefore be carried out by comparing with experimental data the calculated
''trends'' of $\rho _c(T)$ close to $T_c,$ rather than the absolute values
of $\rho _c(T)$.

A closer look at the data on to BSCCO \cite{BMV96} shows indeed that the fit
with (\ref{AB1}) tends to slightly overestimate the experimental data at
both high and low temperatures, while underestimating it at intermediate
temperatures (except for the oxygen annealed sample, which has a weak peak).
Apparently, the fit is forced to increase as much as possible the curvature
of the theoretical function to try to match the steep increase in $\rho
_c(T) $ just above $T_c$. It is likely therefore that a good description of
the region just above $T_c$ (where fluctuation theory predicts a singular
behavior) cannot be given by this model. To better understand this point in 
\cite{BMV96} the curvatures of the experimental data and of the fitting
functions above $T_m$ were compared. The result was that in spite of the
apparently good fit of $\rho _c(T)$ in this temperature region, the
curvature was strongly underestimated (up to a factor of about 3) by the
fitting function for the as grown and argon annealed samples, even if the
experimental curvature is partly depressed by the AL fluctuation
contribution. Moreover, this underestimation disappears (within experimental
resolution) for the sample having the highest oxygen content, which shows
only a very weak peak (Fig. 20).

This seems to be a clear indication that the experimental data show a
divergent trend at $T_c$ which cannot be reproduced by any normal state
theory. Such a divergence is predicted by the DOS contribution and is
counterbalanced by the AL fluctuation contribution very near (a few $K$) to $%
T_c$. The gradual reduction in the curvature as the oxygen content is
increased is consistent with the higher AL fluctuation contribution expected
in less anisotropic samples. For the oxygen annealed sample the divergence
is so weak that, within the experimental error, it is indistinguishable from
a steep, non divergent behavior and the theory of Abrikosov works well even
close to $T_c$.

It has been pointed out \cite{BMV96} that when $T_0,T_1>>T$ as is the case
for the fits on both BSCCO\ and YBCO, (\ref{AB1}) reduces to an activated
behavior:

\begin{equation}
\rho _c(T)/\rho _{ab}(T)=A\left[ 1+\exp \left( \Delta /T\right) \right]
\label{AB2}
\end{equation}
Here $\Delta $ = $T_0-T_1$ is the energy of the lowest resonant impurity
level relative to the Fermi level. Only the difference between $T_0$ and $%
T_1 $ is therefore important and not their separate values, to which the fit
is insensitive when $T_0,T_1>>T$. Eq. ( \ref{AB2}) is very similar to the
semi-phenomenological formula for $\rho _c$ proposed in \cite{YMH95}: 
\begin{equation}
\rho _c(T)=A+BT+(C/T)\exp (\Delta /T)  \label{ONG}
\end{equation}
(where $\Delta $ is some kind of pseudogap). Therefore, in spite of the
additional $1/T$ dependence and the presence of four phenomenological
parameters which slightly reduce the discrepancy with experimental data, ( 
\ref{ONG}) faces the same difficulties as (\ref{AB1}) in describing the
curvature of $\rho _c(T)$ just above $T_c$ in 2212 BSCCO.

The above discussion leads to the conclusion that in models based on an
''activated'' behavior for $\rho _c(T)$ the strong divergence of the
exponential at $T=0$ $K$ simulates the weaker divergence of the experimental
data at $T_c$, except in the temperature region very close to $T_c$. It is
then to be expected that these models will give a curvature which is too
high when applied to the low $T_c$ compound of the BSCCO family (2201 BSCCO)
for which $T_c$ is only about 15 $K$. Applying (\ref{AB1}) to data \cite
{MFF88} on 2201 BSCCO crystals indeed gives a very bad fit, as shown in \cite
{BMV96}. Moreover, while Eqn.(\ref{AB1}) predicts a lower curvature at $T_c$
than that measured in 2212 BSSCO, the opposite is true for 2201 BSSCO.

In conclusion, when $T_c\simeq 100$ $K$, the curvature of the exponential,
which diverges at $T=0$ $K$, is not high enough to account for the apparent
divergence of the experimental $\rho _c$ just above $T_c$. On the other
hand, when $T_c$ is closer to $0$ $K$, the exponential divergence becomes
much higher than the experimental one.

The analysis carried out in the previous paragraph suggests that it will be
very difficult for any normal state theory, with a divergence of any kind at 
$T=0$ $K$, to reconcile both situations. The same is not of course true of
course if the peak in $\rho _c$ has a superconducting origin such as the DOS
fluctuation contribution, whose divergence shifts with $T_c$, while
remaining logarithmic in shape in all cases. Although the
semi-phenomenological formula for $\rho _c$ proposed in \cite{YMH95}, i.e. ( 
\ref{ONG}) correctly describes the $\rho _c(T)$ curves of the low $T_c$ 2201
BSCCO phase, this is only true for $\Delta =0$ $K$, when the exponential
divergence is canceled and substituted by the $1/T$ one. In this case,
however, the formula loses its significance.

If the DOS fluctuation correction to conductivity is responsible for the $%
\rho _c(T)$ peak just above $T_c$, it is clear that the goal of a normal
state theory of electrical conductivity along the c axis in layered
superconductors is not to give a detailed description of this peak. The
normal state resistivity $\rho _{Nc}(T)$ curve must instead lie somewhere
below the measured $\rho _c(T)$ one. To find out what a reasonable behavior
of $\rho _{Nc}(T)$ would be, compatible with the fluctuation origin of the
transverse $\rho _c(T)$ peak, in \cite{BMV96} the calculated contribution
due to fluctuations was subtracted from the experimental $\rho _c(T)$, using
(\ref{d30}) with values for the parameters taken from literature \cite
{BMV93, NBMLV96, BMV95, LLM95, L95} ($\tau =2\cdot 10^{-14}\ s$, $\tau _\phi
=2\cdot 10^{-13}\ s$, and $J=40\ K$), while the critical temperature $T_c$
is taken from experimental data. Three simulated $\rho _{Nc}(T)$ curves were
calculated for three different values of the Fermi energy $E_F$ (i.e. $0.8\
eV$, $1.0\ eV$ and $1.25\ eV$). The Fermi energy is just a scale factor for
the global fluctuation contribution to conductivity $\sigma _{fl}$, and in
order to keep the latter within the limits of validity of the theory
underlined in section 6 ($\sigma _{fl}\ll \sigma _N$) it must be assumed to
be of the order of $1\ eV$, somewhat higher than expected in these
materials. These curves are plotted in Fig. 21 together with
the experimental $\rho _c(T)$ for $T<150\ K$ (for $T>100\ K$ the theory is
however no longer very accurate since the limit $\varepsilon \ll 1$ is not
fulfilled). It is interesting that because of the less divergent behavior of
the simulated $\rho _{Nc}(T)$ curve as compared to $\rho _c(T)$, simpler
functional dependences for $\rho _{Nc}(T)$ could be compatible with it. Some
of the theories for normal state transverse conductivity which failed to
describe the c-axis resistivity peak could be in this context reconsidered. 

We conclude that using the fluctuation theory to describe the
transverse resistivity peak in HTS in zero external magnetic field is well
justified. A further check of the theory must  be sought by adding
another parameter, besides temperature, on which the resistivity depends.
This can be done by applying an external magnetic field, as described in the
next section.

\newpage

\section{ The effects of magnetic field}

\subsection{Introduction}

The behavior of the resistivity peak under an external c-axis oriented
magnetic field \cite{BCZ91} is certainly one of the intriguing anomalies of
HTS. As the field intensity is increased, the position of the peak in $\rho
_c(T)$ is shifted towards lower temperatures. However above the peak
temperature $T_m(B)$, the resistivity curve $\rho _c(T)$ retains the
temperature dependence shown in zero field above the peak temperature $%
T_m(B=0)$. As a result, the magnitude of the peak in $\rho _c(T)$ increases,
and a very strong positive magnetoresistance is observed below $T_m(B=0)$,
as shown in Fig. 22.

The c-axis magnetoresistance shows an even more characteristic behavior above $T_{c0}$. In
contrast to the ab-plane magnetoresistance which is positive at all
temperatures, along the c-axis the magnetoresistance has been
found to have negative sign not too close to $T_{c0}$ in many HTS compounds
- i.e. BSSCO \cite{YMH95, NTHKT94, HASH, HLW96}, LSSCO \cite{KMT96}, YBCO 
\cite{AHE} and TlBCCO \cite{WAHL98} - and turn to positive values at lower
temperatures. We will show how these behaviors find their explanation within the
fluctuation theory in the magnetic-field induced suppression of the AL
contribution along the c-axis.

In this section we will mainly discuss the relevance of fluctuations in the
explanation of c-axis magnetoresistance data in HTS. The in-plane
magnetoresistance is less interesting in the framework of this review, since
it has already been shown in section 7.2 that in zero magnetic field the DOS
contribution to in-plane transport properties is small, at least not too far
from $T_c$. Many authors (see e.g. Refs. \cite
{SH91,SIKYH92,HARKM93,LLM95,LHKS95,HRJH95,L95,WM96,L95B,VRQ97,SLP95}) have
therefore successfully explained in-plane magnetoresistance data in HTS
(including BSCCO) using the AL and MT contributions only \cite
{HL88,AHL89,BM90,BMT91}.

After a review of the theoretical predictions in section 8.2, the
experimental study of magnetoresistance above $T_{c0}$
 and the role of the DOS fluctuation contribution in it will be
discussed in section 8.3. In section 8.3.1 a short survey of in-plane
magnetoresistance results is given, then the
peculiarities of the c-axis magnetoresistance are analyzed in section 8.3.2.
Section 8.4 will be focused on magnetoresistance effects below $T_{c0}$. 
A summary and discussion of the informations given by both zero-field and
in-field electrical transport measurements presented in sections 7 and 8
will conclude this section.

\subsection{Theory of {\bf c}-axis conductivity in magnetic field}

\subsubsection{General expressions}

The full theoretical treatment of the effect of magnetic field on the
fluctuation conductivity of layered superconductors above $T_c$ has been
given in \cite{BDKLV93}. The AL, DOS, regular and anomalous Maki-Thompson
fluctuation contributions to the c-axis conductivity are considered there in
detail. We recall here the qualitative aspects of the problem and present
the results necessary to understand the following sections.

The effect of a magnetic field parallel to the $c$-axis is considered. For
this particular field direction the current vertices do not depend upon the
magnetic field, and both the quasiparticles and the pairs form Landau orbits
within the layers. The $c$-axis dispersion remains unchanged from the
zero-field form. For this simple field direction, it is elementary to
generalize the zero-field results reported in section 6.3 to finite field
strengths. One replaces 
\begin{eqnarray}
\eta {\bf q}^2\rightarrow \eta (\overrightarrow{\nabla }/i-2e{\bf A})^2
\label{landau}
\end{eqnarray}
in each of the integral expressions for the contributions to the fluctuation
conductivity. The two-dimensional integration over ${\bf q}$ is replaced by
a summation over the Landau levels (indexed by $n$), taking account of the
Landau degeneracy factor in the usual way \cite{BDKLV93}, 
\begin{eqnarray}
\int {\frac{{d^2{\bf q}}}{{(2\pi )^2}}}\rightarrow {\frac B{\Phi _0}}\sum_n={%
\frac \beta {{4\pi \eta }}}\sum_n,  \label{intsum}
\end{eqnarray}
where $n=0,1,2,\ldots $. So the general expressions for all fluctuation
corrections to c-axis conductivity in magnetic field can be simply written
in the form: 
\begin{eqnarray}
\sigma _{zz}^{AL} &=&\frac{e^2sr^2\beta }{128\eta }\sum_{n=0}^\infty \frac
1{[(\varepsilon _B+\beta n)(\varepsilon _B+\beta n+r)]^{3/2}}  \label{AL} \\
\sigma _{zz}^{DOS} &=&-\frac{e^2sr\kappa \beta }{16\eta }\sum_{n=0}^{1/\beta
}\frac 1{[(\varepsilon _B+\beta n)(\varepsilon _B+\beta n+r)]^{1/2}}
\label{DSzz} \\
\sigma _{zz}^{MT(reg)} &=&-\frac{e^2s\tilde{\kappa}\beta }{8\eta }%
\sum_{n=0}^\infty \left( \frac{\varepsilon _B+\beta n+r/2}{[(\varepsilon
_B+\beta n)(\varepsilon _B+\beta n+r)]^{1/2}}-1\right)  \label{MTrg} \\
\sigma _{zz}^{MT(an)} &=&\frac{e^2s\beta }{16\eta (\varepsilon -\gamma
_\varphi )}\sum_{n=0}^\infty \left( \frac{\gamma _B+\beta n+r/2}{[(\gamma
_B+\beta n)(\gamma _B+\beta n+r)]^{1/2}}\right.  \nonumber \\
&-&\left. \frac{\varepsilon _B+\beta n+r/2}{[(\varepsilon _B+\beta
n)(\varepsilon _B+\beta n+r)]^{1/2}}\right)  \label{sumcor}
\end{eqnarray}
where $\beta =4\eta eB = B/[2T_c\,|dH_{c2}/dT|_{Tc}]$,
$\varepsilon _B=\varepsilon +\beta /2$,
and $\gamma_B=\gamma _\varphi +\beta/2$.

For the in-plane component of the fluctuation conductivity tensor the only
problem appears in the AL diagram, where the matrix elements of the harmonic
oscillator type, originating from the $B_{\parallel }$ $(q_{\parallel })$
blocks, have to be calculated. The other contributions are essentially analogous
to their c-axis counterparts:

\begin{eqnarray}
\sigma _{xx}^{AL} &=&{\frac{{e^2}}{{4s}}}\sum_{n=0}^\infty (n+1)\biggl( {%
\frac 1{{[(\epsilon _B+\beta n)(\epsilon _B+\beta n+r)]^{1/2}}}}-  \label{l1}
\\
&&{\frac 2{{\{[\epsilon _B+\beta (n+1/2)][\epsilon _B+\beta
(n+1/2)+r]\}^{1/2}}}}+  \nonumber \\
&&{\frac 1{{\{[\epsilon _B+\beta (n+1)][\epsilon _B+\beta (n+1)+r]\}^{1/2}}}}%
\biggr),  \nonumber
\end{eqnarray}

\begin{eqnarray}
\sigma _{xx}^{DOS}+\sigma _{xx}^{MT(reg)}=-{\frac{{e^2\beta (\kappa +\tilde{%
\kappa})}}{{4s}}}\sum_{n=0}^{1/\beta }{\frac 1{{[(\epsilon _B+\beta
n)(\epsilon _B+\beta n+r)]^{1/2}}}},  \label{l2}
\end{eqnarray}

and 
\begin{eqnarray}
\sigma _{xx}^{MT(an)}={\frac{{e^2\beta }}{{8s(\epsilon -\gamma_\varphi )}}}%
\sum_{n=0}^\infty \biggl ({\frac 1{{[(\gamma _B+\beta n)(\gamma _B+\beta
n+r)]^{1/2}}}}-\\
\nonumber
{\frac 1{{[(\epsilon _B+\beta n)(\epsilon _B+\beta n+r)]^{1/2}%
}}}\biggr).  \label{l3}
\end{eqnarray}

Note that we have included the DOS and regular MT terms together, as they
are proportional to each other for this conductivity tensor element, and the
DOS diagram differs from that for the transverse conductivity by the factor $%
(v_F/sJ)^2$, which measures the square of the ratio of the effective Fermi
velocities. The same factor enters into the normal state conductivity,
leading to the standard $\sigma _{xx}^N={\frac 12}N(0)e^2v_F^2\tau =E_F\tau
e^2/(2\pi s)$. We note that (\ref{l1}) was given previously in \cite{K74},
using standard procedures \cite{MS70}, and rederived in \cite{MT89}; the
formula (\ref{l3}) was also given previously \cite{MT89,HL88}.

These results can in principle be already used for numerical evaluations
and fitting of the experimental data.
Resistivity curves calculated by (\ref{AL}) (\ref{sumcor}) using reasonable
parameter values are shown in Fig. 23. It can be seen that the
simulated behavior is similar to the experimental one reported in Fig. 22.

Nevertheless, it is useful to manipulate formula (\ref{sumcor}) and (\ref{l3})
algebraically, in order to remove the apparent (but spurious) singularity at 
$\epsilon =\gamma_\varphi $. The low field expansions of all results (\ref{AL}) - (%
\ref{l3}) can be calculated in a straightforward way. The analysis of the ``strong'' field regime $\varepsilon <<\beta <<1$ turns out
much more sophisticated. In this case, the sums over $n$ for
the AL and MT terms converge rapidly, and it is enough to keep only the $n=0$
term in the sums. For the DOS contribution however the formal logarithmic
divergence of the sum requires a little bit more careful treatment. We will deal with
these expansions in the next sections.

\subsubsection{Weak magnetic field}

In the weak field regime ($\beta <<\varepsilon $), we expand the various
conductivity contributions in powers of $\beta $ \cite{BDKLV93}. Such
expansions are simplified by using the Euler-Maclaurin approximation
formula, 
\begin{eqnarray}
\sum_{n=0}^Nf(n)=\int_0^Nf(x)dx+{\frac 12}[f(N)+f(0)]+{\frac 1{{12}}}%
[f^{^{\prime }}(N)-f^{^{\prime }}(0)]+\ldots .  \label{euler}
\end{eqnarray}
If one writes the expressions in terms of $\varepsilon _B$, terms linear in $%
B$ will appear. However, writing the expressions in term of the zero-field $%
\varepsilon $, all terms linear in $B$ vanish identically, leaving leading
terms of order $B^2$. To order $B^2(\beta ^2)$, we find 
\begin{eqnarray}
\sigma _{zz}^{AL} &=&{\frac{{e^2s}}{{32\eta }}}\biggl[\Bigl({\frac{{\
\varepsilon +r/2}}{{[\varepsilon (\varepsilon +r)]^{1/2}}}}-1\Bigr)-{\frac{{%
\ \beta ^2r^2(\varepsilon +r/2)}}{{32[\varepsilon (\varepsilon +r)]^{5/2}}}}%
\biggr ], \\
\sigma _{zz}^{DOS} &=&-{\frac{{e^2sr\kappa }}{{16\eta }}}\biggl[\ln \Bigl({\
\frac 2{{\varepsilon ^{1/2}+(\varepsilon +r)^{1/2}}}}\Bigr)^2-{\frac{{\beta
^2(\varepsilon +r/2)}}{{24[\varepsilon (\varepsilon +r)]^{3/2}}}}\biggr], \\
\sigma _{zz}^{MT({\rm reg})} &=&-{\frac{{e^2sr\tilde{\kappa}}}{{16\eta }}}%
\biggl[\Bigl({\frac{{(\varepsilon +r)^{1/2}-\varepsilon ^{1/2}}}{{r^{1/2}}}}%
\Bigr)^2-{\frac{{\beta ^2r}}{{48[\varepsilon (\varepsilon +r)]^{3/2}}}}%
\biggr], \\
\sigma _{zz}^{MT({\rm an})} &=&{\frac{{e^2s}}{{16\eta }}}\biggl[\Bigl({\frac{%
{\varepsilon +\gamma_\varphi }_{{\varphi }}{+r}}{{[\varepsilon (\varepsilon
+r)]^{1/2}+[\gamma_\varphi (\gamma_\varphi +r)]^{1/2}}}}_{}-1\Bigr)  \label{XY} \\
&-&{\frac{{\beta ^2r^2(\varepsilon +\gamma_\varphi }_\varphi {+r)\bigl[\varepsilon
(\varepsilon +r)+\gamma }_{{\varphi }}{(\gamma }_{{\varphi }}{%
+r)+[\varepsilon (\varepsilon +r)\gamma }_{{\varphi }}(\gamma _\varphi +r){%
]^{1/2}\bigr]}}{{96[\varepsilon (\varepsilon +r)\gamma }_{{\varphi }}{%
(\gamma }_{{\varphi }}{+r)]^{3/2}\bigl( [\varepsilon (\varepsilon
+r)]^{1/2}+[\gamma }_{{\varphi }}{(\gamma }_{{\varphi }}{+r)]^{1/2}\bigr)}}}%
_{}\biggr].  \nonumber
\end{eqnarray}
In (\ref{XY}), we have assumed $\varepsilon <<1$ and $\beta <<1$. Typically, 
$1~T$ corresponds to $\beta \simeq 10^{-2}$ so that the low field expansion
is often realized in practice.

Using (\ref{XY}), one can find the position of the resistive maximum: 
\begin{eqnarray}
\varepsilon _m/r\approx {\frac 1{{[8r\kappa ]^{1/2}}}}(1-{\frac{{5\beta
^2\kappa }}{{3r}}})-{\frac{{\tilde{\kappa}}}{{8\kappa }}}+{\frac 1{{\
16\gamma }_{{\varphi }}{\kappa }}}.  \label{maxiB}
\end{eqnarray}
Note that the weak magnetic field reduces $T_m$ by an amount proportional to 
$B^2$.

We now present the low-field expansions for the contributions to the
fluctuation conductivity parallel to the layers. Using the same
Euler-Maclaurin approximation formula, we obtain 
\begin{eqnarray}
\sigma _{xx}^{AL}={\frac{{e^2}}{{16s}}}\biggl[{\frac 1{{[\epsilon (\epsilon
+r)]^{1/2}}}}-{\frac{{\beta ^2[8\epsilon (\epsilon +r)+3r^2]}}{{32[\epsilon
(\epsilon +r)]^{5/2}}}}\biggr],  \label{A11}
\end{eqnarray}

\begin{eqnarray}
\sigma _{xx}^{DOS}+\sigma _{xx}^{MT({\rm reg})} &=&-{\frac{{e^2(\kappa +%
\tilde{\kappa})}}{{4s}}}\biggl[2\ln \biggl({\frac 2{{\epsilon
^{1/2}+(\epsilon +r)^{1/2}}}}\biggr)  \label{A12} \\
&&-{\frac{{\beta ^2(\epsilon +r/2)}}{{24[\epsilon (\epsilon +r)]^{3/2}}}}%
\biggr],  \nonumber
\end{eqnarray}
and 
\begin{eqnarray}
\sigma _{xx}^{MT({\rm an})} &=&{\frac{{e^2}}{{8s(\epsilon -\gamma )}}}%
\biggl[ 2\ln \biggl({\frac{{\epsilon ^{1/2}+(\epsilon +r)^{1/2}}}{\gamma {%
^{1/2}+(\gamma +r)^{1/2}}}}\biggr) -{\frac{{\beta ^2}}{{24}}}\biggl({\frac{{%
\gamma +r/2}}{{[\gamma (\gamma +r)]^{3/2}}}}  \nonumber \\
&&-{\frac{{\epsilon +r/2}}{{[\epsilon (\epsilon +r)]^{3/2}}}}\biggr)\biggr].
\label{A13}
\end{eqnarray}
The zero-field term in $\sigma _{xx}^{AL}$ was first given by Lawrence and
Doniach \cite{LD70}, and the term of order $\beta ^2$ was first obtained
explicitly in \cite{HL88} by inverting the order of the summation over $n$
and the integration over $q_z$.

\subsubsection{The AL and anomalous MT contributions in intermediate and
strong fields}

We start the discussion of the non-weak magnetic field ($\beta
>>\varepsilon $) from the simplest case of the well converging AL
contribution. Performing the summation in (\ref{sumcor})-(\ref{l3}) one
finds 
\begin{equation}
\sigma _{zz}^{AL}(\beta >>\max \{\varepsilon ,r\})={\frac{7\zeta (3){e^2s}}{{%
128\eta }}\cdot }\frac{r^2}{\beta ^2}
\end{equation}
in the case of strong field and

\begin{equation}
\sigma _{zz}^{AL}(\epsilon \ \ll \beta \ll r)={\frac{{e^2s}}{{128\eta }}}%
\sqrt{\frac r\beta }\sum_{n=0}^\infty {\frac 1{{[(n+1/2)]^{3/2}}}}={\frac{%
4.57{e^2s}}{{128\eta }}}\sqrt{\frac r\beta }
\end{equation}
for the intermediate regime which can be realized in the $3D$ case.

An analogous treatment for the in-plane components results in:

\begin{equation}
\sigma _{xx}^{AL}(\beta >>\max \{\varepsilon ,r\})={\frac{{e^2}}{{4s\beta }}}
\end{equation}
and

\begin{equation}
\sigma _{xx}^{AL}(\epsilon \ll \beta \ll r)\approx {\frac{{e^2}}{{2s}}}\frac
1{\sqrt{\beta r}}.
\end{equation}

The anomalous MT contribution can be analyzed in the same way, but the
situation is a little bit more cumbersome. Here is necessary to distinguish
the cases of strong ($\beta >>\max \{\varepsilon ,r,\gamma _\varphi \}$) and
several intermediate field regimes ($\epsilon ,\gamma _\varphi \ll \beta \ll
r;\epsilon \ll \beta \ll \gamma _\varphi ,r;$) which can be realized in the $3D$
situation. The first limit can be studied in the same way as was done
above: the expansion of (\ref{sumcor}) over $\beta ^{-1}$permits to evaluate
the sum and results in the high field asymptotic: 
\begin{equation}
\sigma _{zz}^{MT({\rm an})}(\epsilon ,\gamma _\varphi ,r\ll \beta )=\frac{%
3\pi ^2e^2s}{128\eta }\cdot {\frac{{(r+\epsilon +\gamma _\varphi )}}{{\beta }%
}},
\end{equation}

In the intermediate cases one finds:

\begin{equation}
\sigma _{zz}^{MT({\rm an})}(\epsilon ,\gamma _\varphi ,\ll \beta \ll r)=%
\frac{4.57e^2s}{64\eta }\sqrt{\frac r\beta },
\end{equation}

\begin{equation}
\sigma _{zz}^{MT({\rm an})}(\epsilon \ll \gamma _\varphi \ll \beta \ll r)=%
\frac{e^2s}{32\eta }\sqrt{\frac r{\gamma _\varphi }}
\end{equation}

and

\begin{equation}
\sigma _{zz}^{MT({\rm an})}(\epsilon \ll \beta \ll \gamma _\varphi \ll
r)\sim \frac{e^2s}\eta
\end{equation}

The evaluation of the in-plane component gives:

\begin{equation}
\sigma _{xx}^{MT({\rm an})}(\epsilon ,\gamma _\varphi ,r\ll \beta )=\frac{%
3\pi ^2e^2}{16s}\frac 1\beta ,
\end{equation}

\begin{equation}
\sigma _{xx}^{MT({\rm an})}(\epsilon ,\gamma _\varphi ,\ll \beta \ll r)={%
\frac{4.57{e^2}}{16{s}}}\frac 1{\sqrt{\beta r}}
\end{equation}
and

\begin{equation}
\sigma _{xx}^{MT({\rm an})}(\epsilon \ll \beta \ll \gamma _\varphi ,r)=\frac{%
e^2}{8s}\cdot \frac 1{\gamma _\varphi }\cdot \ln \frac{\sqrt{\max \{\gamma
_\varphi ,r\}}}{\sqrt{\beta }+\sqrt{\beta +r}}.
\end{equation}

\subsubsection{Renormalization of the DOS contribution divergency in
intermediate and strong fields}

As mentioned in \cite{AHE}, the fit of experimental data with the theory
based on the fluctuation renormalization of the one-electron density of
states \cite{ILVY93,BDKLV93} is excellent for weak magnetic fields but meets
noticeable difficulties in the region of strong fields. This is due to
the formal divergence of the DOS contribution to conductivity and to the
dependence of the cut-off parameter on the magnetic field itself \cite
{BDKLV93}. In this section we clarify the problem of the regularization of
the DOS contribution in an arbitrary magnetic field

To avoid the problem of the ultraviolet divergence of the DOS contribution
with the badly defined cut-off depending on the magnetic field \cite{BDKLV93}
we calculate the cut-off independent difference \cite
{Buz96}: 
\begin{eqnarray}
\Delta \sigma _{zz}^{DOS}({\beta },\epsilon )=\sigma _{zz}^{DOS}({\beta }%
,\epsilon )-\sigma _{zz}^{DOS}(0,\epsilon ).  \label{delta}
\end{eqnarray}

For this purpose the zero-field value $\sigma _{zz}^{DOS}(0,\epsilon )$ \cite
{BDKLV93} may be rewritten in the form:

\begin{eqnarray}
  &\sigma _{zz}^{DOS}&({0},\epsilon )=-\lim_{\beta\rightarrow 0}\frac{%
e^2sr\kappa }{16\eta }\beta \int_{-1/2}^{1/\beta +1/2}\frac{dn}{\sqrt{%
\epsilon +\beta (n+1/2)}\sqrt{\epsilon +r+\beta (n+1/2)}}=  \nonumber \\
&=&-\lim_{\beta\rightarrow 0}\frac{e^2sr\kappa }{16\eta }\beta 
\sum_{n=0}^{1/\beta } \int_{-1/2}^{1/2}\frac{dx}{\sqrt{\epsilon +\beta
(n+x+1/2)}\sqrt{\epsilon +r+\beta (n+x+1/2)}}  \nonumber \\
&=&\lim_{\beta\rightarrow 0}\frac{e^2sr\kappa }{16\eta } \sum_{n=0}^{1/{\beta}}
\ln {\frac{\sqrt{\epsilon +\beta n+\beta }+\sqrt{\epsilon +r+\beta n+\beta }%
}{\sqrt{\epsilon +\beta n}+\sqrt{\epsilon +r+\beta n}},}  \label{main1}
\end{eqnarray}

Substituting expression (\ref{main1}) in (\ref{delta}), for the
difference $\sigma _{zz}^{DOS}({\beta },\epsilon )-\sigma
_{zz}^{DOS}(0,\epsilon )$ we may write the following formula, where the
summation may be extended up to $N\longrightarrow \infty $ \ since the sum
is convergent (the n-th term of the sum is proportional to $n^{-3/2}$ for
large $n$): 
\begin{eqnarray}
\Delta \sigma _{zz}^{DOS}({\beta },\epsilon ) &=&\sigma _{zz}^{DOS}({\beta }%
,\epsilon )-\sigma _{zz}^{DOS}(0,\epsilon )=  \nonumber \\
&=&\frac{e^2sr\kappa }{16\eta }\beta {\bf \sum_{n=0}^\infty }\{\frac 2\beta
\ln {\frac{\sqrt{\epsilon +\beta n+\beta }+\sqrt{\epsilon +r+\beta n+\beta }%
}{\sqrt{\epsilon +\beta n}+\sqrt{\epsilon +r+\beta n}}}-  \nonumber \\
&&-\frac 1{\sqrt{\epsilon +\beta n+\beta /2}\sqrt{\epsilon +r+\beta n+\beta
/2}}\}  \label{main}
\end{eqnarray}

This expression is very suitable for numerical calculation to analyze
experimental data. It permits to obtain easily the asymptotic behavior of
magnetoconductivity in the case of non-weak fields (note the inaccuracy of
the analysis of this asymptotic in \cite{BDKLV93}). The case of very
strong fields $h\gg \max \{\epsilon ,r\}$, in contrast to \cite{BDKLV93},
becomes now trivial: it is mainly determined just by the
logarithmically large contribution of the $n=0$ term in (\ref{main})( the
contribution of $n\geq 1$ terms is the second in the parentheses below ) :

\begin{eqnarray}
\Delta \sigma _{zz}^{DOS}(h &\gg &\max \{\epsilon ,r\})=\frac{e^2sr\kappa }{%
8\eta }\{\ln {\frac{2\sqrt{\beta }}{ e (\sqrt{\epsilon }+\sqrt{\epsilon +r})}}%
.+0,02\}\approx  \nonumber \\
&\approx &\frac{e^2sr\kappa }{8\eta }\cdot \ln {\frac{\sqrt{\beta}}{\sqrt{%
\epsilon }+\sqrt{\epsilon +r})},}  \label{4*}
\end{eqnarray}
where in the logarithm  $e  = 2,71...$

Further analysis of (\ref{main}) shows that for the intermediate fields
in the temperature range of three dimensional fluctuations ($\epsilon \ll
\beta \ll r$) the DOS contribution shows the magnetic field dependence $\sim 
\sqrt{\frac \beta r}$ in contrast to $\sqrt{\frac r\beta }$ in the above discussed
AL and MT contributions (nevertheless the magnitude of the former
remains smaller that the latters up to $\beta \sim r$):
\begin{eqnarray}
\Delta \sigma _{zz}^{DOS}(\epsilon &\ll &\beta \ll r)=\frac{e^2sr\kappa }{%
16\eta }\sum_{n=0}^\infty \{\ln {\frac{1+\sqrt{\frac \beta {2r}}\sqrt{n+1}}{%
1+\sqrt{\frac \beta {2r}}\sqrt{n}}}-  \label{inter} \\
-\sqrt{\frac \beta {2r}}\frac 1{\sqrt{2n+1}}\} &=&0.428\frac{e^2sr\kappa }{%
16\eta }\sqrt{\frac \beta {2r}}.  \nonumber
\end{eqnarray}

In addition to the DOS contribution it is necessary to take into account the
regular Maki-Thompson one, which in the case of in-plane component, as we
know, has the same sign and functional dependence as the overall $\Delta \sigma
_{xx}^{DOS}$and differs only by the impurity concentration dependent factor $%
\tilde{\kappa}$ (instead of $\kappa$).

In weak fields $\Delta \sigma _{zz}^{MT(reg)}(\beta \ll r,\epsilon )$
becomes comparable with (\ref{main1}) only in 3D case ($\epsilon \ll r$) and
for the dirty or intermediate case ($T\tau \preceq 1$) when $\tilde{\kappa}$
is of the order of $\kappa $. In the dirty limit ($T\tau \ll 1$) $\kappa =2%
\tilde{\kappa}$ and $\Delta \sigma _{zz}^{MT(reg)}(\beta \ll r,\epsilon )\ $%
reaches a half of $\Delta \sigma _{zz}^{DOS}(\beta {\ll r},\epsilon ).$

The evaluation of the regular Maki-Thompson contribution to
magnetoconductivity may be done by the same procedure as in (\ref{main1})
for the analysis of non-weak fields :

\begin{eqnarray}
\Delta \sigma _{zz}^{MT(reg)}(\beta ,\epsilon ) &=&-\frac{e^2s\tilde{\kappa}%
}{8\eta }\beta \sum_{n=0}^\infty \{\frac{\epsilon +\beta (n+1/2)+r/2}{\sqrt{%
\epsilon +\beta (n+1/2)}\sqrt{\epsilon +r+\beta (n+1/2)}}-  \nonumber \\
&&-\frac 1\beta [\sqrt{\epsilon +\beta (n+1)}\sqrt{\epsilon +r+\beta (n+1)}-
\label{mtr} \\
&&\sqrt{\epsilon +\beta n\ }\sqrt{\epsilon +r+\beta n}]\}  \nonumber
\end{eqnarray}
For the $3D$ case in the region of intermediate fields it leads to 
\begin{eqnarray}
\Delta \sigma _{zz}^{MT(reg)}(\epsilon \ll h\ll r) &=&\frac{e^2sr\tilde{%
\kappa}}{8\eta }\sqrt{\frac \beta {2r}}\sum_{n=0}^\infty \{\sqrt{2n+2}-\sqrt{%
2n}-  \label{intmak} \\
-\frac 1{\sqrt{2n+1}}\} &=&0.428\frac{e^2sr\tilde{\kappa}}{8\eta }\sqrt{%
\frac \beta {2r}}.  \nonumber
\end{eqnarray}
One can see that in this region too the contribution $\Delta \sigma
_{zz}^{MT(reg)}(\epsilon \ll \beta \ll r)$ has the same sign and differs
from $\Delta \sigma _{zz}^{DOS}(\epsilon \ll \beta \ll r)$ (see (\ref{inter}%
)) only by the substitution of $\kappa $ by $2\tilde{\kappa}$ .

This means that for the $3D$ situation in the case of strong or intermediate ($%
T\tau \preceq 1$) impurity concentration $\Delta \sigma
_{zz}^{MT(reg)}(\beta )$ has to be taken into account side by side with $%
\Delta \sigma _c^{DOS}(\beta )$ for all fields up to $\beta \sim \max
\{\epsilon ,r\}=r.$ In the case of strong scattering ($T\tau \ll 1$) both
contributions coincide, while as the impurities
concentration increases the role of the regular part of the MT contribution diminishes.

The analysis of (\ref{mtr}) in the case of strong field( $\beta \gg \max
\{\epsilon ,r\}$ ) leads to

\begin{eqnarray}
\Delta &\sigma _{zz}^{MT(reg)}&(\beta \gg \max \{\epsilon ,r\})= \\
&=&\frac{e^2sr\tilde{\kappa}}{16\eta }(\frac{\sqrt{\epsilon +r}-\sqrt{%
\epsilon }}{\sqrt{r}})^2-\frac{\pi ^2e^2s\tilde{\kappa}}{128\eta }\cdot 
\frac{r^2}\beta .  \label{regn}
\end{eqnarray}
The comparison of (\ref{regn}) with (\ref{4*}) demonstrates, that in the $3D$
dirty case it can at least contribute as a constant of the order of $1$ in
comparison with the large field dependent logarithm of (\ref{4*}). In the $2D$ case its
contribution is negligible at all both in the absolute value and the
magnetic field dependence.

\subsubsection{Discussion}

The results obtained for the c-axis magnetoconductivity are collected in Table 1.

  Let us start the analysis from the $2D$ case ($r\ll \epsilon $) (to
visualize them is enough to skip the third column in Table 1). One can
see that the positive DOS contribution in the magnetoconductivity turns out
to be dominant. It growth as $B^2$ up to $B_{c2}(\epsilon )$ and then the
crossover to a slow logarithmic asymptote takes placed. At
$B\sim B_{c2}(0)$ the value of $\Delta \sigma _{zz}^{DOS}(\beta \sim
1,\epsilon )=-\sigma _{zz}^{DOS}(0,\epsilon )$ which means the total
suppression of the fluctuation correction in such a strong field. The
regular part of the Maki-Thompson contribution does not manifest itself in
this case while the AL term can compete with the DOS one in the immediate
vicinity of $T_c, $where the additional small anisotropy factor $r$ can be
compensated by the additional $\epsilon ^3$ in denominator. The MT contribution
can contribute in the case of small pairbreaking only, which is an opposite
to the expecting one in HTS.

In the $3D$ case ($\epsilon \ll r$) the behavior of the magnetoconductivity is
more complex. In weak and intermediate fields the main, negative,
contribution to magnetoconductivity occurs from the AL and the MT terms. At $%
B\sim B_{c2}(\epsilon )(\beta \sim \epsilon )$ the paraconductivity is
already considerably suppressed by the magnetic field and the $\beta ^2-$ dependence
of the magnetoconductivity changes through the $\sqrt{\frac r\beta }$ tendency
to the high field asymptote $-\sigma _{zz}^{(fl)}(0,\epsilon ).$ In this
intermediate region of fields ($\epsilon \ll \beta \ll r),$ side by side
with the decrease ($\sim \sqrt{\frac r\beta }$) of the main AL and MT
contributions, the growth of the still relatively small DOS
term takes place. At the upper border of this region $(\beta \sim r)$ its
positive contribution is of the same order as the AL one and at high fields $%
(r\ll \beta \ll 1)$ the positive DOS contribution determines the slow
logarithmic decay of the fluctuation correction to conductivity which is
completely suppressed only at $B\sim B_{c2}(0).$ The regular part of the
Maki- Thompson contribution is not of special importance in $3D$ case. It
remains comparable with the DOS contribution in the dirty case at fields $\beta
\preceq r$ , but decreases rapidly $(\sim \frac r\beta )$ at strong fields ( 
$\beta \succeq r),$ in the only region where the robust $\Delta
\sigma _c^{DOS}(\beta ,\epsilon )\sim \ln \frac \beta r$
shows up surviving up to $\beta \sim 1$.

The analogous formulae for the in-pane magnetoconductivity are
presented in Table 2.

Analyzing this table one can see that almost in all regions the negative AL
and MT contributions determine the behaviour of in-plane
magnetoconductivity. Nevertheless, similarly to the c-axis case, the high
field behavior is again determined by the positive logarithmic $\Delta
(\sigma _{xx}^{DOS}+\sigma _{xx}^{MT(reg)})$ contribution, which the only
one to survive in strong field.

It is important to stress that the suppression of the DOS contribution by
magnetic field takes place very slowly. Such robustness with respect to the
magnetic field is of the same physical origin as the slow logarithmic
dependence of the DOS-type corrections on temperature. This difference
between the DOS and the Aslamazov-Larkin and Maki-Thompson contributions \cite{AM78}%
 makes the former noticeable in a wide range of temperatures (up to $%
\sim 2-3T_c$) and magnetic fields ($\sim B_{c2}(0)$). The temperature scale of the
suppression of DOS contribution is given by the value of the experimentally
observed pseudogap. It is of the order of $\Delta
_{pseudo}\sim 2-3T_c$ for magnetoconductivity and NMR, $\Delta _{pseudo}\sim
\pi T_c$ for tunneling and $\Delta _{pseudo}\sim \tau ^{-1}$ for optical
conductivity.

\subsection{Magnetoresistance above $T_c$}

\subsubsection{In-plane magnetoresistance}

Soon after the discovery of HTS many investigations of the in-plane
paraconductivity have been performed. It turned out that the early results
obtained on bulk ceramic samples could be well reproduced in highly oriented
thin films and single crystals \cite{APC92}. A major drawback in the
analysis of superconducting fluctuations from the paraconductivity is the
need for separating fluctuation and normal-state conductivity contributions
in the analysis. Usually, a linear temperature dependence of the resistivity
in the normal state is postulated, but such assumption is lacking an
undisputed theoretical background. To overcome the above-mentioned problems
with the unknown transport properties of the normal-state in HTS, an
analysis of the in-plane fluctuation magnetoconductivity above $T_c$ has
been proposed \cite{HL88,MHK88}. Early studies of the in-plane fluctuation
magnetoconductivity on YBCO ceramics, thin films, and single crystals \cite
{MHK88, MHKT89, HS89, WK91, OH92, SLP95, SIM91, SIKYH92, HARKM93, RPV96,
GRBG94, SM97, HER95}, in oxygen-deficient YBCO with $T_c=55{\rm \ K}$ \cite
{LGKS95}, 2212-BSCCO \cite{FBWM92, DBSY94, GRBG94, LLM95, HLW96, WM96},
2223-BSCCO \cite{LHKS95, L95}, ${\rm Tl_2Ba_2CaCu_2O_x}$ \cite{KGKM91}, and $%
{\rm La_{2-x}Sr_xCuO_4}$ \cite{KMT96} were interpreted within the then
available phenomenological dirty-limit \cite{HL88, MT89, AHL89, T91} and
clean-limit theories \cite{BM90} and reanalyzed including non-local
corrections \cite{BMT91}.

Within the phenomenological theoretical context, four different
contributions to $\Delta \sigma_{xx}(B)$ were predicted. As discussed
previously, the orbital AL and anomalous MT contributions result from the
suppression of the paraconductivity by orbital interaction with a magnetic
field. Due to the anisotropic nature of the cuprate superconductors, those
effects are prominent with $B \parallel c$ (transverse orientation), but are
severely suppressed with $B \parallel j \perp c$ (longitudinal orientation).
In addition, an isotropic Zeeman effect on the spins of the fluctuating
pairs has been proposed \cite{AHL89}, which could dominate the
magnetoconductivity of the anisotropic HTS with the magnetic field oriented
parallel to the planes and result in two additional contributions,
the AL-Zeeman and the anomalous MT-Zeeman terms.

The problem of the amplitude of the anomalous MT contribution in HTS, which
could not be satisfactory resolved with paraconductivity analysis, was
re-addressed by several authors with the analysis of the in-plane
magnetoconductivity. Most authors observed that $\Delta \sigma_{xx}(B)$ with 
$B \parallel c$ is dominated by the orbital AL process for $\epsilon < 0.1$,
but in optimally-doped YBCO and 2212-BSCCO an additional contribution
appears at higher temperatures which was at first associated with a
considerable weight of the anomalous MT contribution. An analysis within
both the dirty and clean-limit phenomenological models revealed several
inconsistencies regarding the resulting value for $\tau_\phi$ \cite{BM90,
BMT91, SLP95, SM97}. This additional magnetoconductivity is suppressed in
Zn-doped YBCO \cite{AHE,SM97}, partially Pr-substituted YBCO \cite{MS91}
and in the more impure compounds 2223-BSCCO \cite{LHKS95, L95} and
oxygen-depleted YBCO \cite{LGKS95}. The results obtained by various authors
from the analysis of the in-plane magnetoconductivity of YBCO vary for $%
\xi_{ab}(0)$ from 11 to 18 \AA, for $\xi_c(0)$ from 1 to 4.6 \AA \ and, in
those cases, where the anomalous MT contribution was detected, for $%
\tau_\phi $ from ${\rm 2.2 \times 10^{-16} \ s}$ to ${\rm 5.7 \times
10^{-13} \ s}$. For YBCO in the 60-K phase the respective values are: $%
\xi_{ab}(0) = {\rm 25 \ \AA}$, $\xi_c(0) = {\rm 0.9 \ \AA}$. In 2212-BSCCO
several authors found $\xi_{ab}(0) = {\rm 10 \ to \ 38 \ \AA}$, $\xi_c(0) = 
{\rm 0.1 \ to \ 2.3 \ \AA}$, and $\tau_\phi = {\rm 0.13 \ to \ 7.5 \times
10^{-14} \ s}$. Only few results are available for other compounds, like
2223-BSCCO, with $\xi_{ab}(0) = {\rm 16 \ \AA}$, $\xi_c(0) = {\rm 1.4 \ \AA}$%
, ${\rm Tl_2Ba_2CaCu_2O_x}$ with $\xi_{ab}(0) = {\rm 11.8 \ \AA}$, $\xi_c(0)
= {\rm 0.12 \ \AA}$, $\tau_\phi = 2.9 \times 10^{-14} \ s$, and ${\rm %
La_{1.85}Sr_{0.15}CuO_4}$ with $\xi_{ab}(0) = {\rm 28 \ \AA}$ and $\xi_c(0)
= {\rm 1.4 \ \AA}$. In general the values obtained from the in-plane
magnetoconductivity analysis agree well with those from other experimental
methods, but the scatter in the results for $\tau_\phi$ is quite large.

On the other hand, measurements of the in-plane magnetoresistance in YBCO
single crystals near and above room temperature revealed an unconventional $%
\Delta \rho / \rho_0 \equiv (\rho_{xx} (B) - \rho_{xx} (0)) / \rho_{xx} (0)
\sim T^{-4}$ temperature dependence, which the authors attributed to the
cyclotronic motion of normal-state quasiparticles. They proposed that in a
fourfold-symmetric, two-dimensional metal the orbital magnetoresistance and
the Hall effect measure the variance and the mean of the local Hall angle
along the Fermi surface, respectively. Hence, the temperature dependence of
the normal-state magnetoresistance is $\Delta \rho / \rho_0 = A \ \tan ^2
\theta_{{\rm H}}$, where $\tan \theta_{{\rm H}}$ is the Hall angle in the
material and $A=1.7$ for YBCO \cite{HYMO95}. However, superconducting
fluctuations were entirely neglected in this analysis. Using these results
and magnetoresistance data from YBCO thin films it was shown that neither
the fluctuation picture alone nor the normal-state magnetoresistance can
account for the data from about $T_c$ to above room temperature \cite
{LHLWW97} and it was proposed that the quasiparticle magnetoresistance
resembles the anomalous MT contribution \cite{SLP95, L95B}. In Fig. 24
it can be noticed that in fact the combination of the four fluctuation
contributions, the orbital AL and anomalous MT, and the respective Zeeman effects
(ALO+ALZ+MTO+MTZ) fails to fit the data above $2 T_c$. It should be noted
that using the full theory outlined in the previous chapters (including the
DOS and regular MT expressions) instead of the phenomenological model would
not resolve this discrepancy. As an alternate scenario it was proposed that
the magnetoconductivity close to $T_c$ is dominated by the orbital AL term,
with a crossover to the normal-state magnetoconductivity $\Delta \sigma_{xx}
\cong -\Delta \rho / \rho^2$ at $\epsilon > 0.2$ (see Fig. 24). In
this case the AL Zeeman and the DOS contributions, which are expected to gain
importance relative to the orbital AL term at higher temperatures are masked
by the normal-state magnetoresistance. Similar results were recently
obtained on YBCO single crystals \cite{SM97}. In this context the large
scatter of the results for $\tau_\phi$ and the apparent absence of the
anomalous MT contribution in several materials rather reflect different shapes of
the Fermi surface, resulting in a variation of the parameter $A$, i.e. a
different ratio between the normal-state magnetoresistance and the Hall
angle.

As can be seen from the previous paragraph, the DOS contribution is unlikely
to have significance for the in-plane magnetoresistance of HTS as far as $B
\parallel c$ is concerned. In the longitudinal orientation however, the
orbital terms are suppressed due to the anisotropy of the cuprates and the
normal-state magnetoresistance due to the absence of the Lorentz force on
the quasiparticles, respectively, and, thus, the Zeeman terms become
dominant. The results available for this geometry \cite{MHKT89, SIM91,
HRJH95, SM97, LGKS95, HLW96, WM96, KGKM91} are not entirely conclusive,
probably due to the required very accurate orientation of the magnetic field
parallel to the ${\rm CuO_2}$ planes. Some authors report a good accordance
with the AL Zeeman term only. The anomalous MT Zeeman effect can be supposed to be
negligible if the corresponding orbital anomalous MT contribution is not strong, a
fact which is now well agreed for the cuprates. As mentioned in section 8.3.2,
a DOS Zeeman process is needed to explain the longitudinal out-of-plane
magnetoresistance data above $T_c$. Accordingly, the DOS Zeeman term can be
expected to induce a sign change from negative to positive
magnetoconductivity at temperatures considerably above $T_c$. Apart from a
single report \cite{WM96} this has not been observed so far. Certainly the
application of the DOS effect to the in-plane magnetoresistance needs more
attention in future.

Finally, two other effects commonly associated with the in-plane
magnetoresistance are worth mentioning. It was shown that the sample shape,
in particular the usually almost square geometry of single crystals, has
considerable influence on the in-plane magnetoresistance at higher
temperatures due to bending of the carrier trajectories \cite{GLS98}. Thus,
evaluations of the magnetoresistance in this temperature region have to be
regarded with some caution. On the other hand, a large increase of the
in-plane magnetoconductivity above theoretical predictions close to $T_c$,
which was observed by several authors, can be naturally explained by a
non-uniform $T_c$ in the samples \cite{L94}. It was shown that measurements
of the magnetoresistance in fact can serve as a very sensitive probe for $%
T_c $ inhomogeneities in HTS \cite{L94B}.

\subsubsection{Out-of-plane magnetoresistance}

We will now address the problem of the origin of the negative out of plane
magnetoresistance above $T_{c0}$, discussed in the Introduction. The
negative $c-$axis magnetoresistance of $BSCCO$ single crystals was first
observed in \cite{NTHKT94} and then carefully measured in \cite{YMH95} (see
Fig. 25) at temperatures above $95\ K$, where it was qualitatively
interpreted in terms of the holon and spinon model by Anderson \cite{AND}.
However, this analysis is not quantitative and is based on the
phenomenological assumption of an ''activated'' behavior for $\rho _c(T)$,
which in the previous section has been shown to be unsatisfactory. 

On the other hand, from the data reported in \cite{YMH95} it can be seen
that the effect becomes significant below approximately $140\ K$ and its
magnitude increases dramatically as the temperature goes down to $95\ K$.
This temperature range is the same as that observed for the fluctuation
induced {\it positive} ab-plane magnetoresistance, which naturally leads 
\cite{BMV95} to the hypothesis that fluctuations are also responsible for
the negative c-axis magnetoresistance. Since the DOS fluctuation
contribution is held responsible for the peak in $\rho _c(T)$, its
contribution to magnetoresistance cannot be neglected and may determine its
sign. This contribution is indeed expected to be {\it negative} in sign,
since a suppression of the DOS contribution by the magnetic field would give
a decrease of the resistivity. In the temperature region where the DOS
contribution dominates over the AL one, a negative fluctuation induced
c-axis magnetoresistance is therefore conceivable. All the features of the
observed magnetoresistance are therefore consistent with its attribution to
fluctuations, with a key role played by the DOS contribution. Starting from (%
\ref{XY}) the following expression for the fluctuation c-axis
magnetoresistivity close to $T_c$ in the presence of weak magnetic fields
(this assumption is fulfilled in the experiment reported in \cite{YMH95})
has been found \cite{BMV95}: 
\begin{eqnarray}
\frac{\rho _c(B,T)-\rho _c(0,T)}{\rho _c(0,T)}=2.92\cdot 10^{16}\rho
_c(B,T)f(T)B^2,  \label{mr}
\end{eqnarray}
The temperature dependence is mainly included in the factor $f(T)$: 
\begin{eqnarray}
f(T) &=&\frac{s\eta r^2}{[\varepsilon (\varepsilon +r)]^{3/2}}\left\{ \frac{%
3(\varepsilon +r/2)}{\varepsilon (\varepsilon +r)}-8k\left[ \frac
\varepsilon r+\frac 12\left( 1+\frac{\tilde{k}}k\right) \right] +\right. 
\nonumber  \label{9} \\
&& \\
&&+\left. \frac{2\left( \varepsilon +\gamma _\varphi +r\right) \left\{
\varepsilon (\varepsilon +r)+\gamma _\varphi (\gamma _\varphi +r)+\left[
\varepsilon (\varepsilon +r)\gamma _\varphi (\gamma _\varphi +r)\right]
^{1/2}\right\} }{\left[ \gamma _\varphi (\gamma _\varphi +r)\right]
^{3/2}\left[ (\varepsilon (\varepsilon +r))^{1/2}+(\gamma _\varphi (\gamma
_\varphi +r))^{1/2}\right] }\right\}  \nonumber
\end{eqnarray}

Here cgs units are used except for the magnetic field $B$ (measured in
Tesla) and the resistivity $\rho _c(B,T)$ (measured in $\Omega \,cm$).

The first term in (\ref{9}) represents the AL contribution, the second is
the sum of DOS and regular MT contributions and the third is the anomalous
MT one. The fit of (\ref{mr}) with the experimental data was performed in 
\cite{BMV95} using as adjustable parameters $v_F,\tau $ and the phase
pair-breaking lifetime $\tau _\phi $. The values of the interlayer spacing $%
s\approx 10^{-7}cm$ and of the hopping integral $J\approx 40\ K$ were taken
from literature data \cite{BMV93}, since they are not likely to vary
strongly from sample to sample (at least for BSCCO samples with metallic
behavior far from $T_c$), while $\rho _c(B,T)$ and $T_c\cong 85\ K$ have
been extracted from the experimental curves.

The results of the fit performed using (\ref{mr}) for the magnetoresistance
curves are shown in (Fig. 26). The curves measured at $%
T=95$ $K$ and $T=100$ $K$ were fitted simultaneously (i.e. using the same
values of the fitting parameters for both curves) to put more constraints on
the fitting procedure. Those measured at $T=105$ $K$ and higher temperatures
were not considered in the fit because they lie outside the temperature
region $\varepsilon \ll 1$ in which the theory is quantitatively accurate
(at $105$ $K,\varepsilon =0.21$). However, the theoretical curve at $105$ $K$
has been drawn in Fig. 26 using the values of $v_F$, $%
\tau $ and $\tau _\phi $ found for the curves at $95$ $K$ and $100$ $K$ in
order to show that, even at higher temperatures, the calculated temperature
dependence of the transverse magnetoresistance is in substantial agreement
with the experimental data. 

The values of the fitting parameters were $v_F=3.1\times 10^6\ cm/s$, ${\tau 
}=1.0\cdot 10^{-14}\ s$ and ${\tau _\phi }=8.7\cdot 10^{-14}\ s$.

While the field dependence of the magnetoconductivity is simply $B^2$, at
least for moderate fields, its behavior with temperature, given by (\ref{9}), is much more interesting. On the basis of (\ref{9}) and using the above
values for the fitting parameters, the expected temperature dependence of
the transverse magnetoresistance has been calculated in \cite{BMV95} (see
Fig. 27). The existence of an inversion temperature $T_r$ at
which a reversal of the sign of magnetoconductivity occurs is predicted. The
physical origin of this change of sign is the same as for the appearance of
the peak discussed in Section 6.1 : relatively far from $T_c$ the AL
negative magnetoconductivity is suppressed by its dependence on the square
of the transparency and the positive DOS contribution dominate, while very
close to $T_c$ the very singular temperature dependence of the negative AL
contribution ($\sim \varepsilon ^{-4}$) makes it prevail over the less
singular DOS one ($\sim \varepsilon ^{-2}$), despite the latter's linear
dependence upon transparency.

Last years the problem of the out of plane magnetoresistivity has been studied
experimentally very carefully \cite{HASH,HLW96,LHLWW97,AHE}. Hashimoto et
al. \cite{HASH} measured the c-axis magnetoresistance of BSCCO single
crystals in fields up to $30\ T$ and found results similar to \cite{YMH95},
except that at very high fields/low temperatures the field behavior is no
longer concave, as can be expected from \cite{BDKLV93}). They fitted the
experimental data to the original theory of Ioffe et al. \cite{ILVY93}
(which is valid only in the clean case) but including magnetic field and
renormalization effects on the mass term. The agreement between theory and
experiment is good for fields up to about $10\ T\ $, although at higher
fields the measured magnetoresistance is higher than expected. Fitting
parameters were the ratio $T_c/T_{c0}=0.84$ ($T_{c0}$ being the mean field
critical temperature), the in-plane coherence length $\xi _{\Vert }(0)=16\ 
\hat{A}$ and the effective mass $m_{ab}=3.6$, $\tau $ having been fixed to $%
1.2\cdot 10^{-13}\ s$ to match the clean limit assumption.

In \cite{HLW96} the negative c-axis magnetoresistance of BSCCO single crystals was
again observed at temperatures higher than $100\ K$. The fit of experimental
data was performed using AL and DOS terms only, with parameters $T_c=93.5\ K$
, ${\tau }=1.5\cdot 10^{-14}\ s$, $\xi _{\Vert }(0)=14\ \hat{A}$ and $\xi
_{\bot }(0)=0.12\ \hat{A}$. The unphysically low value of $\xi _{\bot }(0)$
was attributed to the possible overestimation of the intrinsic resistivity
due to the influence of microcracks perpendicular to the crystal c-axis. The
fit was anyway very good, except for a slight overestimation of the effect
at the lowest temperature used $(T=100\ K)$. In a later paper \cite{LHLWW97}
the authors extended their analysis to include both longitudinal and
transverse in-plane magnetoresistance data measured on the same sample in
the fit. With the addition of an AL Zeeman term (necessary to describe
in-plane longitudinal magnetoresistance, in which the other terms are
suppressed), the experimental data were described very well for $B=12\ T$ in
an extremely wide temperature range (up to about $2T_c$) using as parameters 
$v_F=2.2\times 10^7\ cm/s$, ${\tau }=1\cdot 10^{-14}\ s$ and ${J} = 4\ K$.

In \cite{AHE} the transverse magnetoresistance was measured down to
temperatures close to $T_c$ in two YBCO single crystals in different
oxygenation states and with different twin densities. After corrections for
the inhomogeneous current distribution the measured magnetoconductivity for $%
B=12\ T$ was fitted with (\ref{sumcor}), employing a weighted cutoff for the
sum in the DOS contribution to smooth the field dependence of the calculated
values. Its temperature dependence was found to follow the
behavior shown in Fig. 28 and the existence of a sign reversal
temperature was unambiguously confirmed at about $10\ K$ above $T_c$.
Assuming $v_F=2\cdot 10^7\ cm/s$ and the temperature dependence $\tau =\tau
_\phi \sim 1/T$ the fitting parameters were $\tau (100\ K)=\tau_\phi (100\
K)=(4\pm 1)\cdot 10^{-15}\ s$, $J=(215\pm 10)\ K$.

We point out that although YBCO generally shows a weak transverse
resistivity peak because of its relatively small anisotropy, the different
temperature dependences of the AL and DOS terms allow for a negative total
magnetoresistance. This occurs at higher temperatures than for BSCCO, making
the observation of the change of sign easier.

In \cite{AHE} the sign reversal of magnetoresistance in HTS was proved, but
the smallness of the negative magnetoresistance did not allow for a precise
quantitative check of the predictions of the fluctuation theory. Very
recently, similar measurements have been performed on single crystals of the
very anisotropic Tl 2223 compound \cite{WAHL98}. A strong negative
magnetoresistance is observed a few degrees above $T_c$, initially
increasing as temperature is decreased and then turning to positive as $T_c$
is approached. In this extremely anisotropic compound this effect is very
pronounced, so that the relative experimental errors on the data are very
small. It turns out that the fit of the data with the fluctuation theory
including the DOS contribution is excellent, the theoretical behavior
being perfectly reproduced with quite reasonable values of
the fitting parameters.

There is therefore now increasing evidence that the negative c-axis
magnetoresistance observed in many HTS is really connected with the DOS
fluctuation contribution. The competition among the DOS and AL fluctuation
contributions leads to the negative c-axis magnetoresistance observed above $%
T_c$ and to its change of sign in the vicinity of $T_c$.

\subsection{Fluctuation magnetoresistance below $T_c$}

In this section we analyze the role of the DOS fluctuation contribution in
the behavior of $\rho _c(H,T)$ below the zero field critical temperature $%
T_{c0}$, and the associated increase of the c-axis resistance peak with
magnetic field. This effect was first measured by Briceno et al. \cite{BCZ91}
and then analyzed by several authors.

The main features observed are:

1) a well pronounced maximum at zero field in the $\rho _c(T)$ behavior;

2) a very strong shift of the zero resistance temperature in external
magnetic field ($\Delta T_c\sim 30\ K$ already for $B\sim 1\ T$)

3) a large broadening of the peak for nonzero magnetic fields with the
appearance of a long resistivity tail for higher fields and a large
associated positive magnetoresistance.

Kim and Gray \cite{KG93} explained the broadening of the peak
with increasing magnetic field in terms of Josephson coupling, describing a
layered superconductor as a stack of Josephson junctions. The growth of the
resistance above the transition was attributed to the DOS fluctuation
contribution. The results obtained by the Kim and Gray model are
interesting, and its agreement with experimental data is good, as seen in
Fig. 29, but their approach introduces another element in the
discussion beyond fluctuations and lacks a full microscopic foundation. The
field dependence of $\rho _c$ is described assuming that phase slips in
layered superconductors are identical to phase slips in a single Josephson
junction having area $\Phi _0/B$. It has been remarked \cite{Cho94} that
this model is inadequate to describe the field dependence of the activation
energy extracted from resistivity curves in the low temperature region.

In \cite{NBMLV96} the first attempt to describe the experimental $\rho
_c(T,H)$ curves for perpendicular magnetic fields up to $8\ T$ using only
the fluctuation theory was undertaken. The first step was to fit the
zero-field data in the temperature range from $92\ K$ to $115\ K$ (which
corresponds to $0.02\ \leq \ \varepsilon \ \le \ 0.25$ ) with the
fluctuation theory using (\ref{sumcor}). The values obtained for the fitting
parameters were ${\tau }=(5.6\pm 0.6)\cdot 10^{-14}\ s$, ${\tau _\phi }%
=(8.6\pm 1.4)\cdot 10^{-13}\ s$, $E_F=(1.07\pm 0.12)\ eV$ and $J=(43\ \pm \
4)\ K$ ($T_{c0}=89.8\ K$ was taken as the midpoint of the transition).
Keeping these parameters fixed, the data measured in a magnetic field were
then fitted using the field parameter $\beta $ as a fitting parameter. It
was found that to reproduce the shape of the peak the critical temperature $%
T_c(B)$ also had to be used as a fitting parameter, because the mean-field $%
T_c(B)$ incorporated with (\ref{sumcor}) appeared to be unable to describe
the strong shift of the peak in magnetic field. With this correction the
theory satisfactorily describes the resistivity behavior for weak fields $%
B<1\ T$, while at higher fields the strong broadening of the peak cannot be
reproduced. It appears from Fig. 30 that (\ref{sumcor}) are able to
fit the experimental data only for temperatures above the temperature $%
T_m(B) $ corresponding to the maximum in $\rho _c$. For $T<T_m(B)$ the
experimental data decrease much more slowly than expected from the theory,
and this discrepancy increases with increasing field strength.

This lack of agreement is not surprising. The first reason is
the absence of critical fluctuations in the above treatment. As it was shown
in Refs. \cite{IOT89, UD91}, the extension of fluctuation theory beyond the
Gaussian approximation results in a shift of $T_c(B)$ to zero temperature
with a corresponding broadening of the transition, which increases strongly
with increasing magnetic field strength in qualitative agreement with
experimental data. Another reason for peak broadening is the presence of a
complex vortex structure which leads to an additional dissipation in the
mixed state. However, it can be noticed that the broadening of the superconducting
transition in HTS in the presence of an external magnetic field is
quantitatively very similar for in-plane and c-axis experiments. This in spite
of the quite pronounced qualitative difference, i.e. the presence of the
resistivity peak in c-axis measurements as opposed to the monotonic decrease
of in-plane resistivity with decreasing temperature. This suggests a common,
intrinsic origin for this broadening, which in view of the results reported
above and in the previous sections, could be attributed to fluctuations.

The possibility of describing the shape of the transverse resistivity
peak in presence of an external magnetic field within the fluctuation theory
using the Hartree approximation instead of the Gaussian one was analyzed in 
\cite{BLM97}. Here, (\ref{sumcor}) were modified within the self-consistent
Hartree approximation developed by Ullah and Dorsey \cite{UD91}, in which $%
\varepsilon _B$ is renormalized according to self-consistent equation \cite
{UD91}: 
\begin{equation}
\varepsilon _B=\tilde{\varepsilon}_B-\frac 14\ Gi^2\ t\ \beta
\sum_{n=0}^{1/\beta }\frac 1{[(\tilde{\varepsilon}_B+\beta n)(\tilde{
\varepsilon}_B+\beta (n+1)+r)]^{1/2}}  \label{self}
\end{equation}
where $Gi=2\pi T_c(0)/H_c(0)s\xi _{ab}^2(0)$ is the Ginzburg number ($H_c(0)$
is the zero-temperature thermodynamical critical field and $\xi _{ab}(0)$ is
the zero-temperature Ginzburg-Landau coherence length in the $ab$-plane).

In this paper the authors report simultaneous measurements of both the
in-plane and the out-of-plane resistivities of BSCCO\ films in magnetic
fields up to $1\ T$. Measuring both $\rho _{ab}(T,B)$ and $\rho _c(T,B)$ on
the same films allows them to use the same set of microscopical parameters
for all curves, thus putting more constraints on the fitting procedure. They
found that both $\rho _{ab}(T,B)$ and $\rho _c(T,B)$ are in remarkable
agreement with the fluctuation theory in a very wide range of temperatures
both below and above $T_c$. This gives a strong evidence in favor of the
fluctuation origin of the transverse resistance peak.

Figures 31(a) and 31(b) show the $\rho_{ab}(T)$ and $%
\rho_c(T)$ dependences respectively for several applied magnetic fields in
Ref.(\cite{BLM97}). The curves were fitted with the formula $\rho =1/(\sigma
_N+\sigma _{fl})$. Here $\sigma _{fl}=\sigma ^{AL}+\sigma ^{DOS}$ is the
fluctuation conductivity obtained replacing $\varepsilon _B$ with $\tilde{%
\varepsilon}_B$ in ( \ref{sumcor}) and neglecting the Maki-Thompson
contribution, while $\rho _N=1/\sigma _N$ is the normal-state resistivity
which was assumed to be linear and extrapolated from the temperature range
150-200 K for both components of resistivity. The zero-field curves $\rho
_{ab}(T,0)$ and $\rho _c(T,0)$ were fitted by the above equation with two
free parameters, $s$ and $r$. This procedure gave $s=(15\pm 2)\ \AA $ and $%
r=(5.0\pm 0.6)\cdot 10^{-3} $ for both directions. These parameters were
then kept fixed while the fit was performed for finite-field data using as
free parameters $\beta $ (field dependent) and $Gi $ (field independent).
This procedure provides a uniquely determined set of parameters for both
components of the resistivity tensor. It was found that $Gi=0.12\pm 0.01$
for all curves, in agreement with estimates from microscopic parameters,
while $\beta =B/[2T_c\,|dH_{c2}/dT|_{Tc}]$ (Eq.(35) in \cite{BDKLV93} must
be corrected) was correctly found to vary linearly with the applied field,
as expected, yielding a slope of the upper critical field of about $1\, T/K$
at $T_c$. The agreement between theory and experimental data is remarkable,
given the number of curves fitted the small number of free parameters used
and the considerable qualitative and quantitative differences among in-plane
and transverse resistivity behaviors. The fit is unexpectedly good even down
to temperatures at which the resistivity falls below about 10\% of normal
state value, i.e. a region where vortex dynamics is generally assumed to be
the main source of dissipation. 

Soon after similar conclusions were reached using a different approach \cite
{SNSFG97}. It was reported that the ab-plane resistivity of YBCO\ films and
crystals in external magnetic field could be well fitted down to low
resistance values using the fluctuation theory developed by Ikeda et al. 
\cite{IOT91} in the Hartree approximation. In the same region a scaling law
of the resistivity curves for different orientations of the magnetic field
is valid. This, and the comparison of the similarities and differences among
the behavior of films and crystals, leads to the conclusion that in a wide
region of the H-T plane the dissipation is due only to intrinsic properties
(fluctuations) and only at lower temperatures pinning-related dissipation
processes take place.

There is convincing evidence, therefore, that the fluctuation theory in the
Hartree approximation is able to describe very well the experimental $\rho
_{ab}(T,B)$ and $\rho _c(T,B)$ curves, using consistent values for the
fitting parameters for both current flow directions. There are clues that
highly anisotropic layered superconductors can be described as a normal,
strongly fluctuating phase rather than a superconducting mixed phase down to
very low temperatures, possibly leading to a reconsideration of the role of
vortex dynamics in these compounds.

To conclude this section, we want to mention again the paper by Nakao et al. 
\cite{NTHKT94} in which the resistance of a BSCCO single crystal was
measured by sweeping an external magnetic field up to $40\ T$ at a constant
temperature. Below $T_{c0}$ a peculiar behavior was observed, the resistance
initially increasing steeply with field, then levelling off and finally
decreasing in very high fields (Fig. 32(a)). This behavior (later
observed also in \cite{YMH95,CMS96}) can be ascribed to the combined effect
of the magnetic field induced destruction of the phase coherence between
layers and the suppression of the DOS. Indeed, curves simulated using the
fluctuation theory in \cite{NTHKT94} show a behavior very similar to the
observed one (Fig. 32(b)).

\subsection{Final remarks}

At the end of this survey of conductivity and magnetoconductivity
measurements explained in terms of the DOS fluctuation contribution, we want
to make some comments about the values of the parameters extracted from the
fits by the various authors.

The variety of experiments on c-axis electrical transport properties which
can be quantitatively explained by fluctuation theory alone convinces one of
the substantial correctness of this approach. It must also be stressed that
in this fluctuation theory all parameters have a well defined physical
meaning, no phenomenological parameter ever being used. The reasonable
values obtained for the physical parameters involved in the theory are an
important check of the consistency of the theory itself. However, the values
of the microscopic parameters found in the literature are not always fully
compatible.

There are many obvious reasons for a certain scatter of the values of
microscopic parameters found by comparing the results of several papers,
apart from the intrinsic uncertainties. It can be ascribed both to
differences among the samples used (single crystals vs. films, different
oxygenation states) and to differences in the methods used to compute $\rho
_c $ from voltage drop measurements (see e.g. \cite{BMV93, AHE}). Some
uncertainty in the values of the fitting parameters is due to the rather
arbitrary choice of additional parameters which are simply deduced from
experimental data or taken from literature data and, apart from
magnetoresistance measurements, the linearly extrapolated normal state resistance behavior. Also,
the weak temperature dependencies of $\tau $ and $\tau_\phi $ in the narrow
temperature range considered have been neglected by some authors but
estimated as $\sim 1/T$ by others, and formulae containing different
approximations have been used in different papers. Finally, other physical
effects can partially contribute to the observed phenomena.

In spite of this, we generally find a good overall agreement among the sets
of parameters. For BSCCO (there are not enough papers concerning YBCO), $%
\tau $ is the range $1\div 5\cdot 10^{-14}\ s$, while $\tau _\phi $ varies
over one order of magnitude from $8\cdot 10^{-14}\ s$ to $8\cdot 10^{-13}\ s$%
. However, $\tau _\phi $ only enters in the small Maki-Thompson term, so
that the uncertainty on this parameter is likely to be rather large. $E_F$
and $v_F$ have a similar role in the fits (they define only the scale for
the fluctuation conductivity) but cannot be directly compared. To provide
indicative reference values for $E_F$ we repeated the fits performed in Refs 
\cite{BMV93, BMV95} using the same formulae as in \cite{NBMLV96}. $E_F$ is
introduced in the formulae instead of $v_F$ through the expression for the
normal state conductivity, thus avoiding this source of error too. It turns
out that for \cite{BMV93} (resistivity data on films) ${\tau }=3.0\cdot
10^{-14}\ s$, ${\tau _\phi } =3.6\cdot 10^{-13}\ s$ and $E_F=1.07\ eV$,
while for the magnetoresistance data of single crystals analyzed in \cite
{BMV95} ${\tau }=0.9\cdot 10^{-14}\ s$, ${\tau _\phi }=7.8\cdot 10^{-14}\ s$
and $E_F=0.25\ eV$, values to be compared with those found in \cite{NBMLV96}%
. A big discrepancy is instead found among the values for $J$. However, it
must be noticed that the estimate $J=40\ K$ used by some authors was deduced
from in-plane measurements in overdoped samples \cite{BMV93, H97}, while the
fitting value $J=4\ K $ refers to c-axis measurements on samples optimally
doped for the highest $T_c$ and showing a very pronounced peak.

We have collected the parameters following from the analysis of the c-axis
magnetoresistance in Table 3 \footnote{$*$ means that the parameters where
recalculated by the authors of the review on the basis of the complete
theory presented here;
(x) means that this parameter was not used as a fitting one but was
taken by authors from literature;{x} means that this parameter has been
recalculated by the authors of the review from the original parameters
presented by the authors of the article}.

It can be seen that $\tau $ is almost always close to $2\cdot 10^{-14}\ s$, which
is a widely accepted value in the intermediate region between clean and
dirty limits, while the ratio $\tau _\phi /\tau $ is always about 10, in
good agreement with the expected one \cite{BDKLV93}. The $\tau _\phi $
values imply that the pair-breaking in $BSCCO$ has a moderate strength,
which is consistent with several papers (see section 8.2) where it was shown
that the AL term alone describes adequately the fluctuation contribution to
the in-plane conductivity in $BSCCO$. The regular MT contribution is
negligible in the clean and intermediate cases, while the relative
importance of the anomalous MT contribution depends upon the relationship
between $\gamma _\varphi $ and $r$. For the parameters found above $\gamma
_\varphi \approx 5\ r$ which still corresponds to moderate pair-breaking,
but closer to the weak limit. This is the reason why it usually suffices to
consider only the AL contribution. As for the Fermi energy value, which
being a scale factor is most influenced by the experimental conditions
(oxygenation state, methods for calculating resistivity, etc.), it lies
within a factor of 2 around $0.5\ eV$ , in good agreement with the estimate
given in \cite{BMV96} and not too far from that calculated using band
structure calculations and photoemission measurements.

Comparing the values of parameters extracted using (\ref{sumcor}), it must
also be noticed that these equations are used to describe both the
zero-field corrections leading to the existence of the resistivity peak
(i.e. a change in conductivity of the order of 20\% in typical BSCCO\ films)
and the small magnetoresistance corrections (i.e. an effect of the order of
0.1\%), thus spanning more than two orders of magnitude. The consistency of
the values for the fitting parameters, and especially those which only give
the scale of the corrections (e.g. $E_F$) is much better than that.

We can conclude that the number of unexpected experimental facts concerning
the electrical transport properties of HTS for which the fluctuation theory
provides a quantitative explanation without any other additional assumption
is impressive. Moreover, the theory uses no phenomenological parameter and,
apart from a reasonable scatter of the values of the physical parameters
involved, these values are certainly acceptable. No other theory at present
has achieved so many successes in this field.

\newpage

\section{The fluctuation induced pseudogap in HTS}

\subsection{Introduction}

The observations of pseudo-gap like phenomena in the normal state of HTS is
currently one of the most debated problems. Among these an important place
is occupied by measurements \cite{BTD94,BTD95} of the $c$-axis reflectivity
spectra, in the FIR region on $YBa_2Cu_4O_8$ single crystals. With the
decrease of temperature the $c$-axis optical conductivity decreases showing
around $180K$ a transition from a Drude-like to a pseudogap-like behavior.
The value of gap is $\omega _0\sim 180cm^{-1}$ and it seems weakly dependent
on temperature. The decrease of temperature does the gap structure more
pronounced without any abrupt change at the superconducting transition
temperature $T_c=80K$.

There are many hypotheses concerning the origin of this pseudo-gap. We
will show below that it can also be explained by the suppression of the
one-electron density of states on the Fermi level due to the interelectron
interaction. It will be demonstrated that the relatively low electromagnetic
wave frequencies suppress the main positive AL and MT terms, while leaving
almost unaffected the negative DOS contribution. This means that noticeable
changes will occur in the electromagnetic wave reflectivity at frequencies
well before the value $\omega \sim \tau ^{-1}$expected for a Drude term. The
positive AL and anomalous MT contributions show frequency dependence when $%
\omega \sim \min \{T-T_c,\tau _\varphi ^{-1}\}$\cite{AV80,S68}, whilst the
negative DOS contribution shows dependence on scale $\omega \sim \min
\{T,\tau ^{-1}\}.$ This competition results in the rapid decay of the
dissipative processes at frequencies of the order of $\omega \sim T-T_c$ and
in the appearance of a transparency window up to $\omega \sim \min \{T,\tau
^{-1}\}$. The high frequency behavior of ${\rm Re}\left[ \sigma (\omega
)\right] $ is mostly governed by $\sigma ^{{n}}(\omega )$ which decreases,
in agreement with the Drude law, for $\omega \geq \tau ^{-1} $.

Below we will study the {\bf a.c.} fluctuation conductivity tensor for a
layered superconductor taking into account all contributions and paying
attention to the most interesting case of $c$-axis polarization of the
field. Nevertheless, for completeness, the $ab$-plane results of the old
paper of Aslamazov and Varlamov \cite{AV80} will be re-examined and
discussed in application to the novel HTS layered systems.

\subsection{Paraconductivity}

The optical conductivity of a layered superconductor can be expressed by the
same analytical continuation of the current-current correlator
(electromagnetic response operator) $Q_{\alpha \beta }^{(R)}(\omega )$ as in
the section 6 but in contrast to d.c. conductivity case, calculated without
the assumption $\omega \rightarrow 0$: 
\begin{eqnarray}
{\rm Re}\left[ \sigma _{\alpha \beta }(\omega )\right] =-\frac{{\rm Im}%
\left[ Q_{\alpha \beta }^{(R)}(\omega )\right] }\omega  \label{Resig}
\end{eqnarray}
One can see that the calculations are analogous to those for the d.c.
fluctuation conductivity, discussed in section 6.4, but here we cannot treat
the external frequency as a small parameter. We will use the same model and
definitions introduced in section 6.

Let us start with the AL contribution (diagram 1 of Fig. 9) to
the {\bf a.c.} fluctuation conductivity. The general expression for the
appropriate contribution to the electromagnetic response operator as a
function of the Matsubara frequencies of the external electromagnetic field $%
\omega _\nu $ is defined by (\ref{d9}).

In the vicinity of $T_c$, for frequencies $\omega \ll T$, the leading
singular contribution to the response operator $Q_{\alpha \beta }^{AL\,(R)}$
arises from the fluctuation propagators in (\ref{d9}) rather than from the
frequency dependencies of the $B_\alpha $ blocks, so it suffices to neglect
its frequency dependencies \cite{AV80}, as was done for the d.c. case.
Carrying out the same calculations as in section 6.4 but without the limit $%
\omega \rightarrow 0$, one can find the explicit expression for the
imaginary part of the retarded electromagnetic response operator \footnote{%
It is necessary to mention that the direct calculation of the expression ( 
\ref{d9}) leads to the appearance of divergent expressions in the ${\rm Re}
Q $. Nevertheless, as it was shown in \cite{AV80}, the thorough summation of
all diagrams from the first order of the perturbation theory for $Q$ before
momentum integration leads to the exact cancellation of ${\rm Re}\left[ Q^{%
{\rm fl}}(0)\right]$. This fact justifies the possibility of the
further calculation of different diagrammatic terms to $Q^{{\rm fl}}$
separately.} for real frequencies $\omega \ll T$: 
\begin{eqnarray}
&&{\rm Im}\left[ Q_{\perp }^{AL(R)}(\omega )\right] =\frac{e^2T}{4\pi s}%
\left( \frac{s^2}\eta \right) \left( \frac{16T_c}{\pi \omega }\right) {\rm Re%
}\Bigg\{ \left( \frac{\pi \omega }{16T_c}\right) ^2-\left( \varepsilon -%
\frac{i\pi \omega }{16T_c}+\frac r2\right) \times  \nonumber  \label{QALperp}
\\
&& \\
&\times &\left[ \Delta D_2\left( \varepsilon -\frac{i\pi \omega }{16T_c}%
\right) -\left( \frac r2\right) ^2\Delta D_1\left( \varepsilon -\frac{i\pi
\omega }{16T_c}\right) \right] \Bigg\}  \nonumber
\end{eqnarray}
\begin{eqnarray}
&&{\rm Im}\left[ Q_{\parallel }^{AL(R)}(\omega )\right] =\frac{2e^2T}{\pi s}%
{\rm Im}\Bigg\{\left[ 1+i\left( \frac{16T_c}{\pi \omega }\right) \left(
\varepsilon +\frac r2\right) \right] \times  \nonumber  \label{QALparal} \\
&& \\
&\times &\left[ \Delta D_1\left( \varepsilon -\frac{i\pi \omega }{16T_c}%
\right) \right] +i\left( \frac{16T_c}{\pi \omega }\right) \left[ \Delta
D_2\left( \varepsilon -\frac{i\pi \omega }{16T_c}\right) \right] \Bigg\} 
\nonumber
\end{eqnarray}
where $D_1(z)=2\ln \left[ \sqrt{z}+\sqrt{\left( z+r\right) }\right] $, $%
D_2(z)=-\sqrt{z(z+r)}$, $\Delta D_1\left( z\right) =D_1\left( z\right)
-D_1\left( \varepsilon \right) $, $\Delta D_2\left( z\right) =D_2\left(
z\right) -D_2\left( \varepsilon \right) $and $r=4\eta J^2/v_F^2$.

The expressions presented above solve the problem of the frequency
dependence of the paraconductivity tensor in the general form for $%
\varepsilon \ll 1$and $\omega \leq T$for an arbitrary relation between $%
\varepsilon, r$and $\omega$, but they are too cumbersome.

We concentrate therefore on the most interesting case of $2D$ fluctuations where $%
\xi _c(T)\ll s$ ($r\ll \varepsilon $). $\sigma _{\perp }^{AL}$ turns out to
be suppressed by the necessity of the independent tunneling of each electron
participating in the fluctuation pairing from one $CuO_2$ layer to the
neighboring one. The approximation $r\ll \varepsilon $ simplifies
considerably the expressions (\ref{QALperp}) and ( \ref{QALparal}) \footnote{%
The second expression coincides with that one obtained in \cite{AV80,S68}}
while still giving valid results for frequencies comparable to $T_c$: 
\begin{eqnarray}
\sigma _{\perp }^{AL(2D)}(\varepsilon ,\omega ) &=&\frac{e^2s}{64\eta }%
\left( \frac r{2\varepsilon }\right) ^2\displaystyle{\frac 1{\tilde{\omega}%
^2}\ln \left( 1+\tilde{\omega}^2\right) }=  \nonumber  \label{perp} \\
&& \\
&=&\sigma _{\perp }^{AL(2D)}(\varepsilon ,0)\cases{ \displaystyle{
1-\frac{\tilde \omega ^2}{2}} & for $\tilde \omega \ll 1$ \cr \displaystyle{
\frac{2}{\tilde \omega ^2}\ln {\tilde \omega}} & for $\tilde \omega \gg 1$} 
\nonumber
\end{eqnarray}
and 
\begin{eqnarray}
\sigma _{\parallel }^{AL(2D)}(\varepsilon ,\omega ) &=&\frac{e^2}{16s}\frac
1\varepsilon \left\{ \frac 2{\tilde{\omega}}\arctan \tilde{\omega}-\frac 1{%
\tilde{\omega}^2}\ln (1+\tilde{\omega}^2)\right\} =  \nonumber
\label{abplane} \\
&& \\
&=&\sigma _{\parallel }^{AL(2D)}(\varepsilon ,0)\cases{\displaystyle{
1-\frac{\tilde \omega ^2}{6}} & for $\tilde \omega \ll 1$ \cr \displaystyle{
\frac{\pi }{\tilde \omega }} & for $\tilde \omega \gg 1$}  \nonumber
\end{eqnarray}
where $\tilde{\omega}=\displaystyle{\frac{\pi \omega }{16(T-T_c)}}$.

Two facts following from the expressions obtained should be stressed. First, the
paraconductivity begins to decrease rapidly with the increase of frequency
for $\omega \geq T-T_c$ (the critical exponents of this power decrease
coincide with those of the $\varepsilon $-dependence of d.c. -conductivity
tensor components: $\nu _{\parallel }=1$($2D$ fluctuations) and $\nu _{\perp
}=2$($0D$ fluctuations)). Secondly the assumption that one can neglect the $%
\omega $-dependence of the Green's functions blocks evidently breaks down at
frequencies $\omega \geq T$ and has the effect of accelerating the decrease
of paraconductivity.

\subsection{Density of States contribution}

The four main diagrams for the DOS contribution to the electromagnetic
response tensor are diagrams 5--8 of Fig. 9. The general
expression for the DOS contribution to $Q_{\alpha \beta }(\omega )$ from
diagram 5 is given by expression (\ref{d14}). The external frequency $\omega
_\nu $ enters in this expression (\ref{d14}) only by means of the Green's
function $G({\bf p},\omega _{n+\nu })$ and it is not involved in $q$
integration. So, near $T_c$, even in the case of an arbitrary external
frequency, we can choose the propagator frequency $\omega _\mu =0$. The
contribution of diagram 7 of Fig. 9 can be treated in the same
manner.

The diagrams 5 and 6 of Fig. 9 are topologically equivalent and
this fact would lead one to believe that they give equal contributions to $%
\sigma (\omega )$. Nevertheless the thorough analysis of the analytical
continuation over the external frequency shows that their contributions
differ slightly and for the total DOS contribution to conductivity one can
find: 
\begin{eqnarray}
{\rm Re}\left( 
\begin{array}{c}
\sigma _{\perp }^{{\rm DOS}}(\omega ) \\ 
\sigma _{\parallel }^{{\rm DOS}}(\omega )
\end{array}
\right) =-\displaystyle{\frac{e^2}{2\pi s}}\left( 
\begin{array}{c}
\displaystyle{\ \frac{s^2J^2}{v_F^2}} \\ 
1
\end{array}
\right) \ln \left[ \frac 2{\sqrt{\varepsilon +r}+\sqrt{\varepsilon }}\right] 
\hat{\kappa}\left( \omega ,T,\tau \right) ,  \label{ReDOSzzxx}
\end{eqnarray}
where 
\begin{eqnarray}
&&\hat{\kappa}\left( \omega ,T,\tau \right) =\displaystyle{\ \frac{Tv_{{\rm F%
}}^2}\eta }\frac 1{(\tau ^{-2}+\omega ^2)^2}\left\{ \frac 4\tau \left[ \psi
\left( \frac 12\right) -{\rm Re}\psi \left( \frac 12-\frac{i\omega }{2\pi T}%
\right) \right] +\right.   \nonumber  \label{kappa} \\
&&  \nonumber \\
&+&\left. \frac{\tau ^{-2}+\omega ^2}{4\pi T\tau }\frac 1\omega {\rm Im}\psi
^{\prime }\left( \frac 12-\frac{i\omega }{2\pi T}\right) +\right.   \nonumber
\\
&& \\
&+&\left. (\tau ^{-2}-\omega ^2)\frac 1\omega \left[ {\rm Im}\psi \left(
\frac 12-\frac{i\omega }{2\pi T}\right) -2\,{\rm Im}\psi \left( \frac 12-%
\frac{i\omega }{4\pi T}+\frac 1{4\pi T\tau }\right) \right] \right\} . 
\nonumber
\end{eqnarray}

In contrast to (\ref{QALperp}) and (\ref{QALparal}),
this result has been found with only the assumption $\varepsilon \ll 1$, so
it is valid for any frequency, any impurity concentration and any
dimensionally of the fluctuation behavior. The function $\hat{\kappa}\left(
\omega ,T,\tau \right) $ can be easily used to fit experimental data.
Nevertheless we also present the asymptotics of the expression (\ref{kappa}) for
clean and dirty cases. In the dirty case 
\begin{eqnarray}
\hat{\kappa}_{{\rm d}}\left( \omega ,T\ll \tau ^{-1}\right) =\displaystyle{\ 
\frac{Tv_F^2}{2\eta }}\cases{\displaystyle{\frac{\tau } {(2\pi T)^2}}\left
|\psi''\left(\displaystyle{\frac12}\right)\right | & for $\omega \ll T\ll
\tau^{-1}$ \cr \displaystyle{\frac{\tau}{\omega ^2}} & for $T\ll \omega \ll
\tau ^{-1}$ \cr -\displaystyle{\frac{\pi}{\omega^3}} & for $T\ll \tau
^{-1}\ll \omega $}  \label{dirty}
\end{eqnarray}
and in the clean case 
\begin{eqnarray}
\hat{\kappa}_{{\rm cl}}\left( \omega ,T\gg \tau ^{-1}\right) =\displaystyle{%
\ \frac{Tv_F^2}{2\eta }}\cases{\displaystyle{\frac{\pi \tau ^2}{4T}} & for
$\omega \ll \tau ^{-1}\ll T$ \cr -\displaystyle{\frac{\pi }{4\omega ^2T}} &
for $\tau^{-1}\ll \omega \ll T$ \cr -\displaystyle{\frac{\pi}{\omega^3}} &
for $\tau ^{-1}\ll T\ll \omega $}  \label{clean}
\end{eqnarray}

\subsection{Maki-Thompson contribution}

The total contribution of the MT-like diagrams to $\sigma _{\alpha \beta
}(\omega )$ has been analyzed in \cite{BDKLV93} for the case of zero
frequency and the frequency dependence of $\sigma _{\parallel }(\omega )$
has been studied in \cite{AV80}. In section 6.4 we have demonstrated that
the {\it regular} part of the MT diagram can almost always be omitted. So we
will not discuss it in this section and will concentrate instead on the {\it %
anomalous} MT contribution only.

In \cite{AV80} it was demonstrated that in the case of quasi two-dimensional
electron motion (\ref{d1}) there is no formal necessity to introduce the
pair-breaking time $\tau _\varphi $ because the Maki-Thompson logarithmic
divergence is automatically cut off by interlayer hopping. Nevertheless, all
evidences show that the intrinsic pair-breaking in HTS is strong (at least
one of its sources may be identified as thermal phonons) and an estimate of
the appropriate $\tau _\varphi \sim 2\div 5\cdot 10^{-13}s$ is only several
times larger than $T_c^{-1}$.So we are actually in the overdamped region of
the MT contibution, and the latter does not noticeably affect the $\epsilon $
dependence of conductivity \cite{BDKLV93} (the major part of experimental
results is explained in terms of AL or AL and DOS contributions). However we
are still interested in the MT contribution because of its frequency
dependence which evidently determines another characteristic scale in
addition to the previous three ($T-T_c$, $T$, $\tau ^{-1}$) we have
introduced: $\omega _{MT} \sim \tau _\varphi ^{-1}$.

We start, as usual, from the general expression for the {\it anomalous} MT
contribution to the electromagnetic operator tensor (\ref{d21}) and after
the integration over momentum $p$ and the summation over $\omega _n$ in the
range $\omega _n\in [-\omega _\nu ,0[$ ({\it anomalous} part), one finds: 
\begin{eqnarray}
&&\left( 
\begin{array}{c}
Q_{\perp }^{{\rm MT(an)}}(\omega _\nu ) \\ 
Q_{\parallel }^{{\rm MT(an)}}(\omega _\nu )
\end{array}
\right) =e^2T\tau \left[ \psi \left( \frac 12+\frac{\omega _\nu }{2\pi T}%
\right) -\psi \left( \frac 12\right) \right] \times  \nonumber  \label{MTint}
\\
&& \\
&\times &\int \frac{d^3q}{(2\pi )^3}\left( 
\begin{array}{c}
J^2s^2\cos {q_{\perp }s} \\ 
v_F^2
\end{array}
\right) \frac 1{\left( \omega _\nu +\tau _\varphi ^{-1}+\hat{{\bf D}}%
q^2\right) \left( \varepsilon +\eta q^2+r\sin ^2(q_{\perp }s/2)\right) } 
\nonumber
\end{eqnarray}

At this stage of calculation we artificially introduce the phase-breaking
time in the ``Cooperon''

vertices. Carrying out the integration and separating the real and the
imaginary parts we have: 
\begin{eqnarray}
&&{\rm Re}\left( 
\begin{array}{c}
\sigma _{\perp }^{{\rm MT(an)}}(\omega ) \\ 
\sigma _{\parallel }^{{\rm MT(an)}}(\omega )
\end{array}
\right) =\frac{e^2}{2\pi s}\left( 
\begin{array}{c}
\displaystyle{\ \frac{s^2}{2\eta }} \\ 
1
\end{array}
\right) \frac T\omega {\rm Im}\left\{ \displaystyle{\ \frac{\psi \left(
\frac 12-\frac{i\omega }{2\pi T}\right) -\psi \left( \frac 12\right) }{\frac{%
i\pi \omega }{8T_c}+\varepsilon -\gamma _\varphi }}\times \right.  \nonumber
\label{sigmaMT} \\
&& \\
&\times &\left. \left( 
\begin{array}{c}
-\Delta D_2\left( -\frac{i\pi \omega }{8T_c}+\gamma _\varphi \right) +\left( 
\frac{i\pi \omega }{8T_c}+\varepsilon -\gamma _\varphi \right) \\ 
\Delta D_1\left( -\frac{i\pi \omega }{8T_c}+\gamma _\varphi \right)
\end{array}
\right) \right\}  \nonumber
\end{eqnarray}
where$\gamma _\varphi =\displaystyle{\frac \pi {8T_c\tau _\varphi }}$. In
the two-dimensional overdamped regime ($r\ll \varepsilon \leq \gamma
_\varphi $) the expression (\ref{sigmaMT}) gives the following limits: 
\begin{eqnarray}
\sigma _{\perp }^{MT(an)(2D)}(\omega )=\frac{e^2s}{2^7\eta }\frac{r^2}{%
\gamma _\varphi \varepsilon }\cases{1 & for $\omega \ll \tau_{\varphi}^{-1}$
\cr \displaystyle{\left(\frac{8T_c\gamma_\varphi } {\pi \omega }\right)^2} & for
$\omega \gg \tau_{\varphi}^{-1}$}  \label{MTzzlim}
\end{eqnarray}
\begin{eqnarray}
\sigma _{\parallel }^{MT(an)(2D)}(\omega )=\frac{e^2}{8s}\cases{
\displaystyle{\frac{1}{\gamma_\varphi }}\ln
\left(\displaystyle{\frac{\gamma_\varphi}{\varepsilon}}\right) & for $\omega \ll
\tau_{\varphi}^{-1}$ \cr \displaystyle{\frac{4T_c}{\omega}} & for $\omega
\gg \tau_{\varphi}^{-1}$}  \label{MTxxlim}
\end{eqnarray}

We point out that the expression (\ref{sigmaMT}) has been obtained
without any limitation on frequency. Nevertheless we have made the
assumption $\tau \hat{{\bf D}}q^2\ll 1$ as in section 6 and it turns out
that, while this condition doesn't restrict our results for the AL and the
DOS contributions over the full range of frequency, temperature and impurity
concentration, this is not so for the MT contribution. In fact, as already
mentioned in section 6.4, in the ultra-clean (or non-local) limit, when $%
T\tau >1/\sqrt{\varepsilon }$, the assumption $\tau \hat{{\bf D}}q^2\ll 1$
is violated for the MT contribution and the results obtained are not valid
there. Nevertheless one can see that this non-local situation can be
realized only in the clean case ($T\tau \gg 1$) for temperatures in the
range $1/(T\tau )^2\ll \varepsilon \ll 1$. We suppose $T\tau \sim 1$, as in
the case of HTS, so we postpone the discussion of the non-local limit up to
section 11.

\subsection{Discussion}

We will now analyse each fluctuation contribution separately and then
discuss their interplay in ${\rm Re}\left[ \sigma _{\perp }(\omega )\right] $.
Because of the large number of parameters entering the expressions we
restrict our consideration to the $c$-axis component of the conductivity
tensor in the region of 2D fluctuations (above the Lawrence-Doniach crossover
temperature). In purpose to make the discussion more transparent we present
the asymptotics of the results obtained in the Table 4. The in-plane
component will be discussed in the end of this section.

The AL contribution describes the fluctuation condensate response to the
applied electromagnetic field. The currents associated with it can be
treated as the precursor phenomenon of the screening currents in the
superconducting phase. Above $T_c$ the binding energy of virtual Cooper
pairs gives rise to a pseudo-gap of the order of $T-T_c$, so it is not
surprising that the AL contribution decreases with the further increase of $%
\omega $\footnote{%
experimentally such decrease was observed recently on the in-plane
conductivity measurements in the vicinity of $T_c$ on YBCO samples \cite
{BDA96}}. Actually $\omega ^{AL}\sim T-T_c$is the only relevant scale for $%
\sigma ^{AL}$: its frequency dependence doesn't contain $T,\tau _\phi $ and $%
\tau $. The independence from the latter is due to the fact that elastic
impurities do not represent obstacles for the motion of Cooper pairs. The
interaction of the electromagnetic wave with the fluctuation Cooper pairs
resembles, in some way, the anomalous skin-effect where its reflection is
determined by the interaction with the free electron system.

The anomalous MT contribution also involves the formation of fluctuation
Cooper pairs, this time on self-intersecting trajectories. Being the
contribution related with the Cooper pairs electric charge transfer it
doesn't depend on the elastic scattering time but it turns out to be
extremely sensitive to the phase-breaking mechanisms. So two characteristic
scales turn out to be relevant in its frequency dependence: $T-T_c$and $\tau
_\phi ^{-1}$. In the case of HTS, where $\tau _\phi ^{-1}$ has been
estimated as at least $0.1T_c$ for temperatures up to $5\div 10K$ above $T_c$%
, the MT contribution is overdamped, is determined by the value of $\tau
_\phi $ and is almost temperature independent.

The DOS contribution to ${\rm Re}\left[ \sigma (\omega )\right] $ is quite
different with respect to those above. At low frequencies ($\omega \ll \tau
^{-1}$) the lack of electron states at the Fermi level leads to an opposite
effect in comparison with the AL and MT contributions: ${\rm Re}\left[
\sigma ^{DOS}(\omega )\right] $ turns out to be negative and this means the
increase of the surface impedance, or, in other words, the decrease of
reflectance. Nevertheless, the applied electromagnetic field affects the
electron distribution and at high frequencies $\omega \sim \tau^{-1}$ the
DOS contribution changes its sign. It is interesting that the DOS
contribution, as a one-electron effect, depends on the impurity scattering
in a similar manner to the normal Drude conductivity. The decrease of ${\rm %
Re}\left[ \sigma ^{DOS}(\omega )\right] $ starts at frequencies $\omega \sim
\min \{T,\tau ^{-1}\}$ which for HTS are much higher than $T-T_c$ and $\tau
_\varphi ^{-1}$.

The $\omega $-dependence of ${\rm Re}[\sigma _{\perp }^{{\rm tot}}]$ with
the most natural choice of parameters ($r\ll \varepsilon \leq \tau _\varphi
^{-1}\ll \min \{T,\tau ^{-1}\}$) is presented in Fig. 33.

The positive AL and MT contributions are pronounced at low frequencies on
the background of the DOS contribution which remains in this region a
negative constant. Then at $\omega \sim T-T_c$ the former decays and the $%
{\rm Re}\sigma _{\perp }$ remains negative up to $\omega \sim \min \{T,\tau
^{-1}\}$. The DOS contribution changes its sign at $\omega \sim \tau ^{-1}$
and then it rapidly decreases. The following high frequency behavior is
governed by the Drude law. So one can see that the characteristic
pseudo-gap-like behavior in the frequency dependence of the optical
conductivity takes place in the range $\omega \in [T-T_c,\tau ^{-1}]$. The
depth of the window increases logarithmically with $\varepsilon $ when $T$
tends to $T_c$, as shown in Fig. 34.

In case of $ab$-plane optical conductivity the two first positive
contributions are not suppressed by the interlayer transparency, and exceed
considerably the negative DOS contribution in a wide range of frequencies.
Any pseudo-gap like behavior is therefore unlikely in $\sigma _{\parallel
}^{fl}(\omega )$: the reflectivity will be of the metallic kind. The
comparison between (\ref{abplane}) and (\ref{ReDOSzzxx}) shows that the
compensation of the two contributions could only take place at $\omega
_0\sim T/\ln \varepsilon $ which is out of the range of validity of the AL
contribution.

Let us compare now the results of these calculations with the experiments
available. The recent measurements \cite{BTD94,BTD95} of the $c$-axis
reflectivity spectra, in FIR region on $YBa_2Cu_4O_8$ single crystals, show
the response of a poor metal with the additional contributions from $IR$
active phonon modes (which we do not discuss here).

With the decrease of temperature the $c$-axis optical conductivity decreases
showing around $180K$ a transition from a Drude-like to a pseudogap-like
behavior. The value of gap is $\omega _0\sim 180cm^{-1}$ and it seems weakly
dependent on temperature. The decrease of temperature does the gap structure
more pronounced without any abrupt change at the superconducting transition
temperature $T_c=80K$ (see Fig. 35).

Such experimentally observed behavior of the optical conductivity is in
qualitative agreement with our results. The suppression of the density of
states due to the superconducting fluctuations in the vicinity of $T_c$leads
to a decrease of reflectivity in the range of frequencies up to $\omega \sim
\tau ^{-1}$. The magnitude of this depression slowly (logarithmically)
increases with the decrease of temperature but at the edge of the transition
it reaches some saturation because of the crossover to the $3D$ fluctuation
regime (where instead of $\ln (1/\varepsilon )$ one has $\ln (1/r)-\sqrt{%
\varepsilon }$, see (\ref{ReDOSzzxx} )). So no singularity is expected in
the value of the minimum even in the first order of perturbation theory.
Below $T_c$ the fluctuation behavior of $\langle \Psi _{fl}^2\rangle$ is
mostly symmetrical to that one above $T_c$ (see \cite{VD83}) and, with the
further decrease of temperature, the fluctuation pseudo-gap minimum in the
optical conductivity smoothly transforms itself into the real
superconducting gap, which opens very sharply in HTS. We stress that the
temperature independence of the pseudo-gap threshold appears naturally in
the theory: it is determined by $\omega _0\sim \tau ^{-1}$(see (\ref{dirty}%
), (\ref{clean}) and Fig. 36). Comparing Fig. 33(b)
and Fig. 36, one can easily see that
the threshold doesn't move when the temperature changes but varies when the
inverse of the scattering rate changes. As far as far as the numerical
value of $\omega _0$ is concerned, assuming $T_c\tau =0.35$ (which is the
value for the scattering rate of the sample used in the experiment under
consideration \cite{BTD95}, that is also in the experimental range of the
inverse of the scattering rate $T\tau \approx 0.3\div 0.7$\cite{TT92,C96}),
one can see that the pseudo-gap threshold is predicted to be of the order of 
$200cm^{-1}$, in quantitative agreement with the experimental data \cite
{BTD94,BTD95}.

The outlined theory is, strictly speaking, valid only in the vicinity of the critical
temperature, where $\varepsilon \ll 1$. Nevertheless the logarithmic
dependence on $\varepsilon $ of the result obtained gives grounds to believe
that qualitatively the theory can be valid up to $\varepsilon =\ln
(T/T_c)\sim 1$, i.e. for temperatures up to $200K$ in the experiment
discussed. So the theory is again in agreement with the experimental value
of temperature $180K$ up to which the pseudo-gap is observable.

In conclusion we have calculated the optical conductivity tensor for layered
superconductors. A pseudo-gap-like minimum of its $c$-axis component in a
wide range of frequencies for temperatures in the vicinity of $T_c$ is
found. It is due to the fluctuation density of states renormalization which
can be treated as the opening of a fluctuation pseudo-gap. These result are
qualitatively, and in some aspects quantitatively, in agreement with 
recent experiments on $YBa_2Cu_4O_8$ samples. Further experiments, with more
anisotropic samples like $BSSCO$ single crystals, would be useful, because
the effect should be more pronounced.

\newpage

\section{Thermoelectric power above the superconducting transition}

\subsection{Introduction}

Thermoelectric effects are difficult both to calculate and measure if
compared to electrical transport properties. At the heart of the problem
lies the fact that thermoelectric coefficients in metals are the small
resultant of two opposing currents which almost completely cancel. In
calculating thermoelectric power one finds that electrons above the Fermi
level $E_F$ carry a heat current that is nearly the negative of that carried
by electrons below $E_F$. In the model of a monovalent metal in which band
structure and scattering probabilities are symmetric about $E_F$, this
cancellation would be exact; in a real metal small asymmetry survives. In
calculating this small effect one cannot with impunity ignore any possible
correction or renormalization merely on the grounds that it has been shown
to be negligible in calculations of the conductivity alone. This small
effect is proportional to the electron-hole asymmetry factor $f_{as}$, which
is defined as the ratio of the difference between numbers of electrons and
holes to the total number of particles. Apart from this factor, to estimate
the thermoelectric coefficient, one should consider also the characteristic
energy involved in thermoelectric transport, $\epsilon ^{*}$. The
thermoelectric coefficient $\vartheta $ may then be obtained from electrical
conductivity $\sigma $, if transport coefficients are defined through the
electric current flowing on the system as a response on electric field and
temperature gradient: 
\[
{\bf J}_e=\sigma {\bf E}+\vartheta \nabla T,\ \ \ \ \vartheta \sim (\epsilon
^{*}/eT)f_{as}\sigma 
\]
For non-interacting electrons $\epsilon ^{*}\sim T$ and $f_{as}\sim T/E_F$,
and therefore $\vartheta _0\sim (T/eE_F)\sigma _0$. Because of their
compensated nature, thermoelectric effects are very sensitive to the
characteristics of the electronic spectrum, the presence of impurities and
peculiarities of scattering mechanisms. The inclusion of many-body effects,
such as electron-phonon renormalization, multi-phonon scattering, a drag
effect, adds even more complexity to the problem of calculating the
thermoelectric power. Among such effects, there is also the influence of
thermodynamical fluctuations on thermoelectric transport in superconductor
above the critical temperature. This problem has been attracting the
attention of theoreticians for more than twenty years, since the paper of
Maki \cite{M74} appeared. After this pioneering work about ten contradictory
papers appeared and up to now not only the magnitude, but also the
temperature dependence and even the sign of the fluctuation correction to
thermoelectric coefficient ($\vartheta $) are under discussion.

The main question which should be answered is whether or not the correction
to $\vartheta $ has the same temperature singularity in the vicinity of the
critical temperature $T_c$ as the correction to electrical conductivity $%
\sigma $. In the paper of Maki \cite{M74} the only logarithmically divergent
contribution was predicted in the two-dimensional case and its sign was
found to be opposite to the sign of the normal state thermoelectric
coefficient $\vartheta _0$. Later on, in a number of papers \cite
{VL90,R91,K93,YP95,KB97} it was claimed that the temperature singularity of
the fluctuation correction to $\vartheta $ is the same as it is for $\sigma $
($\propto (T-T_c)^{-1}$ in 2D). Finally, Reizer and Sergeev \cite{RS94} have
recently revisited the problem using both quantum kinetic equation and
linear response methods and have finally shown that, in the important case
of an isotropic electronic spectrum, strongly divergent contributions \cite
{VL90,R91,K93} are canceled out for any dimensionally, with the final result
having the same logarithmic singularity as found by Maki, but of opposite
sign. We should emphasize that in all papers cited above only the
Aslamazov-Larkin (AL) contribution was taken into account, because the
anomalous Maki-Thompson (MT) term was shown to be absent in the case of
thermoelectric transport. It was mentioned \cite{RS94} that the reason for
discrepancy between different authors lies in the difficulties connected
with the introduction of many-body effects in the heat-current operator. As
a matter of fact, the incorrect evaluation of interaction corrections to
heat-current operator can produce erroneously large terms which are really
canceled out within the consistent procedure. Due to this strong
cancellation the AL term turns out to be less singular compared with the
corresponding correction to conductivity \cite{RS94}.

On the other hand, in every case where the main AL and MT fluctuation
corrections are suppressed for some reason, the contribution connected with
fluctuation renormalization of the one-electron density of states (DOS) can
become important. In this section we show that the analogous situation also
occurs in the case of the thermoelectric coefficient \cite{VLR92,VLF97}. In what
follows we study the DOS contribution to the thermoelectric coefficient of
superconductors with an arbitrary impurity concentration above $T_c$. We
will be mostly interested in 2D case, but the generalization to the case of
layered superconductor will be done at the end. We show that, although the
DOS term has the same temperature dependence as the AL contribution \cite
{RS94}, it turns out to be the leading fluctuation contribution in both the
clean and dirty cases due to its specific dependence on electron mean free
path.

\subsection{DOS contribution to thermoelectric coefficient}

We introduce the coefficient $\vartheta $ in the framework of linear
response theory as: 
\begin{eqnarray}
\vartheta =\lim_{\omega \rightarrow 0}\frac{{\rm Im}[Q^{{\rm (eh)R}}(\omega
)]}{T\omega }
\end{eqnarray}
where $Q^{{\rm (eh)R}}(\omega )$ is the Fourier representation of the
retarded correlation function of two current operators 
\begin{eqnarray}
Q^{{\rm (eh)R}}(X-X^{\prime })=-\Theta (t-t^{\prime })\langle \langle \left[
J^{{\rm h}}(X),J^{{\rm e}}(X^{\prime })\right] \rangle \rangle .
\end{eqnarray}
Here $J^{{\rm h}}$ and $J^{{\rm e}}$ are heat-current and electric current
operators in Heisenberg representation, $X=({\bf r},t)$ and $\langle \langle
\cdots \rangle \rangle $ represents both thermodynamical averaging and
averaging over random impurity positions. We use units with $\hbar =c=k_{%
{\rm B}}=1$.

The correlator in the diagrammatic technique is represented by the bubble
with two exact electron Green's functions and two external field vertices,
the first, $e{\bf v}$, associated with the electric current operator and the
second one, $\frac i2(\epsilon _n+\epsilon _{n+\nu }){\bf v}$, associated
with the heat current operator ($\epsilon _n=\pi T(2n+1)$ is fermionic
Matsubara frequency and ${\bf v}=\partial \xi ({\bf p})/\partial {\bf p}$
with $\xi $ being the quasiparticle energy). Working to the first order of
perturbation theory in Cooper interaction and averaging over impurity
configurations one finds the ten diagrams presented in Fig.37.

As usual, the solid lines represent the single-quasiparticle normal state
Green's function averaged over impurities. The shaded objects are the vertex
impurity renormalization $\lambda $ which, neglecting the ${\bf q}$
-dependence (see Section 6) that is unimportant here, are given by 
\begin{eqnarray}
\lambda ({\bf q}=0,\epsilon _n,\epsilon _{n^{\prime }})=\lambda (\epsilon
_n,\epsilon _{n^{\prime }})=\frac{|\tilde{\epsilon}_n-\tilde{\epsilon}%
_{n^{\prime }}|}{|\epsilon _n-\epsilon _{n^{\prime }}|}  \label{lambda}
\end{eqnarray}
Finally the wavy line represents the fluctuation propagator $L({\bf q}%
,\Omega _k)$.

The first diagram describes the AL contribution to thermoelectric
coefficient and was calculated in \cite{RS94} with the electron-hole
asymmetry factor taken into account in the fluctuation propagator. Diagrams
2-4 represent the Maki-Thompson contribution. As was mentioned in Refs.\cite
{VL90,RS94}, neither anomalous nor regular parts of these diagrams contribute
to $\vartheta $ in any order of electron-hole asymmetry. In what follows we
will discuss the contribution from diagrams 5-10 which describes the
correction to $\vartheta $ due to fluctuation renormalization of
one-electron density of states.

Let us start from the contribution of the diagrams 5 and 6. We have 
\begin{eqnarray}
Q^{(5+6)}(\omega _\nu ) &=&2eT\sum_{\Omega _k}\int (d{\bf q})L({\bf q}%
,\Omega _k)T\sum_{\epsilon _n}\frac{i\left( \epsilon _{n+\nu }+\epsilon
_n\right) }2\int (d{\bf p})v^2\times  \nonumber  \label{Q12} \\
&&  \nonumber \\
&\times &\left[ \lambda ^2(\epsilon _n,-\epsilon _n)G^2\left( {\bf p}%
,\epsilon _n\right) G\left( {\bf q}-{\bf p},-\epsilon _n\right) G\left( {\bf %
p},\epsilon _{n+\nu }\right) +\right. \\
&&  \nonumber \\
&+&\left. \lambda ^2(\epsilon _{n+\nu },-\epsilon _{n+\nu })G^2\left( {\bf p}%
,\epsilon _{n+\nu }\right) G\left( {\bf q}-{\bf p},-\epsilon _{n+\nu
}\right) G\left( {\bf p},\epsilon _n\right) \right] .  \nonumber
\end{eqnarray}
(We use the shorthand notation $(d{\bf q})=d^Dq/(2\pi )^D$, where D is
dimensionality). Evaluating Eq. (\ref{Q12}) one naturally obtains a
vanishing result if electron-hole asymmetry is not taking into account. The
first possible source of this factor is contained in the fluctuation
propagator and was used in \cite{RS94} for the AL diagram. Our calculations
show that for the DOS contribution this correction to the fluctuation
propagator results in non-singular correction to $\vartheta $ in 2D case and
can be neglected. That's why we ignored such correction in Eq. (\ref{d6}).
Another source of electron-hole asymmetry is connected with expansion of
energy-dependent functions in power of $\xi /E_F$ near Fermi level
performing ${\bf p}$ -integration in Eq. (\ref{Q12}) ($E_F$ is the Fermi
energy). In the case under discussion we have to perform such expansion of
the following product: 
\begin{eqnarray}
N(\xi ){\bf v}^2(\xi )=N(0){\bf v}^2(0)+\xi \left[ \frac{\partial (N(\xi )%
{\bf v}^2(\xi ))}{\partial \xi }\right] _{\xi =0}.  \label{exp}
\end{eqnarray}
Only second term in Eq. (\ref{exp}) contributes to thermoelectric
coefficient. Contribution of diagrams 7 and 8 is described by the
expression: 
\begin{eqnarray}
&&Q^{(7+8)}(\omega _\nu )=\frac{eT}{\pi N(0)\tau }\sum_{\Omega _k}\int (d%
{\bf q})L({\bf q},\Omega _k)T\sum_{\epsilon _n}\frac{i\left( \epsilon
_{n+\nu }+\epsilon _n\right) }2\int (d{\bf p}){\bf v}^2\times  \nonumber
\label{Q12cross} \\
&&  \nonumber \\
&&\times \left[ \lambda ^2(\epsilon _n,-\epsilon _n)G^2\left( {\bf p}%
,\epsilon _n\right) G\left( {\bf p},\epsilon _{n+\nu }\right) \int (d{\bf k}%
)G^2\left( {\bf k},\epsilon _n\right) G\left( {\bf q}-{\bf k},-\epsilon
_n\right) +\right. \\
&&  \nonumber \\
&&+\left. \lambda ^2(\epsilon _{n+\nu },-\epsilon _{n+\nu })G^2\left( {\bf p}%
,\epsilon _{n+\nu }\right) G\left( {\bf p},\epsilon _n\right) \int (d{\bf k}%
)G^2\left( {\bf k},\epsilon _{n+\nu }\right) G\left( {\bf q}-{\bf k}%
,-\epsilon _{n+\nu }\right) \right] .  \nonumber
\end{eqnarray}
The calculation of this expression is analogous to Eq. (\ref{Q12}), but the
expansion of $N(\xi ){\bf v}(\xi )$ should be performed either in ${\bf p}$
or in ${\bf k}$ momentum integration. Diagrams 9-10 do not give any singular
contribution to thermoelectric coefficient due to the vector character of
external vertices and as a result an additional $q^2$ factor appears after $%
{\bf p}$-integration. The same conclusion concerns the MT-like diagram 3-4.

Performing integration over $\xi $ we find the contribution of the important
diagrams 5-8 in the form 
\begin{eqnarray}
Q^{(5-8)}(\omega _\nu )=-\frac{eT^2}4\left[ \frac{\partial (N(\xi ){\bf v}%
^2(\xi ))}{\partial \xi }\right] _{\xi =0}\int (d{\bf q})L({\bf q},0)(\Sigma
_1+\Sigma _2+\Sigma _3)  \label{Q12_c}
\end{eqnarray}
In the last equation we have taken into account that near $T_c$ only the
term with $\Omega _k=0$ is important and we have separated sums over
semi-infinite ($]-\infty ,-\nu -1]$, $[0,\infty [$) and finite ($[-\nu ,-1]$
) intervals : 
\begin{eqnarray}
\Sigma _1 &=&2\sum_{n=0}^\infty \frac{2\epsilon _n+\omega _\nu }{2\tilde{%
\epsilon}_n+\omega _\nu }\left( \frac{\tilde{\epsilon}_n+\omega _\nu }{%
(\epsilon _n+\omega _\nu )^2}+\frac{\tilde{\epsilon}}{\epsilon _n^2}\right) ,
\nonumber  \label{sums} \\
&&  \nonumber \\
\Sigma _2 &=&\frac 1{(1/\tau +\omega _\nu )^2}\sum_{n=-\nu }^{-1}(2\epsilon
_n+\omega _\nu )^2\left( \frac{\tilde{\epsilon}_{n+\nu }}{\epsilon _{n+\nu
}^2}-\frac{\tilde{\epsilon}_n}{\epsilon _n^2}\right) \\
&&  \nonumber \\
\Sigma _3 &=&(1+\omega _\nu \tau )\sum_{n=-\nu }^{-1}(2\epsilon _n+\omega
_\nu )\left( \frac 1{\epsilon _{n+\nu }^2}-\frac 1{\epsilon _n^2}\right) 
\nonumber
\end{eqnarray}
$\Sigma _1$ and $\Sigma _2$ are associated with diagram 5-6, while $\Sigma
_3 $ with diagram 7-8. Calculating the sums (\ref{sums}) we are interested
in terms which are linear in external frequency $\omega _\nu $. The sum $%
\Sigma _1$ turns out to be an analytical function of $\omega _\nu $ and it
is enough to expand it in the Taylor series after analytical continuation $%
\omega _\nu \rightarrow -i\omega $. The last two sums over finite intervals
require more attention because of their nontrivial $\omega _\nu $-dependence
and before analytical continuation they have to be calculated rigorously. As
a result: 
\begin{eqnarray}
\Sigma _1^R=\frac{i\omega }{4T^2}\ ;\ \Sigma _2^R=-\frac{2i\omega \tau }{\pi
T}\ ;\ \Sigma _3^R=-\frac{i\omega }{2T^2}
\end{eqnarray}

Finally, we perform integration over ${\bf q}$. In 1D case the total
contribution associated with density of states renormalization takes the
form 
\begin{eqnarray}
\vartheta _{{\rm 1D}}^{{\rm DOS}}=\frac e4\frac 1{p_F}\left( \frac{T_c}{T-T_c%
}\right) ^{1/2}\left[ \left( 1+\frac \pi {8T_c\tau }\right) T_c\tau \kappa
(T_c\tau )\right] ^{1/2},  \label{1D}
\end{eqnarray}
where 
\begin{eqnarray}
\kappa ^{*}(T\tau ) &=&-\displaystyle{\frac{1+\displaystyle{\frac \pi
{8T\tau }}}{T\tau \left[ \psi \left( \displaystyle{\frac 12}+\displaystyle{%
\frac 1{4\pi T\tau }}\right) -\psi \left( \displaystyle{\frac 12}\right) -%
\displaystyle{\frac 1{4\pi T\tau }}\psi ^{\prime }\left( \displaystyle{\frac
12}\right) \right] }}  \nonumber  \label{kappa*} \\
&& \\
&=&\cases{\displaystyle{\frac{8\pi^2}{7\zeta(3)}} T\tau \approx 9.4 T\tau\ \
\ & for $T\tau \gg 1$ \cr \displaystyle{\frac1{T\tau}}\ \ \ & for $T\tau \ll
1$}  \nonumber
\end{eqnarray}

In 2D case: 
\begin{eqnarray}
\vartheta ^{{\rm DOS}}=\frac 1{4\pi ^2}\frac{eT_c}{{\bf v}_F^2N(0)}\left[ 
\frac{\partial ({\bf v}^2N(0))}{\partial \xi }\right] _{\xi =0}\ln \left( 
\frac{T_c}{T-T_c}\right) \kappa ^{*}(T_c\tau ),  \label{final1}
\end{eqnarray}
The generalization of this result to the important case of layered
superconductor is straightforward. One has to replace $\ln (1/\varepsilon
)\rightarrow \ln [2/(\sqrt{\varepsilon }+\sqrt{\varepsilon +r})]$ and to
multiply Eq. (\ref{final1}) by $1/p_Fs$, where $s$ is the interlayer
distance. In the limiting case of 3D superconductor ($r\gg \varepsilon $)
both the AL \cite{RS94} and the DOS contributions are not singular.

\subsection{Discussion}

Comparing Eq. ({\ref{final1}) with the
results of \cite{RS94} for the AL contribution, we conclude, that in both
limiting cases of clean and dirty systems the decrease of $\vartheta $ due
to fluctuation the DOS renormalization dominates the thermoelectric
transport due to the AL process. Really, the total relative correction to
thermoelectric coefficient in the case of 2D superconducting film of
thickness $s$ can be written in the form: 
\begin{eqnarray}
\frac{\vartheta ^{{\rm DOS}}+\vartheta ^{{\rm AL}}}{\vartheta _0}=-0.17\frac
1{E_\tau }\frac 1{p_Fs}\ln \left( \frac{T_c}{T-T_c}\right) \left[ \kappa
^{*}(T_c\tau )+5.3\ln \frac{\Theta _D}{T_c}\right] ,
\end{eqnarray}
where the first term in square brackets corresponds to the DOS contribution
( \ref{final1}) and the second term describes the AL contribution from \cite
{RS94} ($\Theta _D$ is the Debye temperature). Assuming $\ln (\Theta
_D/T_c)\approx 2$ one finds that the DOS contribution dominates the AL one
for any value of impurity concentration: $\kappa ^{*}$ as a function of $%
T\tau $ has a minimum at $T\tau \approx 0.3$ and even at this point the DOS
term is twice larger. In both limiting cases $T\tau \ll 1$ and $T\tau \gg 1$
this difference strongly increases. In one-dimensional case 
\begin{eqnarray}
\frac{\vartheta _{1D}^{{\rm DOS}}+\vartheta _{{\rm 1D}}^{{\rm AL}}}{%
\vartheta _0} &=&~~~~~~~~~~~~~~~~~~~~~~~~~~~~~~~~~~~~~~~~~~  \nonumber \\
&=&-\frac 1{(p_Fs)^2}\left( \frac{T_c}{T-T_c}\right) ^{1/2}\cases{
1.15\left[ 1+\displaystyle{\frac{0.47}{T\tau}}\ln\left(
\displaystyle{\frac{\Theta_{ D}}{T_c}} \right) \right] & for $T\tau\gg 1$
\cr \displaystyle{\frac{0.24}{\left(T\tau\right)^{1/2}}} \left[
\displaystyle{\frac1{T\tau}}+8.86\ln\left (\displaystyle{\frac{\Theta_{
D}}{T_c}}\right) \right] & for $T\tau\ll 1$}  \label{1D-TOT}
\end{eqnarray}
and the DOS correction again turns out to be dominant. }

The temperature and impurity concentration dependencies of fluctuation
corrections to $\vartheta $ in important 2D case can be evaluated through a
simple qualitative consideration. The thermoelectric coefficient may be
estimated through the electrical conductivity $\sigma $ as $\vartheta \sim
(\epsilon ^{*}/eT)f_{{\rm as}}\sigma $, where $\epsilon ^{*}$ is the
characteristic energy involved in thermoelectric transport and $f_{{\rm as}}$
is the electron-hole asymmetry factor, which is defined as the ratio of the
difference between numbers of electrons and holes to the total number of
particles. Conductivity can be estimated as $\sigma \sim e^2{\cal N}\tau
^{*}/m$, where ${\cal N}$, $\tau ^{*}$ and $m$ are the density, lifetime and
mass of charge (and heat) carriers, respectively. In the case of the AL
contribution the heat carriers are nonequlibrium Cooper pairs with energy $%
\epsilon ^{*}\sim T-T_c$ and density ${\cal N}\sim p_F^d\frac T{E_F}\ln 
\frac{T_c}{T-T_c}$ and characteristic time, given by Ginzburg-Landau time $%
\tau ^{*}\sim \tau _{GL}=\frac \pi {8(T-T_c)}$. Thus in 2D case $\Delta
\vartheta ^{{\rm AL}}\sim (T-T_c)/(eT_c)f_{{\rm as}}\Delta \sigma ^{{\rm AL}%
}\sim ef_{{\rm as}}\ln \frac{T_c}{T-T_c}$. One can easily get that the
fluctuation correction due to the AL process is less singular (logarithmic
in 2D case) with respect to the corresponding correction to conductivity and
does not depend on impurity scattering \cite{RS94}.

The analogous consideration of the single-particle DOS contribution ($%
\epsilon ^{*}\sim T$, $\tau ^{*}\sim \tau $) evidently results in the
estimate $\vartheta \sim ef_{{\rm as}}T_c\tau \ln {\frac{T_c}{T-T_c}}$ which
coincides with (\ref{final1}) in clean case. The dirty case is more
sophisticated because the fluctuation density of states renormalization
strongly depends on the character of the electronic motion, especially in
the case of diffusive motion \cite{CCRV90}. The same density of states
redistribution in the vicinity of Fermi level directly enters into the
rigorous expression for $\vartheta $ and it is not enough to write the
fluctuation Cooper pair density ${\cal N}_{c.p.}$ but is necessary to take
into account some convolution with $\delta N_{{\rm fl}}(E)$. This is what
was actually done in the previous calculations.

Experimentally, although the Seebeck coefficient $S=-\vartheta /\sigma $ is
probably the easiest to measure among thermal transport coefficients, the
comparison between experiment and theory is complicated by the fact that $S$
cannot be calculated directly; it is rather a composite quantity of
electrical conductivity and thermoelectric coefficient. As both $\vartheta $
and $\sigma $ have corrections due to superconducting fluctuations, total
correction to the Seebeck coefficient is given by 
\begin{equation}
\Delta S=S_0\left( \frac{\Delta \vartheta }{\vartheta _0}-\frac{\Delta
\sigma }{\sigma _0}\right)  \label{S}
\end{equation}
Both these contributions provide a positive correction $\Delta \vartheta $,
thus resulting in the decrease of the absolute value of $S$ at the edge of
superconducting transition ($\Delta \vartheta /\vartheta _0<0$). As for
fluctuation correction to conductivity $\Delta \sigma /\sigma _0>0$, we see
from Eq. (\ref{S}) that thermodynamical fluctuations above $T_c$ always
reduce the overall Seebeck coefficient as temperature approaches $T_c$. So
the very sharp maximum in the Seebeck coefficient experimentally observed in
few papers \cite{H90,ZSY92,KB97} seems to be unrelated to the fluctuation
effects within our simple model even leaving aside the question about the
experimental reliability of these observations. This conclusion is supported
by recent analysis of temperature dependence of thermoelectric coefficient
close to transition in Refs. \cite{MVF94}.

\newpage

\section{DOS fluctuations on NMR characteristics in HTS}

\subsection{Introduction}

In this section we discuss the contribution of superconducting fluctuations
to the spin susceptibility ${\chi _s}$ and the NMR relaxation rate $1/T_1$.
We base on the work of M.Randeria and A.Varlamov \cite{RV94} were the effect
of fluctuations on ${\chi _s}$ was examined first time. The fluctuation
contribution to NMR relaxation rate $1/T_1$ has been previously studied in
the set of papers \cite{AM76,KF89,H} focusing mainly on the most singular
contribution, the anomalous Maki-Thompson (MT) term in the dirty limit $%
T\tau \ll 1$. Nevertheless in \cite{C94,RV94} the problem was reexamined and
the important role of the DOS contribution was underlined. The matter of
fact that in conditions of the unusually strong pair breaking in high $T_c$
materials, one might expect the main MT contribution to be suppressed and $%
1/T_1$ to be dominated by the less singular DOS contributions.

It has recently been suggested \cite{RTMS} that dynamic \cite{SRE} pairing
correlations beyond the perturbative weak coupling regime are responsible
for the spin gap anomalies \cite{SPINGAP} observed well above $T_c$ in the
underdoped cuprates. The analysis presented here, which only treats static
fluctuations very close to $T_c$, nevertheless constitutes the first
correction \cite{RAINER} to Fermi liquid behavior, to order $%
\max(1/E_F\tau,T_c/E_F)$, arising from pairing correlations above $T_c$.

The main results of this section, valid for $\varepsilon \ll 1$, can be
summarized as follows:

(1) Fluctuations lead to a suppression of the spin susceptibility ${\chi_s}$
, due to the combined effect of the reduction of the single particle density
of states (DOS) arising from self energy contributions, and of the regular
part of the Maki-Thompson (MT) process.

(2) ``Cooperon'' impurity interference terms, involving impurity ladders in
the particle-particle channel, are crucial for the ${\chi_s}$ suppression in
the dirty limit.

(3) The processes which dominate the results in (1) and (2) above have
usually been ignored in fluctuation calculations (conductivity, $1/T_1$,
etc.). The spin susceptibility is unusual in that the Aslamazov-Larkin, and
the anomalous MT terms, which usually dominate, are absent.

(4) For weak pair-breaking ($1/\tau_\varphi \ll T_c$), we find an
enhancement of $1/T_1T$ coming from the anomalous MT term. We recover known
results \cite{KF89} in the dirty limit, and extend these to arbitrary
impurity scattering.

(5) In the clean limit ($T_c\tau \gg 1$) we find a different asymptotic
behavior of $1/T_1T$ depending on whether one has $T_c\tau$ is greater or
smaller than $1/{\sqrt\varepsilon}$.

(6) Finally, strong dephasing suppresses the anomalous MT contribution, and $%
1/T_1T$ is then dominated by the less singular DOS and the regular MT terms.
These contributions lead to a suppression of spectral weight and a decrease
in $1/T_1T$.

\subsection{Definitions}

We begin with the dynamic susceptibility $\chi _{+-}^{(R)}({\bf k},\omega
)=\chi _{+-}({\bf k},i\omega _\nu \rightarrow \omega +i0^{+})$ where 
\begin{equation}
\chi _{+-}({\bf k},\omega _\nu )=\int_0^{1/T}d\tau e^{i\omega _\nu \tau
}\langle \langle \hat{T}\left( \hat{S}_{+}({\bf k},\tau )\hat{S}_{-}(-{\bf k}%
,0)\right) \rangle \rangle
\end{equation}
with the Bose frequency $\omega _\nu =2\pi \nu T$. Here $\hat{S}_{\pm }$ are
the spin raising and lowering operators, $\hat{T}$ is the time ordering
operator, and the brackets denote thermal and impurity averaging in the
usual way. The uniform, static spin susceptibility is given by ${\chi _s}%
=\chi _{+-}^{(R)}({\bf k}\rightarrow 0,\omega =0)$ and the NMR relaxation
rate by 
\begin{equation}
{\frac 1{{T_1T}}}=\lim_{\omega \rightarrow 0}{\frac A\omega }{\int \,(d{\bf k%
})\,\Im \chi _{+-}^{(R)}({\bf k},\omega )}
\end{equation}
where $A$ is a positive constant involving the gyromagnetic ratio.

For non-interacting electrons $\chi_{+-}^0({\bf k}, \omega_{\nu})$ is
determined by the loop diagram presented in Fig. 38. Simple
calculations lead to the well known results for $T \ll E_F$: $\chi_s^{0} =
N(0)$ (Pauli susceptibility) and $\left(1 /T_1T\right)^{0} = A \pi [N(0)]^2$
(Korringa relaxation), where $N(0)$ is the DOS at the Fermi level. We shall
present the fluctuation contributions in a dimensionless form by normalizing
with the above results.

To leading order in $\max(1/E_F\tau,T_c/E_F)$ the fluctuation contributions
to $\chi_{+-}$ are given by the diagrams shown in Fig. 38. The
diagrams are constructed from fermion lines, fluctuation propagators
(denoted by the wavy lines) and impurity vertex corrections (represented by
shaded objects). It is important to note that the two fermion lines attached
to the external vertex have opposite spin labels (up and down) for $%
\chi_{+-} $. Consequently, the Aslamazov-Larkin diagram (1) does not exist
since one cannot consistently assign a spin label to the fermion line marked
with a `?' for spin-singlet pairing.

The next set of diagrams to consider is the Maki-Thompson (MT) diagram (2a
and 2b), and the MT with the Cooperon impurity corrections (3) and (4).
While the MT diagrams for $\chi _{+-}$ appear to be identical to the well
known MT diagrams for conductivity, there is an important difference in
topology which arises from the spin structure. It is easy to see, by drawing
the fluctuation propagator explicitly as a ladder of attractive interaction
lines (diagram (2b)), that the MT diagram is a non-planar graph with a
single fermion loop. In contrast the MT graph for conductivity is planar and
has two fermion loops. The number of loops, of course, affects the sign of
the contribution.

The diagrams (5) and (6) represent the effect of fluctuations on the
one-particle self energy, leading to a decrease in the DOS. The DOS diagrams
(7) and (8) include impurity vertex corrections. (Note that these have only
a single impurity scattering line as additional impurity scattering, in the
form of a ladder, has a vanishing effect.) Finally (9) and (10) are the DOS
diagrams with the Cooperon impurity corrections.

The fermion lines represent the one electron Green function $G({\bf p},
\omega_{n})=( i\tilde{\epsilon}_n-\xi({\bf p}))^{-1}$, where $\tilde{%
\epsilon }_n=\epsilon_n+{\rm sgn}(\epsilon_n)/2\tau$ describes the
self-energy effects of impurity scattering. The momentum relaxation rate $%
1/\tau \ll E_F$ , however, $T_c\tau$ is arbitrary. For simplicity, we will
discuss the two-dimensional (2D) case for the most part, indicating at the
end how the results are modified for the layered system, and the 2D to 3D
crossover.

Pairing fluctuations above $T_c$ are described in the usual way by means of
the vertex part of electron-electron interaction in the Cooper channel \cite
{AL68,AV80}, or the fluctuation propagator (shown in Fig. 38 by the
shaded wavy line) $L({\bf q},\Omega _\mu )$ where the arguments refer to the
total momentum and frequency of the pair. The full $({\bf q},\Omega _\mu )$
-dependence of $L$ is important far from $T_c$; see ~\cite{AV80}. We
restrict our attention here to the vicinity of transition and thus it
suffices to focus on long wavelength, static ($\Omega _\mu =0$)
fluctuations. In this regime we have 
\begin{equation}
L^{-1}({\bf q},\Omega _\mu =0)=-N(0)\left[ \varepsilon +\eta _Dq^2\right] ,
\end{equation}

We now turn to vertex corrections due to impurity scattering. First, it can
be shown \footnote{%
The external vertex correction equals zero for the case of ${\chi_s}$ since
the external frequency $\omega_\nu = 0$. For $1/T_1$, the large external
momentum ${\bf k}$ leads to a suppression of this correction by the
parameter $1/E_F\tau$.} that the external vertices do not need to be
renormalized by impurity lines. Next, the (three-legged) impurity vertex $%
\lambda ({\bf q},-\epsilon _n,\epsilon _{n+\nu })$ is defined as the sum of
impurity ladders dressing the bare vertex consisting of two fermion lines,
with frequencies $-\epsilon _n$ and $\epsilon _{n+\nu }=\epsilon _n+\omega
_\nu $, and a fluctuation propagator with momentum and frequency $({\bf q}%
,\omega _\nu )$. Having in mind to discuss in this section the non-local
limit ($l\gg \xi _{ab}(\varepsilon )$ ) and to show how this modifies the MT
contribution (the same situation takes place in conductivity, but we skipped
this in the view of cumbersome calculations of the section 6) we use below
the purely 2D presentation of the vertex, calculated for an arbitrary value
of $\tau ^{-1}\hat{{\bf D}}q^2$. The result differs from (\ref{d3}) (but
naturally coincides with the latter in the local limit) 
\begin{equation}
\lambda ({\bf q},-\epsilon _n,\epsilon _{n+\nu })=\left( 1-{\frac{\Theta
(\epsilon _n\epsilon _{n+\nu })}{{\tau \sqrt{(\tilde{\epsilon}_n+\tilde{%
\varepsilon}_{n+\nu })^2+v^2q^2}}}}\right) ^{-1}.
\end{equation}
where $\Theta (x)$ is the Heaviside step function.

Finally, the Cooperon $C({\bf q},-\epsilon_n,\epsilon_{n+\nu})$ is defined
as the sum of impurity ladders in the particle-particle channel where $%
-\epsilon_n$ and $\epsilon_{n+\nu}$ are the frequencies of the two particle
lines and ${\bf q}$ is their {\it total} momentum. It is given by 
\begin{equation}
C({\bf q},-\epsilon_n,\epsilon_{n+\nu}) = {\frac{1}{2\pi N(0)\tau}} \left[{\ 
\frac{1 }{\tau}}{\frac{\Theta(\epsilon_n\epsilon_{n+\nu}) }{{|2\epsilon_n +
\omega_\nu| +{\bf D}q^2}}} + \Theta(-\epsilon_n\epsilon_{n+\nu})\right].
\end{equation}
Note that the diffusion pole involves the sum of the momenta and the
difference of frequencies.

\subsection{Spin Susceptibility}

We note that, when the external frequency and momentum can be set to zero at
the outset, as is the case for ${\chi_s}$, there is no anomalous MT piece
(which as we shall see below is the most singular contribution to $1/T_1$).
The MT diagram (2) then yields a result which is identical to the sum of the
DOS diagrams (5) and (6). We have: 
\begin{equation}  \label{dos.chi}
{\frac{{\chi_s}_5 + {\chi_s}_6 }{{\chi_s}^{0}}} = - {\frac{T_c }{E_F}} \ln
(1/\varepsilon)
\end{equation}
and this result is valid for any impurity concentration.

We first discuss the clean limit, where the fluctuation contribution is
given by $\chi_s^{{\rm fl}} = {\chi_s}_2 + {\chi_s}_5 + {\chi_s}_6$; all
other diagrams are negligible for $T_c\tau \gg 1$. The final result is 
\begin{equation}  \label{clean.chi}
{\frac{\chi_s^{{\rm fl}} }{{\chi_s}^{0}}} = - {\frac{2T_c }{E_F}} \ln
(1/\varepsilon) \ \ \ {\rm for}\ T_c\tau \gg 1.
\end{equation}

In the dirty limit ($T_c\tau \ll 1$), the DOS diagrams (5) and (6), together
with the regular part of MT diagram (2), yield the same result (\ref
{clean.chi}) of the order $T_c/E_F$. One can see, that this contribution is
a negligible in comparison with the expected for the dirty case dominant
one, of the order ${\cal {O}}(1/E_F\tau )$. The thorough study of all
diagrams shows that the important graphs in dirty case are those with the
Cooperon impurity corrections MT (3) and (4), and the DOS ones (9) and (10).
This is the first example known to us where the Cooperons, which play a
central role in the weak localization theory, give the leading order result
in the study of superconducting fluctuations. Diagrams (4) and (5) give the
one half the final result given below; diagrams (9) and (10) provide the
other half. The total fluctuation susceptibility $\chi _s^{{\rm fl}}={\chi _s%
}_3+{\chi _s}_4+{\chi _s}_9+{\chi _s}_{10}$, is 
\begin{equation}
{\frac{\chi _s^{{\rm fl}}}{{\chi _s}^{(0)}}}=-{\frac{7\zeta (3)}{\pi ^3}}{\
\frac 1{E_F\tau }}\ln (1/\varepsilon )\ \ \ {\rm for}\ T_c\tau \ll 1.
\label{dirty.chi}
\end{equation}

It is tempting to physically understand the negative sign of the fluctuation
contribution to the spin susceptibility in eqns.~(\ref{clean.chi}) and (\ref
{clean.chi}) as arising from a suppression of the DOS at the Fermi level.
But one must keep in mind that only the contribution of diagrams (5) and (6)
can strictly be interpreted in this manner; the MT graphs and the coherent
impurity scattering described by the Cooperons do not permit such a simple
interpretation.

\subsection{Relaxation Rate}

The calculation of the fluctuation contribution to $1/T_1$ requires rather
more care than ${\chi_s}$ because of the subtleties of analytic
continuation. We define the local susceptibility $K(\omega_\nu) = \int
(d {\bf k})\chi_{+-}({\bf k},\omega_\nu)$. The ``anomalous'' MT
contribution, denoted by a subscript ${({\rm an})}$, comes from that part of
the Matsubara sum which involves summation over the interval $%
-(\omega_\nu/2\pi T) = -\nu \le n \le -1$. The reason this piece dominates,
and has to be treated separately, is that $S_\an$ has a singular ${\bf q}$%
-dependence as we shall see. The remaining terms in $S$, i.e., those with $n
< -\nu$ and $n \ge 0$, are called the ``regular'' MT contribution.

We evaluate $S_\an$ using standard contour integration techniques and then
make the analytic continuation $i\omega_\nu \rightarrow \omega + i0^+$ to
obtain 
\begin{equation}  \label{eqn.a}
\lim_{\omega \rightarrow 0} {\frac{1 }{\omega}} \Im K_{2{({\rm an})}
}^{(R)}(\omega) = - {\frac{\pi N(0)^2 }{8}} \int (d{\bf q}) L({\bf q},0) 
{\cal K}({\bf q}),
\end{equation}
\begin{equation}  \label{eqn.b}
{\cal K}({\bf q}) = 2\tau\int^\infty_{-\infty}{\frac{dz }{\cosh^2(z/4T\tau)}}
{\frac{1 }{\left(\sqrt{l^2q^2 - (z-i)^2} - 1\right) \left(\sqrt{l^2q^2 -
(z+i)^2} - 1\right)}}
\end{equation}
where $l = v\tau$ is the mean free path.

The first simple limiting case for (\ref{eqn.b}) is $l q \ll 1$, for which $%
{\cal K}({\bf q}) = 2\pi/{\bf D}q^2$. Since (\ref{eqn.a}) involves a ${\bf q}
$-integration, we need to check when the above approximation is justified,
the characteristic $q$-values being determined from the fluctuation
propagator $L$. In the dirty limit, we have ${\bf D}q^2 \sim \varepsilon T_c$
, thus leading to $l^2q^2 \sim \varepsilon T_c\tau \ll 1$. In the clean
case, on the other hand, $v^2q^2/T_c \sim \varepsilon T_c$ and $l^2q^2 \sim
\varepsilon (T_c\tau)^2 \ll 1$ only when $1 \ll T_c \tau \ll 1/\sqrt{
\varepsilon}$.

For the above conditions (either $T_c\tau \ll 1$ or $1\ll T_c\tau \ll 1/%
\sqrt{\varepsilon }$) we obtain the singular MT contribution $\int
d^2q\left[ ({\bf D}q^2+1/\tau _\varphi )(\varepsilon +\eta _2q^2)\right]
^{-1}$ where we have introduced the pair breaking rate $1/\tau _\varphi $ as
an infrared cutoff. We define the dimensionless pair-breaking parameter $%
\gamma _\varphi =\eta _2/{\bf D}\tau _\varphi \ll 1$; in the dirty limit $%
\gamma _\varphi =\pi /8T_c\tau _\varphi $ while for the clean case $\gamma
_\varphi =7\zeta (3)/16\pi ^2T_c^2\tau \tau _\varphi $ . The ``bare''
transition temperature ${T_c}_0$ is shifted by the pair breaking, so that $%
\varepsilon =\varepsilon _0+\gamma _\varphi $, with $\varepsilon _0=(T-{T_c}%
_0)/{T_c}_0$, and we obtain the final result 
\begin{equation}
{\frac{(1/T_1T)^{{\rm fl}}}{(1/T_1T)^0}}={\frac \pi {8E_F\tau }}{\frac
1{\varepsilon -\gamma _\varphi }}\ln (\varepsilon /\gamma _\varphi ).
\label{mt.dirty}
\end{equation}

The other limiting case of interest is the ``ultra-clean limit'' when the
characteristic $q$-values satisfy $l q \gg 1$. This is obtained when $%
T_c\tau \gg 1/\sqrt{\varepsilon} \gg 1$. From (\ref{eqn.b}) we then find $%
{\cal K}({\bf q}) = 4\ln(l q)/vq$, which leads to 
\begin{equation}
{\frac{(1/T_1T)^{{\rm fl}} }{(1/T_1T)^{0}}} = {\frac{\pi^3 }{\sqrt{%
14\zeta(3) }}} {\frac{T_c }{E_F}} {\frac{1 }{\sqrt{\varepsilon}}}
\ln(T_c\tau \sqrt{ \varepsilon}).
\end{equation}

We note that in all cases the anomalous MT contribution leads to an {\it %
enhancement} of the NMR relaxation rate over the normal state Korringa
value. In particular, the superconducting fluctuations above $T_c$ have the 
{\it opposite} sign to the effect for $T\ll T_c$ (where $1/T_1$ drops
exponentially with $T$). One might argue that the enhancement of $1/T_1T$ is
a precursor to the coherence peak just below $T_c$. Although the physics of
the Hebel-Slichter peak (pile-up of the DOS just above gap edge and
coherence factors) appears to be quite different from that embodied in the
MT process, we note that both effects are suppressed by strong inelastic
scattering \cite{RV94,AF95,AFK94}.

We now discuss the DOS and the regular MT contributions which are important
when strong dephasing suppresses the anomalous MT contribution discussed
above. The local susceptibility arising from diagrams (5) and (6) can be
easily evaluated. The other remaining contribution is from the regular part
of the MT diagram. This corresponds to terms with $n<-\nu $ and $n\ge 0$ in
the Matsubara sum. It can be shown that this regular contribution is exactly
one half of the total DOS contribution from diagrams (5) and (6). All other
diagrams either vanish (as is the case for graphs (7) and (8)) or contribute
at higher order in $1/E_F\tau $ (this applies to the graphs with the
Cooperon corrections). The final results are given by 
\begin{equation}
{\frac{(1/T_1T)^{{\rm fl}}}{(1/T_1T)^0}}=-{\frac{6T_c}{E_F}}\ln
(1/\varepsilon )
\end{equation}
for $T_c\tau \gg 1$, and 
\begin{equation}
{\frac{(1/T_1T)^{{\rm fl}}}{(1/T_1T)^0}}=-{\frac{21\zeta (3)}{\pi ^3}}{\
\frac 1{E_F\tau }}\ln (1/\varepsilon )
\end{equation}
for $T_c\tau \ll 1$.

The negative sign of the result indicates a suppression of low energy
spectral weight as in the ${\chi _s}$ calculation, however in contrast to ${%
\ \chi _s}$ we note that the same graphs (2), (5) and (6) dominate in both
the clean and dirty limits.

It is straightforward to extend the above analysis to layered systems by
making the following replacement in the 2D results given above: 
\begin{eqnarray}
\ln\left(\frac{1 }{\varepsilon }\right) \rightarrow 2\ln\left({{\frac{2 }{{\ 
\sqrt{\varepsilon} + \sqrt{\varepsilon + r}}}}}\right).  \label{quasi2D}
\end{eqnarray}

\subsection{Discussion}

The negative DOS contribution to the NMR relaxation rate is evident from the
Korringa formula and it sign seems very natural while the sign of the
positive Maki-Thompson contribution can generate a questions about its
physical origin. It is why we present here the qualitative consideration of
the result obtained above from the microscopic theory in the spirit of the
section 6.2.

In the case of nuclear magnetic relaxation rate calculations the electron
interaction with nuclei with spin flip is considered. If one would try to
imagine the AL process of this type he were in troubles, because the
electron-nuclei scattering with the spin-flip evidently transforms the
initial singlet state of the fluctuation Cooper pair in the triplet-one,
what is forbidden in the scheme discussed. So the formally discovered
absence of the AL contribution to relaxation rate is evident enough.

The negative density of states contribution in $\frac 1{T_1}$ has the same
explanation given above for the conductivity: the number of normal electrons
decreases at the Fermi level and as a result the relaxation rate diminishes
with respect to the Korringa law.

The positive Maki-Thompson contribution can be treated in terms of the
pairing on the self-intersecting trajectories like this was done with that
one in conductivity (see section 6.2). This consideration \cite{RV97} clears
up the situation with the sign and explains why the MT type of pairing is
possible when the AL one is forbidden. Nevertheless the principle of the
electron pairing on the self-intersecting trajectory in the case of the NMR
relaxation rate has to be changed considerably with respect to the case of
conductivity.

Let us consider a self-intersecting trajectory and the motion of the
electron along it with fixed spin orientation (e.g. ''spin up''). If,
after passing of the full turn, the electron interacts with the nucleus and
changes its spin state and momentum to the opposite value it can pass again
the previous trajectory moving in the opposite direction (see Fig. 39).
The interaction of the electron with itself on the previous
stage of the motion is possible due to the retarded character of the Cooper
interaction and such pairing process, in contrast to the AL one, turns out
to be an effective mechanism of the relaxation near $T_c$. This purely
quantum process opens a new mechanism of spin relaxation, so contributes
positively to the relaxation rate $\frac 1{T_1}$.

In conclusion of this section we want to say several words about the
experimental difficulties in the observation of fluctuation effects in NMR.
Much of the existing work on fluctuations as probed by NMR has been
restricted to small particles (zero-dimensional limit) of conventional
superconductors; see \cite{MAC} for a review. There has been resurgence of
interest in superconducting fluctuations since the high $T_c$ cuprates show
large effects above $T_c$ due to their short coherence length and layered,
quasi-two-dimensional structure. However, in order to extract the
``fluctuation contributions'' from experiments one needs to know the normal
state backgrounds, which in a conventional metal would simply be the Pauli
susceptibility for ${\chi _s}$ and the Korringa law for $1/T_1T$. The
problem for the high $T_c$ materials is that the backgrounds themselves have
nontrivial temperature dependencies \cite{PS90} above $T_c$: for example,
the non-Korringa relaxation for the Cu-site in YBa$_2$Cu$_3$O$_{7-\delta }$,
and the spin-gap behavior \cite{SPINGAP} with $d{\chi _s}/dT>0$ and the O
and Y $1/T_1T\sim {\chi _s}(T)$. The attempt to overpass these difficulties
was done in \cite{CLRV96} which we discuss in the next section.

\newpage

\section{D versus S pairing scenario: {\it pro} and {\it contra} from the
fluctuation phenomena analysis}

\subsection{Introduction}

One of the hottest debate in solid state physics is the question of the
pairing state symmetry in HTS. The determination of the order parameter
symmetry is the crucial first step in the identification of the pairing
mechanism in HTS and subsequent development of a microscopic theory of high
temperature superconductivity. Recently much of the attention of both
theoreticians and experimentalists has been focused on a pairing state with $%
d-$wave symmetry, although many other pairing symmetries are also possible.
The challenge for investigators is to derive tests that are capable to
distinguish the different possible pairing states and to make the
determination of pairing state symmetry. The direct way to study the
anisotropy of the phase of the order parameter is investigation of phase
coherence of Josephson and tunnel junctions incorporating HTS. Recent phase
coherence experiments support a scenario of superconducting pairing state
with $d_{x^2-y^2}$ symmetry in HTS \cite{VanHar,B94}. Nevertheless the
studies of the amplitude of the order parameter may be also fruitful and the
simplest experiments are studies of fluctuation effects above
superconducting transition temperature. In view of the fact that the theory
of fluctuation effects in different physical properties of superconductor is
now well established, the main question of the present section is how the
interpretation of experimental data on fluctuation effects in HTS alters if
one assumes $d-$ rather than $s-$ pairing state and which experiment on
fluctuation effect would be a test for a pairing state in HTS.

In the theoretical studies of fluctuation effects in $d-$wave
superconductors two models were used. First is a model spherically symmetric
superconductor with the bare electron-electron interaction with the
interaction constant 
\begin{equation}
V({\bf k},{\bf k}^{\prime})=g_0 P_2({\bf k k}^{\prime})
\end{equation}
($P_2$ is Legendre polynomial) \cite{Y90}. The second model considers
tight-binding electrons with nearest-neighbor transfer with underlying 2D
square lattice \cite{MW94,KF89,NG95}. In this case 
\begin{equation}
V({\bf k},{\bf k^{\prime}})=g_0(\cos k_xa-\cos k_ya)(\cos k_x^{\prime}a-\cos
k_y^{\prime}a)
\end{equation}
with $a$ being a lattice constant. Both models lead essentially to the same
results. Further we will use the second model which provides a more simple
calculations. The main difference between $s-$ and $d-$ wave superconductors
lies in the different manifestation of non-magnetic impurities, which are
pair-breaking for the last. Further, in order to avoid cumbersome equations,
we restrict ourselves to establish only dependencies of fluctuation
contributions on temperature and impurity scattering thus ignoring numerical
factors of order unity.

\subsection{The fluctuation propagator}

In the case of $d-$wave superconductor the fluctuation propagator depends
upon directions of ${\bf k}$ and ${\bf k}^{\prime}$, but using the ansatz: 
\begin{equation}
L_{{\bf k k}^{\prime}}({\bf q})=\hat L({\bf q}) (\cos k_xa-\cos k_ya)(\cos
k_x^{\prime}a-\cos k_y^{\prime}a)
\end{equation}
(${\bf q}={\bf k}-{\bf k}^{\prime}$), one can easily solve the Dyson
equation for a quantity $\hat L({\bf q})$. One has: 
\begin{eqnarray}  \label{propag}
{\hat L({\bf q})}^{-1}&=&{g_0}^{-1}-P({\bf q},\omega_k) , \\
P({\bf q},\omega_k)&=&T\sum_{\omega_n}\int \frac{d^2 k}{(2\pi)^2} (\cos k_xa
-\cos k_y a)^2 G({\bf k},\omega_n)G({\bf q}-{\bf k},\omega_k -\omega_n) 
\nonumber
\end{eqnarray}
Here 
\begin{equation}
{G({\bf k},\omega_n)}^{-1}=i\omega_n+i{\rm sgn}\omega_n/2\tau-\xi_k\ \ \ \
,\ \ \ \ \ \xi_k=-t(\cos k_xa + \cos k_ya)-\mu
\end{equation}
($t$ and $\mu$ are the transfer integral between the nearest-neighbor sites
and the chemical potential, respectively). Note, that in contrast to $s-$
wave pairing state, the impurity renormalization of quantity $P({\bf q}
,\omega_k)$ is simply absent \cite{Y90,KF89} due to dependence of bare
interaction on the momentum directions.

For temperatures close to transition point, one can solve (\ref{propag}) for
small ${\bf q}$ and $\omega _k$. In this case 
\begin{equation}
\xi _{q-k}\approx \xi _k+\varsigma \ \ \ ,\ \ \ \ \varsigma =at(q_x\sin
k_xa+q_y\sin k_ya)
\end{equation}
Then one can replace the ${\bf k}$ by ${\nu }/{(2\pi )}d\xi \langle
...\rangle $ , where $\nu $ is 2D density of states and $\langle ...\rangle $
means averaging over angles of ${\bf k}$ taken under condition $\xi _k=0$.
After straightforward calculations one has: 
\begin{eqnarray}
\frac{P({\bf q},\omega _k)}{\alpha _1\nu } &=&4\pi T\sum_{n=0}^{4t}\frac
1{2\omega _n+\tau ^{-1}}-\frac{|\omega _k|}{4\pi T}\psi ^{\prime }\left(
\frac 12+\frac 1{4\pi T\tau }\right) -  \nonumber  \label{Polariz} \\
&-&\frac{\alpha _2(q)}{\alpha _14\pi T^2}\psi ^{\prime \prime }\left( \frac
12+\frac 1{4\pi T\tau }\right)
\end{eqnarray}
Here we defined: 
\begin{eqnarray}
\alpha _1 &=&\langle (\cos k_xa-\cos k_ya)^2\rangle \approx 1  \nonumber \\
\alpha _2(q) &=&\langle (\cos k_xa-\cos k_ya)^2\eta ^2\rangle \approx
a^2t^2q^2  \nonumber
\end{eqnarray}
The divergence of $\hat{L}({\bf q},\omega _k)$ at ${\bf q}=0,\omega _k=0$
determines the critical temperature. Finally: 
\begin{eqnarray}
{\hat{L}({\bf q},\omega _k)}^{-1}=\alpha _1\nu \left[ \varepsilon +\frac{%
|\omega _k|}{4\pi T}\psi ^{\prime }\left( \frac 12+\frac 1{4\pi T\tau
}\right) -\frac{\alpha _2(q)}{\alpha _14\pi T^2}\psi ^{\prime \prime }\left(
\frac 12+\frac 1{4\pi T\tau }\right) \right]  \label{prop2}
\end{eqnarray}
where 
\begin{eqnarray}
-\ln \frac{T_c}{{T_c}_0}=\psi \left( \frac 12+\frac 1{2\pi T\tau }\right)
-\psi \left( \frac 12\right)
\end{eqnarray}
is a shifted by impurities critical temperature and 
\begin{equation}
{T_c}_0=\frac{8\gamma t}\pi \exp \left( -\frac 2{g_0\nu \alpha _1}\right)
\end{equation}
(where $\gamma $ is the Euler constant) is the hypothetical critical
temperature in the absence of impurities. One can see that usual
non-magnetic impurities result in the reducing of critical temperature for $%
d-$wave superconductor in the same manner as magnetic impurities in the case
of $s-$wave superconductor. It is interesting to note, that taking into
account typical for HTS values of parameter $T\tau \approx 1$ one obtains a
huge suppression of $T_{c0}$. Thus $T_{c0}/T_c\ =\ 2.86,\ 1.85,$ and $1.43$
for $T\tau \ =\ 0.5,\ 1,$ and $2$, respectively, resulting in embarrassing
prediction of critical temperature in pure HTS materials at room
temperature. It seems that such prediction rules out completely the $d-$wave
scenario. Nevertheless, one should realize that HTS materials are obtained
by the isovalent substitutions from the parent compounds and structural
disorder here is an intrinsic feature. Thus considering the ``pure'' limit
for these materials has no sense.

\subsection{ d.c. conductivity}

The contribution of AL process to in-plane conductivity was calculated by
Yip \cite{Y90}. Current-current response function, ignoring factors of order
unity, is: 
\begin{equation}
Q^{{\rm AL}}_{xx} \approx e^2\int\frac{d^2q}{(2\pi)^2}T\sum_{\omega_k}
\left( \frac{\partial P({\bf q},\omega_k)}{\partial q_x}\right)^2 \hat L(%
{\bf q} ,\omega_k)\hat L({\bf q},\omega_k+\omega_{\nu})
\end{equation}
As a result: 
\begin{equation}  \label{ALA}
\sigma_{xx}^{{\rm AL}}\approx \psi^{\prime}\left(\frac{1}{2}+\frac{1}{4\pi
T\tau}\right) \frac{e^2}{d \varepsilon}
\end{equation}
Here $\psi-$function represents the pair-breaking effect of impurities which
reduces the magnitude of AL conductivity in the dirty case with respect to $%
s-$wave superconductor. As impurity vertex corrections do not contribute, the
anomalous MT contribution is simply absent \cite{Y90}. Other terms (DOS and
regular MT) give: 
\begin{eqnarray}  \label{DOS}
Q_{xx}^{{\rm DOS}}&\approx& e^2T\int\frac{d^2q}{(2\pi)^2} \hat L({\bf q}
,0)T\sum_{\omega_n} \int d\xi_k \nu \left(\frac{\partial \xi_k} {\partial 
{\bf k}}\right)^2 G({\bf k},\omega_n)^3 G(-{\bf k},-\omega_n)  \nonumber \\
\sigma_{xx}^{{\rm DOS}}&\approx& -\frac{e^2}{d}\ln\frac{1}{\varepsilon} \ \
\ \ \ \ \ \ \ \ \ \ \ 
\end{eqnarray}
In contrast to what happens in the $s-$wave case, the DOS contribution does not
depend on $\tau$. It is easy to extend (\ref{ALA}) and (\ref{DOS}) to the
case of layered superconductors. In this case one can calculate also
out-of-plane component of conductivity tensor. The results are similar to
those found above: both relevant contributions have the same temperature
dependence as for $s-$ wave case, but other dependence on $\tau$. Namely, AL
contribution is suppressed for dirty case, while the DOS contribution does not
depend upon $\tau$.

\subsection{NQR-NMR relaxation}

For $s-$wave the main contribution to spin relaxation originates from MT
process, if the latter is not suppressed by strong intrinsic pair-breaking.
Instead, the hierarchy of fluctuation contributions changes essentially if
one consider the $d-$wave superconductor. As both AL and MT processes are
absent, the negative DOS term becomes the only present. Corresponding
results for the d-wave superconductor with quasi-two-dimensional spectrum
are 
\begin{eqnarray}  \label{DOS2a}
\frac{\chi_s^{DOS}}{\chi_s^{(0)}}\approx\frac{(1/T_1T)^{DOS}_{B=0}}
{(1/T_1T)^{(0)}} \approx -\frac{T_c}{t}\ln\frac{2}{\varepsilon^{1/2}+(
\varepsilon+r)^{1/2}}
\end{eqnarray}
By comparing (\ref{DOS2a}) with the appropriate result for s-wave case from 
\cite{RV94} one can see that the type of pairing does not effect on the
magnitude and temperature dependence of the DOS contribution.

Finally, one can conclude that the only essential difference between $s-$
and $d-$ pairing states in context of fluctuation theory consists in the
absence of anomalous MT process in the latter. Therefore, studies based on
measurements of thermodynamical character do not give the possibility to
distinguish the type of pairing on the basis of fluctuation effects
experiments. In-plane and out-of-plane conductivities, the experimental data
in zero field do not manifest the signs of MT term. Nevertheless the
measurements of fluctuation conductivity cannot provide reliable tests for
possible $d-$ or $s-$pairing because the high values of $T_c$ determine
strong pair-breaking (at least due to the electron-phonon scattering the
expected minimal value of $\tau_\phi^{-1} \sim \frac{T}{\hbar} (\frac{T}{
\Theta_D})^2$). Thus the MT process is ineffective even in the case of $s-$
pairing scenario. Additionally, as the MT contribution to conductivity
temperature dependence is similar to AL one, they can hardly be
distinguished. The additional comparison with $s-$wave case $\psi -$function
gives the numerical factor of order unity, which cannot be tested
experimentally. Therefore, the only physical property related to
superconducting fluctuations which is due to the MT contribution is the NMR
relaxation rate, in which AL process does not contribute at all \cite{RV94}.

Within the context of the $s-$wave scenario both the positive MT singular
contribution \cite{KF89} and the negative DOS \cite{RV94} (independent on
phase-breaking) fluctuation renormalization exist. The concurrence of these
two effects should be observable in the relaxation measurements. In the case
of $d-$pairing, vice versa, the MT anomalous process in accordance with the
consideration presented above does not contribute at all and in the NMR
relaxation rate above $T_c$ a monotonous decrease with respect to Korringa
law has to be observed. Consequently, from point of view of fluctuation
theory, the sign of the correction to NMR relaxation rate above $T_c$ could
be a test for the symmetry of the order parameter. Further, as the form of
background (normal state) relaxation is unknown, the better choice for
experimental verification of the existence of the MT contribution is the study of
relative change of $1/T_1$ in the vicinity of transition induced by external
magnetic field. The main results for the fluctuation contributions to the
relaxation rate in external field in the case of $s-$wave superconductor are
the following. The DOS contribution has the same form as calculated above.
The MT contribution is given by 
\begin{equation}
\frac{(1/T_1T)_{B=0}^{MT}}{(1/T_1T)^{(0)}}\approx \frac 1{E_F\tau }\frac
1{\varepsilon -\gamma _\varphi }\ln \frac{\varepsilon ^{1/2}+(\varepsilon
+r)^{1/2}}{\gamma _\varphi ^{1/2}+(\gamma _\varphi +r)^{1/2}}  \label{MT}
\end{equation}
While the MT contribution is very sensitive to the presence of pair-breaking,
the simplest way to discriminate DOS and MT contributions is to apply an
external magnetic field, a relative small value being expected practically
to suppress the MT contribution.

\subsection{Experiment and discussion}

As it is clear from discussion above the most appropriate physical property
those measurement would be used as a test for a pairing state symmetry in
HTS is NQR-NMR relaxation rate. Thus below we present the result of recent
experimental work \cite{CLRV96} performed in order to get a clear evidence
of the role of superconducting fluctuations and meantime to achieve some
conclusion about the pairing mechanism in HTS.

At first, we discuss experimental details which are important for
interpretation of a such kind of experiments in HTS. From the recovery of
the $^{63}$Cu signal after RF saturation in both type of NQR and NMR
experiments one arrives at $1/T_1=2W$ given by 
\begin{equation}
2W={\frac{\Gamma^2}{2}}\int \langle h_+(t)h_-(0)\rangle e^{-i\omega_Rt}dt
\label{7}
\end{equation}
where $\Gamma$ is the gyromagnetic ratio of $^{63}$Cu nucleus. In (\ref{7}) $%
h_{\pm}$ are the components of the field at the nuclear site transverse with
respect to the $c$ axis both for NQR where the quantization axis is the $Z$
one of the electric field gradient tensor as well as for NMR when the
external field is along the $c$ axis itself. $\omega_R$, the resonance
frequency, is $\omega_R(0)= 31$ MHz in NQR (zero field) and $\omega_R(H)= 67$
MHz in NMR for $H=5.9$ T. In the following this difference will be
neglected. The fictitious field $\vec h$ can be related to the electron spin
operators $S_{\pm}$ through the electron-nucleus Hamiltonian and the
relaxation rate can formally be written in terms of a generalized
susceptibility. One can write 
\begin{eqnarray}
2W={\frac{\Gamma^2}{2}}k_BT\sum_{\vec k}A_k{\frac{\chi^{\prime\prime}(\vec
k,\omega_R)}{\omega_R}}\simeq  \nonumber \\
{\frac{\Gamma^2}{2}}k_BT\langle A_k\rangle_{BZ}\sum_{\vec k}{\frac{
\chi^{\prime\prime}(\vec k,\omega_R)}{\omega_R}}  \label{8}
\end{eqnarray}
where $A_k$ is a term involving the square of the Fourier transform of the
effective field $\vec h$, which can be averaged over the Brillouin zone. In
a Fermi gas picture in (\ref{8}) one can introduce, in the limit $%
\omega_R\rightarrow 0$, the static spin susceptibility $\chi^o(0,0)$ and the
density of states $N(0)$, namely $W\propto T[\chi^o(0,0)/(1-\alpha)]\hbar
N(0)$ ($\alpha$ is a Stoner-like enhancement factor) and the Korringa law $%
1/T_1\propto T$ is thus recovered. Having to discuss only the effect of the
external field $H$ on $W$ around $T_c^+$, in the assumption that the field
does not change appreciably the electron-nucleus Hamiltonian (as it is
proved by the equality $W(0)=W(H)$ for $T\gg T_c$) we will simply write $%
W/T\propto\chi_{ab}\equiv\chi$, where $\chi$ is then the $k$-integrated, $%
\omega_R\rightarrow 0$ contribution. The fact that in YBCO the Korringa law
is not obeyed above $T_c$ for $^{63}$Cu NQR-NMR relaxation rates (most
likely because of correlation effects) should not invalidate the comparison
of $W(0)$ to $W(H)$ in the relatively narrow temperature range of $10$ K
above $T_c$.

The $^{63}$Cu NQR and NMR relaxation measurements have been carried out in
an oriented powder of YBa$_2$Cu$_3$O$_{6.96}$. From the $^{63}$Cu NMR
line-width (FWHI$\simeq 40$ kHz) the spread in the direction of the $c$ axis
was estimated within $1-2$ degrees. The superconducting transition was
estimated $T_c=90.5$ K in zero field and $T_c=87.5$ K in the field of $5.9$
T used for NMR relaxation. In $^{63}$Cu NQR measurements the recovery of the
amplitude $s(t)$ of the echo signal at the time $t$ after complete
saturation of the $\pm 1/2\rightarrow \pm 3/2$ transition was confirmed of
exponential character, thus directly yielding the relaxation rate. The
temperature was measured with precision better than $25$ mK and the long
term stabilization during the measurements was within $100$ mK. In the
presence of the magnetic field the sample was aligned with $c\parallel \vec
H $. The echo signal for the central transition was used to monitor the
recovery of the nuclear magnetization after fast inversion of the $\pm 1/2$
populations. From the solution of the master equations one derives for the
recovery law $y(t)= 0.9exp(-12Wt) + 0.1exp(-2Wt)$ This law was observed to
be very well obeyed and the relaxation rate was extracted. The experimental
error in the evaluation of $W$ was estimated well within $5\ \%$. In Fig. 40
the experimental results for $2W(0)$ and $2W(H)$ around $T_c$
are reported. It is noted that $W(0)=W(H)$ for $T^{>}_{\sim} T_c + 15$ K
while the effect of the field, namely a decrease of the relaxation rate, is
present in a temperature range where paraconductivity and anomalies in the $%
c $ axis transport are observed.

Let us discuss the interpretation of the experimental results in terms of
the contributions related to superconducting fluctuations. The relative
decrease of $W$ around $T_c(H)$ and $T_c(0)$ induced by the field is about $%
20\ \%$. We can evaluate if this decrease is quantitatively consistent with
the picture of superconducting fluctuations. First one should estimate the
strength of pair-breaking. By using the reasonable values, $\tau \simeq
10^{-14}\ s$ and $\tau _\phi \simeq 2\times 10^{-13}\ s$, one finds that the
pair-breaking parameter $\gamma _\varphi $ is about 0.15. The effective
anisotropy parameter $r$ in YBCO can be estimated around 0.1. Finally one
can observe that in zero field the relative MT contribution is larger than
the absolute value of the DOS contribution by a factor 1.5, thus providing
the positive fluctuation correction to $W$. The effect of the field on these
two contributions can be expected as follows. In the case of strong
pair-breaking $\gamma _B>\{\varepsilon _B,r\}$ ($\varepsilon _B=\varepsilon
+\beta /2,\ \gamma _B=\gamma _\varphi +\beta /2,\ \beta =2B/H_{c2}(0)$)
there are two different regimes for the fluctuation corrections in magnetic
field \cite{BDKLV93}. Low-field regime ($\beta \ll \varepsilon $)
corresponds to a decrease quadratic in $\beta $ of the fluctuation
correction. In the high-field regime ($\varepsilon \ll \beta $) one can use
lowest Landau level approximation. In our experiment $\varepsilon \approx
0.05$, while $\beta $ is about 0.2. Since relative corrections to $W$
coincide with relative corrections to conductivity \cite{BDKLV93}, one
easily finds

\begin{eqnarray}
\frac{(W)_B^{DOS}}{(W)^{(0)}}(\beta \succeq \varepsilon )=- \frac{21\zeta (3)%
}{\pi^3} \frac 1{\epsilon_F\tau} \ln (1/\sqrt{\beta})  \label{9a} \\
\frac{(W)_B^{MT}}{(W)^{(0)}}(\beta \succeq \varepsilon ) = \frac{\pi}{%
8\epsilon _F\tau} \frac 1{\beta}.  \nonumber  \label{10}
\end{eqnarray}
Direct calculations according to these equations show that the MT contribution
is much more affected by the magnetic field. In fact from (\ref{9}) the
modification in $W^{DOS}$ induced by the field is small, of the order of $%
15\ \%$. On the contrary, according to (\ref{10}) the field reduces the MT
term to $1/4$ of its zero field value. In the case of d-wave pairing
symmetry, as it was already noted above, the MT contribution is absent and
the applied magnetic field results only in the slight reducing of the DOS
contribution in accordance with the equation similar with (\ref{9}). If the
decrease of $W$ by magnetic field is due to the reduction of the MT term,
then one deduces that zero-field total fluctuation correction to NMR
relaxation rate is positive and is within $10\ \%$ of background, due to the
partial cancellation of MT and DOS contribution. In a field of 6\ T the
total fluctuation correction becomes negative with an absolute value about $%
15\ \%$ of the background. The lack of detailed information on the
normal-state NMR relaxation rate in HTS does not allow one to achieve more
quantitative estimates. However, it should be remarked that the observation
of the decrease of $W$ in a magnetic field cannot be accounted for in the
case of $d$-wave pairing. Finally we would like to emphasize the following.
It has been recently pointed out by M\"{u}ller \cite{M95} that two type of
condensates, with different symmetry but the same transition temperature,
can exist in oxide superconductors. It is conceivable that also the spectrum
of the fluctuations of the order parameter above $T_c$ could reflect both
components, if present. Since the effect of the magnetic field discussed in
our paper works only on the component of $s$ symmetry, our conclusion is not
in contrast with the experimental evidences indicating $d-$wave pairing and
it could be considered a support to the hypothesis of order parameter having
simultaneously $s$ and $d$ symmetry.

Summarizing, from the accurate comparison of $^{63}$Cu NQR and NMR
relaxation rates in zero field and in a field of 5.9 T around $T_c^+$ in $%
YBa_2Cu_3O_x$, we have provided evidence of a contribution to $W$ related to
the superconducting fluctuations, in a temperature range of about $10$ K.
The experimental observation of a decrease in $W$ induced by the field is
consistent with the hypothesis of a strong reduction of the MT contribution 
\footnote{%
Concerning the behaviour of $W(H)$ there is still some controversy in
literature. Results analogous to those presented above, were obtained by the
Urbana group \cite{Urbana} and indicate that the magnetic field reduces the
relaxation rate in accordance with our consideration. However a comparison
of NQR and NMR measurements in the $YBa_2Cu_4O_x$ phase yields an opposite
behaviour of $W(H)$, which turns out to be larger than $W(0)$ \cite{B95,B92}%
. The reason of this discrepancy could be due either to the presence of a
negligible $s$-component in the orbital pairing of the $YBa_2Cu_4O_x$ or to
the role of diamagnetic terms \cite{RV97}.}. Since the MT contribution does
not exist in the case of a $d$-wave scenario, the interpretation of the
experimental finding is an indication in favor of the presence of an $s$
symmetry component in the orbital pairing.

\newpage

\section{Conclusions}

Several comments should be made in the conclusion. We have demonstrated that
the strong and narrow in energy scale renormalization of the one-electron
density of states in the vicinity of the Fermi level due to the Cooper
channel interelectron interaction manifests itself in experiments as the
wide enough pseudo-gap-like structures. The scale of these anomalies, as we
have seen above, is different for various phenomena ($eV=\pi T$ for tunnel
conductance, $\omega \sim \tau ^{-1}$ for optical conductivity, $T\sim T_c$
in NMR and conductivity measurements).

The results presented above are based on the Fermi liquid approach which is
formally expressed through the presence of a small parameter of the theory $%
Gi_2 \ln{\frac1{Gi_2}} \approx \frac{T_c}{E_F} \ln{\frac{E_F}{T_c}}\ll 1$ in
obtained results. Moving along the phase diagram of HTS from the metal
region (overdoped or optimally doped samples) to poor metals (underdoped
compounds) one can see that the small parameter of the perturbation theory
grows ($E_F \rightarrow 0$) causing the effects discussed to be more
pronounced.  Nevertheless, analysing the rapid growth of the normal state
anomalies with the decrease of the oxygen content below the optimal doping
concentration one can notice that it strongly overcomes our theoretical
prediction. We can attribute this discrepancy to the simplicity of the Fermi
surface model supposed above to be isotropic in the ab-plane. The ARPES
study of HTS shows the presence of the strong anisotropy of the Fermi
surface of such type and even the existence of two characteristic
energy scales $E_F \approx 0.3 eV$  and $\Delta \approx 0.01 eV$
(extended saddle point of the spectrum). So one can suppose that as the
oxygen concentration decreases below the optimal one, by some reasons the
massive part of the Fermi surface is "obliterated" and the crossover in
properties related with the special role of "slow" electrons of extended
saddle points takes place. Formally in this case the large value
$E_F\approx 0.3 eV$ in the denominator of the Ginzburg-Levanyuk parameter
has to be substituted by small $\Delta \approx 0.01 eV$ making rapidly the
perturbation approach to be unapplicable.

The existence of the nonequilibrium Cooper pairs in the normal metal phase
of HTS (the state with $<\Psi^2> \neq 0, <\Psi>=0, <\phi>=0$) resembles the
state of preformed Cooper pairs in the underdoped phase \cite{MLL96,GIL97}
($<\Psi^2> \neq 0,<\Psi> \neq 0, <\phi>=0$). Both of them are determined by
the presence of interelectron interaction and it would be interesting to
study the plausible "condensation" of the fluctuation pairs in the preformed
ones in the unknown land {\it ''hic sunt leones''} where $Gi_2 \approx 1$.
Such crossover qualitatively was discussed by M.Randeria \cite{R94}, but its
systematic study requires the formulation of the appropriate model.

In conclusion, we have presented here what we hope to be a comprehensive
overview of the theoretical and experimental facts which allow us to
attribute an important role in the behaviour of HTS (the metal part of its
phase diagram) to the fluctuation theory, and especially, to the frequently
neglected DOS contribution. We did show that a large number of non-trivial
experimental observed anomalies of HTS normal state properties can be
qualitatively, and often quantitatively, described under this model. Of
course, as a rule, this is not the only explanation which can be given for
these behaviours. However, in our opinion, alternative approaches often lack
the internal coherence and self consistency of the picture based on the
fluctuation theory. In particular, it is important that all the experimental
facts reviewed here can be explained by the fluctuation theory alone,
without need for any other {\it ad hoc} assumption or phenomenological
parameters, and that the values of all physical parameters extracted from
the fits are always in a good agreement with independent measurements. Even
more important is the fact that within the same approach one can explain so
many different properties of HTS, in the wide temperature range from tens of
degrees below to tens of degree above the critical temperature. It is very
unlikely that these circumstances are purely fortunate.

Therefore we believe that, although almost for all experiments discussed the
alternative explanations can be proposed, no other model, besides the
fluctuation theory, has the same appeal embracing in a single microscopic
view the wide spectrum of unusual properties which have been discussed in
this review.

\newpage

\section{ Acknowledgments}

The authors are grateful to all colleagues in collaboration with whom the
original results which constitute the background of this review were
obtained. We would like to thank A.A.Abrikosov, B.L.Altshuler, M.Ausloos,
A.Barone, A.I.Buzdin, C.Castellani, C.Di Castro, F.Federici, R.A.Klemm,
A.Koshelev, A.I.Larkin, K.Maki, D.Rainer, M.Randeria, A.Rigamonti,
Yu.N.Ovchinnikov, S.Pace, A.Paoletti, O.Rapp, C.Strinati, V.Tognetti,
R.Vaglio, L.Yu for the numerous discussions of the problems discussed above.
We would like to thank A.M.Cucolo for the elucidating discussion of the
tunneling experiments on HTS and permission to include in section 5.3 some
unpublished results. We are grateful to W.Lang for numerous discussions and
valuable contribution in section 8. Special thanks to R.A.Smith for the
attentive reading of the manuscript and valuable comments and suggestions.

Two of us (A.A.V. and D.V.L.) acknowledge the financial support of the
Collaborative NATO Grant N 941187 in the framework of which this article was
partially accomplished. All authors acknowledge support from the INTAS Grant
N 96 - 0452 and inter-university collaboration programme between "Tor
Vergata" University and Moscow Steel and Alloys Institute.

\newpage

\section{Appendix A: Calculation of the impurity vertex $\lambda(\vec
q,\omega_1,\omega_2)$}

Let us demonstrate the details of calculations of the impurity vertex $%
\lambda (\vec{q},\omega _1,\omega _2)$ which appears in the
particle-particle channel as the result of the averaging over the impurities
configuration. These calculations can be done in the frameworks of the
Abrikosov-Gorkov approach to the diagrammatic description of the alloys \cite
{AGD}.

In the assumption of relatively small impurity concentration $p_Fl\gg 1$
(what in practice means the mean free path up to tens of interatomic
distances) and for the spectra with dimensionality $D>1$ one can neglect the
contribution of the diagrams with intersecting impurities lines. In these
conditions the renormalized vertex $\lambda (\vec{q},\omega _1,\omega _2)$
is determined by the graphical equation of the ladder type (see Fig. 41):

Analytically the last equation can be written in the form 
\begin{eqnarray}
\lambda ^{-1}(\vec{q},\omega _1,\omega _2)=1-\frac 1{2\pi N(0)\tau }{\cal P}
( \vec{q},\omega _1,\omega _2),  \label{a1}
\end{eqnarray}
where 
\begin{eqnarray}
{\cal P}(\vec{q},\omega _1,\omega _2) &=&\int (d{\vec{p}})G({\vec{p}+\vec{q}}
,\omega _1)G({-\vec{p}},\omega _2)=  \label{a2} \\
&=&N(0)\langle {\ \int_{-\infty }^\infty \frac{d\xi (\vec{p})}{\left( \xi ( 
\vec{p})-i\tilde{\omega}_1+\Delta \xi (\vec{q},\vec{p})|_{|\vec{p}
|=p_F}\right) \left( \xi (\vec{p})-i\tilde{\omega}_2\right) }}\rangle
_{F.S.}=  \nonumber \\
&=&2\pi N(0)\Theta (-\omega _1\omega _2)\langle {\ \frac 1{|\tilde{\omega}
_1- \tilde{\omega}_2|+i~sign(\omega _1)\Delta \xi (\vec{q},\vec{p})|_{|\vec{%
p } |=p_F}}}\rangle _{F.S.}.  \nonumber
\end{eqnarray}
Here $\Theta (x)$ is the Heaviside step-function and 
\begin{eqnarray}
\Delta \xi (\vec{q},\vec{p})|_{|\vec{p}|=p_F}=[\xi ({\vec{p}+\vec{q}})-
\xi(- \vec{p})]|_{|\vec{p}|=p_F}.  \label{a3}
\end{eqnarray}

Now one has to reduce the formal averaging of the general expression (\ref
{a2}) over the Fermi surface ($\langle {\ ...}\rangle _{F.S.}$), to the
particular one of the expression (\ref{a3}). For an arbitrary $\vec{q}$ it
is possible to accomplish this program for some simple spectra types only.
Below we will show such exact solution for the important for HTS case of the
2D isotropic spectrum, but let us start before from the analysis of the
practically important calculation of $\lambda (\vec{q},\omega _1,\omega _2)$
for small momenta in the case of an arbitrary spectrum.

The angular averaging of (\ref{a2}) over the Fermi surface can be carried
out in general form for values of 
\begin{eqnarray}
\Delta \xi (\vec{q},\vec{p})|_{|\vec{p}|=p_F}\ll |\tilde{\omega}_1-\tilde
{\omega}_2|,  \label{a4}
\end{eqnarray}
what in the case of corrugated cylinder spectrum (\ref{d1}) means $%
q_{\parallel }\ll \min \{\xi _{ab}^{-1},l^{-1}\}$. Expanding the denominator
of (\ref{a2}) over the appropriate small parameter one can find: 
\begin{eqnarray}
\frac 1{2\pi N(0)}{\cal P}(\vec{q},\omega _1,\omega _2)=\frac{\Theta
(-\omega _1\omega _2)}{|\tilde{\omega}_1-\tilde{\omega}_2|}-\frac{\langle
(\Delta \xi (\vec{q},\vec{p})|_{|\vec{p}|=p_F})^2\rangle _{F.S.}} {|\tilde{
\omega}_1-\tilde{\omega}_2|^3}\Theta (-\omega _1\omega _2)  \label{a5}
\end{eqnarray}
and 
\begin{eqnarray}
\lambda (\vec{q},\omega _1,\omega _2)=\frac{|\tilde{\omega}_1-\tilde{\omega}
_2|}{|\omega _1-\omega _2|+\frac{\langle (\Delta \xi (\vec{q},\vec{p})|_{| 
\vec{p}|=p_F})^2\rangle _{F.S.}}{\tau |\tilde{\omega}_1-\tilde{\omega}_2|^2}
\Theta (-\omega _1\omega _2)}.  \label{a6}
\end{eqnarray}

It is easy to see that the restriction (\ref{a4}) is not too severe and
almost always can serve as a reasonable approximation for temperatures near $%
T_c$ (the exclusion is the very clean case, when the non-local fluctuation
effects take place (see section 11)). For instance, from the calculations of
section 6 one can see that not far from $T_c$ the effective propagator
momenta are determined by $|{\bf q}|_{eff}\sim [\xi _{ab}^{GL}(T)]^{-1}=\xi
_{ab}^{-1}\sqrt{\epsilon }\ll \xi _{ab}^{-1}$, while in Green functions the $%
\vec{q}$-dependence becomes important for much larger momenta $q\sim \min
\{\xi _{ab},l^{-1}\}$ (what is equivalent to the condition (\ref{a4})).

The last average in (\ref{a6}) can be easily calculated for some particular
types of spectra. For example in the cases of $2D$ and $3D$ isotropic
spectra it is expressed in terms of the diffusion coefficient ${\bf D}_D$ : 
\begin{eqnarray}
\langle (\Delta \xi (\vec{q},\vec{p})|_{|\vec{p}|=p_F})^2\rangle
_{F.S.D}=\tau ^{-1}{\bf D}_Dq^2=\frac{v_F^2q^2}D,  \label{a7}
\end{eqnarray}
while in the case of the quasi-two-dimensional electron motion (\ref{d1})
expression (\ref{a3}) may be presented by means of the action of the
diffusion operator $\hat{D}$ on the momentum $\vec{q}$: 
\begin{eqnarray}
\langle (\Delta \xi (\vec{q},\vec{p})|_{|\vec{p}|=p_F})^2\rangle _{F.S.}={\
\frac 12}(v_F^2{\bf q}^2+4J^2\sin ^2(q_zs/2))\equiv \tau ^{-1}\hat{{\bf D}}
q^2,  \label{a8}
\end{eqnarray}
The vertex $\lambda (\vec{q},\omega _1,\omega _2)$ in this case can be
written as 
\begin{eqnarray}
\lambda (\vec{q},\omega _1,\omega _2)=\frac{|\tilde{\omega}_1-\tilde{\omega}
_2|}{|\omega _1-\omega _2|+\frac{(v_F^2{\bf q}^2+4J^2\sin ^2(q_zs/2))}{2\tau
|\tilde{\omega}_1-\tilde{\omega}_2|^2}\Theta (-\omega _1\omega _2)}.
\label{a9}
\end{eqnarray}
and namely this expression was used through the review. In the dirty case $%
(T\tau \ll 1)$ this expression is reduced to (\ref{d3}).

Finally we discuss the exact calculation of the vertex $\lambda (\vec{q}
,\omega _1,\omega _2)$ in the case of $2D$ electron spectrum which turns out
to be necessary for the consideration of the Maki-Thompson contribution in
the extra-clean limit $l\gg \xi _{ab}^{GL}(T)$. In this case the $\vec{q}$
-dependence of the vertices in the domain of small momenta becomes of the
same importance than that one which comes from the propagator, but because
of the large value of $l$ the condition (\ref{a4}) is violated. This is why
we express (\ref{a3}) in the exact for $2D$ case form

\begin{eqnarray}
\Delta \xi (\vec{q},\vec{p})|_{|\vec{p}|=p_F}=[\xi ({\vec{p}+\vec{q}})-\xi
(- \vec{p})]|_{|\vec{p}|=p_F}.=v_Fq\cos \varphi ,  \label{a10}
\end{eqnarray}
which is valid for any $v_Fq\ll E_F$ , and average (\ref{a2}) without any
expansion. The angular average in this case is reduced to the calculation of
the integral 
\begin{eqnarray}
{\cal P}(\vec{q},\omega _1,\omega _2)=\frac{2\pi N(0)\Theta (-\omega
_1\omega _2)sign(\omega _1)}{iv_Fq}{\ \int_0^{2\pi }\ }\frac{d\varphi }{2\pi 
}{\frac 1{\cos \varphi -\frac{i\cdot sign(\omega _1)|\tilde{\omega}_1-\tilde{
\omega}_2|}{v_Fq}.}}  \label{a11}
\end{eqnarray}
which can be easily carried out. Really, the integral in (\ref{a11}) is
calculated by means of the substitution $z=e^{i\varphi }$: 
\begin{eqnarray}
{\ \int_0^{2\pi }\ }\frac{d\varphi }{2\pi }{\frac 1{\cos \varphi .-ia}=} 
\frac{i\cdot sign(a)}{\sqrt{1+a^2}}  \label{a12}
\end{eqnarray}
what gives for ${\cal P}(\vec{q},\omega _1,\omega _2):$ 
\begin{eqnarray}
{\cal P}(\vec{q},\omega _1,\omega _2)=\frac{2\pi N(0)\Theta (-\omega
_1\omega _2)}{\sqrt{v_F^2q^2+(\tilde{\omega}_1-\tilde{\omega}_2)^2}}.
\label{a13}
\end{eqnarray}
The proper exact expression for $\lambda (\vec{q},\omega _1,\omega _2)$ can
be written as

\begin{equation}
\lambda (\vec{q},\omega _1,\omega _2)=\left( 1-{\frac{\Theta (-\omega
_1,\omega _2)}{{\tau \sqrt{(\tilde{\omega}_1-\tilde{\omega}_2)^2+v_F^2q^2}}}}
\right) ^{-1}.  \label{a14}
\end{equation}
which was used for the study of non-local limit in section 11. One can see
that this expression can be reduced to (\ref{a6})- (\ref{a7}) in the case of 
$v_Fq\ll |\tilde{\omega}_1-\tilde{\omega}_2|$.

\newpage

\section{ Appendix B: Calculation of the fluctuation propagator $L(\vec{q}
,\omega_\mu)$}

In this section we discuss the calculation of the fluctuation propagator $%
L(q,\omega _\mu )$ which is nothing else that the two-particle Green
function of electrons interacting in the Cooper channel. Graphically it is
presented by the sum of the diagrams with two entrances and two exits. In
the BCS theory it is demonstrated that the ladder type diagrams with the
accuracy of $\ln ^{-1}(\frac{\omega _D}{2\pi T})$ turn out to be of first
importance in the calculation of $L(q,\omega _\mu ).$ This statement can be
applied for any type of weak interaction with characteristic energy $%
\varepsilon _0\gg T_c$ in the range of temperatures $T_c\leq T\ll
\varepsilon _0,$ so in purpose to calculate the fluctuation propagator $%
L(q,\omega _\mu )$ one has to solve the ladder type Dyson equation presented
in the Fig. 42: 

Analytically this equation can be written in the form
\begin{eqnarray}
L^{-1}(q,\omega _\mu )=g^{-1}-\Pi (q,\omega _\mu )  \label{b1}
\end{eqnarray}
where $g$ is the effective constant of the electron-electron interaction in
the Cooper channel and the {\it polarization operator} $\Pi (q,\omega _\mu )$
($\omega _\mu \geq 0$) is determined as the two one-electron Green's
functions correlator averaged over impurities positions, which can be
expressed in terms of the introduced above functions $\lambda (q,\omega
_{n+\mu },\omega _{-n})$ and ${\cal P}(\vec{q},\omega _{n+\mu },\omega
_{-n}) $ (see Appendix 1) :

\begin{eqnarray}
\Pi (q,\omega _\mu ) &=&T\sum_{\omega _n}\lambda (q,\omega _{n+\mu },\omega
_{-n})\int {\frac{{d^3p}}{{(2\pi )^3}}}G(p+q,\omega _{n+\mu })G(-p,\omega
_{-n})=  \label{b2} \\
&=&T\sum_{\omega _n}\lambda (q,\omega _{n+\mu },\omega _{-n}){\cal P} (\vec{
q },\omega _{n+\mu },\omega _{-n}),  \nonumber
\end{eqnarray}

One can easily calculate $L(q,\omega _\mu )$ for the most important case of
small momenta $\Delta \xi (\vec{q},\vec{p})|_{|\vec{p}|=p_F}\ll |\tilde{
\omega}_{n+\mu }-\tilde{\omega}_{-n}|\sim \max \{T,\tau ^{-1}\}.$ Using the
proper expressions for $\lambda (q,\omega _{n+\mu },\omega _{-n})$ and $%
{\cal P}(\vec{q},\omega _{n+\mu },\omega _{-n})$ it is possible to find:

\begin{eqnarray}
{\Pi }(\vec{q},\omega _\mu ) &=&2\pi TN(0)\sum_{\omega _n}\Theta (-\omega
_{n+\mu }\omega _{-n})\frac{1-\frac{\langle (\Delta \xi (\vec{q},\vec{p}
)|_{| \vec{p}|=p_F})^2\rangle _{F.S.}}{|\tilde{\omega}_{n+\mu }-\tilde{%
\omega } _{-n}|^2}}{|\omega _{-n}-\omega _{n+\mu }|+\frac{\langle (\Delta
\xi (\vec{ q} ,\vec{p})|_{|\vec{p}|=p_F})^2\rangle _{F.S.}}{\tau |\tilde{%
\omega}_{n+\mu }- \tilde{\omega}_{-n}|^2}}=  \nonumber  \label{b4} \\
&=&4\pi TN(0)[\sum_{\omega _n=0}\frac 1{2\omega _n+\omega _\mu }{\ -}
\label{b4a} \\
&&-\langle (\Delta \xi (\vec{q},\vec{p})|_{|\vec{p}|=p_F})^2\rangle
_{F.S.}\sum_{\omega _n=0}\frac 1{(2\omega _n+\omega _\mu )(2\omega _n+\omega
_\mu +\frac 1\tau )^2}]  \nonumber  \label{b5}
\end{eqnarray}
The calculation of sums can be accomplished in terms of the logarithmic
derivative of $\Gamma $-function $\psi (x)$ what leads to the explicit
expression for $\Pi (q,\omega _\mu ):$ 
\begin{eqnarray}
\frac 1{N(0)}\Pi (q,\omega _\mu ) &=&\psi (\frac 12+\frac{\omega _\mu }{4\pi
T}+\frac{\varepsilon _0}{2\pi T})-\psi (\frac 12+\frac{\omega _\mu }{4\pi T}
)-  \label{b6} \\
- &&\tau ^2\left[ \psi (\frac 12+\frac 1{4\pi T\tau })-\psi (\frac 12)-\frac
1{4\pi T\tau }\psi ^{^{\prime }}(\frac 12)\right] \langle (\Delta \xi (\vec{
q },\vec{p})|_{|\vec{p}|=p_F})^2\rangle _{F.S.}  \nonumber
\end{eqnarray}
The term proportional to $\langle (\Delta \xi (\vec{q},\vec{p})|_{|\vec{p}
|=p_F})^2\rangle _{F.S.}$ may be expressed in terms of the $\eta $
-coefficient of Ginzburg-Landau theory or diffusion operator $\hat{{\bf D}}:$
\begin{equation}
\frac 1{N(0)}\Pi (q,\omega _\mu )=\psi (\frac 12+\frac{\omega _\mu }{4\pi T}
+ \frac{\varepsilon _0}{2\pi T})-\psi (\frac 12+\frac{\omega _\mu }{4\pi T}
)- \hat{\eta}\vec{q}^2.  \nonumber
\end{equation}

The definition of the critical temperature as the temperature $T_c$ at which
the pole of $L(0,0,T_c)$ takes place 
\begin{eqnarray}
L^{-1}(0,0,T_c)=g^{-1}-\Pi (0,0,T_c)=0  \label{b7}
\end{eqnarray}
permits us to express the fluctuation propagator in terms of the reduced
temperature $\epsilon =\ln (\frac T{T_c}):$ 
\begin{eqnarray}
L(q,\omega _\mu )=-\frac 1{N(0)}\frac 1{\ln (\frac T{T_c})+\psi (\frac 12+ 
\frac{\omega _\mu }{4\pi T})-\psi (\frac 12)+\hat{\eta}\vec{q}^2}  \label{b8}
\end{eqnarray}
One has to remember that this expression was carried out in the assumption
of the small momenta $\Delta \xi (\vec{q},\vec{p})|_{|\vec{p}|=p_F}\ll | 
\tilde{\omega}_{n+\mu }-\tilde{\omega}_{-n}|\sim \max \{T,\tau ^{-1}\},$ so
the range of its applicability is restricted by the Ginzburg-Landau region
of temperatures $\epsilon =\ln (\frac T{T_c})\ll 1,$ where the integrated
functions in diagrammatic expressions have the singularities at small
momenta.

Nevertheless it is possible to generalize the last expression for arbitrary
momenta. The first hint appears in the dirty case when $\eta \sim \psi
^{^{\prime }}(\frac 12)$ and one can suppose that the term $\hat{\eta}\vec{q}
^2$ is nothing else as the first term of the expansion of the function $\psi
(\frac 12+\frac{\omega _\mu +\hat{{\bf D}}\vec{q}^2}{4\pi T}).$ One can find
the confirmation of this hypothesis in the fact that the analytically
continued expansion of the propagator

\begin{eqnarray}
L^R(q,-i\omega )=-\frac 1{N(0)}\frac 1{\epsilon -\frac{i\pi \omega }{8T}+ 
\hat{\eta}\vec{q}^2},  \label{b9}
\end{eqnarray}
which is valid for the region $\Delta \xi (\vec{q},\vec{p})|_{|\vec{p}
|=p_F}\ll \max \{T,\tau ^{-1}\}$ and $\omega \ll T,\ $ is nothing else as
the Time Dependent Ginzburg-Landau (TDGL) equation's fundamental solution.
As it is well known \cite{parks69} the Maki-De Gennes equation, which in the
case discussed has the form

\begin{eqnarray}
\left( \ln (\frac T{T_c})+\psi (\frac 12+\frac{\omega _\mu +\hat{{\bf D}}( 
\vec{q}-\frac{2e}cA)^2}{4\pi T})-\psi (\frac 12)\right) \Phi =0  \label{b10}
\end{eqnarray}
serves as the generalization of TDGL equation . So it is natural to suppose
that the most general form of the propagator, valid for arbitrary momenta
and frequencies, has the form

\begin{eqnarray}
L(q,\omega _\mu )=-\frac 1{N(0)}\frac 1{\ln (\frac T{T_c})+\psi (\frac 12+ 
\frac{\omega _\mu +\Omega _L+\hat{{\bf D}}\vec{q}^2}{4\pi T})-\psi (\frac
12)}  \label{b11}
\end{eqnarray}
(The Larmour frequency $\Omega _L$ appeared in (\ref{b11}) as electron
eigen-energy of the Landau state in magnetic field).

The second confirmation of the correctness of the hypothesis proposed is the
direct calculation of the polarization operator from the formula (\ref{b2})
for the case of a $2D$ spectrum, when it is possible to carry out exactly
the angular average over the Fermi surface in ${\cal P}(\vec{q},\omega
_{n+\mu },\omega _{-n})$ . Really, using the definition of $\lambda
(q,\omega _{n+\mu },\omega _{-n})$ we can rewrite the polarization operator
in the form:

\begin{equation}
\Pi (q,\omega _\mu )=T\sum_{\omega _n}\frac 1{\left[ {\cal P}(\vec{q},\omega
_{n+\mu },\omega _{-n})\right] ^{-1}-\frac 1{2\pi N(0)\tau }}  \nonumber
\end{equation}
In Appendix 1 for $2D$ spectrum case it was found$:$

\begin{eqnarray}
{\cal P}(\vec{q},\omega _1,\omega _2)=\frac{2\pi N(0)\Theta (-\omega
_1\omega _2)}{\sqrt{v_F^2q^2+(\tilde{\omega}_1-\tilde{\omega}_2)^2}},
\label{b12}
\end{eqnarray}
what gives for the polarization operator

\begin{equation}
\Pi (q,\omega _\mu )=N(0)\sum_{n=0}\frac 1{\sqrt{\left( n+\frac 12+\frac{%
\omega _\mu }{4\pi T}+\frac 1{4\pi T\tau} \right) ^2+\frac{v^2\vec{q}^2}{16\pi ^2T^2}}%
-\frac 1{4\pi T\tau}}  \nonumber
\end{equation}
 For $\max \{\frac{v^2\vec{q}^2}{16\pi
^2T^2},\frac{\omega _\mu }{4\pi T}\}\gg 1$ the summation can be substituted
by integration and one can find (for simplicity we put $ T\tau \rightarrow
\infty $): 
\begin{eqnarray}
L(q,\omega _\mu )=-\frac 1{N(0)}\times \frac 1{\ln (\frac T{T_c})+\ln \left(
\frac 12+\frac{\omega _\mu }{4\pi T}+\sqrt{(\frac 12+\frac{\omega _\mu }{%
4\pi T})^2+\frac{v^2\vec{q}^2}{16\pi ^2T^2}}\right) -\psi (\frac 12)}.
\label{b13}
\end{eqnarray}
Taking into account that $\psi (x\gg 1)\rightarrow \ln x$ one can see that ( 
\ref{b13}), in the limit of $\max \{\frac{v^2\vec{q}^2}{16\pi ^2T^2},\frac{%
\omega _\mu }{4\pi T}\}\gg 1,$ coincides with (\ref{b11}).

\newpage

\section{Glossary}

AL term ( paraconductivity, $\delta \sigma _{AL})$ is Aslamazov-Larkin
contribution;

ARPES is angular resolved photo-emission;

$A_\alpha $ is a vector-potential;

$a$ is the interatomic distance;

$\alpha =\frac 1{4m\eta _D}$ is the coefficient of the Ginzburg-Landau
theory;

$\alpha _1=\langle (\cos k_xa-\cos k_ya)^2\rangle \approx 1$ is the
coefficient of the Ginzburg-Landau theory for d-pairing;

$\alpha _2(q)=\langle (\cos k_xa-\cos k_ya)^2\eta ^2\rangle \approx
a^2t^2q^2 $ is the kinetic energy part of the Ginzburg-Landau theory for
d-pairing;

$\alpha _q={\frac{{4\eta }_2{\hat{{\bf D}}q^2}}{{\pi ^2v_F^2\tau }}}\ $ is
dimensionless kinetic energy in the fluctuation propagator;

$B_\alpha (q,\omega _\mu ,\omega _\nu )$ is the integrated three Green
functions block of AL contribution;

$\beta =B/[2T_c\,|dB_{c2}/dT|_{Tc}]=4\eta eB=\frac B{B_{c2}}\ $is reduced
magnetic field;

$B$ is magnetic field;

$B_{c2}$ is upper critical magnetic field at zero temperature;

$\Gamma $ is the gyromagnetic ratio;

$\gamma $ is the Euler constant;

$\gamma _\varphi ={\frac{{2\eta }}{{v_F^2\tau \tau _\phi }}}\rightarrow {%
\frac{{\pi }}{{8T\tau _\phi }}}$ is the phase-breaking rate related with $%
\tau _\phi ;$

$\gamma _B=\gamma _\varphi +\beta /2$ is the phase-breaking rate related
with $\tau _\phi $ in the presence of a magnetic field;

${\bf D}$ is diffusion coefficient; ${\bf D}\sim \frac{p_Fl}m\sim \frac{%
E_F\tau }m\rightarrow \frac{E_F}{mT_c}$ is its generalization from dirty to
clean case;

${\bf D}_{\alpha \beta }$ is the diffusion tensor;

DOS term ($\delta \sigma _{DOS})$ is the density of states fluctuation
contribution;

$D$ is the space dimensionality;

$D_1(z)=2\ln \left[ \sqrt{z}+\sqrt{\left( z+r\right) }\right] $ is the
function of complex variable;

$D_2(z)=-\sqrt{z(z+r)}$ is the function of complex variable;

$\Delta D_1\left( z\right) =D_1\left( z\right) -D_1\left( \varepsilon
\right) $ is the function of complex variable;

$\Delta D_2\left( z\right) =D_2\left( z\right) -D_2\left( \varepsilon
\right) $ is the function of complex variable;

$d$ is the films thickness;

$E_F$ is the Fermi energy;

$E({\bf p})={\bf p}^2/(2m)$, is the kinetic energy of $2D$ free electron;

$E_0(T)$ is the energy scale at which the DOS renormalization occurs; $%
E_0^{(cl)}\sim \sqrt{T_c(T-T_c)},$

$E_0^{(d)}\sim T-T_c;$

$\varepsilon =\ln (T/T_c)$ is reduced temperature;

$\varepsilon _B=\varepsilon +\beta /2$ is reduced temperature in magnetic
field;

$\zeta (x)$ is the Riemann zeta function;

$f_{as}\sim T/E_F$, the symmetry factor entering in the thermo-epf;

$f(\varepsilon )$ is a universal function entering in $2D$ AL
paraconductivity;

$Gi_{(D)}=Gi_{(D)}(4\pi T\tau )$ is Ginzburg-Levanyuk parameter;

$G(V=0,T)$ is the tunnel junction zero-bias conductance;

$G(V)$ is the differential tunnel conductance;

$G_n(0)$ is the background value of the Ohmic conductance supposed to be
bias independent;

$\delta G(V)=G(V)-G_n(0)$;

$G(p,\omega _n)={\frac 1{{i\tilde{\omega}_n-\xi (p)}}}$ is the single
quasiparticle normal state Green's function;

$I_{qp}(V)$ is tunneling current;

$\int (d{\bf q})=\int d^Dq/(2\pi )^D;$

$\int d^3q\equiv \int d^2{\bf q}\int_{-\pi /s}^{\pi /s}dq_z$ is momentum
space integral transformation for a layered superconductor;

$I_{\alpha \beta }(q,\omega _\mu ,\omega _\nu )$ is the integrated four
Green functions block entering in the MT contribution;

$\vec{i},\vec{j},\vec{l}$ are the unit vectors along the axes;

$J$ is a hopping integral describing the Josephson interaction between
layers;

$J^{{\rm h}}$ and $J^{{\rm e}}$ are heat-current and electric current
operators in Heisenberg representation;

$\eta _D=-\frac{v_F^2\tau ^2}D\psi \left( {\frac 12}+{\frac 1{{4\pi \tau T}}}%
\right) -\psi \left( {\frac 12}\right) -{\frac 1{{4\pi \tau T}}}\psi
^{^{\prime }}({\frac 12)}$ is the coefficient of the gradient term of $D$%
-dimensional Ginzburg-Landau theory; $\eta \equiv \eta _2;$

HTS - high temperature superconductors;

$H_c(0)$ is the zero-temperature thermodynamical critical field;

$h_{\pm }$ are the components of the magnetic field at the nuclear site
transverse with respect to the $c$ axis both for NQR as well as for NMR;

$K(\omega _\nu )=\int (d{\bf k})\chi _{+-}({\bf k},\omega _\nu )$;

$\kappa ={\frac{{r_1+r_2}}{{r}},}\tilde{\kappa}$ are the functions of the
impurity concentration entering in the DOS contribution in the case of an
arbitrary impurity concentration;

$\hat{\kappa}\left( \omega ,T,\tau ^{-1}\right) $ is the function
determinating the frequency and temperature dependence of the pseudogap in
optical conductivity;

$L_{x,y,z}$ are the sample dimensions in appropriate directions;

$l$ is the intralayer mean-free path;

$L_T=\sqrt{\frac{{\bf D}}T}$ is the diffusion length;

$L(q,\omega _\mu )$ is the fluctuation propagator; in the absence of the
magnetic field and in the vicinity of $T_c$, it has the form : $%
L^{-1}(q,\omega _\mu )=-N(0)\left[ \varepsilon +\psi \left( {\frac 12}+{%
\frac{{\omega _\mu }}{{4\pi T}}}+\alpha _q\right) -\psi \left( {\frac 12}%
\right) \right] ;$

$L_{{\bf kk}^{\prime }}({\bf q})=\hat{L}({\bf q})(\cos k_xa-\cos k_ya)(\cos
k_x^{\prime }a-\cos k_y^{\prime }a)$ is the fluctuation propagator in the
case of d-pairing;

$\lambda (q,\omega _n,\omega _{n^{\prime }})$ is the impurity vertex in the
Cooper channel;

$\mu $ is a chemical potential;

MT term, anomalous and regular ($\delta \sigma _{MT}=$ $\delta \sigma
^{MT(reg)}$+$\delta \sigma ^{MT(an)}$), is the Maki-Thompson contribution;

${\cal N}_{c.p.}$ is the concentration of Cooper pairs;

$\delta {\cal N}_{s.i.}$ is the concentration of interfering Cooper pairs;

${\cal N}_e^{(2)}=\frac m{2\pi }E_F$ is the one-electron concentration in 2D
case;

$N_{(2)}=\frac m{2\pi }$ is the density of states for 2D electron gas;

$N_L(0)$, $N_R(0)$ are densities of states at the Fermi levels in each of
electrodes of tunnel junction in the absence of interaction;

$\delta N_{fl}^{(2)}(E,\varepsilon )$ is the fluctuation correction to the
density of states of the $2D$ electron gas;

$n_i$ is the impurity concentration;

$\xi (p)=E({\bf p})+J\cos (p_zs)-E_F$ is the spectrum of corrugated cylinder
type;

$\xi _0$ is the superconductor coherence length at zero temperature: $\xi
_{0,cl}^2=\frac{7\zeta (3)}{12\pi ^2T_c^2}\frac{E_F}{2m}$ in the clean case

and $\xi _{0,d}^2=\frac{\pi {\bf D}}{8T_c}$ in the dirty case;

$\xi (T)=\xi _0\left( \frac{T_c}{T-T_c}\right) ^{1/2}$ is the temperature
dependent coherence length of the Ginzburg-Landau

theory;

$\xi _{ab}$ is the in-plane BCS coherence length of layered superconductor;

$\xi _c$ is the out of plane BCS coherence length of layered superconductor;

$\langle [\xi (p)-\xi (q-p)]^2\rangle \equiv \tau ^{-1}\hat{{\bf D}}q^2={%
\frac 12}\Bigl(v_F^2{\bf q}^2+4J^2\sin ^2(q_zs/2)\Bigr);$

${\cal P}(\vec{q},\omega _1,\omega _2)$ is the two one-electron Green's
function loop integrated over internal electron momentum;

$\Pi (q,\omega _\mu )$ is the polarization operator defined as the two
one-electron Green's function correlator averaged over impurities positions;

$p\equiv ({\bf p},p_z)$ is a vector of momentum space;

${\bf p}\equiv (p_x,p_y)$ is a two-dimensional intralayer wave-vector;

$Q^{{\rm (eh)R}}(\omega )$ is the Fourier representation of the retarded
correlation function of two current operators;

$Q^{{\rm (eh)R}}(X-X^{\prime })=-\Theta (t-t^{\prime })\langle \langle
\left[ J^{{\rm h}}(X),J^{{\rm e}}(X^{\prime })\right] \rangle \rangle :X=(%
{\bf r},t)$ and $\langle \langle \cdots \rangle \rangle $ represents both
thermodynamical averaging and averaging over random impurity positions;

$Q_{\alpha \beta }(\omega _\nu )$ is the electromagnetic response operator;

$R_n$ is the Ohmic resistance for unit area;

$\rho _{ab}(T)$ and $\rho _c(T)$ the components of the resistivity tensor in 
$ab$-plane and along $c$-axis;

$r=4\eta _2J^2/v_F^2=4\xi _{\perp }^2(0)/s^2$ is the Lawrence-Doniach
anisotropy parameter;

$r_1, r_2 $ are the functions of impurity concentration accounting for the
contributions of DOS diagrams;

$S=-\vartheta /\sigma $ is the Seebeck coefficient;

$\sigma _{\alpha \beta }$ is the conductivity tensor;

$\sigma ^{{\rm n}}(\omega )$ is Drude conductivity;

$T_{c0}$ is the BCS value of critical temperature;

$T_c^{*}$ is the critical temperature reduced by the effect of fluctuations;

$\delta T_c$ is the fluctuation shift of the critical temperature;

$T_0$ and $T_1$ are respectively the mean energy and the half width of the
energy spread of the resonant defects referred to Fermi level;.

${\cal T}$ is the period of electron Bloch oscillations;

$1/T_1=2W={\frac{\Gamma ^2}2}\int \langle h_{+}(t)h_{-}(0)\rangle
e^{-i\omega _Rt}dt$ is the NMR relaxation rate;

$\nabla T$ is a temperature gradient;

$t_\xi ^{-1}={\bf D}\xi ^{-2}\sim \tau _{GL}^{-1}\sim T-T_c$ is the inverse
of the time necessary for the electron to diffuse over a distance equal to
the coherence length $\xi (T);$

$t_\xi ^{-1}\sim v_F\xi ^{-1}\sim \sqrt{T_c(T-T_c)}$ the same value for the
ballistic motion;

$t$ is the transfer integral between the nearest-neighbor sites in the
theory of d-pairing;

$\tau _\phi (\varepsilon )$ is the one-electron phase-breaking time;

$\tau _{GL}=\pi \hbar /{8k_B(T-T_c)}$ is the characteristic time of the
Time-Dependent Ginzburg-Landau Theory; plays the role of a fluctuation
Cooper pair lifetime in the vicinity of $T_c;$

$\tau $ is the quasiparticle scattering time;

$\tau _{hop}$ is the characteristic time of anisotropic diffusion from one
layer to the neighboring one;

$\Theta (x)$ is the Heaviside step function;

{$\Theta _D$ is the Debye temperature;}

$\vartheta $ is the thermoelectric coefficient;

$\Phi _0$ is the elementary magnetic flux;

$V$ is the volume of the sample;

$V({\bf k},{\bf k^{\prime }})=g_0(\cos k_xa-\cos k_ya)(\cos k_x^{\prime
}a-\cos k_y^{\prime }a)$ is the interaction potential for the case of
d-pairing;

$v_\alpha (p)={\frac{{\partial \xi (p)}}{{\partial p_\alpha }}}$.is the
quasiparticle velocity; $v_F$ is the Fermi velocity in $ab$-plane;

$v_z(p)={\frac{{\partial \xi (p)}}{{\partial p_z}}}=-Js\sin (p_zs)$ is the
electron velocity along the $c$-axis direction;

$\chi _{+-}^{(R)}({\bf k},\omega )=\chi _{+-}({\bf k},i\omega _\nu
\rightarrow \omega +i0^{+})$ is the dynamic susceptibility;

${\chi _s}=\chi _{+-}^{(R)}({\bf k}\rightarrow 0,\omega =0)$ is the spin
susceptibility;

$\Psi ({\vec{r}})$ - superconducting order parameter, $\Psi _{\vec{k}}={%
\frac 1{\sqrt{V}}}\int \Psi (\vec{r})\exp ^{-i\vec{k}\vec{r}}dV$ is its
Fourier transform;

$\psi (x)$ and $\psi ^{(n)}(x)$ are the digamma function and its derivatives
respectively;

$\Omega _{(fl)}$ is the fluctuation part of the thermodynamical potential;

$\Omega _L$ is the Larmor frequency;

$\omega _\nu =(2\nu +1)\pi T$ are the Matsubara frequencies;

$\tilde{\omega}_n=\omega _n[1+1/(2|\omega _n|\tau )]$ are the Matsubara
frequencies renormalized by the impurity scattering;

$\omega _{n\pm \nu }=\omega _n\pm \omega _\nu $, is the sum of Matsubara
frequencies;

$\tilde{\omega}=\displaystyle{\frac{\pi \omega }{16(T-T_c)}}$ is
dimensionless electromagnetic field frequency;

$\omega _R$ is the resonance frequency.

\newpage

{\bf FIGURE CAPTIONS}

Figure 1. The temperature dependence of the averaged order parameter after
renormalisation by fluctuations. The dash-dotted line represents the BCS
curve for the unperturbed transition temperature, $T_c$. The dashed line is
the BCS curve for the fluctuation-renormalised transition temperature, $%
T_c^* $. The solid line is the fluctuation-renormalised average of the
squared order parameter, $\left<|\Psi_{fl}|^2\right>(T)$, as given in Eq. (%
\ref{delta2}).

Figure 2. The theoretical prediction of Eq. (\ref{ds2Dc}) for the normalised
correction $\delta N(E)$ to the single-particle density of states vs energy $%
E$ (measured in units of $T_c$) for a clean two-dimensional superconductor
above $T_c$. $\tau_{GL}^{-1}$ assumes the values $0.02T_c$, $0.04T_c$ and $%
0.06T_c$. In the inset the dependence of the energy at which $\delta N(E)$
is a maximum, $E_0$, on $\tau_{GL}^{-1}$ is shown \cite{CCRV90}. 

Figure 3. The theoretical prediction of Eq. (\ref{flcon1}) for the
fluctuation-induced zero-bias anomaly in tunnel-junction resistance as a
function of voltage for reduced temperatures $t=1.05$ (top curve), $t=1.08$
(middle curve) and $t=1.12$ (bottom curve). The insert shows the
experimentally observed differential resistance as a function of voltage in
an Al-I-Sn junction just above the transition temperature \cite
{Khachat}.

Figure 4. Theoretical fit (solid line) of the experimentally observed
temperature dependence zero-bias conductance of the YBaCuO/Pb junction of
Ref. \cite{AMC96}. The theory used is that of Eqs. (\ref{flcon1})
and (\ref{q2D}) with $r=0.07$ and $T_c=90K$. The inset shows the same
results in a wider region of temperature.

Figure 5. Normalised tunneling conductance data of the BSSCO-2212/Pb
junctions of Ref. \cite{tao97}. Pseudo-gap type non-linearities are
seen in the temperature range from $T_c=87-89K$ to $110K$.

Figure 6. Theoretical fit of the $G(V,T)$ data from the BSSCO-2212 junctions
of Ref. \cite{SKN97}. The solid lines are the experimental data at $%
T=90K$ and $100K$ for a sample where $T_c=87K$. The thin lines are the
theory of Eq. (\ref{flcon1}) with parameters $r=0$ and $Gi_{(2)}=0.008$.

Figure 7. The electronic spectrum for our model of HTSC materials as written
in Eqn. (\ref{d1}). The Fermi surface takes the form of a corrugated
cylinder.

Figure 8. The origin of the Aslamazov-Larkin and Maki-Thompson contributions
to fluctuation conductivity in terms of electron-electron pairing. (a)
Correlations between electrons of opposite spin and momenta moving in a
straight line in opposite directions lead to the Aslamazov-Larkin terms. (b)
Correlations between electrons of opposite spin moving in opposite
directions along a self-intersecting trajectory in real space lead to the
Maki-Thompson terms.

Figure 9. Feynman diagrams for the leading-order contributions to
fluctuation conductivity. Wavy lines are fluctuation propagators, thin solid
lines with arrows are impurity-averaged normal-state Green's functions,
shaded semicircles are vertex corrections arising from impurities, dashed
lines with central crosses are additional impurity renormalisations and
shaded rectangles are impurity ladders. Diagram 1 is the Aslamazov-Larkin
term; diagrams 2--4 are the Maki-Thompson terms; diagrams 5--10 arise from
corrections to the normal state density of states; diagram 11 is an example
of a higher-order contribution.

Figure 10. The minimal self-intersecting trajectory for the Maki-Thompson
contribution to the $c$-axis conductivity. A quasiparticle hops from one
layer to neighbouring layer, diffuses within that layer, hops back to the
first layer, and finally diffuses back to its starting point.

Figure 11. The graphical form of the Maki-Thompson contribution showing that
it is the precursor phenomenon of the Josephson effect. Starting from the
diagram for the Josephson current for $T<T_c$, we expand the Gorkov
F-functions to linear order in $\Delta$ as $T\rightarrow T_c$, and for $%
T>T_c $ replace $\Delta$, $\Delta^*$ by the $\Delta\Delta^*$ propagator
which is just the fluctuation propagator $L$.

Figure 12. Theoretical predictions \cite{BDKLV93} for the zero-field resistivities $%
\rho_{xx}/\rho_{xx}^N$ (dashed curves) and $\rho_{zz}/\rho_{zz}^N$ (solid
curves) vs reduced temperature $T/T_{c0}$. In each plot $\tau T_{c0}=1$ and $%
\tau_{\phi}T_{c0}=1$ (top curves), $\tau_{\phi}T_{c0}=10$ (middle curves)
and $\tau_{\phi}T_{c0}=100$ (bottom curves). All temperatures shown are in
the region $1\le T/T_{c0}\le 1.06$. Plots (a)--(c) differ in the values of
the parameters $r$ and $E_F$: (a) $r(T_{c0})=0.1$, $E_F/T_{c0}=300$, (b) $%
r(T_{c0})=0.01$, $E_F/T_{c0}=300$, (c) $r(T_{c0})=0.001$, $E_F/T_{c0}=500$. 

Figure 13. Theoretical predictions \cite{BDKLV93} for zero-field resistivities $%
\rho_{xx}/\rho_{xx}^N$ (dashed curves) and $\rho_{zz}/\rho_{zz}^N$ (solid
curves) vs reduced temperature $T/T_{c0}$ for the parameters $r(T_{c0}=0.01$%
, $E_F/T_{c0}=300$, $\tau T_{c0}=0.1$ and $\tau_{\phi}T_{c0}=1$ (top
curves), $\tau_{\phi}T_{c0}=10$ (middle curves) and $\tau_{\phi}T_{c0}=100$
(bottom curves). Changing the value of $\tau T_{c0}$ into the dirty limit
has suppressed the peak in $\rho_{zz}/\rho_{zz}^N.$

Figure 14. Theoretical predictions \cite{BDKLV93} for zero-field resistivity $\rho
_{xx}/\rho _{xx}^N$ with (top curves) and without (bottom curves) the
density of states contributions. The parameters used are $E_F/T_{c0}=300$, $%
\tau _\phi T_{c0}=10$ and $r(T_{c0})=0.1$ (dashed curves) or $r(T_{c0})=0.01$
(solid curves). We see that the DOS contibutions lead to an overall increase
in resistivity.

Figure 15. Resistance $R(t)$ vs reduced temperature $t=T/T_c$ for the three
BSCCO-2212 samples utilised in \cite{BMMRVV92} for fluctuation
measurements. Resistances are normalised to their values at $T=1.33T_c$.

Figure 16. Temperature dependence of the excess conductivity of 4 YBCO
samples as measured in \cite{Freitas87}. The solid lines are the
predictions of the AL theory for 3D and 2D cases. The dashed line is the
modified theory of \cite{FPFV97}.

Figure 17. The normalised excess conductivity $f(\epsilon)=(16\hbar
s/e^2)\Delta\sigma$ of the BSCCO samples in Fig. 15 plotted against $%
\epsilon=\ln{(T/T_c)}$ in a ln-ln plot as described in \cite
{BMMRVV92}. The solid line is the extended theory of Reggiani, Vaglio and
Varlamov \cite{RVV91}. The dashed line is the 2D AL theory.

Figure 18. The normalised excess conductivity $f(\epsilon)=(16\hbar
s/e^2)\Delta\sigma$ for samples of YBCO-123 (triangles), BSSCO-2212
(squares) and BSSCO-2223 (circles) plotted against $\epsilon=\ln{(T/T_c)}$
in a ln-ln plot as described in \cite{FPFV97}. The dotted and
solid lines are the AL theory in 3D and 2D respectively. The dashed line is
the extended theory of \cite{RVV91}.

Figure 19.  Fit of the temperature dependence of the transverse resistance of
an underdoped BSCCO c-axis oriented film with the results of the fluctuation
theory \cite{BMV93}. The inset shows the details of the fit in the
temperature range between $T_c$ and $110K$.

Figure 20.  Comparison of the curvatures of experimental curves of $\rho
_c/\rho _{ab}$ vs temperature for a BSSCO film \cite{BMV96} with the
predictions of the resonant tunneling model of Abrikosov (\ref{AB1}). The
solid lines are the theoretical fits; the points are the data for a reduced
sample (open squares), as-grown sample (closed circles) and oxidised sample
(open diamonds). These fits show that the experimental data cannot be fit by
a theory which predicts singular behaviour at $T=0$ instead of at $T=T_c$.
The dashed line is the simulated behavior according to the fluctuation
theory for the argon annealed sample. 

Figure 21.  Plots of the possible behaviour \cite{BMV96}
of the normal-state $c$-axis
resistance of a BSSCO-2212 sample after the predicted fluctuation
contributions have been subtracted from the experimental data (open
circles). The solid lines are the subtracted data for three values of the
Fermi energy: $E_F=0.8eV$ (bottom curve), $E_F=1.0eV$ (middle curve) and $%
E_F=1.25eV$ (upper curve). We see that after subtraction of the fluctuation
contribution, there may still be a weaker temperature dependence of the
normal-state resistance to explain.

Figure 22.  Experimental data showing the effect of a magnetic field parallel
to the $c$-axis on the in-plane resistivity $\rho _{ab}(T)$ and transverse
resistivity $\rho _c(T)$ of BSCCO single crystals \cite{BCZ91}. The
insets show the zero field behaviour. We see that the size of the peak
increases and moves to lower temperature. 

Figure 23.  Theoretical prediction \cite{BDKLV93}
for the magnetic field dependence of
resistivities $\rho_{xx}(T)/\rho_{xx}^N$ (dashed curves) and $%
\rho_{zz}(T)/\rho_{zz}^N$ (solid curves). The parameters used are $\tau
T_{c0}=1$, $\tau_{\phi}T_{c0}=10$, $r(T_{c0}=0.01$, and $E_F/T_{c0}=300$,
with field strengths corresponding to $\beta(T_{c0})=0,0.05$ and $0.1$.
 
Figure 24.  Transverse magnetoconductivity of YBCO thin films ($\circ$) (from 
\cite{SLP95}) and single crystals ($\bullet$) (calculated from 
\cite{HYMO95}) as a function of the reduced temperature $\epsilon$
(after \cite{LHLWW97}). The broken line is a fit to the clean-limit
theory \cite{BM90,T91} including all four contributions and the
solid line represents the sum of the orbital AL contribution and the
quasiparticle magnetoconductivity estimated from the Hall effect with $A =
1.7$. The latter two are also shown separately by dash-dotted and dotted
curves, respectively.

Figure 25.  Negative c-axis magnetoresistance at various temperatures in 2212
BSCCO single crystals after \cite{YMH95}.

Figure 26.  Fit \cite{BMV95} of the experimental $c$-axis magnetoresistance in the
BSCCO-2212 samples of Ref. \cite{YMH95} to the fluctuation theory
predictions of Eqs. (99)--(102). The fitting parameters used were $%
v_F=3.1\times 10^6$ cm/s, $\tau=1.0\times 10^{-14}$s and $%
\tau_{\phi}=8.7\times 10^{-14}$s.

Figure 27.  Theoretical prediction of Eq. (\ref{9}) \cite{BMV95}
for the temperature
dependence of the c-axis magnetoresistivity of BSSCO-2212. The fitting
parameters are the same as those used in Fig. 26 to fit the data of Ref. 
\cite{YMH95}.

Figure 28.  Measured temperature dependence of the $c$-axis
magnetoconductivity of two YBCO single crystals at $B=12T$ \cite{AHE}%
. The solid lines are fits to the fluctuation theory prediction of Eq. (\ref
{9}) with fitting parameters $v_F=2\times 10^7$cm/s, $\tau(100K)=\tau_{%
\phi}(100K)=(4\pm 1)\times 10^{-15}$s and $J=(215\pm 10)K$. The inset is an
enlarged view of the temperature region close to $T_c$.

Figure 29.  Fit of c-axis resistance curves of BSCCO-2212 single crystals in a
magnetic field with the Josephson coupling theory of Kim and Gray 
\cite{KG93}. 

Figure 30.  Fit of the measured temperature dependence of the $c$-axis
magnetoresistance of a BSSCO-2212 film to the fluctuation theory 
\cite{BMV96}. The points are experimental data at different magnetic
fields: $B=0T$ (circles), $B=0.2T\equiv\beta=0.003$ (squares) and $%
B=0.4T\equiv\beta=0.006$ (triangles). The lines are the predictions of Eqs.
(99)--(102) with fitting parameters $\tau=(5.6\pm0.6)\times 10^{-14}$s, $%
\tau_{\phi}=(8.6\pm 1.4)\times 10^{-13}$s, $E_F=(1.07\pm 0.12)eV$ and $%
J=(43\pm 4)K$. For temperatures below the resistance maximum the fit
underestimates the resistance, the discrepancy becoming larger with
increasing field.

Figure 31.  (a). Fit of experimental $\rho_{ab}(T)$ curves of a 2212 BSCCO film
in various external magnetic fields (see inset) with the fluctuation theory
in the Hartree approximation after \cite{BLM97} (b). Fit of
experimental $\rho_c(T)$ curves of the same film as in (a) in various
external magnetic fields with the fluctuation theory in the Hartree
approximation after \cite{BLM97}. Fitting parameters (see text) are
the same as for (a).

Figure 32.  (a). c-axis resistance vs. magnetic field in 2212 BSCCO single
crystals. Temperatures are 63.9 K, 68.1 K, 70.4 K, 74.0 K, 77.8 K, 80.1
K, 84.4 K, 91.4 K, 95.7 K, 99.5 K from above. $T_c$ was 79 K. (b).
Calculated c-axis magnetoresistance at several temperatures according to the
fluctuation theory including the DOS contribution. After \cite{NTHKT94}.

Figure 33.  The theoretical dependence \cite{FV96} of the real part of conductivity,
normalized on the Drude normal conductivity, on $\omega /T$, $\Re \left[
\sigma ^{\prime }(\omega )\right] ={\rm Re}\left[ \sigma (\omega )\right]
/\sigma ^{{\rm n}}$. The dashed line refers to the $ab$-plane component of
the conductivity tensor whose Drude normal conductivity is $\sigma
_{\parallel }^{{\rm n}}=N(0)e^2\tau v_F^2$. The solid line refers to the $c$%
-axis component whose Drude normal conductivity is $\sigma _{\perp }^{{\rm n}%
}=\sigma _{\parallel }^{{\rm n}}J^2s^2/v_F^2$. In this plot we have put $%
T\tau =0.3,E_F/T=50,r=0.01,\varepsilon =0.04,T\tau _\varphi =4$.

Figure 34.  The theoretical behavior \cite{FV96}
of the $c$-axis component of conductivity frequency
dependence, for different values of temperature, is shown. The solid line
refers to $\varepsilon =0.04$; the dashed line refers to $\varepsilon=0.06$;
the dot-dashed line refers to $\varepsilon =0.08$. $T\tau=0.2$ for all the
curves. The other parameters of this plot are the same used in Fig. 33

Figure 35.  The opening of the pseudo-gap in the c-axis conductivity
measurements on $YBa_2Cu_4O _8$ samples \cite{BTD94}

Figure 36.  The theoretical dependence \cite{FV96}
of ${\rm Re}\left[\sigma_{\perp}(\omega)/\sigma_{\perp}^{{\rm n}}\right]$
on $\omega/T$ for different values
of $T\tau$. The solid line refers to $T\tau =0.4$; the dot-dashed line
refers to $T\tau =0.3$; the dashed line refers to $T\tau =0.2$. The other
parameters of this plot are the same used in Fig. 33.

Figure 37.  The Feynman diagrams for the fluctuation correction to
thermoelectric coefficient are shown. Shaded partial circles are impurity
vertex corrections (\ref{lambda}), dashed curves with central crosses are
additional impurity renormalizations, and shaded thick lines are additional
impurity vertex corrections.

Figure 38.  Diagrams for the fluctuation contribution to the dynamic spin
susceptibility $\chi_{+-}$. 

Figure 39.  The transformation of the MT type diagram for the NMR relaxation
rate in self-intersecting trajectory.

Figure 40.  The $^{63}$Cu relaxation rates in zero field $2W(0)$ (from NQR
relaxation) and $2W(H)$ in a field of 5.9 T (from NMR relaxation of the $%
-1/2\rightarrow 1/2$ line) in the oriented powders of YBCO, with $%
T_c(0)=90.5 $ K and $T_c(H)=87.5$ K. In the inset the relaxation rates,
normalized with respect to $W(H)=W(0)$ for $T\gg T_c$, are reported as a
function of $T/T_c$.

Figure 41.  The Dyson equation in the ladder approximation for the particle
particle channel vertex $\lambda (\vec{q},\omega _1,\omega _2)$ renormalized
by impurities (shaded partial circles). Here the dashed line, as usually,
presents the ''impurity propagator'' $1/2\pi N(0)\tau$ (which is nothing
else as the square of the scattering amplitude $|U|^2$ avaraged over the
solid angles), solid lines are one-electron Green functions (\ref{d2})
already avareged over impurities configuration.

Figure 42.  The Dyson equation in the ladder approximation for the fluctuation
propagator.

\newpage

Table 1

\begin{tabular}{|c|c|c|c|}
\hline
& $\beta \ll \epsilon $ & $\epsilon \ll \beta \ll r\quad (3D)$ & $\max
\{\epsilon ,r\}\ll \beta \quad (2D)$ \\ \hline
$\Delta \sigma _{zz}^{DOS}$ & ${\frac{{e^2s\kappa }}{3\cdot 2^7{\eta }}\frac{%
r{(\varepsilon +r/2)}}{{[\varepsilon (\varepsilon +r)]^{3/2}}}\beta ^2}$ & $%
0.428\frac{e^2s\kappa }{16\eta }\cdot r\sqrt{\frac \beta {2r}}$ & $\frac{%
e^2s\kappa }{8\eta }\cdot r\cdot \ln {\frac{\sqrt{\beta /2}}{\sqrt{\epsilon }%
+\sqrt{\epsilon +r})}}$ \\ \hline
$\Delta \sigma _{zz}^{MT(reg)}$ & ${\frac{{e^2s\tilde{\kappa}}}{{3\cdot
2^8\eta }}\frac{r^2}{{[\varepsilon (\varepsilon +r)]^{3/2}}}\beta ^2}$ & $%
0.428\frac{e^2s\tilde{\kappa}}{8\eta }\cdot r\sqrt{\frac \beta {2r}}$ & $%
\sigma _{zz}^{MT(reg)}(0,\epsilon )-\frac{\pi ^2e^2s\tilde{\kappa}}{2^7\eta }%
\cdot \frac{r^2}\beta $ \\ \hline
$\Delta \sigma _{zz}^{AL}$ & $-{\frac{{e^2s}}{2^{10}{\eta }}\frac{{%
r^2(\varepsilon +r/2)}}{{[\varepsilon (\varepsilon +r)]^{5/2}}}\beta ^2}$ & $%
-\sigma _{zz}^{AL}(0,\epsilon )+{\frac{4.57{e^2s}}{{\eta }}}\sqrt{\frac
r\beta }$ & $-\sigma _{zz}^{AL}(0,\epsilon )+{\frac{7\zeta (3){e^2s}}{2^7{%
\eta }}\cdot }\frac{r^2}{\beta ^2}$ \\ \hline
\begin{tabular}{c}
$\Delta \sigma _{zz}^{MT(an)}$ \\ 
$\min \{\epsilon ,r\}\ll \gamma _\varphi $%
\end{tabular}
& $-{\frac{{e^2s}}{{3\cdot 2^9\eta }}\frac{r^2}{{[\varepsilon (\varepsilon
+r)]^2}}\beta ^2}$ & $-\sigma _{zz}^{MT(an)}(0,\epsilon )+\frac{e^2s}{32\eta 
}\sqrt{\frac r{\gamma _\varphi }}$ & $-\sigma _{zz}^{MT(an)}(0,\epsilon )+%
\frac{3\pi ^2{e^2s}}{2^7{\eta }}{\frac{{\max \{r,\gamma _\varphi \}}}{{\beta 
}}}$ \\ \hline
\begin{tabular}{c}
$\Delta \sigma _{zz}^{MT(an)}$ \\ 
$\gamma _\varphi \ll \min \{\epsilon ,r\}$%
\end{tabular}
& $-{\frac{{e^2s}}{{3\cdot 2^9\eta }}\frac{\sqrt{r}}{{\varepsilon \gamma
^{3/2}}}\beta ^2}$ & $-\sigma _{zz}^{MT(an)}(0)+\frac{4.57e^2s}{64\eta }%
\sqrt{\frac r\beta }$ & $-\sigma _{zz}^{MT(an)}(0,\epsilon )+\frac{3\pi ^2{%
e^2s}}{2^7{\eta }}{\frac{{(r+\epsilon )}}{{\beta }}}$ \\ \hline
\end{tabular}

\newpage

Table 2

\begin{tabular}{|c|c|c|}
\hline
& $\beta \ll \epsilon $ & 
\begin{tabular}{cc}
$\epsilon \ll \beta \ll r$ & $\;\qquad \max \{\epsilon ,r\}\ll \beta $%
\end{tabular}
\\ \hline
$\Delta \sigma _{xx}^{AL}$ & $-{\frac{{e^2}}{2^9{s}}\frac{{[8\epsilon
(\epsilon +r)+3r^2]}}{{[\epsilon (\epsilon +r)]^{5/2}}}\beta ^2;}$ & 
\begin{tabular}{cc}
$-\sigma _{xx}^{AL}(0,\epsilon )+{\frac{{e^2}}{{2s}}}\frac 1{\sqrt{\beta r}};
$ & $-\sigma _{xx}^{AL}(0,\epsilon )+{\frac{{e^2}}{{4s}}}\frac 1\beta {;}$%
\end{tabular}
\\ \hline
\begin{tabular}{c}
$\Delta \sigma _{xx}^{MT(an)}$ \\ 
$(\min \{\epsilon ,r\}\ll \gamma _\varphi )$%
\end{tabular}
& $-{\frac{{e^2}}{3\cdot 2^7{s}}\frac{{(\epsilon +r/2)}}{{[\epsilon
(\epsilon +r)]^{3/2}}}}\beta ^2;$ & $-\sigma _{xx}^{MT}(0,\epsilon )+\frac{%
e^2}{8s}\frac 1{\gamma _\varphi }\ln \frac{\sqrt{\gamma _\varphi }}{\sqrt{%
\beta }+\sqrt[=]{\beta +r}};$ \\ \hline
\begin{tabular}{c}
$\Delta \sigma _{xx}^{MT(an)}$ \\ 
$(\gamma _\varphi \ll \min \{\epsilon ,r\})$%
\end{tabular}
& $-{\frac{{e^2}}{3\cdot 2^7{s}}\frac{{1}}{\epsilon {\gamma ^{3/2}r^{1/2}}}}%
\beta ^2;$ & 
\begin{tabular}{cc}
$-\sigma _{xx}^{MT}(0,\epsilon )+{\frac{4.57{e^2}}{16{s}}}\frac 1{\sqrt{%
\beta r}};$ & $-\sigma _{xx}^{MT}(0,\epsilon )+\frac{3\pi ^2e^2}{16s}\frac
1\beta ;$%
\end{tabular}
\\ \hline
\begin{tabular}{c}
$\Delta (\sigma _{xx}^{DOS}+$ \\ 
$\sigma _{xx}^{MT(reg)})$%
\end{tabular}
& ${\frac{{e^2(\kappa +\tilde{\kappa})}}{3\cdot 2^9s}\frac{{(\varepsilon
+r/2)}}{{[\varepsilon (\varepsilon +r)]^{3/2}}}\beta ^2;}$ & 
\begin{tabular}{cc}
$0.428\frac{e^2{(\kappa +\tilde{\kappa})}}{2^6s}\sqrt{\frac \beta {2r}};$ & $%
\qquad \frac{e^2{(\kappa +\tilde{\kappa})}}{32s}\cdot \ln {\frac{\sqrt{\beta
/2}}{(\sqrt{\epsilon }+\sqrt{\epsilon +r})}.}$%
\end{tabular}
\\ \hline
\end{tabular}

\newpage

Table 3

\begin{tabular}{|c|c|c|c|c|c|}
\hline
& $\tau (s)$ & $\tau _\varphi (s)$ & J(K) & E$_F(eV)$ & v$_F(cm/s)$ \\ \hline
\begin{tabular}{c}
Balestrino et al.[32] \\ 
$\ast $%
\end{tabular}
& 
\begin{tabular}{c}
$5\ 10^{-14}$ \\ 
$3\ 10^{-14}$%
\end{tabular}
& 
\begin{tabular}{c}
- \\ 
$3.6\ 10^{-13}$%
\end{tabular}
& 
\begin{tabular}{c}
$(40)$ \\ 
$(40)$%
\end{tabular}
& 
\begin{tabular}{c}
- \\ 
$1.07$%
\end{tabular}
& 
\begin{tabular}{c}
$1.4\ 10^7$ \\ 
-
\end{tabular}
\\ \hline
\begin{tabular}{c}
Balestrino et al.[40] \\ 
$\ast $%
\end{tabular}
& 
\begin{tabular}{c}
$1\ 10^{-14}$ \\ 
$9\ 10^{-15}$%
\end{tabular}
& 
\begin{tabular}{c}
- \\ 
$7.8\ 10^{-14}$%
\end{tabular}
& 
\begin{tabular}{c}
$(40)$ \\ 
$(40)$%
\end{tabular}
& 
\begin{tabular}{c}
- \\ 
$0.25$%
\end{tabular}
& 
\begin{tabular}{c}
$3.1\ 10^6$ \\ 
-
\end{tabular}
\\ \hline
Heine et al. [186] & $1.5\ 10^{-14}$ & - & $\{10\}$ & - & - \\ \hline
Lang et al. [187] & $1\ 10^{-14}$ & - & $4$ & - & $2.2\ 10^7$ \\ \hline
Nygmatulin et al.[35] & $5\ 10^{-14}$ & $8.6\ 10^{-13}$ & $43$ & $1.07$ & -
\\ \hline
Axnas et al. [6] (YBCO) & 
\begin{tabular}{c}
$5.0\ 10^{-15}$ \\ 
$3.1\ 10^{-15}$%
\end{tabular}
& 
\begin{tabular}{c}
$(=\tau )$ \\ 
$(=\tau )$%
\end{tabular}
& 
\begin{tabular}{c}
$225$ \\ 
$205$%
\end{tabular}
& - & $(2\ 10^7)$ \\ \hline
\end{tabular}

\newpage

Table 4

\begin{tabular}{|c|c|c|}
\hline
& $\omega \ll \min \{T,\tau ^{-1}\}$ & $
\begin{tabular}{cc}
$\min \{T,\tau ^{-1}\}\ll \omega \ll \max \{T,\tau ^{-1}\}$ & $\max \{T,\tau
^{-1}\}\ll \omega $%
\end{tabular}
$ \\ \hline
$%
\begin{tabular}{c}
$\Delta \sigma _{\perp }^{DOS(2D)}(\omega )$ \\ 
$T\ll \tau ^{-1}$%
\end{tabular}
$ & $-\frac{7\zeta (3)}{2^3\pi ^3}\frac{e^2s}\eta \frac{J^2\tau }T\ln \frac
2{\sqrt{\varepsilon +r}+\sqrt{\varepsilon }}$ & 
\begin{tabular}{cc}
$-\frac{e^2s}{4\pi \eta }\frac{J^2}{\omega ^2}(T\tau )\ln \frac 2{\sqrt{%
\varepsilon +r}+\sqrt{\varepsilon }}$ & $+\frac{e^2s}{4\eta }\frac{J^2T}{%
\omega ^3}\ln \frac 2{\sqrt{\varepsilon +r}+\sqrt{\varepsilon }}$%
\end{tabular}
\\ \hline
\begin{tabular}{c}
$\Delta \sigma _{\perp }^{DOS(2D)}(\omega )$ \\ 
$\tau ^{-1}\ll T$%
\end{tabular}
& $-\frac \pi {2^4}\frac{e^2s}\eta (J\tau )^2\ln \frac 2{\sqrt{\varepsilon +r%
}+\sqrt{\varepsilon }}$ & $
\begin{tabular}{cc}
$+\frac{e^2s}{2^4\eta }\frac{J^2}{\omega ^2}\ln \frac 2{\sqrt{\varepsilon +r}%
+\sqrt{\varepsilon }}$ & $\;\;\quad +\frac{e^2s}{4\eta }\frac{J^2T}{\omega ^3%
}\ln \frac 2{\sqrt{\varepsilon +r}+\sqrt{\varepsilon }}$%
\end{tabular}
$ \\ \hline
\begin{tabular}{c}
$\Delta \sigma _{\perp }^{AL(2D)}(\varepsilon ,\omega )$ \\ 
$r\ll \varepsilon $%
\end{tabular}
& 
\begin{tabular}{c}
$\omega \ll T$ \\ 
$+\frac{e^2s}\eta \left( \frac r{\pi \omega }\right) ^2{\ln }[{1+(}\frac{\pi
\omega }{16\epsilon })^2]$%
\end{tabular}
& $-$ \\ \hline
$
\begin{tabular}{c}
$\Delta \sigma _{\perp }^{MT(2D)}(\varepsilon ,\omega )$ \\ 
$r\ll \varepsilon \ll \gamma _\varphi $%
\end{tabular}
$ & 
\begin{tabular}{cc}
$\omega \preceq \tau _\varphi ^{-1}$ & $\tau _\varphi ^{-1}\preceq \omega
\ll T$ \\ 
$+\frac{e^2s}{2^7\eta }\frac{r^2}{\gamma _\varphi \varepsilon }$ & $+\frac{%
e^2s}{2\pi ^2\eta }\frac{r^2\gamma _\varphi T_c^2}{\varepsilon \cdot \omega
^2}$%
\end{tabular}
& $-$ \\ \hline
\end{tabular}

\end{document}